\documentclass[twoside,twocolumn,9pt]{article}
\usepackage{extsizes}
\usepackage[super,sort&compress,comma]{natbib} 
\usepackage[version=3]{mhchem}
\usepackage[left=1.5cm, right=1.5cm, top=1.785cm, bottom=2.0cm]{geometry}
\usepackage{balance}
\usepackage{times,mathptmx}
\usepackage{sectsty}
\usepackage{graphicx} 
\usepackage{lastpage}
\usepackage[format=plain,justification=justified,singlelinecheck=false,font={stretch=1.125,small,sf},labelfont=bf,labelsep=space]{caption}
\usepackage{float}
\usepackage{fancyhdr}
\usepackage{fnpos}
\usepackage[english]{babel}
\addto{\captionsenglish}{%
  
}
\usepackage{array}
\usepackage{droidsans}
\usepackage{charter}
\usepackage[T1]{fontenc}
\usepackage[usenames,dvipsnames]{xcolor}
\usepackage{setspace}
\usepackage[compact]{titlesec}
\usepackage{hyperref}
\usepackage{amsfonts}
\usepackage{bm}
\usepackage{amsmath}
\usepackage{tabularx}
\usepackage{supertabular}
\usepackage{epstopdf}

\usepackage{amssymb}
\newcommand{\kt}{k_B T}
\newcommand{\UU}{\mathcal{U}}

\makeatletter
\let\Hy@linktoc\Hy@linktoc@none
\makeatother

\newcommand{\LB}[1]{{#1}}
\newcommand{\LBNEW}[1]{{#1}}
\newcommand{\SMM}[1]{{#1}}
\newcommand{\SM}[1]{{#1}}
\newcommand{\sopp}[1]{{#1}}
\newcommand{\review}[1]{{#1}}

\newcommand{\qote}[1]{``#1''}
\newcommand{\citefull}[1]{Ref.\citenum{#1}}
\newcommand{\citefulls}[1]{Refs.\citenum{#1}}
\renewcommand{\eqref}[1]{Eq.~(\ref{#1})}
\newcommand{\eqrefs}[1]{Eqs.~(\ref{#1})}
\newcommand{\noeqref}[1]{(\ref{#1})}

\usepackage{epstopdf}%This line makes .eps figures into .pdf - please comment out if not required.

\definecolor{cream}{RGB}{222,217,201}

\begin{document}

\pagestyle{fancy}
\thispagestyle{plain}
\fancypagestyle{plain}{

%%%HEADER%%%
\fancyhead[C]{\includegraphics[width=18.5cm]{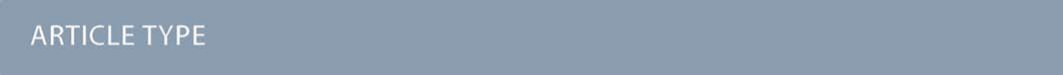}}
\fancyhead[L]{\hspace{0cm}\vspace{1.5cm}\includegraphics[height=30pt]{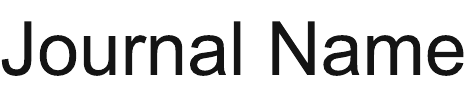}}
\fancyhead[R]{\hspace{0cm}\vspace{1.7cm}\includegraphics[height=55pt]{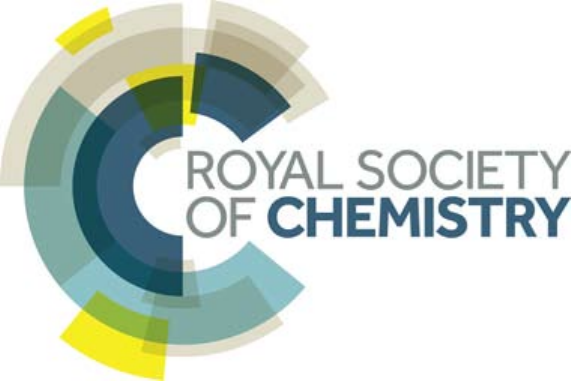}}
\renewcommand{\headrulewidth}{0pt}
}
%%%END OF HEADER%%%

%%%PAGE SETUP - Please do not change any commands within this section%%%
\makeFNbottom
\makeatletter
\renewcommand\LARGE{\@setfontsize\LARGE{15pt}{17}}
\renewcommand\Large{\@setfontsize\Large{12pt}{14}}
\renewcommand\large{\@setfontsize\large{10pt}{12}}
\renewcommand\footnotesize{\@setfontsize\footnotesize{7pt}{10}}
\makeatother

\renewcommand{\thefootnote}{\fnsymbol{footnote}}
\renewcommand\footnoterule{\vspace*{1pt}% 
\color{cream}\hrule width 3.5in height 0.4pt \color{black}\vspace*{5pt}} 
\setcounter{secnumdepth}{5}

\makeatletter 
\renewcommand\@biblabel[1]{#1}            
\renewcommand\@makefntext[1]% 
{\noindent\makebox[0pt][r]{\@thefnmark\,}#1}
\makeatother 
\renewcommand{\figurename}{\small{Fig.}~}
\sectionfont{\sffamily\Large}
\subsectionfont{\normalsize}
\subsubsectionfont{\bf}
\setstretch{1.125} %In particular, please do not alter this line.
\setlength{\skip\footins}{0.8cm}
\setlength{\footnotesep}{0.25cm}
\setlength{\jot}{10pt}
\titlespacing*{\section}{0pt}{4pt}{4pt}
\titlespacing*{\subsection}{0pt}{15pt}{1pt}
%%%END OF PAGE SETUP%%%

%%%FOOTER%%%
\fancyfoot{}
\fancyfoot[LO,RE]{\vspace{-7.1pt}\includegraphics[height=9pt]{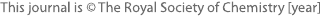}}
\fancyfoot[CO]{\vspace{-7.1pt}\hspace{13.2cm}\includegraphics{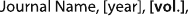}}
\fancyfoot[CE]{\vspace{-7.2pt}\hspace{-14.2cm}\includegraphics{head_foot/RF}}
\fancyfoot[RO]{\footnotesize{\sffamily{1--\pageref{LastPage} ~\textbar  \hspace{2pt}\thepage}}}
\fancyfoot[LE]{\footnotesize{\sffamily{\thepage~\textbar\hspace{3.45cm} 1--\pageref{LastPage}}}}
\fancyhead{}
\renewcommand{\headrulewidth}{0pt} 
\renewcommand{\footrulewidth}{0pt}
\setlength{\arrayrulewidth}{1pt}
\setlength{\columnsep}{6.5mm}
\setlength\bibsep{1pt}
%%%END OF FOOTER%%%

%%%FIGURE SETUP - please do not change any commands within this section%%%
\makeatletter 
\newlength{\figrulesep} 
\setlength{\figrulesep}{0.5\textfloatsep} 

\newcommand{\topfigrule}{\vspace*{-1pt}% 
\noindent{\color{cream}\rule[-\figrulesep]{\columnwidth}{1.5pt}} }

\newcommand{\botfigrule}{\vspace*{-2pt}% 
\noindent{\color{cream}\rule[\figrulesep]{\columnwidth}{1.5pt}} }

\newcommand{\dblfigrule}{\vspace*{-1pt}% 
\noindent{\color{cream}\rule[-\figrulesep]{\textwidth}{1.5pt}} }

\makeatother
%%%END OF FIGURE SETUP%%%

%%%TITLE, AUTHORS AND ABSTRACT%%%
\twocolumn[
  \begin{@twocolumnfalse}
\vspace{3cm}
\sffamily
\begin{tabular}{m{4.5cm} p{13.5cm} }

\includegraphics{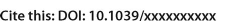} & \noindent\LARGE{\textbf{Osmosis, from molecular insights to large-scale applications}} \\%Article title goes here instead of the text "This is the title"

\vspace{0.3cm} & \vspace{0.3cm} \\

 & \noindent\large{Sophie Marbach\textit{$^{a,\dagger}$} and Lyd\'eric Bocquet\textit{$^{a,\ast}$}} \\%Author names go here instead of "Full name", etc.

%%%%%%%%%%%%%%%%%%%%%%%%%%%%%%%%%%%%%%%
%%%%%%%%% ABSTRACT %%%%%%%%%%%%%%%%%%%%%%%
%%%%%%%%%%%%%%%%%%%%%%%%%%%%%%%%%%%%%%%

\includegraphics{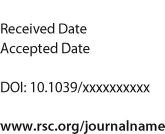} & \noindent\normalsize{Osmosis is a universal phenomenon occurring in a broad variety of  processes and fields. It is the archetype of entropic forces, both trivial in its fundamental expression -- the van\,'t Hoff perfect gas law -- and highly subtle in its physical roots. 
While osmosis is intimately linked with transport across membranes, it also manifests itself as an interfacial transport phenomenon: 
 %as raises fundamental questions in the context of interfacial transport, with phenomena 
 the so-called diffusio-osmosis and -phoresis, whose consequences are presently actively explored for example for the manipulation of colloidal suspensions or the development of active colloidal swimmers.  %and emerging aspects of "active separation". 
Here we give a global and unifying view of the phenomenon of osmosis and its consequences with a multi-disciplinary perspective. 
Pushing the fundamental understanding of osmosis allows one to propose new perspectives for different fields and we highlight a number of examples along these lines, for example introducing the concepts of osmotic diodes, active separation and far from equilibrium osmosis, raising in turn fundamental questions in the thermodynamics of separation. The applications of osmosis are also obviously considerable and span very diverse fields. Here we discuss a selection of phenomena and applications where osmosis shows great promises:  osmotic phenomena in membrane science (with recent developments in separation, desalination, reverse osmosis for water purification thanks in particular to the emergence of new nanomaterials); applications in biology and health (in particular discussing the kidney filtration process); osmosis and energy harvesting (in particular, osmotic power and blue energy as well as capacitive mixing); applications in detergency and cleaning, as well as for oil recovery in porous media. } %It also provides some insight for processes occurring in newly discovered channels in the brain or in the lungs \LB{[LB: ???]}, but also in various biologically related processes, highlighting how powerful this phenomenon is.} %The abstract should be a single paragraph which summarises the content of the article. Any references in the abstract should be written out in full \textit{e.g.}\ [Surname \textit{et al., Journal Title}, 2000, \textbf{35}, 3523].

\end{tabular}

 \end{@twocolumnfalse} \vspace{0.6cm}
 ]

%%%END OF TITLE, AUTHORS AND ABSTRACT%%%

%%%FONT SETUP - please do not change any commands within this section
\renewcommand*\rmdefault{bch}\normalfont\upshape
\rmfamily
\section*{}
\vspace{-1cm}

%%%%%%TABLE OF CONTENTS %%%%%%%%%%%%%
%\newpage
%\null\newpage
%
%\tableofcontents
%\newpage

%%%%%%%%%%%%%%%%%%%%%%%%%%%%%%%

%%%FOOTNOTES%%%

\footnotetext{\textit{$^{a}$~Laboratoire de Physique , Ecole Normale Sup\'{e}rieure, PSL Research University, 24 rue Lhomond, 75005, Paris}}
\footnotetext{\textit{$^{\dagger}$~now at Courant Institute, New York University, New York}}
\footnotetext{\textit{$^{\ast}$~ Corresponding author e-mail: lyderic.bocquet@lps.ens.fr}}

%Please use \dag to cite the ESI in the main text of the article.
%If you article does not have ESI please remove the the \dag symbol from the title and the footnotetext below.
%\footnotetext{\dag~Electronic Supplementary Information (ESI) available: [details of any supplementary information available should be included here]. See DOI: 10.1039/b000000x/}
%additional addresses can be cited as above using the lower-case letters, c, d, e... If all authors are from the same address, no letter is required

%\footnotetext{\ddag~Additional footnotes to the title and authors can be included \textit{e.g.}\ `Present address:' or `These authors contributed equally to this work' as above using the symbols: \ddag, \textsection, and \P. Please place the appropriate symbol next to the author's name and include a \texttt{\textbackslash footnotetext} entry in the the correct place in the list.}

%%%END OF FOOTNOTES%%%

%%%%%%%%%%%%%%%%%%%%%%%%%%%%%%%%%%%%%%%
%%%%%%%%% ABSTRACT %%%%%%%%%%%%%%%%%%%%%%%
%%%%%%%%%%%%%%%%%%%%%%%%%%%%%%%%%%%%%%%

\section{Introduction}

%\cite{guell1996physical} I really like this paper actually; though it's not especially interesting 

\begin{figure}[h!]
\centering
  \includegraphics[width=0.45\textwidth]{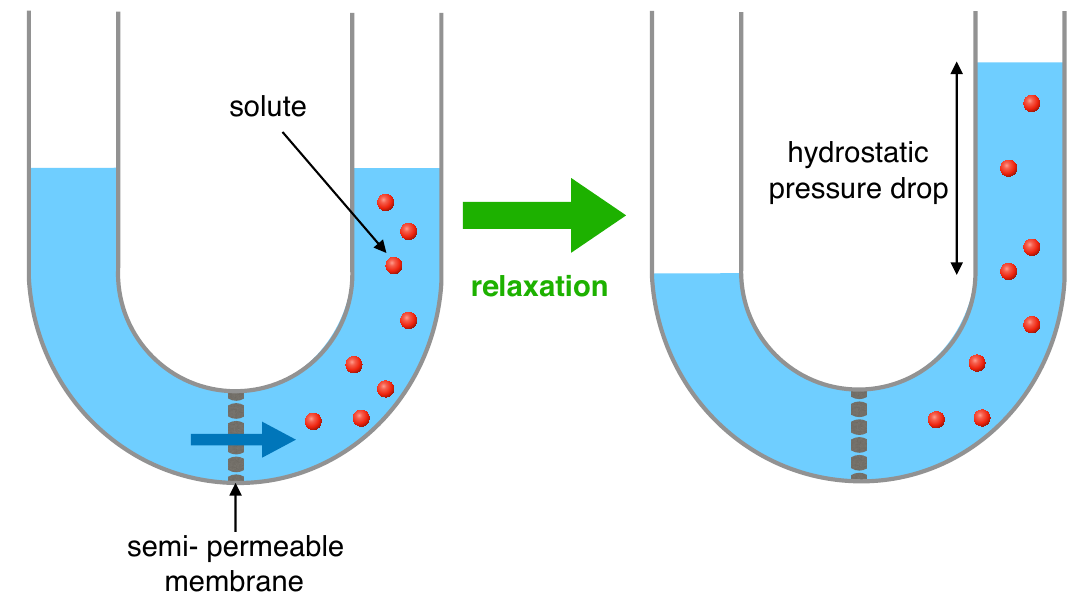}
  \caption{\textbf{: Key manifestation of osmosis.} A semi-permeable membrane allows transport of water upon a solute concentration difference (in red). The flow of water is directed from the fresh water reservoir to the concentrated reservoir. }
  \label{fig:principle}
\end{figure}

From the etymological point of view, osmosis denotes a \qote{push} and indeed osmosis is usually associated with the notion of force and pressure. 
Osmosis is a very old topic, it was first observed centuries ago with reports by Jean-Antoine Nollet in the 18$^{th}$ century. It was rationalized more than one century later by van\,'t Hoff, who showed that the osmotic pressure took the form of a perfect gas equation of state. 
\SM{In practice, an osmotic pressure is typically expressed across %in the presence of %a membrane, more commonly 
a \textit{semi-permeable membrane}, \textit{e.g.} a membrane that allows only the solvent to pass while retaining solutes}. If two solutions of a liquid containing different solute concentrations are put into contact through such a semi-permeable membrane, the fluid will undergo a driving force pushing it towards the reservoir with the highest solute concentration, see Fig.~\ref{fig:principle}.
Reversely, in order to prevent the fluid from passing through the membrane, a pressure has to be applied to the fluid to counteract the flow: the applied pressure is then equal to the \textit{osmotic pressure}. 
%  The phenomenon It is usually associated with the motion of a liquid across a semi-permeable membrane separating two fluid reservoir containing  contrasting 
%
%Osmosis is commonly described as the phenomenon occurring when two reservoirs of liquid, say water, are put in contact by a membrane permeable to the fluid but not to a solute. When a solute is added to one reservoir, the presence of the solute -- that cannot cross the membrane -- induces a water flow from the clear reservoir to the concentrated one, see Fig.~\ref{fig:principle}. A sensible definition of osmotic pressure may be found in \cite{guell1996physical} as  %"osmotic pressure is 
%"the macroscopic, mechanical pressure difference existing at equilibrium between two solutions separated by a membrane". The membrane being "any barrier (separating two bulk phases) which is permeable to at least one species present in the bulk phases and which also distinguishes between the several species by their respective rates of transport through that barrier". 

Osmosis is therefore extremely simple in its expression. Yet it is one of the most subtle physics phenomenon in its roots -- it resulted in many debates over years~\cite{meyer1890pression,van1890wesen}. Osmosis also implies subtle phenomena, in particular as a prototypical illustration for the explicit conversion of entropy of mixing into mechanical work. In spite of centuries of exploration, osmosis as a field remains very lively, with  a number of recent breakthroughs both in its concepts and applications as we shall explore in this review. A simple reason for the importance of osmosis is that it is a very {\it powerful} phenomenon:
giving just one illustrative number, it is amazing to realize that a concentration difference of $\sim 1.2$ molar, which corresponds roughly to the difference between sea and fresh water (and can be easily achieved in anyone's kitchen), yields an osmotic pressure of $\sim$ 30 atmospheres. This is the hydrostatic pressure felt under a 300m  water column ! Osmosis has  potentially a destructive power, in particular in soft tissues and membranes, with possible fatal consequences \cite{kung2010mechanosensitive}. This explains actually why it is also an efficient asset for food preservation \SM{(such as fish and meat curing with dry salt)}.

Osmosis is accordingly also a key and universal phenomenon occurring in many processes, ranging from biological transport in plants, trees and cells, to water filtration, reverse and forward osmosis, energy harvesting and osmotic power, capacitive mixing, oil recovery, detergency and cleaning, active matter, to quote just a few. 

{The litterature on osmosis and its consequences is accordingly absolutely huge~\footnote{The word \qote{osmosis} in {\it Web of Science} results in tens of thousands of referenced papers on this topic.}, and it may seem hopeless to cover in a single review all aspects of the topic with an exhaustive discussion of all possible applications. Also, such a comprehensive list would probably be useless for readers who want to catch up with the topics related to osmosis. \SM{In writing this review, we thus decided to rather present a tutorial and unified perspective of osmosis, obviously with personal views,} avoiding exhaustiveness to highlight a number of significant %chosen 
questions discussed in the recent literature. The review will therefore explore the fundamental foundations of osmosis, emphasizing in particular  the -- sometimes subtle -- mechanical balance at play; then report on more recent concepts and applications related to osmosis which - in our opinion - prove promising for future perspectives. 
We will accordingly put in context phenomena like diffusio-osmosis and -phoresis, as well as \qote{active} (non-equilibrium) counterparts of osmosis, which were realized lately to play a growing role in numerous applications in filtration and energy harvesting.}

{
The review is organized as follows. We start with some basic reminder of the fundamentals of osmosis in terms of equilibrium and non-equilibrium thermodynamics of the underlying process. We further highlight simplistic views clarifying the mechanical aspects of osmosis. We then discuss membrane-less osmosis and the so-called diffusio-osmotic flows. We then show how such phenomena may be harnessed to go beyond the simple views of  
%osmosis beyond 
van\,'t Hoff. We then explore the transport of particles under solute gradients, diffusio-phoresis, and discuss how this phenomenon can be harnessed to manipulate colloidal assemblies. And we finally illustrate a number of applications for the introduced concepts, from desalination, water treatment, the functioning of the kidney, blue energy harvesting, \textit{etc}. We conclude with some final, brief, perspectives.}

%Accordingly we will first  focus on basic undertanding of osmosis, recalling  the concepts and some pedagogical interpretations; then discuss some more advanced forms of osmosis, in particular osmosis 'without membranes', namely diffusio-osmosis and -phoresis which - although first explored by the russian literature decades ago - came back on the scene lately with a variety of novel applications; then explore how these concepts and some advanced extensions of osmosis - in particular towards active counterparts of osmosis - may be harvested for various applications. Again we will choose specific ones.

%While we will come back to applications in the last section of this review, and
%We first focus on its working mechanism and its implications for advanced osmotic transport. 

\vskip0.5cm
\section{Osmosis : the van\,'t Hoff legacy}

\subsection{A quick history of osmosis}

%Il y a aussi les expériences de Henri Dutrochet (dans son livre sur les phénomènes qui ont lieu dans les végétaux et les animaux, 1826). Il s'intéresse à différents phénomènes dans les arbres (montée de la sève, mais aussi dans les plantes) et également aux limaces et à l'intestin. Il introduit les termes d'\textit{endosmose} et \textit{exosmose}.  (dedans/dehors et du grec (ancien je pense) $\omega\sigma\mu o ?$, (« poussée »)/ impulsion), associés au fait qu'il observe des quantités d'eau importantes entrer et sortir de compartiments lorsque ceux ci contiennent différents ingrédients dissous (le sperme pour la limace, il y a surement d'autres exemples). 
% p.115 de https://archive.org/details/bub_gb_WAIAAAAAQAAJ/page/n117
%Dutrochet's most valuable contributions to science were his emphasis on the similarity of basic processes in all living organisms and his belief that all such processes can be explained in terms of physical and chemical forces. - ça c'est bien vrai, surtout quand on lit son bouquin
%\cite{dutrochet1995new} initial confusion with capillarity, trying to disentangle effects;

%\cite{wisniak2013thomas} review on the work of graham, who is considered as the father of the word "osmosis", though still exactly what that meant at the time was highly controversial;
We start this review with 
%The following section is dedicated to 
a short and non-exhaustive journey through time in order to highlight how a complete understanding of osmosis emerged over time. We refer \textit{e.g.} to \citefull{guell1996physical} for a more detailed historical review. The first occurence of the term "osmosis" and clear observation of its effects -- \SM{beyond the seminal work of Nollet} -- is reported at least as early as in the works of Henri Dutrochet in the 1820s~\cite{dutrochet1826agent,dutrochet1995new}. He observed swelling events or emptying of pockets driven by the presence of various dissolved components in water (different sugars in plants, sperm in slugs...). In reference to the greek term "osmose" (meaning "impulsion" or "push") he introduced the vocabulary "\textit{endosmose}" and "\textit{exosmose}". Interestingly, Dutrochet served as a pioneer in linking these different topics by claiming that the \textit{same physical force} could be used to describe all these events~\cite{dutrochet1826agent}, which is indeed a unique and fascinating feature of osmosis. Yet, the mechanisms driving osmotic flow were still unclear, and entangled (or believed to be entangled) with capillary and electrical effects. In 1854 T. Graham introduced the word "osmosis" building on the work of Dutrochet~\cite{graham1854vii}. 

\begin{figure}[h!]
\centering
  \includegraphics[width=0.35\textwidth]{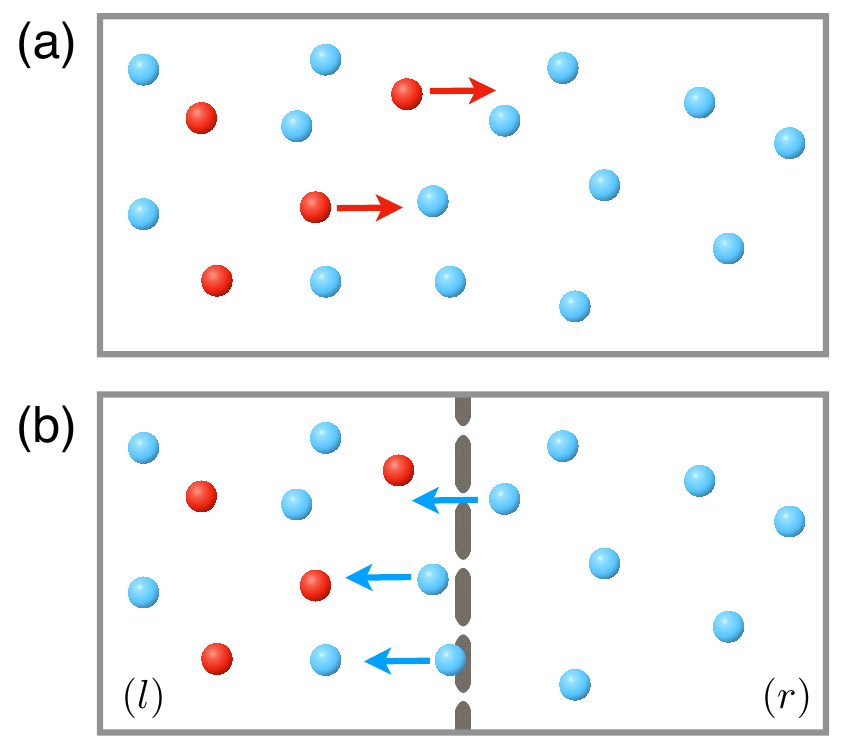}
  \caption{\textbf{: Osmosis versus diffusion.} (a) situation where motion of the red solute particles is governed by diffusion alone (b) osmosis situation, where motion of the blue solvent particles is driven by osmosis; the solute particles being "repelled" by the membrane, the membrane exerts an effective force on the liquid (solute + solvent) that drives solvent flow towards the highly concentrated reservoir.}
  \label{fig:diffusion}
\end{figure}

Interestingly, the distinction between \textit{osmosis} and pure diffusion -- \textit{without} a membrane, see Fig.~\ref{fig:diffusion} -- is not clear from the beginning. The confusion will grow stronger with the work of \SM{Adolf} Fick in 1855~\cite{fick1855v}, where he claims that diffusive motion (Fickian diffusion) is the driver for osmotic flow (the water concentration imbalance between the two compartments drives the water flow). The question of finding whether osmotic flow is diffusion-driven or not will be an ongoing debate for a century. 
%That osmosis is not equivalent to pure diffusion is still a rather puzzling statement. 
That diffusion alone cannot account for osmosis is not widely appreciated.
In 1957, the debate is definitely closed by an experimental visualization of water flow, using radioactively labeled water molecules~\cite{mauro1957nature} and verified in~\citefull{robbins1960experimental}. The flows measured were significantly higher than that expected by pure diffusion. 

\begin{figure}[h!]
\centering
  \includegraphics[width=0.45\textwidth]{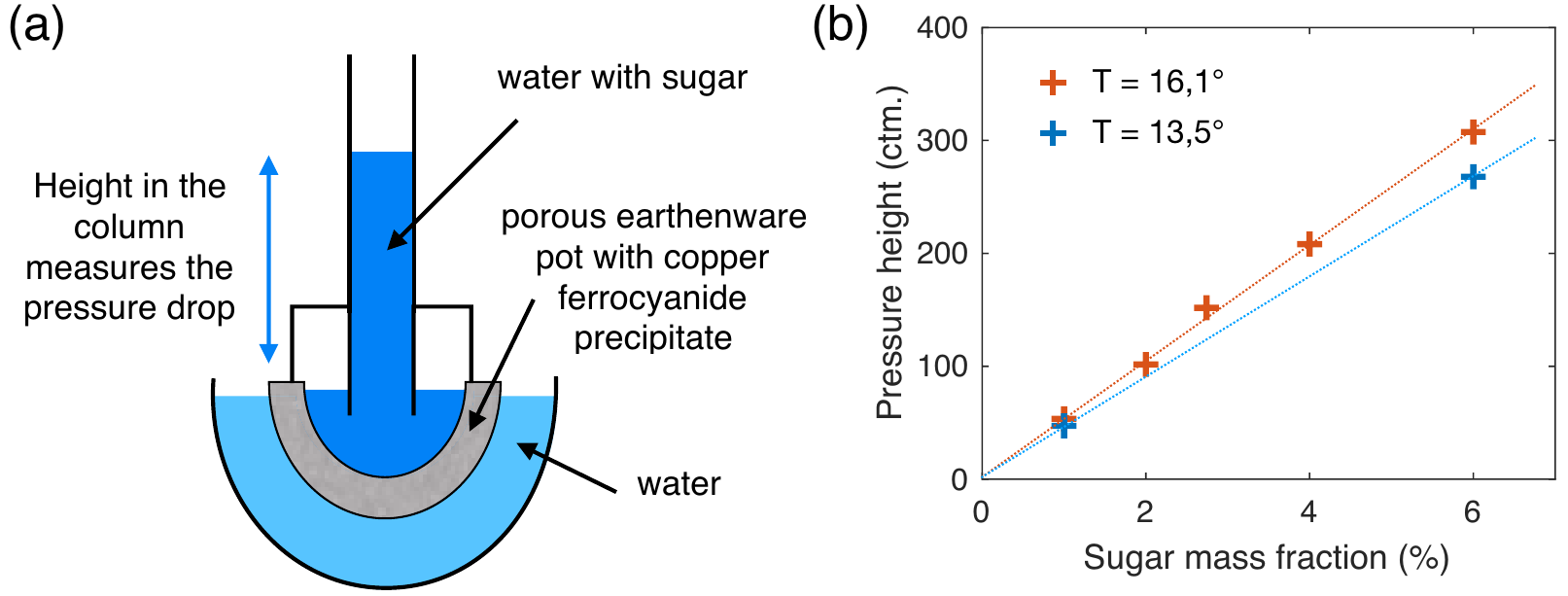}
  \caption{\textbf{: First measurements of osmotic pressure.}(a) Schematic of a Pfeffer cell, the device used by W. Pfeffer to perform the first measurements of osmotic pressure; (b) Osmotic pressure as a function of solute concentration at two temperatures, with experimental data points from W. Pfeffer~\cite{pfeffer1877osmotische} verifying a linear relation (the lines are a guide for the eye). }
  \label{fig:history}
\end{figure}

In 1877, \SM{Wilhelm} Pfeffer made the first measurements of osmotic pressure~\cite{pfeffer1877osmotische}, see Fig.~\ref{fig:history}. At equilibrium, he measured a rise in the concentrated solution, corresponding to a hydraulic pressure drop that is equal to the osmotic pressure. He measured a linear relation between the osmotic pressure and the concentration difference. \sopp{But also, Pfeffer measured that for each degree rise in temperature, the pressure would go up by 1/270~\cite{wald1986theory}. This fact was reported to  \SM{Jacobus Henricus} van\;'t Hoff by the botanist Hugo de Vries and van\;'t Hoff immediately recognized that 270 was an approximation of 273 K. %corresponding to 0 $^{\circ}$C. 
} Intrigued by this result, he attempted in 1887 to rationalize this linear dependence~\cite{van1887rolle} and suggested to interpret that the osmotic pressure $\Delta \Pi$ was exerted by the solute particles and equal to the partial pressure that they would have in gas phase (therefore the term "osmotic pressure"):
\begin{quote}\sopp{\textit{[...] it occurred to me that with the semipermeable barrier all the reversible transformations that so materially ease the application of thermodynamics to gases, become equally available for solutions... That was a ray of light; and led at once to the inescapable conclusion that the osmotic pressure of dilue solutions must vary with temperature entirely as does gas pressure [...].~\cite{wald1986theory}}}\end{quote}
then writing
\begin{equation}
\Delta \Pi = k_B T \Delta c_s
\label{eq:vantHoff}
\end{equation}
\SM{with $k_B$ the Boltzmann constant, $T$ temperature and $\Delta c_s$ the solute imbalance between reservoirs}.

 \eqref{eq:vantHoff} is today referred to as the van\;'t Hoff law, and gives in practice good agreement for the osmotic pressure measured between two solutions separated by a membrane permeable only to the solvent. For a solute imbalance of $\Delta C_s = 1.2$ mol.L$^{-1}$ (corresponding to the ionic strength difference between fresh and sea water, which is twice -- two ions for salt -- the typical concentration 0.6 mol.L$^{-1}$), we find an osmotic pressure of $\Delta \Pi = k_B T \Delta C_s \mathcal{N}_A \simeq 30$ bar.
%\textcolor{blue}{maybe comment at this stage that spaghetti in water, or more formally colloids may not generate a sufficient osmotic pressure (very very small)}.

At the time, the interpretation of van\;'t Hoff gave rise to a number of debates~\cite{meyer1890pression,van1890wesen}.
 %showing that still a number of subtleties had to be explained. 
 In the following decades a great number of theories were invented to describe the osmotic phenomenon and a detailed review of these theories can be found in~\citefull{borg2003osmosis}. Among all these theories, two of them caught a lot of attention. One of them was the proof of van\;'t Hoff's law using the kinetic theory of gas to describe the two solutions~\cite{einstein1905movement} (which was later improved for multicomponent systems~\cite{mcmillan1945statistical}). The other one is  acknowledged today as the common description of osmosis, and makes use of the concept of chemical potentials first introduced by \SM{Josiah Willard} Gibbs~\cite{gibbs1897semi} (actually introduced as a physical descriptor required to understand osmosis), that we recall in the next section.  

%(\textcolor{blue}{Actually I think that osmotic pressure allowed Gibbs to introduce this notion of chemical potentials; I think the idea is more that osmotic pressure may not be explained without adding a physical descriptor, that he named the chemical potential})
%\cite{morse1905osmotic} improved version of the ad-hoc law by adapting the volume being not the volume of the total solution but that of the solvant only; Not sure it's super relevant
%formalisation complète en termes d'activité chimique par Lewis quelques années plus tard; 
%\cite{lewis1908osmotic}

\subsection{Thermodynamic equilibrium}

%\textcolor{blue}{Je trouve que c'est vraiment instructif de comprendre d'ou vient l'egalite des potentiels chimiques de part et d'autre de la membrane et en fait ca ne me parait pas evident avec G. A adapter ? et consider shortening the following ?}

We start with the thermodynamic derivation of the osmotic pressure as proposed by Gibbs. We follow here the clear-cut presentation proposed in the textbook by Callen~\cite{callen1998thermodynamics}, which we recall here for the purpose of settling properly the foundations. \sopp{In addition to Callen it is also worth reading the rigorous thermodynamic treatment by Guggenheim~\cite{guggenheim1985thermodynamics}}. We consider a composite system made of two simple reservoirs (left and right) separated by a rigid wall permeable to component $w$ (usually the solvent) and totally impermeable to all other components (labelled $s_1$, $s_2$ and so on). The whole system is in contact with a thermal bath at temperature $T$. 
The solvent is in equilibrium over the whole system, {\it i.e.} over the two reservoirs, while solutes cannot equilibrate between the reservoirs.
Due to the imbalance of solute fraction between both reservoirs, the solvent cannot keep a homogeneous pressure across the two reservoirs while ensuring
the equality of chemical potential at equilibrium. An (osmotic) pressure drop builds up, which the membrane withstands.

Assuming that the solute concentration is dilute, 
%We start with 
the Gibbs free energy of a binary system of solvent and dilute solute can be written as 
%(with no membrane and for simplicity just one dilute specie) that may be written as
\begin{equation}
\begin{split}
G(T,p,N_w,N_{s_1}) = &N_w \mu_w^0(p,T) + N_{s} \mu_s^0(p,T) \\
&+ N_w k_B T \ln \frac{N_w}{N_w + N_{s} } + N_{s} k_B T \ln \frac{N_{s}}{N_w + N_{s} } 
\end{split}
\end{equation}
where $ \mu_w^0(p,T)$ and $ \mu_s^0(p,T)$ are the chemical potentials of the pure solvent and solute and
%, $\phi$ is an unspecified function, and 
the last two terms correspond to the entropy of mixing terms. In the dilute regime where $N_{s} \ll N_w$ the Gibbs free energy simplifies to
\begin{equation}
%\begin{split}
G(T,p,N_w,N_{s}) = N_w \mu_w^0(p,T) + N_{s} \mu_s^0(p,T) - k_B T N_{s}  + N_{s} k_B T \ln \frac{N_{s}}{N_w}
%\end{split}
\end{equation}
and the chemical potential of the solvent may be obtained as $\mu_w = \partial_{N_w} G$
\begin{equation}
\mu_w(T,p,X) \simeq  \mu_w^0(p,T) -k_B T\,  X. % \frac{N_{s}}{N_w}.
\end{equation}
with $X\simeq N_s/N_w$ the solute molar fraction.
The chemical potential balance, $\mu_w^{(l)}  = \mu_w^{(r)} $, thus writes
%gives $\mu_w(T,P^{(l)},N^{(l)}_w,N_{s_1}^{(l)}) = \mu_w(T,P^{(r)},N^{(r)}_w,N^{(r)}_{s_1},..)$ and thus
\begin{equation}
\mu_w(T,p^{(l)},0) - k_B T\frac{N^{(l)}_{s}}{N^{(l)}_w} = \mu_w(T,p^{(r)},0)- k_B T\frac{N^{(r)}_{s}}{N^{(r)}_w}. 
\end{equation}
Noting then that for small pressure drops, $ \mu_w(T,p^{(r)},0)\simeq\mu_w(T,p^{(l)},0) + (p^{(r)}-p^{(l)})v_w$, 
with $v_w = \partial_p \mu_w(T,p,0)$ the molecular volume, one deduces finally  
%In the case of a sufficiently small pressure drop, with $v_w = \partial_P \mu_w(T,P,0)$ we have
%\begin{equation}
%\mu_w(T,P^{(l)},0) - k_B T\frac{N^{(l)}_{s_1}}{N^{(l)}_w} = \mu_w(T,P^{(l)},0) + (P^{(r)}-P^{(l)})v_w- k_B T\frac{N^{(r)}_{s_1}}{N^{(r)}_w}. 
%\end{equation}
%and finally
\begin{equation}
\Delta \Pi = p^{(r)}-p^{(l)} = k_B T \left[ \frac{N_{s}^{(r)}}{V^{(r)}} - \frac{N_{s}^{(l)}}{V^{(l)}}\right]
\end{equation}
Introducing the concentration as $c_s = N_{s}/V$, one thus  recovers the result of van 't Hoff
\begin{equation}
\Delta \Pi = k_B T \Delta c_s.
\end{equation}
In the case of several dilute solutes, this generalizes simply to
\begin{equation}
\Delta \Pi = k_B T \left[ \frac{N_{s_1}^{(r)}+ N_{s_2}^{(r)} + ...}{V^{(r)}} - \frac{N_{s_1}^{(l)}+ N_{s_2}^{(l)} + ...}{V^{(l)}}\right].
\end{equation}

The derivation above is limited to dilute solutes. For arbitrary molar fractions $X$ of solute/solvent mixtures, the osmotic pressure
is given in terms of the general expression for the pressure, namely~\cite{barrat2003basic}
%\LB{Beyond dilute solutes, 
% thermodynamics predicts that the thermodynamic force driving motion is a generalized osmotic pressure
%taking the formal expression:~\cite{barrat2003basic}
%
\begin{equation}
\Pi(X)=X {\partial f\over \partial X} - f[X] + f[X=0],
\label{PiGeneral}
\end{equation}
with $f(X)=F/V$ the Helmoltz free energy density calculated for a solute molar fraction $X$. Deviations from ideality are for example measured for polymers, where the range of validity of the van 't Hoff law decreases with
increasing molecular weight \cite{barrat2003basic}. Deviations are also expected for highly concentrated brines or solvent mixtures, {\it e.g.} in the context of solvophoresis, see below \cite{paustian2015direct}. %and even in dense electrolytes \cite{marbach2016active}.

An interesting, and quite counter-intuitive remark is that -- provided it is semi-permeable -- the membrane characteristics do not appear in this thermodynamic expression for the osmotic pressure. 
%\LB{comment on osmotic polarization instead of osmotic pressure ?}.
Another puzzling remark is that  the osmotic pressure is \sopp{a \textit{colligative} property, \textit{i.e.} it does not depend on the nature of the solute (nor that of the membrane), but only on the concentration of the solute}. This is relevant when the membrane is completely impermeable to the solute, but  when the membrane is only {\it partially} impermeable, or when there are different solutes with different permeation properties, there may be both a solvent and a solute flux driven by the solute concentration imbalance 
(in opposite directions)~\cite{talen1965negative,talen1965osmometry,weinstein1968charge,lee2014osmotic,lee2017nanoscale}.
The osmotic pressure is then usually assumed to be reduced by a  so-called (dimensionless) reflection factor, $\sigma$, which depends on the
specific properties of solvent-membrane interactions and transport. 
This requires to go beyond the thermodynamic equilibrium and consider 
 %Determining $\sigma$ requires to describe 
 the detailed mass and solute transport across the membrane, as we now explore.

%Already discussion of the reflection coefficient $\sigma$, to this end we can probably also discuss negative or anomalous osmosis, measured e.g. in the setup of~\cite{weinstein1968charge}
%and also measurements by Staverman~\cite{talen1965negative}

%Now, when the membrane is only {\it partially} impermeable to the solute, there is still a solvent flux driven
%by the solute concentration imbalance.~

%1960s Still some questions around the details of the membrane itself; and why that does not come into play // and also the physical mechanism behind osmosis, how does it work ?  

%\cite{callen1998thermodynamics} he introduces something that is very very close to the mechanistic view to understand the physical mechanism, namely that the membrane exerts a force on the solute particles, that are repelled and thus "attract" water -- this last sentence is going to be truly confusing for many people. %% En fait ce serait vraiment intéressant à discuter plus loin, avec la vision comme une barrière; 

\subsection{Osmotic fluxes and thermodynamic forces}
\label{fluxforces}
%review \cite{hill1979osmosis} not very happy about this idea of symmetry due to Onsager principles, and says that in full generality it should be verified; something interesting that is brought to light is how 1D models fail to account for radial gradients of concentration and pressure that change eventually what is going on. Then he gives his interpretation of the reflexion coefficients with a derivation. There is some discussion of anomalous osmosis with references to look at; 

Following the work of Staverman~\cite{staverman1951theory}, Kedem and Katchalsky derived a relation between solute and solvent flows through a porous membrane and the corresponding thermodynamic forces~\cite{kedem1958thermodynamic}, based on 
%The objective is to find the proper description of irreversibility (or entropy production) in the system, and from there deduce the relevant flows and forces. 
 Onsager's framework of irreversible processes~\cite{onsager1931reciprocal,degroot2013non}.
% the one staverman paper that inspired the derivation of kedem-katchalsky, the only thing that they add being clarity, and also using the full expression of the osmotic pressure, and a thorough discussion of what the reflexion coefficient is. \cite{staverman1952non} and this one extends to forces being like electrical potentials;
%the one derivation
%

As in the previous section, we consider a composite system made of two simple reservoirs (left and right), containing a solvent $w$ and a solute $s$. The reservoirs are separated by a rigid wall, which is now permeable to all components, but with a differential permeability between the solute and the solvent. Obviously, the objective of the membrane is somehow to reject the solute but the rejection is incomplete here.
%  different interactions 
 %The interactions of the solvent and solute with the membrane are however different. 
 The whole system is put in contact with a thermal bath at temperature $T$. %, and the reasoning is thus performed in the canonical ensemble.  
 
 The entropy production (per unit membrane area $\cal A$) is accordingly written as:
 %determined by the passage of components through the membrane and thus change in chemical potential of components between one side to the other; the dissipation function per unit membrane area $A$ thus writes as:
\begin{equation}
\Phi = \frac{T}{\cal A} \frac{d S}{dt}  = - \left( \mu_w^{(r)} - \mu_w^{(l)} \right) \frac{d N_w^{(r)}}{dt} -  \left( \mu_s^{(r)} - \mu_s^{(l)} \right) \frac{d N_s^{(r)}}{dt}
\label{eq:dissipation}
\end{equation}
with $\frac{dN_i^{(r)}}{dt}$ the flux of molecules of component $i$ per unit area. The dissipation function of \eqref{eq:dissipation} is a product of fluxes $\frac{dN_i^{(r)}}{dt}$ and the corresponding thermodynamic forces, here the differences in chemical potentials.

Now, restricting ourselves to ideal solutions for simplicity, one may write
%We will express these fluxes and forces in a slightly more explicit way. For that, we make in the following the hypothesis that the solutions are ideal such that 
the chemical potential difference as $\mu_i^{(r)} - \mu_i^{(l)} = v_i \Delta p + k_B T \Delta \ln X_i$ where $X_i$ is the molar fraction of component $i$ and $v_i = (\partial \mu_i / \partial p)$ the molar volume of $i$. Accordingly, 
%In the case of dilute solutions, and small concentration gradients, the chemical potential of the solute reduces to:
%\begin{equation}
$\mu_s^{(r)} - \mu_s^{(l)} = v_s \Delta p + k_B T \frac{\Delta c_s}{c_s}$ for the solute and 
%\end{equation}
%and of the solvent:
%\begin{equation}
$\mu_w^{(r)} - \mu_w^{(l)} = v_w \Delta p - k_B T \frac{\Delta c_s}{c_w}$ for the solvent \review{(where we used
%\end{equation}
$c_s \ll c_w$)}. \eqref{eq:dissipation} then rewrites:
\begin{equation}
\Phi = - \left( v_w \frac{d N_w^{(r)}}{dt} + v_s \frac{d N_s^{(r)}}{dt}\right) \Delta p - \left( \frac{1}{c_s}\frac{d N_s^{(r)}}{dt} - \frac{1}{c_w} \frac{d N_w^{(r)}}{dt} \right) k_B T \Delta c_s
\label{eq:dissipation2}
\end{equation}
From the dissipation function in \eqref{eq:dissipation2}, we may thus identify a new set of forces and fluxes: new forces are $- \Delta p$ and $- k_B T \Delta c_s$, respectively the hydrostatic pressure and solute concentration imbalance;  new flows are (a) the total volume flow through the membrane (sum of all flows):
\begin{equation}
Q = v_w \frac{d N_w^{(r)}}{dt} + v_s \frac{d N_s^{(r)}}{dt}
\end{equation}
and (b) the excess solute flow (as compared to the solute flow carried by the solvent) or the exchange flow:
\begin{equation}
J_e =  \frac{1}{c_s}\frac{d N_s^{(r)}}{dt} - \frac{1}{c_w} \frac{d N_w^{(r)}}{dt}
\end{equation}
Under the assumption that the concentration of solute is small $c_s \ll c_w$, one may thus rewrite $J_e \simeq J_s/c_s - Q$ where $J_s$ is the solute flow. 

The framework of irreversible processes assumes a linear relation between fluxes and forces \cite{degroot2013non}, hereby taking the form 
%Transport across the membrane is characterized within the framework of irreversible processes, via a transport matrix ${\mathbb L}$, relating fluxes to thermodynamic forces
%
\begin{equation}
\label{Larray}
\left(\begin{array}{c} Q \\ J_s-c_s Q\end{array}\right)= 
{\mathbb L}
\times \left(\begin{array}{c} - \Delta p \\ - k_B T \Delta \log c_s \end{array}\right).
\end{equation}
where ${\mathbb L}$ is the transport matrix. Importantly, as we  discuss below and in Sec.~\ref{transportmatrix}, 
this matrix is symmetric according to Onsager's principle -- due to microscopic time reversibility -- and definite positive {-- due to the second principle of thermodynamics --}. 

The question then amounts to characterizing the transport coefficients of this matrix. %associated with osmotic gradients. 
By identifying limiting regimes, Kedem and Kachalsky %identified the signs of the coefficients and
rewrote these transport equations in a more explicit form as~\cite{kedem1958thermodynamic,kedem1961physical,kedem1963permeabilitya,kedem1963permeabilityb,kedem1963permeabilityc}
\begin{eqnarray}
&Q= -\mathcal{L}_\text{hyd} \left( \Delta p - \sigma k_B T \Delta c_s\right), \label{kk1}\\
&J_s= - \mathcal{L}_\text{D} \omega_s \Delta c_s + c_s (1-\sigma) Q, \label{kk2}
\end{eqnarray}
where $\mathcal{L}_\text{hyd}=\kappa_\text{hyd} \mathcal{A}/(\eta L)$ is the solvent permeance through the membrane
with $\kappa_\text{hyd}$ the permeability (with units of a length squared), $\mathcal{A}$ the membrane area, $\eta$ the fluid viscosity, and
$L$ the membrane thickness; $\mathcal{L}_\text{D}=\mathcal{A} D_s/L$ is the solute permeability with $D_s$ the diffusion coefficient of the solute.
\eqref{kk1} is often referred to as the Starling equation in the physiology literature\cite{starling1896absorption}, see {\it e.g.} \citefulls{pappenheimer1953passage,adamson2004oncotic}. The osmotic pressure generated by the large scale molecules involved in the body (complex proteins such as albumin and more) is referred to as the \textit{oncotic pressure}. 
These equations introduce two dimensionless (numerical) factors: $\sigma$ is the so-called reflection or selectivity coefficient and $\omega_s$ is a solute \qote{mobility} across the membrane -- both of which we discuss in details below.
%A fully semi-permeable membrane corresponds to the case where $\sigma=1$ and $\omega=0$ while, reversly, a fully permeable membrane to both solute and solvent correspond to $\sigma=0$ and $\omega=1$.
%\LB{check}

The Onsager symmetry relations for \eqref{Larray} can be verified by exploring two limiting cases: (1) The situation where $\Delta p = 0$  yields osmotic flow only as $Q=\sigma \mathcal{L}_\text{hyd} c_s\Delta\mu$ (using  $\Delta \mu= k_B T \Delta c_s/c_s$ in the dilute case); (2) and the situation where $\Delta\mu =0$ yields $J_s-c_sQ=\sigma\mathcal{L}_\text{hyd}c_s\Delta p$. One obtains therefore $\left[Q/\Delta\mu\right]_{\Delta p =0} = \left[(J_s-c_sQ)/\Delta p\right]_{\Delta \mu =0}$ and the symmetry of the transport matrix is indeed verified.

%  terms are equalFrom the symmetry relation principle of Onsager~\cite{onsager1931reciprocal}, we expect thus $\sigma_0 = \sigma_f = \sigma$. \textcolor{blue}{However some discussion exists as to this equality, based on hydrodynamical models of permeation~\cite{anderson1974mechanism,hill1979osmosis}, and so far no measurement has clearly proven that hypothesis.} 
%  For the following we consider that in most situations the Péclet number is small, and thus advection by the flow represents a small contribution to the solute flow. 

{\it The reflection coefficient and the solute mobility --}
The Kedem--Katchalsky equations introduce the reflection coefficient $\sigma$ mentioned previously and first described by Staverman~\cite{staverman1951theory}. This coefficient {\it a priori} depends on the relative interactions of the membrane with the solute and solvent~\cite{talen1965negative,talen1965osmometry,anderson1974mechanism}. The Kedem-Katchalsky framework also introduces the permeability of the solute through the membrane via the combination $\mathcal{L}_\text{D} \omega_s $. 
A fully semi-permeable membrane corresponds to the case where $\sigma=1$ and $\omega_s=0$:  the solute flux vanishes $J_s = 0$ and the  pressure driving the fluid identifies with the van 't Hoff result $\Delta \Pi = k_B T \Delta c_s$. 
Reversely, a \qote{transparent} membrane which is fully permeable to both solute and solvent correspond to $\sigma=0$ and $\omega_s=1$: no osmotic pressure is expressed and the solute flux reduces to Fick's law.

\begin{figure}[h!]
\centering
  \includegraphics[width=0.45\textwidth]{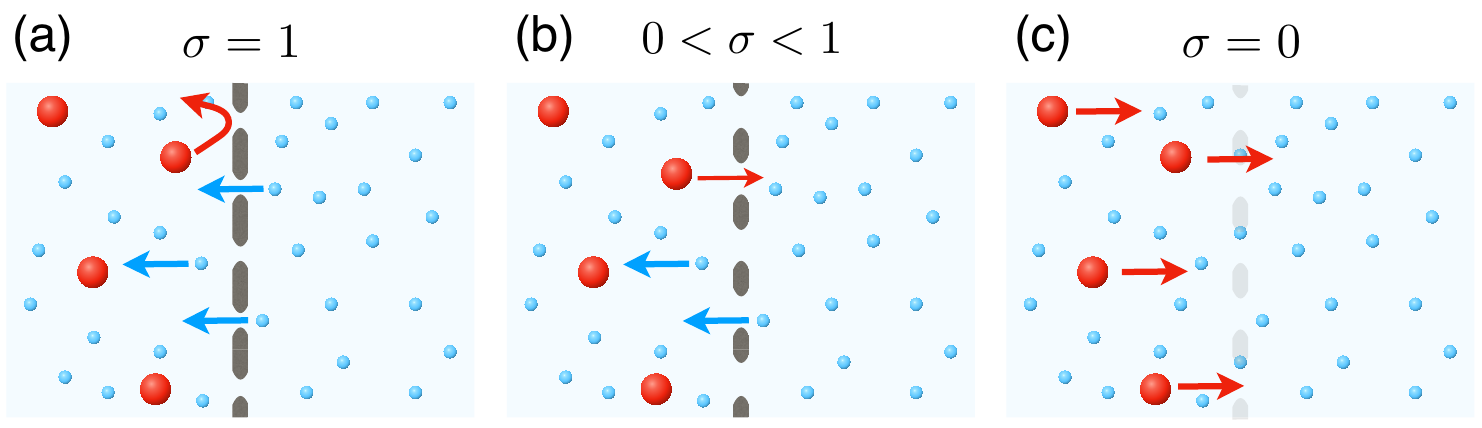}
  \caption{\textbf{: Examples of reflection coefficient based on steric exclusion.} In (a) the blue solvent only may traverse the pores; while in (b) the red solute particles may also traverse; their permeability through the pores is however small since the accessible volume in the pore for the red solute particles is smaller than that for the blue solvent. (c) The membrane is now fully permeable to all species, and therefore diffusion dominates and solute particles move towards the low concentration side.}
  \label{fig:sigma}
\end{figure}

%To recover the result of van\;'t Hoff in the case of a semi-permeable membrane, we find that we need $\sigma = 1$ and $\Delta \Pi = k_B T \Delta c_s$ drives the solvent flow. In this case we naturally have $\omega = 0$ (semi-permeable membrane) and as a result Eq.~\ref{kk2} gives $J_s = 0$ as expected. 
In the intermediate case, the membrane  is  partially permeable to the solute and we expect $0 < \sigma < 1$, see Fig.~\ref{fig:sigma}. 
%This corresponds to a more general and realistic description. 
As an example, in a pure nafion membrane about $18 \mu m$ thick, the reflection coefficient between water and $KCl$ salt was measured as $\sigma = 0.82$ (at concentration $0.25$ mol/L)~\cite{yamauchi2000membrane}.

Interestingly, cases with negative reflection coefficient, $\sigma < 0$, were reported. This situation is often termed {\it anomalous osmosis}~\cite{fujita1968interpretation,hill1979osmosis} and it corresponds to situations where the solute is more permeable than the solvent. 
We will discuss in Sec.~\ref{beyondVO} various examples where such a situation with reversed osmosis occurs.
%We will discuss below a variety of effects  that may also induce a reversed current. 

The specificity of the membrane and its interaction with the solute molecules actually come into play into this reflection coefficient $\sigma$. A number of models have tried to rationalize the dependence of $\sigma$ on the chemical and physical properties of the components. The first models took into account \textit{steric effects} (similar to Fig.~\ref{fig:sigma}), where in fact the volume accessible to the solute inside the pore would differ (because of its typically larger size) than that accessible to the solvent~\cite{ferry1936ultrafilter,ray1960theory,giddings1968statistical}. The next generation of models sought to include as well hydrodynamic interactions, investigating how friction induced by the proximity of the solute to the pore walls would reduce permeability~\cite{renkin1954filtration,anderson1974restricted}. Anderson also studied interactions with the pore walls and adsorption of the solute in the pore~\cite{anderson1974mechanism}. 
Similarly, the \qote{mobility} coefficient $\omega_s$ entering the transport equations will depend both on the solute-membrane interactions and transport parameters. 
A first, naive, estimate is to identify this coefficient with the partition coefficient of the solute between the membrane and the reservoirs at equilibrium, $K_s=c_s^{\rm m}/c_s^{\rm bulk}$, so that $\omega_s = K_s$~\cite{kedem1961physical}. But this estimate does not account for the complex transport processes occurring within the membrane. Interestingly, the non-dimensional coefficients $\omega_s$ and $\sigma$ are expected to be linearly related,~\cite{kedem1961physical} as $1-\sigma \propto \omega_s$, a result that we will recover below in a specific case.
%$\omega$ as the partition function ??\cite{kedem1961physical}
%\LB{In simple situations $\sigma$ and $\omega$ are linearly related.}

{Altogether, a complete determination of the reflection and mobility coefficients requires to implement a microscopic description of the membrane-fluid interactions. We will explore below \SM{and in Sec.~\ref{Sec:OsmosisWithoutMembrane}} various situations highlighting how playing with interactions may lead to advanced osmotic transport behavior.}

\subsection{Mechanical views of osmosis: a tutorial perspective}
\label{mech}

Beyond the general formalism introduced above, it is interesting to get further fundamental insights into the microscopic mechanisms which underlie osmosis.  In particular it is of interest to get some intuition on the {\it mechanical force balance} associated with the osmotic pressure.
% and its mechanical origin.  
%\LB{PEDAGOGICAL !! first, mechanics at equilibrium: high barrier, then with flow}
%We start by exploring some simple views of osmosis, which are based 
%Taking a more mechanical perspective of the underlying processes, we reduce the microscopic ingredients of osmosis to their minimal function. 
{To do so, we will reduce the microscopic ingredients of osmosis to their minimal function and this description has merely a tutorial purpose.} Still it is very enlightening in order to understand %the physical origin of the phenomenon and show 
how the connection between \qote{microscopic} parameters and thermodynamic forces builds up. %As we will highlight in this section, 
Such mechanistic views of osmosis also allow to envision advanced osmotic phenomena, beyond the van\,'t Hoff perspective.
Alternative approaches with similar illustrative objectives were proposed for one-dimensional single file channels, see \citefulls{chou1998fast,chou1999kinetics}.
%A model that physically describes transport and how
%flows depend on microscopic molecular parameters and
%macroscopic thermodynamic constraints would serve as
%a useful benchmark in more sophisticated models

%\LB{ICICICI}

\begin{figure}[h!]
\centering
  \includegraphics[width=0.35\textwidth]{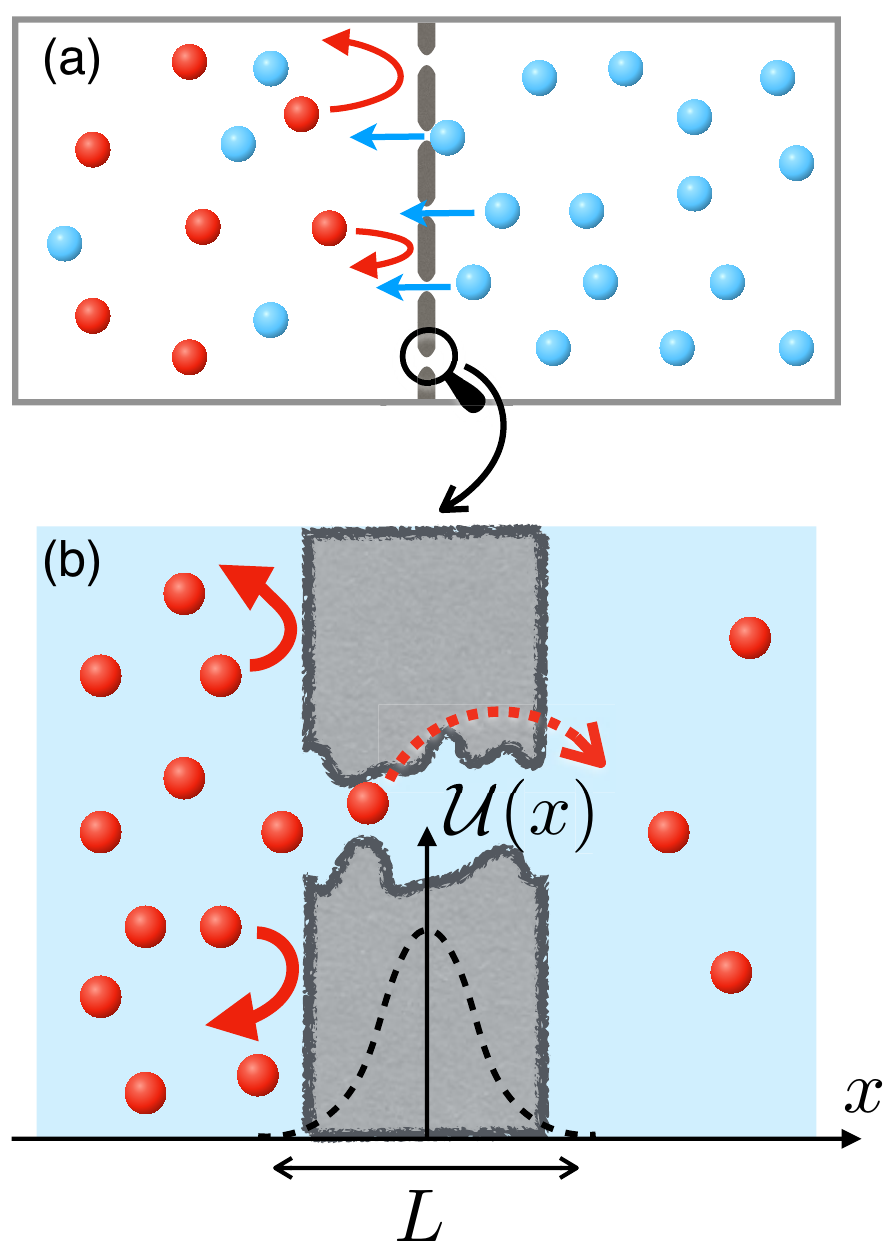}
  \caption{\textbf{: The physical force driving osmosis is interaction of the solute with the membrane.} (a) The membrane exerts a repulsive force (red arrows) on the solute particles (red) that creates a pressure gradient (or a void on the immediate left hand side of the membrane) that drives the flow (blue arrows of blue solvent particles).
  % \textcolor{blue}{is that correct ?} (b) Large scale analogy; a spider repellent sticker prevents spiders from crossing or approaching the center of the box; therefore creating a depletion region where frogs like to jump in, creating an "osmotic frog flow". 
  %\textcolor{blue}{problem: frogs like to eat spiders; we could also just put bugs}
  (b) Mechanical view of osmosis: the partially permeable membrane may be viewed as an energy barrier for the solute molecules (in red) that they have to overcome in order to traverse the membrane.
  }
  \label{fig:barrier}
\end{figure}

We pointed out above that the van\,'t Hoff law for the osmotic pressure does not involve the membrane properties {\it per se}, provided that it is semi-permeable. So it is tempting to replace the membrane by a crude equivalent, namely an energy barrier acting on the solute only, say  $\mathcal{U}(x)$ (assuming for simplicity a unidimensional geometry) -- see Fig.~\ref{fig:barrier}. This approach, which captures the minimal ingredients %of partial or semi-permeability 
at play in osmosis,
%It was first introduced by Manning \cite{manning1968binary} and more recently generalized to charged perm-selective pores\cite{picallo2013nanofluidic} and to high solute concentration \cite{marbach2017osmotic}. 
was first introduced by Manning~\cite{manning1968binary} in the low concentration regime, and generalized more recently to explore the osmotic transport across perm-selective charged nanochannels~\cite{picallo2013nanofluidic} or in non-linear regimes at high solute concentrations~\cite{marbach2017osmotic}.
One may note that such a potential barrier can also be physically achieved; for example, it may be generated from a nonuniform electric field acting on a polar solute in a nonpolar solvent~\cite{debye1954equilibrium}, or it can represent the nonequivalent interactions of solute and of solvent particles with a permeable membrane, {\it e.g.}, charge interactions~\cite{grim1957contributions,picallo2013nanofluidic}.
%In the case of osmosis, this approach was first introduced by Manning~\cite{manning1968binary} in the low concentration regime, and generalized more recently to explore the osmotic transport across perm-selective charged nanopores~\cite{picallo2013nanofluidic} or in non-linear regimes at high concentrations~\cite{marbach2017osmotic}.

Let us first consider the ideal case where the barrier's maximum is high, {\it i.e.} $\mathcal{U}_{\rm max}\gg k_BT$, so that the solute cannot cross the barrier: this is the perfectly semi-permeable case. 
In both reservoirs the solute is at equilibrium
and the solute profile follows accordingly the Boltzmann relation
\begin{equation}
c_s^{(r)/(l)}(x)=c_s^{(r)/(l)} \times e^{-{\mathcal{U}(x)\over k_BT}}
\end{equation}

Now, a key remark is that the force on a {\it fluid element} of volume $d\tau$ (consisting of the solvent and solute mixture) will write
\begin{equation}
df(x)=c_s^{(r)/(l)}(x) \times (-\partial_x\mathcal{U}(x)) d\tau.
\end{equation}
{%Since the interaction potential is only dependent on the coordinate $x$, we may write 
with $d\tau = \mathcal{A} dx$; $\mathcal{A}$ is the membrane area. The total force per unit area acting on the fluid is accordingly integrated over $x$}
%\begin{eqnarray}
%&{F_T\over {\cal A}}=&\int_0^\infty dx\, c_{(r)} e^{-{\mathcal{U}(x)\over k_BT}}\times (-\partial_x\mathcal{U}(x))\nonumber \\
%&&\int_{-\infty}^0 dx\, c_{(l)} e^{-{\mathcal{U}(x)\over k_BT}}\times (-\partial_x\mathcal{U}(x))
%\end{eqnarray}
\begin{equation}
{F_T\over {\cal A}}=\int_0^\infty dx\, c_s^{(r)} e^{-{\mathcal{U}(x)\over k_BT}}\times (-\partial_x\mathcal{U}(x))+
\int_{-\infty}^0 dx\, c_s^{(l)} e^{-{\mathcal{U}(x)\over k_BT}}\times (-\partial_x\mathcal{U}(x))
\end{equation}
(where we arbitrarily put $x=0$ at the position of the maximum of the energy barrier), leading immediately to
\begin{equation}
{F_T\over {\cal A}}=k_BT \times [c_s^{(r)}-c_s^{(l)}]\equiv \Delta \Pi
\end{equation}
where we neglected  terms behaving as $\exp[-\mathcal{U}_{\rm max}/k_BT]$.
Altogether this simple approach allows one to retrieve the van\,'t Hoff law. It highlights the mechanical origin of  osmosis: as is transparent from the previous derivation, the osmotic pressure results from the fact that the reservoir containing more solute particle will generate a higher repelling force on the fluid than from the other reservoir: accordingly a fluid flow will be generated from the low to the high concentrations, hence diluting the more concentrated reservoir.

While the above approach is instrinsically at equilibrium, it can be easily generalized to a non-equilibrium situation by releasing the assumption of an (infinitely) high energy barrier: in this case the solute can cross the \qote{membrane} between the two reservoirs at a finite rate, see Fig.~\ref{fig:barrier}, generating a solute flux. We further assume that 
the membrane is fully permeable to the solvent (no energy barrier acting on it), with a permeance $\mathcal{L}_\text{hyd}$ relating the fluid flux $Q$ to the pressure
drop $\Delta p$ in the absence of a concentration difference: $Q=\mathcal{L}_\text{hyd} (-\Delta p)$. 

The stationary dynamics of the system is described by the coupled set of equations for the solute diffusive dynamics -- Smoluchowski equation -- and fluid transport -- Navier-Stokes equation.
%In the prospect of fully describing the system of Fig.~\ref{fig:barrier} one should write essentially two governing equations (Eqs.~\ref{Smolu} and~\ref{St}), basically momentum and mass conservation equations on the fluid and the solute. 
In the 1D geometry described above, the stationary solute concentration $c_s(x)$ obeys a Smoluchowski  equation:
\begin{align}
0={\partial_t} c_s =& -\partial_x j_s  \nonumber\\
=& -\partial_x\left( -D_s \partial_x c_s + \lambda_s c_s \,(-\partial_x \UU) + v_x c_s \right),
\label{Smolu}
\end{align}
where $j_s = J_s/\mathcal{A}$ is the solute flux per unit surface, $D_s$ is the solute diffusion coefficient, $\lambda_s=D_s/k_B T$ the mobility and $v_x$ the  local fluid velocity.
We will further assume a low P\'eclet number, $\mathrm{Pe}=v_x L/D_s \ll 1$, such that the convective term of \eqref{Smolu} is negligible. This is valid for low permeability (nanoporous) membranes. The full derivation including the convective term was considered in~\citefull{manning1968binary}. 
Since the solute flux across the membrane $J_s$ is constant in time and spatially uniform, \eqref{Smolu} is explicitly solved with respect to the concentration as:
\begin{equation}
c_s(x)= c_s^{(r)}-\Delta c_s\, e^{-\beta\UU(x)} { \int_{x}^{L/2} dx^\prime\, \exp[+\beta\UU(x^\prime)]\over \int_{-L/2}^{L/2} dx^\prime\, \exp[+\beta\UU(x^\prime)]},
\label{profile}
\end{equation}
where $\beta=1/k_B T$. %\LB{attention $c$ est l'exces par rapport a $c_-$}
The solute concentration difference between the two volumes is $\Delta c_s=c_s^{(r)}-c_s^{(l)}$. For simplicity we assumed that the barrier has an extension $L$.

Now turning to the momentum conservation equation %. We search for a momentum conservation equation on 
for the fluid (solvent + solute), the flow field $\bm{v}$ of the fluid obeys a Stokes equation (neglecting inertial terms)
\begin{equation}
%\rho \left( \partial_t \bm{v} + (\bm{v}.\bm{\nabla})\bm{v} \right) 
0= -\bm{\nabla} p + \eta \bm{\nabla}^2 \bm{v} + \bm{f}_{ext},
\label{St}
\end{equation}
where $p$ is the fluid pressure and $\bm{f}_{ext}$ represents the total volume forces acting on the system, $e.g.$ the forces acting on the solvent \textit{and} on the solute, here
\begin{equation}
\bm{f}_{ext} = c_s (-\bm{\nabla} \UU).
\end{equation}
%Therefore on a unit volume the force exerted by the membrane on the solute is the only external force and thus $\bm{f}_{ext} = c (-\partial_x \UU)$. As the systems studied (micro to nano-porous systems) are in the range of small Reynolds number, we may neglect inertial effects and at steady state solve the Stokes equation along the $x$ direction:
%
%\begin{equation}
%0 = -\partial_x p + \eta \nabla^2 u + c(x) (-\partial_x \UU),
%\label{St}
%\end{equation}
%
The driving force inducing the solvent flow along the $x$ axis is accordingly written in terms of an apparent pressure drop, $-\partial_x \mathcal{P}=-\partial_x p+c_s(x)(-\partial_x\UU)$. 
The membrane, via its potential $\UU$, will therefore create an average force on the fluid, which writes per unit surface 
\begin{equation}\label{eq_Fc}
-\Delta\mathcal{P}=-\Delta p+ \int_{-L/2}^{L/2} dx\, c_s \,(-\partial_x \UU) \equiv -\Delta p +\sigma \Delta \Pi,
\end{equation}
%
% -\Delta\mathcal{P}=-\Delta p+{1\over\mathcal{A}}\int d{\mathcal{A}} \int_{-L/2}^{L/2} dx\, c \,(-\partial_x \UU),
%we highlighted in this expression an average over the lateral area $\mathcal{A}$ (which will be dropped in the following to simplify notations).
where $\Delta$ means the difference of a quantity between the two sides.
The second term of \eqref{eq_Fc} can be interpreted as the osmotic contribution. Using the expression for the concentration profile given in \eqref{profile}, one recovers the classical van\;'t Hoff law of the osmotic pressure, $\Delta\Pi= k_B T \Delta c_s$, and furthermore 
obtains an expression for the reflection coefficient $\sigma$ as
\begin{equation}
\sigma= 1-{L \over \int_{-L/2}^{L/2} dx^\prime\, \exp[+\beta\UU(x^\prime)]}.
\label{reflectDilute}
\end{equation}

The above result correctly recovers the case of a completely semi-permeable membrane (no solute flux across the membrane), {\it i.e.}, $\beta \UU\gg 1$ and  $\sigma\rightarrow 1$, yielding $-\Delta \mathcal{P}=-\Delta [p-\Pi]$. 
In the intermediate cases, although the membrane is permeable, a flow arises due to the solute concentration gradient even in the absence of an imposed pressure gradient. When the potential is repulsive and small $\UU \sim \kt$, then $0<\sigma<1$; the flow is in the direction of increasing concentration. %When the potential is attractive, then $\sigma < 0$ and the flow reverses. 
%\revb{ Integrated over the nanochannel, this allows us to express the total water flux Q as $Q=xxx$ with Lhyd the hydrodynamic permeability of the channel}

Integrating \eqref{St} over the membrane area ($\mathcal{A}$) and thickness ($L$) allows the
total flux $Q$  to be expressed as
\begin{equation}
Q = - \mathcal{L}_\text{hyd} \left(\Delta p -\sigma k_B T \Delta c_s \right).
\label{eq:KKbis}
\end{equation}
Here the permeance $\mathcal{L}_{\rm hyd}$ can be expressed in terms of the  permeability, $\kappa_{\rm hyd}$, as $\mathcal{L}_\text{hyd} = {\mathcal{A}\over L} {\kappa_{\rm hyd}\over\eta}$.
The permeability $\kappa_{\rm hyd}$  is defined formally in terms of the flow as 
$\kappa_{\rm hyd}^{-1}= \langle - \nabla^2 v \rangle / \langle v \rangle$, where $\langle\cdot\rangle=\mathcal{V}^{-1} \int\int dx\,d\mathcal{A}\,(\cdot)$ denotes an average over the pore volume, here $\mathcal{V}=\mathcal{A}\,L$.
%$$
%Q\equiv  \kappa_{\rm hyd}  \times {1\over {L} }\int\int dx\,d\mathcal{A}\,\nabla^2 v.
%%\equiv - {Q}/{\kappa_{\rm hyd}}
%$$
%with $\mathcal{V}=\mathcal{A}L$ the channel volume.
%and the corresponding permeability is defined as $\mathcal{L}_\text{hyd} = \mathcal{A}\kappa_{\rm hyd}/(\eta L)$. 
These parameters, $\kappa_{\rm hyd}$ and $\mathcal{L}_\text{hyd}$, take into account  the detailed geometry of the pores in the membrane (pore cross section,  length, etc.). 
%\revr{we have: $({\eta}/{\mathcal{A}}) \int d\mathcal{A} \nabla^2 v = {\Delta\mathcal{P}}/{L}$. We define the volume flux of the solute and the permeability $\kappa$ of the membrane as: $({1}/{\mathcal{A}}) \int d\mathcal{A} \nabla^2 v = - {Q}/{\kappa}$. The permeability $\kappa$ takes into account all the detailed geometric specificities of the membrane. As a result, with $\mathcal{L}_\text{hyd} = \kappa/(\eta L)$}:
Overall \eqref{eq:KKbis} agrees with the Kedem--Kachalsky result in \eqref{kk1}.
%Equation~\eqref{eq:KKbis} is often referred to as the Starling equation in the physiology literature\cite{starling1896absorption}, see {\it e.g.} Refs.~\citenum{pappenheimer1953passage,adamson2004oncotic}. The osmotic pressure generated by the larger scale molecules involved in the body (complex proteins such as albumin and more) is referred to as the \textit{oncotic pressure}. 
While this approach is derived here in the dilute regime for the solute, it can be generalized to arbitrary concentrations, see \citefull{marbach2017osmotic}.

%\LB{Ajouter discussion sur non-ideal solution cf JCP Yoshida/Marbach}
%\LB{Insister plus sur la symetrie entre flux solute et force dans l'osmose, origine de $\sigma$ et active osmosis }

As a last remark, it is interesting to note that the mechanistic approach highlights an underlying fundamental symmetry in the transport phenomenon. Indeed
\eqref{eq_Fc} introduces the osmotic pressure as the driving force on the fluid: $\int_{-L/2}^{L/2} dx\, c_s \,(-\partial_x \UU) = \sigma \Delta \Pi$. Now the Smoluchowski equation for the solute -- integrated over the membrane thickness $L$, \eqref{Smolu} -- contains the very same term and one may accordingly rewrite the solute flux as
%\begin{equation}
%J_s  = -D_s {\Delta c_s\over L} + {D\over k_BT} \times \sigma \Delta \Pi
%\end{equation}
%with $\Delta \Pi = k_BT \Delta c$, so that 
%and
\begin{equation}
J_s= -{D_s \over L} \mathcal{A} \left[ \Delta c_s - \sigma {\Delta \Pi\over k_BT} \right] 
\label{sym}
\end{equation}
The solute flux is therefore intimately related to the osmotic pressure. As is transparent from this equation, the van\,'t Hoff osmotic pressure is fully expressed, {\it i.e.} $\Delta \Pi= k_BT \Delta c_s$,  only when the solute flux vanishes $J_s = 0$ ($\sigma=1$ and $\omega_s=0$). Reversely for a fully permeable membrane $J_s = -{D_s \mathcal{A} \over L} \Delta c_s$, and there is no osmotic pressure ($\sigma=0$ and $\omega_s=1$). Finally this equation can be rewritten 
as $J_s= -D_s \mathcal{A} (1-\sigma) \Delta c_s/L$, so that the \qote{mobility} coefficient $\omega_s$ is related here to the reflection coefficient as $\omega_s = 1-\sigma$.

%\LB{ICICICICI}

\vskip0.5cm

\review{In this first part we have reviewed the basic understanding of osmosis, from the historical discovery of the phenomenon to the precise understanding of the effect in terms of thermodynamic forces. Although simplistic, the previous mechanical/kinetic approach provides a fruitful and complementary perspective on osmotic transport, which suggests a number of generalizations -- that we will discuss below. It also reveals that the key aspect of osmosis is not really the membrane itself, but the existence of differential forces acting separately on the solvent and the solute. This is crucial to understand a number of phenomena related to osmosis that we discuss below.}

%\section{Osmosis beyond van\,'t Hoff}
%\section{Membrane-less osmosis}
\section{Osmosis without a membrane}
\label{Sec:OsmosisWithoutMembrane}

\review{Situations where differential forces act on the solvent and the solute} occur naturally, especially at interfaces: for example a charged surface does act specifically on dissolved ions, repelling co-ions and attracting counter-ions; or a neutral hard wall will repel polymers via excluded volume. As we now discuss in the following sections, these specific forces may be harnessed to induce interfacially-driven osmotic flows. 
%Accordingly a semi-permeable membrane is not  a prerequisite to induce osmosis. 

%Disturbing such Playing with the associated forces allows to . 
%Such configurations are good candidates to induce osmotic forces.
%Such situations is encountered in many phenomena. For example a charged surface generates an inhomogeneous distribution of ions, constituting the so-called electric-double layer (EDL). Co-ions are repelled from the surface while counter-ions are attracted, while having a rather minimal direct effect on the solvent.
%Now as osmosis may be viewed as a net force acting on the fluid originating from a membrane, a similar perspective may be found without a membrane, as long as a force acting on the fluid remains. For example, harvesting surface forces (within \textit{e.g.} fully permeable pores) may lead to osmotically driven flows with different implications. 

%In the present section and the next one, we review several phenomena which harvest in various ways the mechanical nature of osmosis to envision advanced osmotic phenomena. 

%Here we first explore how osmotic may be driven at solid surfaces, in the absence of a semi-permeable membrane (hence membrane-less osmosis). 
The geometry we will consider here involves a solid surface along which a solute gradient, or more generally a thermodynamic force -- an electric field, a temperature gradient... --  is established, as sketched in Figs.~\ref{fig:EO}-\ref{fig:DO}. Under an electric field, the net electric forces occurring within the diffuse interface close to the solid will push the fluid and generate a so-called electro-osmosis flow for the solvent. But as we will show below,  a solute gradient $\nabla c_{\infty}$ parallel to the surface can also generate fluid motion whose amplitude  is proportional to $\nabla c_{\infty}$: %interfacially driven flow of the solvent. 
\begin{equation} v_{DO} \propto \nabla c_{\infty}. \end{equation}
This latter phenomenon is usually coined as {\it diffusio-osmosis}. The phenomenon bears some fundamental analogy with Marangoni effects where a gradient of surface tension at an interface may drive fluid (or reversely particle) motion as $v_f\propto \nabla \gamma_{LV}$ \cite{anderson1989colloid}. 
Now  extending Marangoni flows to solid-fluid interfaces is definitely not obvious, but it was recognized by Derjaguin and collaborators \cite{derjaguin1993kinetic,derjaguin1993diffusiophoresis} that the diffuse nature of the interface may allow the fluid to \qote{slip} over the solid surface under a concentration gradient. Diffusio-osmosis is accordingly an interfacially driven flow, and takes its origin in the interfacial structure of the solute close to the solid surface, within the first few nanometers close to the surface. 

%
%
%\LB{a comment on Marangoni flows, $V\propto \nabla \gamma_{LV}$ \cite{anderson1989colloid}. But initial question: does this extend to liquid-solid interfaces ? yes if the diffuse nature of the interface is taken into account. Derjaguin. Requires to investigate the underlying structure of the interface}.

%\LB{Papiers historiques:\\
%- Dukhin SS, Derjaguin BV. Electrokinetic phenomena. In:
%Matijevic E, editor Surface and Colloid Science. Volume 7.
%New York: John Wiley and Sons; 1974.\\
%- Derjaguin BV, Dukhin SS, Korotkova AA \cite{derjaguin1993diffusio-phoresis} (1993) diffusio-phoresis in electrolyte solutions
%and its role in the mechanism of the formation of films from caoutchouc latexes
%by the ionic deposition method. Prog Surf Sci 43(20):153?158.}

%The study of interfacial transport of fluids at interfaces bares a lot of analogies to the previous mechanical approach to osmosis, since the structure of the solute at the solid surface is required. BLABLA

\subsection{From electro- to diffusio- osmosis}
\label{EO-DO}

\subsubsection{From electro-osmosis...}
%Before exploring diffusio-osmosis, let us first start with a short discussion of electro-osmosis, as a canonical interfacial transport.
%\LB{First EO as a canonical example}

\begin{figure}[h!]
\centering
  \includegraphics[width=0.49\textwidth]{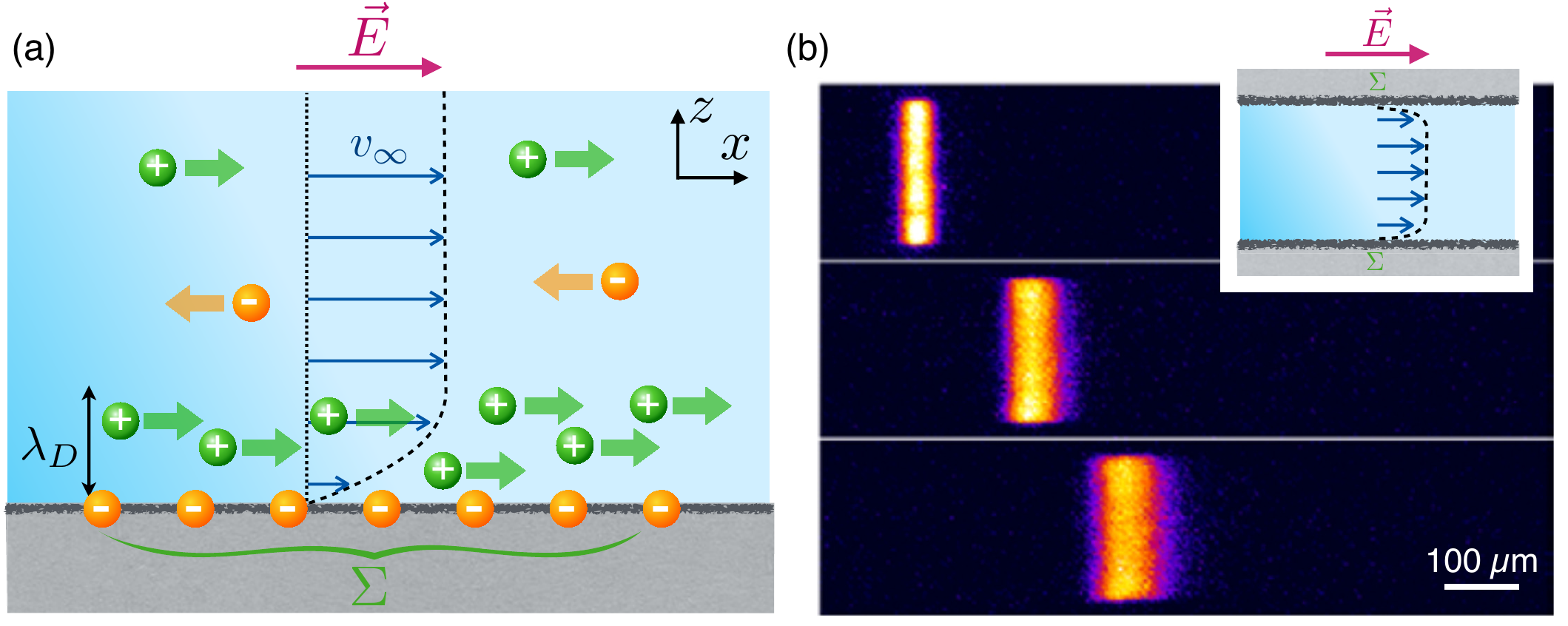}
  \caption{\textbf{: Electro-osmosis.} (a) In the presence of a  surface charge $\Sigma$, an electrical double layer forms extending typically over a distance fixed by the Debye length $\lambda_D$. Under an applied electric field $\vec{E}$ parallel to the surface, a net electric
  force builds up due to the unbalanced charge within the electric double layer. It drives a solvent flow parallel to the surface,  extending as a flat flow profile into the bulk. (b) Visualization of the electro-osmotic flow in a capillary, via the displacement  of (neutral) tagged molecules. In contrast to the parabolic Poiseuille flow, the electro-osmotic flow profile takes the form of a plug flow, which barely disperses the dye.  \SM{Reproduced from \citefull{devasenathipathy2005electrokinetic} with permission from Springer Nature, copyright 2005.}
  %\LB{not sure it is a good idea to show such a linear velocity profile which is very uncommon for EO.}
  }
  \label{fig:EO}
\end{figure}

Let us start with the canonical example of electro-osmosis,  {\it i.e.} the fluid flow close to a solid surface generated under an applied electric field. A solution containing ions will build up a so-called electric double layer (EDL) close to any charged surface: counter-ions are attracted by the surface charge, while co-ions are repelled. The surface charge, say $\Sigma$, is balanced in the fluid by a  density of charge  $\rho_e = e(c_+-c_-)$, defined as the difference between the density of positive and negative ions (assuming monovalent ions here for illustrative purposes). The resulting double layer is diffuse and extends over a finite width, see Fig.~\ref{fig:EO}. The structure of the EDL was amply discussed in many textbooks and reviews, and we refer in particular to \citefulls{hunter2001foundations,andelman1995electrostatic,squires2016particles} for further insights.  As a rule of thumb, the  extension of the EDL is typically given by the Debye screening length %Within the Poisson-Boltzmann framework 
\cite{andelman1995electrostatic,schoch2008transport}, defined as
\begin{equation}
\lambda_D={1\over \sqrt{8\pi\ell_B c_s}}
\end{equation}
where $c_s$ is the (bulk) salt concentration in the bulk and $\ell_B=e^2/4\pi \epsilon k_BT$ is the Bjerrum length ($\epsilon$ is the dielectric permittivity of water). Typically for water at room temperature, $\ell_B=0.7$nm and the Debye length ranges between 30nm for a salt concentration of $10^{-4}$mol.l$^{-1}$ to 0.3nm for a 1mol.l$^{-1}$ salt concentration. %This estimates puts aside further complexities in the detailed structure of the EDL, even at the mean-field level of the non-linear Poisson-Boltzmann framework, see \cite{andelman1995electrostatic,schoch2008transport}. 

%Similarly as for diffusio-osmotic transport, where solvent flow arises under a salinity gradient in the vicinity of a surface interacting with the solute, it is possible to generate electro-osmotic transport. In electro-osmosis, solvent flow arises under a voltage gradient in the vicinity of a surface interacting electrically with the solute, see Fig.~\ref{fig:EO}. Alike diffusio-osmosis, the phenomenon is driven by the interface, taking its origin in the electrical double layer forming close to the surface, due e.g. to a local surface charge $\Sigma$~\cite{lyklema2005fundamentals,anderson1989colloid}. 

Within the EDL, there is a net charge density in the fluid, and whenever an external electric field is applied to the fluid (parallel to the surface), this will generate
a net bulk force $\rho_e E$. 
%To account formally for electro-osmosis, the steps are quite similar as for diffusio-osmosis. \LB{renverser}
%The local electrical concentration $\rho_e$ verifies the Poisson equation:
%%\begin{equation}
%$\rho_e = - \epsilon \partial_{zz} V$
%%\end{equation}
%where $\epsilon$ is the dielectric permittivity of the fluid, $V$ the electrical potential, dependent essentially on the $z$ direction because we have considered slow variations along the $x$ direction. 
The Stokes equation for the fluid velocity writes accordingly in the direction $x$ (parallel to the solid interface)
%The fluid verifies the general equation~\ref{NSt}, that reduces in this case, at steady state and projected on the $x$ axis, to:
\begin{equation}
\eta \partial_{zz} v + \rho_e E = 0
\label{eq:EO}
\end{equation}
where $z$ is the direction perpendicular to the interface. The pressure-gradient term vanishes for the shear-flow considered here.
%In Eq.~\ref{eq:EO} we have neglected the pressure component for it is small compared to the gradient of electrical potential~\cite{anderson1989colloid}. 
Using the Poisson equation $\rho_e = - \epsilon \partial_{zz} V_e$, relating the charge density to the electric potential $V_e$ in the fluid, one
can integrate twice \eqref{eq:EO} to obtain the velocity profile
%Integrating Eq.~\ref{eq:EO} twice with the expression for $\rho_e$, and using the fact that the velocity gradient vanishes far from the surface $\partial_z v|_{z = \infty} = 0$, and also the potential gradient $\partial_z V|_{z = \infty} = 0$, and a no-slip boundary condition at the surface gives:
\begin{equation}
v (z)= - \frac{\epsilon E}{\eta} \left[ V_e(z=0) - V_e(z) \right]
\end{equation}
where a no-slip boundary condition was assumed here. The electrical potential at the interface  is usually identified as
the zeta potential $V_e(z = 0)=\zeta$. 
% where in the absence of slippage at the interface the potential $\zeta$ may be identified with the electrical potential at the interface $V(z = 0)$. It is called the Zeta potential. Far from the surface 
The electro-osmotic velocity is constant beyond the EDL and reaches its asymptotic value %far from the surface:
\begin{equation}
v_{\infty} = \mu_{EO} E
\label{eq:vEO}
\end{equation}
where $ \mu_{EO} = - \frac{\epsilon\, \zeta}{\eta} $ is the electro-osmotic mobility. 
%Formally the electrical field $E$ may be seen as the longitudinal voltage gradient, $E = \partial_x V|_{\infty}$. 
In the presence of hydrodynamic slippage on the surface, the electro-osmotic mobility is typically enhanced by a factor $1+b/\lambda_D$, where $b$
is the slip length, see \citefulls{bocquet2010nanofluidics,khair2009influence,Mouterde2019} for more details. %the relationship between the surface electrostatic potential and the mobility is modified
%The chemical potential of the positively charged solute is therefore modified as $\mu_s^{+} = k_B T \left[ \phi_w (T) + ln \left( c_s k_B T \right) + Z e \mathcal{N}_A V \right]$, and respectively for the negatively charged solute. We therefore have from Gibbs-Duhem, $d\Pi = c_s d\mu$ giving $\partial_x \Pi_{\infty} = Z e \mathcal{N}_A  c_s \partial_{x} V_{\infty}$. 
%Therefore one may define an electro-osmotic mobility as $K_{EO} =  - \frac{\epsilon}{\eta Z e \mathcal{N}_A  c_s} \zeta$ giving $v_{\infty} = K_{EO} \partial_x \Pi_{\infty}$ and electro-osmosis is trully an \textit{osmotic} driven phenomenon. \textcolor{blue}{Is this right ? this is where we have to insist on the fact that electro-osmosis and diffusio-osmosis are really the same in a way}. 
%\LB{Note sure it is a good idea: no osmotic pressure per se, but indeed justify the wording osmosis in electro-osmosis}
We finally note that 
%It is interesting to rewrite 
the $\zeta$-potential may be rewritten as a function of the electrical concentration $\rho_e$ (by integrating twice \eqref{eq:EO})
\begin{equation}
\zeta = - \frac{1}{\epsilon} \int_0^{\infty} z\, \rho_e(z) dz.
\label{eq:zetaPotential}
\end{equation}
%\LB{simplifier}
%In simple cases, one may infer the sign of the $\zeta$ potential and thus the direction of the electro-osmotic current. 
%For example, in the case of Fig.~\ref{fig:EO} where a large negative surface charge is present, one expects the electrical double layer to be essentially positively charged. As a result we expect $\zeta < 0$ and thus $ \mu_{EO} > 0$ and the electro-osmotic flow will be in the direction of the electric field. Electro-osmosis and the $\zeta$ potential are strongly dependent on the properties of the interface, in particular on the slip length, and we refer to Ref.~\citenum{bocquet2010nanofluidics} for more details. 
%Electro-osmosis may also be interpreted in a very simple way~\cite{lyklema2005fundamentals}. In fact it 
From a physical point of view, electro-osmosis may be seen as a force balance between the viscous friction force at the interface and the electrostatic driving force within the EDL. The velocity field is expected to establish over the Debye length $\lambda_D$ and thus the fluid friction force is typically $\sim \eta v_{\infty} /\lambda_D$. Now the body electrical force within the EDL %corresponds to the number of charges to be displaced 
is simply $\Sigma\times E$ where $\Sigma$ is the surface charge. From Gauss' electrostatic boundary condition, we have  $\Sigma = - \epsilon \partial_z V_e |_{z= 0}\approx -\epsilon  \zeta / \lambda_D$. %and typically $\partial_z V_{z=0} = \zeta / \lambda_D$. 
Altogether the force balance thus takes the form  
\begin{equation}
\eta {v_{\infty} \over \lambda_D} \approx \Sigma\times E\approx  -{\epsilon \zeta\over \lambda_D}\times E
\end{equation} 
and this leads accordingly to the expression in \eqref{eq:vEO} for the electro-osmotic mobility. A simple extension of this argument highlights immediately the potential role of hydrodynamic slippage:  with a slip length $b$, the viscous friction force will reduce to $\sim \eta v_{\infty} /(\lambda_D+b)$ while keeping the body force identical, so that the electro-osmotic velocity will be increased by a factor $1+b/\lambda_D$. 
%Note that due to reciprocal relations, upon application of a pressure difference $\Delta P$, one will measure an electric current $I$ in this setting. Since the electrical double layer is not electro-neutral, the pressure difference will induce liquid flow that will drive an electric current in the electrical double layer~\cite{bocquet2010nanofluidics,siria2013giant}. For more details on electro-osmosis, we refer to Ref.~\cite{squires2016particles}.
Altogether, the electro-osmotic flow thus takes its origin within the very few nanometers close to the boundary and can be therefore strongly affected by molecular details: hydrodynamic slippage \cite{bocquet2010nanofluidics}, nanoscale roughness \cite{messinger2010suppression}, contamination \cite{pascall2010induced}, dielectric inhomogeneities \cite{bonthuis2012unraveling}, \textit{etc}. This makes the underlying physics of interfacial transport both complex and very rich.
%Also influence of the surface slip length and roughness influence the global electroosmotic response~\cite{messinger2010suppression} (roughness can however kill electro-osmosis). Actually the enhancement for electro-osmosis is scaling very differently (more like 1 + $\kappa/b$)~\cite{khair2009influence} and the dependence on the slip length of electro-osmosis has nothing obvious. \textcolor{blue}{By the way, did they take that into account when they looked at the streaming currents in carbon nanotubes ?!}

\vskip1cm
\subsubsection{ ... To diffusio-osmosis \SM{...}}
%%%%%%%%%%%%%%%%%%%%%%%%%%%%%%%%%%%
%\LB{DO is an alternative form of EO with solute gradients  as a driving force}

\begin{figure}[h!]
\centering
  \includegraphics[width=0.4\textwidth]{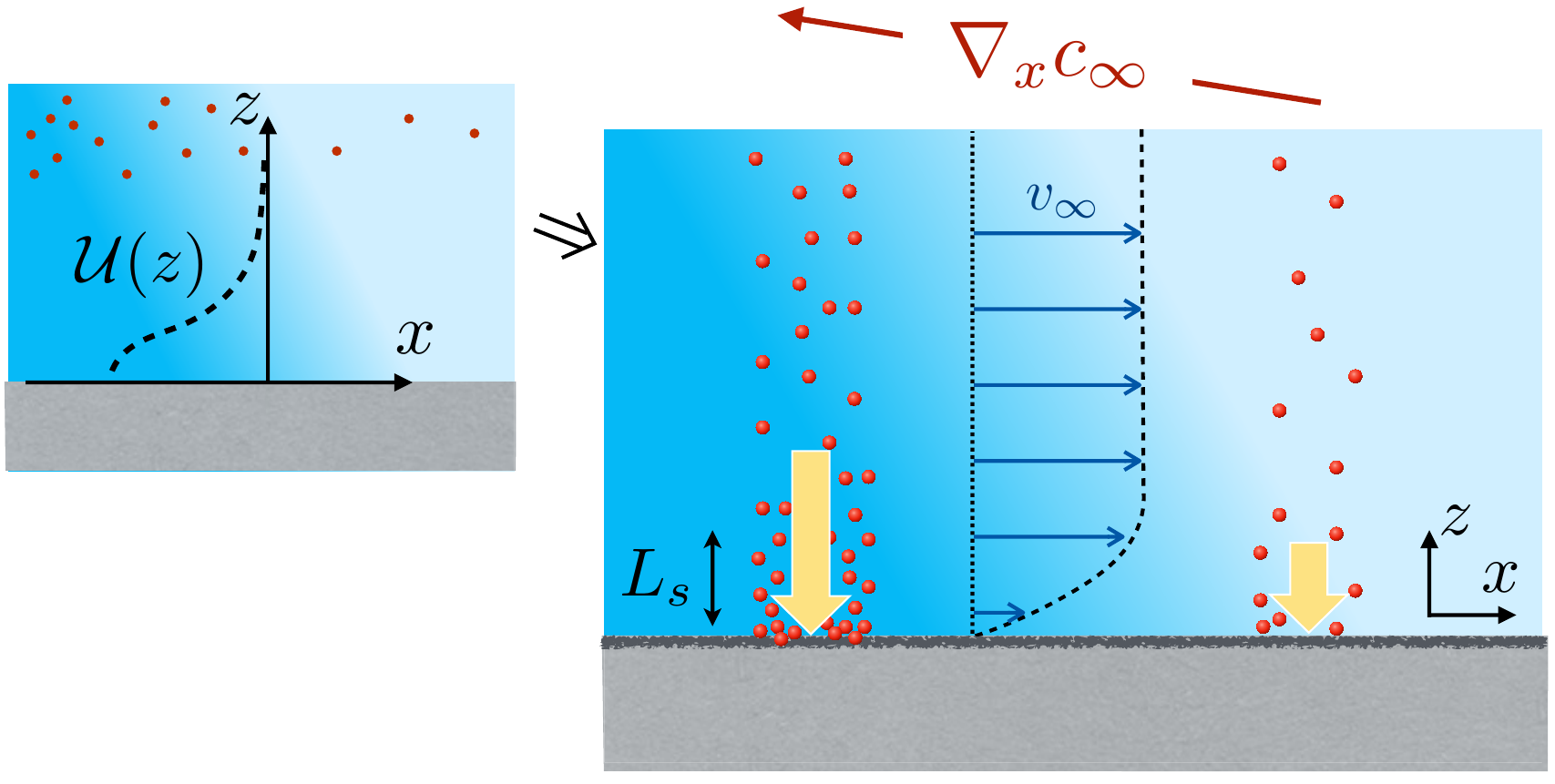}
  \caption{\textbf{: Diffusio-osmosis.} A gradient of solute imposed far from the surface induces a fluid flow. Here the solute is assumed to interact with the surface via an interaction potential $\mathcal{U}(z)$ (an adsorbing profile on the figure). The solute interaction with the surface induces a force on the fluid, here towards the surface, which is higher in the more concentrated area. This normal force converts into a parallel pressure drop  which generates a fluid flow from  high  to  low concentrations (and reversely for a repelling interaction). }
  \label{fig:DO}
\end{figure}

While  electro-osmosis corresponds to interfacially driven fluid motion under an external electric field, diffusio-osmotic motion occurs under the gradient of a solute, $\partial_x c_\infty$,  in the vicinity of a solid surface -- see Fig.~\ref{fig:DO}. %phenomenon in mind, it is 
%We first explore diffusio-osmotic transport corresponding to solvent flow under a salinity gradient, close to a solid surface, see Fig.~\ref{fig:DO}. 
%It is an interfacially driven phenomenon, which takes its origin within the diffuse interfacial layer close to the surface where the solute interacts specifically with the surface~\cite{derjaguin1972capillary,anderson1974mechanism,anderson1989colloid,ajdari2006giant}. The geometry is described in Fig.~\ref{fig:DO}, it can be seen as the inner surface of a porous membrane, or the surface of a colloid (see later). 
Similarly to electro-osmosis, a key ingredient is the specific interaction of the solute with the surface, which occurs within a diffuse layer of finite thickness. 
Reflecting the discussion of osmosis across a model potential barrier in Sec.~\ref{mech}, the solute will be assumed to interact via an external potential $\UU(z)$ with the solid surface. %, which corresponds to specific interactions of the solute with the surface (adsorption, repulsion..); 
One noticeable difference to the previous membrane case though is that this potential now acts {\it perpendicular} to the solid surface and solute gradient ({\it i.e.} depending on $z$ but not on $x$), see Fig.\ref{fig:DO}.

%We consider a flat surface with a solute gradient $\partial_x c_\infty$ along the membrane imposed by an external force; $\partial_x c_\infty$ is %uniform along $x$ and taken far from the surface. $x$ is the coordinate parallel to the surface, and $z$ the orthogonal one.  Similarly to the mechanical approach for osmosis across a membrane, we introduce an external potential $\UU(z)$ from the surface, which acts only on the solute. It corresponds to specific interactions of the surface with the solute (adsorption, repulsion..); one noticeable difference to the previous membrane case is that it now acts perpendicular to the solid surface and solute gradient ({\it i.e.} depending on $z$ but not on $x$). Typically, $\UU$ is strong near the surface within a thin layer and vanishes far from the surface.

{\it Diffusio-osmosis with neutral solutes -- }
We first consider the case of neutral solutes.  The fluid velocity and solute density obey the coupled Stokes and Smoluchowski equations, which write in the stationary state as:
\begin{eqnarray}
&0=& -\bm{\nabla} p + \eta \bm{\nabla}^2 \bm{v} + (-\bm{\nabla} \UU) ,\nonumber \\
&0=& -\bm{\nabla}\cdot\left[ -D_s \bm{\nabla} c_s + \lambda_s c_s \,(-\bm{\nabla} \UU) + \bm{v} c_s \right]
\label{eqDO}
\end{eqnarray}
At infinity, we assume a fixed gradient $\partial_x c_\infty$ along $x$ for the solute concentration. 

These coupled equations are strongly entangled. However
 in the limit of a thin interfacial layer --  corresponding to a range for the potential $\UU(z)$ which is small compared to the lateral variations of the solute gradient --,  one expects the concentration profile to relax quickly to a {\it local} equilibrium across the diffuse layer $c_s(x,z) \simeq c_{\infty}(x) \exp(-\UU(z)/k_B T)$.
%To understand diffusio-osmosis, one should write exactly the same equations as in the previous section, except in this different geometry (\textit{e.g.} equations.~\ref{Smolu} and~\ref{St}). We first focus on the solute distribution. Assuming a thin diffusive layer, ({\it i.e.}, the equilibrium along $z$ is fast, 
%so that the gradient along the surface (along $x$) is small compared with the gradient orthogonal to the surface (along $z$)) we may write in the dilute regime 
%$$c(x,z) = c_{\infty}(x) \exp(-\UU(z)/k_B T),$$ 
%as discussed above from Eq.~\ref{Smolu}. Depending on the interaction potential $\UU(z)$, a surface excess or a surface depletion of the solute will occur at the surface.

Turning now to the fluid transport equation, the Stokes equation projected along the $z$ direction writes simply
\begin{equation} 
0= - \partial_z p + c_s (-\partial_z{\cal U})
\end{equation}
because the $z$ component of fluid velocity is expected to be negligible for thin layers. 
We can integrate this pressure balance to obtain
\begin{equation}
p(x,z)-p_\infty= k_B T c_s(x,z) - k_B T c_{\infty}(x)
\label{pressure}
\end{equation}
which can be seen as an osmotic equilibrium across the diffuse layer~\cite{anderson1991diffusiophoresis}. In simple terms, the existence of a specific solute-wall interaction allows the membrane to \qote{express} the solute osmotic pressure $\Pi(x,z) = k_B T c_s(x,z)$ within the interfacial layer. However the effects of the latter disappear in the bulk ($z\rightarrow \infty$) and there is no bulk osmotic pressure gradient. 
%

%This result then allows to obtain the solvent velocity profile. Indeed, 
Now inserting the pressure from \eqref{pressure} into the Stokes equation projected along $x$, see \eqref{eqDO}, leads to  
\begin{equation}
\eta \partial_z^2 v_x - \partial_x [p(x,z)-p_\infty] = 0. 
\label{eq:StDO}
\end{equation}
Following the same steps as for the electro-osmosis, one obtains the fluid velocity along the $x$ coordinate in the bulk fluid as 
%One then one obtains the velocity field at infinity in terms of the osmotic pressure by integrating twice, and assuming a no slip boundary condition at the surface and a vanishing velocity gradient at infinity:
\begin{equation}
v_{\infty} = \mu_{DO} \times (-k_B T \nabla_x c_{\infty})
\label{vDO}
\end{equation}
%\LB{attention signe invers\'e par LB !!!}
 %$\Pi_{\infty} = k_B T \partial_x c_{\infty}$ and 
%
%\begin{equation}
% v_x(x,z)={k_B T \over \eta} \int_0^z\! dz^\prime\! \int_{z^{\prime}}^\infty\! dz^{\prime\prime} \partial_x c_\infty \left( 1 - \exp\left(-\frac{\UU(z^{\prime})}{RT} \right) \right).
%\end{equation}
%
%
with the diffusio-osmotic mobility $\mu_{DO}$ given by
\begin{equation}
\mu_{DO}=   \frac{1}{\eta}\int_0^\infty  \, z \,\left(\frac{c_s(x,z)}{c_{\infty}}  - 1\right) dz =  \frac{1}{\eta}\int_0^\infty  \, z \, \left(\exp\left(\frac{-\UU(z)}{k_B T}\right) - 1\right) dz.
\label{KDO}
\end{equation}
This expression is similar to \eqref{eq:zetaPotential} for the electro-osmotic mobility.
%\LB{idem signe}
%
%Note that $\eta$ can also be assumed to depend on the concentration $c$. In this case, ${1/\eta}$ in $\mu_{DO}$ has to be integrated along $z$ as well.
The effect of hydrodynamic slippage on the surface can also be taken into account, along the same lines as in \citefulls{ajdari2006giant,huang2008massive} and leads to an enhancement factor of the diffusio-osmotic mobility scaling as  $(1 + b/\lambda)$, where $b$ is the slip length and $\lambda$ is the typical width of the diffuse interface. The amplification effect is expected to be massive on superhydrophobic surfaces \cite{huang2008massive} and amplification by orders of magnitude are predicted. Interestingly for strongly hydrophobic surfaces where the liquid-vapor interface dominates, the diffusio-osmotic velocity takes the physically transparent expression 
$v_{DO}= {b_{\rm eff}\over \eta} \nabla \gamma_{LV}$, where $b_{\rm eff}$ is the effective slip length on the superhydrophobic surface
and $\gamma_{LV}$ is the (solute concentration dependent) surface tension of the liquid-vapor interface.

Similarly as in electro-osmosis, diffusio-osmosis can be interpreted in terms of a simple force balance within the diffuse layer.
A first integration of \eqref{eq:StDO} indeed shows that diffusio-osmotic flow results from the balance between the viscous stress  on the surface and an osmotic pressure gradient integrated over  the diffuse layer:
\begin{equation}
0= \eta \partial_x v_x\vert_{wall} + \int_0^\infty dz\,\partial_x\left[\Pi(x,z)-\Pi_\infty(x)\right]
\end{equation}
%
%Such a dependence may be understood on the basis of a simple scaling argument.
%%Here we give a scaling argument for the observed flow. 
%With a reasoning similar as the one leading to the pressure in diffusio-osmosis, Eq.~\ref{pressure}, one finds that the pressure gradient outside of the electrical double layer will typically scale as $ - k_B T \nabla_x c_{\infty}$ and has to balance the fluid friction as:
Simple estimates of the various terms lead to a more qualitative version of this force balance as
\begin{equation}
\eta\frac{ v_{\infty}}{\lambda} \sim \pm \lambda\times (-k_B T \nabla_x c_{\infty})
\label{lambda2}
\end{equation}
where $\lambda$ is defined here as the range of the potential $\cal U$, and the $\pm$ sign depends on whether the solute is attracted or depleted by the surface. 
This leads to $v_{\infty}\sim \pm{\lambda^2\over \eta}\times (-k_B T \nabla_x c_{\infty})$ in full agreement with \eqrefs{vDO}-(\ref{KDO}).
%It is interesting however that the osmotic pressure is only "expressed" within the diffuse layer, as highlighted by Eq.(\ref{eq:StDO}), but disappears in the bulk of the fluid. 
%Note that this force balance can be 
%For EO:
%\begin{equation}
%0= \eta \partial_x v_w\vert_{wall} + \int_0^\infty dz\,\rho_e\times E
%\end{equation}
%
%for DO:

\begin{figure}[h!]
\centering
  \includegraphics[width=0.49\textwidth]{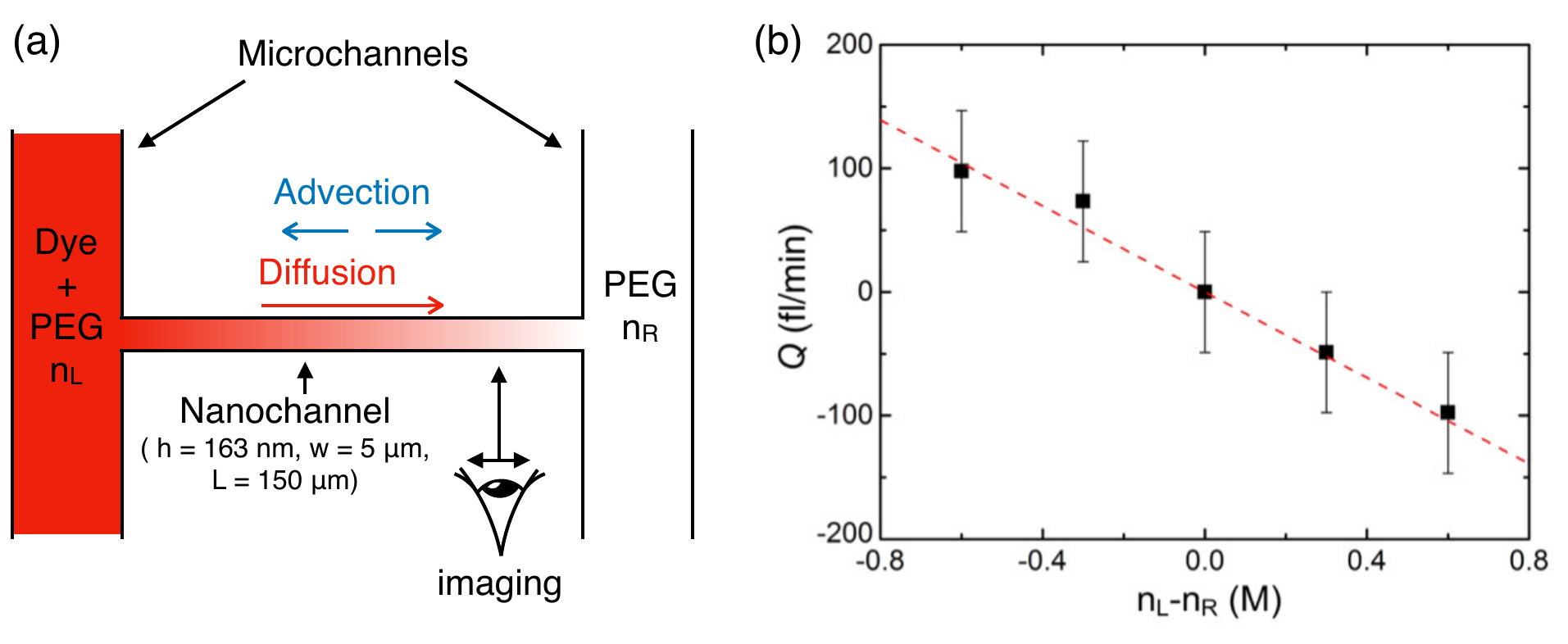}
  \caption{\textbf{: Experimental evidence for diffusio-osmosis.} (a) A gradient of polyethylene glycol polymer PEG is maintained along a nanochannel thanks to lateral microchannels acting as reservoirs. The nanochannel is 160nm in thickness and is
   %made of passivated silicon and glass on one side (potentially acting as the most interactive surface); and 
   fully permeable to PEG. Under a PEG concentration gradient, a diffusio-osmotic flow arises: water moves towards higher concentrations of PEG. The flow rate $Q$ is measured via the concentration profile of a dye. 
   %highlighted by the A dye is inserted in the left channel only. The dye concentration profile inside the nanochannel is determined by diffusion and advection by the diffusio-osmotic flow. From the dye intensity profile along the nanochannel it is possible to reconstitute the magnitude of the diffusio-osmotic flow.  
   (b) Measured diffusio-osmotic flux $Q$ as a function of the PEG concentration difference, showing a velocity proportional to the PEG concentration difference. This behavior and the sign of the effect are consistent with a steric exclusion of PEG on the surfaces, as predicted in Eq.(\ref{musteric}).  (a) and (b) are reproduced and adapted from \citefull{lee2014osmotic} with permission from the American Physical Society (APS), copyright 2014. }
  \label{fig:DO_exp}
\end{figure}

Diffusio-osmosis is definitely an osmotic flow, \textit{e.g.} a flow driven by an osmotic pressure gradient located within the diffuse layer.
However  the direction of the diffusio-osmotic flow can be along or against the gradient of the solute, in strong contrast to bare osmosis which induces a flow towards the highest solute concentration.
That is highlighted in the expression of the diffusio-osmotic mobility, \eqref{KDO}, which can be positive or negative depending on the
attractive or repulsive nature of the interaction potential $\UU(z)$.
%Looking at Eq.(\ref{KDO}), the sign of the diffusio-osmotic mobility can be either positive or negative. 
As a rule of thumb, the sign of the mobility will be dominantly determined by the adsorption $\Gamma = \int_0^\infty dz \left({c(x,z)/ c_\infty(x)} -1\right)$.
If there is a surface excess ($\Gamma>0$ or $\UU(z) < 0$), the solvent flow goes towards the low concentrated area ($\mu_{DO}>0$). 
That may appear as surprising because it amounts to concentrating even more the already concentrated solution; we shall discuss this
apparent paradox in Sec.~\ref{transportmatrix}.
Reversely a surface depletion resulting from a repulsion of the solute from the wall ($\Gamma<0$ or $\UU(z) > 0$) reverses the direction of the solvent flow 
towards the high concentrated zone ($\mu_{DO}<0$). 
An interesting limiting case for this behavior is exemplified by a solute interacting with the wall via steric effect, {\it i.e.} hard-core excluded volumes. 
For a solute particle with radius $R$, the mobility in \eqref{eq:StDO} reduces to 
\begin{equation}
\mu_{DO}^{\rm steric} = - {R^2\over 2\eta}.
\label{musteric}
\end{equation}
This behavior was measured in particular in \citefull{lee2014osmotic} for the diffusio-osmotic flow under a neutral polymer concentration gradient, see Fig.~\ref{fig:DO_exp} for an illustration.
%with $R$ the radius of the solute particle. 
A final remark is that
this simple rule for the correlation between adsorption and the sign of diffusio-osmosis is not exact and may fail for more complex interactions between the solute and the wall, for instance with an oscillatory spatial dependence of the concentration profile due to layering. 
%\cite{lee2017nanoscale}. % the correlation to adsorption is not obvious: ethanol-water mixtures + ethylene glycol
%\textcolor{blue}{There is however no issue with conservation of energy because ??}. 
%But in case of a complex concentration profile, for instance with an oscillatory spatial dependence on $z$ due to layering, 
The sign of $\mu_{DO}$ may  then be expected to differ from the sign of the adsorption $\Gamma$. In this case, no obvious conclusion can be made for the direction of the diffusio-osmotic velocity and a full calculation has to be made, see for example \citefull{lee2017nanoscale}.

\begin{figure}[h!]
\centering
  \includegraphics[width=0.49\textwidth]{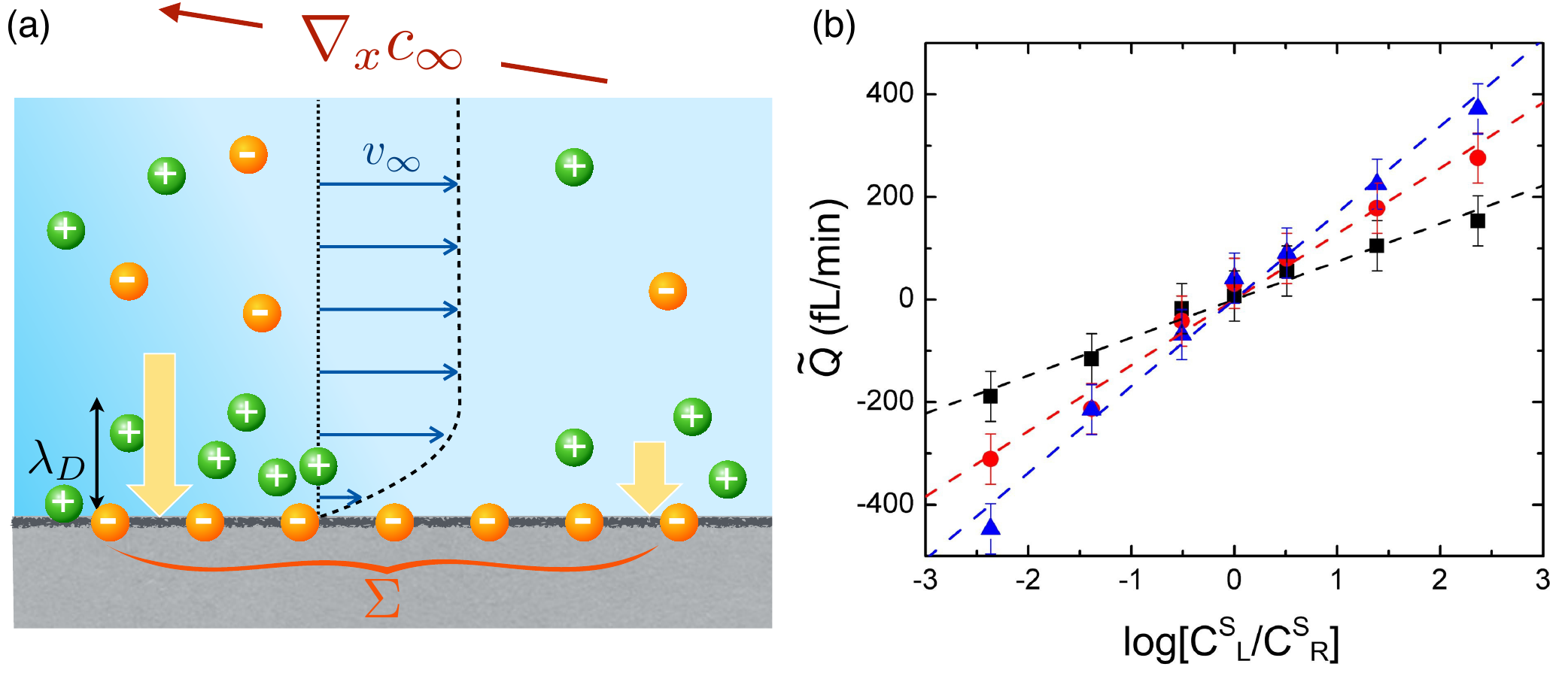}
  \caption{\textbf{: Diffusio-osmosis with charged electrolytes.} (a) Geometry; an electrolyte concentration difference is imposed far from the charged interface. The electrical interaction with the surface induces an (attractive) electrostatic force -- sketched with  arrows-- on the fluid which is larger where the salt concentration is larger, hereby inducing a net flow towards the low salinity region. (b) Measurement of the diffusio-osmotic flux as a function of the difference of the logarithm of the salt concentration between two reservoirs, in a similar way as in Fig.~\ref{fig:DO_exp}. Note the reversal of the sign as compared to Fig.~\ref{fig:DO_exp}-b. Reproduced from \citefull{lee2014osmotic} with permission from the APS, copyright 2014}
  \label{fig:DEO}
\end{figure}

{\it Diffusio-osmosis with electrolytes --} We now discuss specifically the case of diffusio-osmosis under salinity gradients. Here, as for electro-osmosis, the diffuse layer corresponds to the electric double layer created close to a charged surface, see Fig.~\ref{fig:DEO}.
%Now in numerous contexts, one requires a description for the motion of salt near interfaces. This amounts to assembling the diffusio-osmotic and the electro-osmotic geometry. We consider an electrolyte solution with a concentration gradient $\nabla_x c_{\infty}$ far from the charged interface. The charged interface will interact with the electrolyte solution, forming an electrical double layer whose properties will depend on the position along the surface, see Fig.~\ref{fig:DEO}. As a result, the force acting on the fluid will vary along the surface, inducing a net fluid flow~\cite{derjaguin1972capillary}. 
The derivation follows similar steps as above, from \eqrefs{eqDO} to (\ref{KDO}), except that one has to take into account the spatial distribution of both the counter- and co- ions in the EDL that follow a Poisson-Boltzmann distribution, see \citefull{prieve1984motion}.
%To compute the net fluid flow arising in this context, the reasoning follows closely that of the two previous sections, and we refer to Ref.~\cite{prieve1984motion} for the full proper derivation. 
{In this case the diffusio-osmotic velocity is shown to take the form
\begin{equation}
v_{\infty}=D_{DO} (-\nabla \log c_\infty)
\end{equation}
where we introduced a mobility $D_{DO}$ which has now the units of a diffusion coefficient. It takes the expression \cite{anderson1989colloid}
\begin{equation}
D_{DO}= { k_BT\over 2\pi \eta \ell_B} %{\epsilon \over \eta} \left({k_BT\over e}\right)^2 
\times \log \left(\cosh^2{\Phi_0\over 4} \right)
\label{DDO}
\end{equation}
where $\Phi_0=eV_0/k_BT$ is the dimensionless surface potential $V_0$ (usually identified with the zeta potential).
Note that for an electrolyte with unequal diffusion coefficients for the anions and cations ($D_+\ne D_-$), a diffusion electric field
builds up under the gradient of the salt concentration (if no current exists in the bulk). This takes the form $E_{\rm diff}={k_BT\over e} \delta \nabla \log c_\infty$ with $\delta=(D_+-D_-)/(D_++D_-)$ and  adds a supplementary electro-osmotic contribution to the diffusio-osmotic velocity as $v_{\rm diff}=-{\epsilon \zeta\over \eta} \times E_{\rm diff}$. Accordingly this leads to a supplementary contribution to the mobility as:
\begin{equation}
D_{DO}^{\rm diff}=-{\epsilon \zeta\over \eta} \times {k_BT\over e} \delta
\label{DOdiff}
\end{equation}
}

An important remark is that for electrolytes the velocity is proportional to the gradient of the {\it logarithm} of salt concentration, in contrast
to solutes where it is basically linear in the gradient, see \eqref{vDO}.
We will refer to this dependence as \qote{log-sensing} by analogy to behaviors occurring for the chemotaxis of biological entities ({\it e.g.} bacteria). 
%It leads to a wealth of counter-intuitive behaviors.
Such a dependence may be understood on the basis of the simple scaling argument based on the force balance above, see \eqref{lambda2}.
%Here we give a scaling argument for the observed flow. 
Indeed the thickness of the diffuse layer is now given by the Debye length, and
%With a reasoning similar as the one leading to the pressure in diffusio-osmosis, Eq.~\ref{pressure}, one finds that the pressure gradient outside of the electrical double layer will typically scale as $ - k_B T \nabla_x c_{\infty}$ and has to balance the fluid friction as:
%\begin{equation}
$v_{\infty}\simeq{\lambda_D^2\over \eta} \times  (- k_B T \nabla_x c_{\infty})$.
%\end{equation}
Since the Debye length depends on the salt concentration as $\lambda_D^2 = (8 \pi l_B c_{\infty})^{-1}$, %where $l_B$ is the Bjerrum length
%~\cite{bocquet2010nanofluidics}, 
one obtains :
\begin{equation}
v_{\infty} \approx - \frac{k_B T}{8 \pi \eta l_B} \nabla_x \log c_{\infty}.
\label{eq:vDEO}
\end{equation}
%Note that in Eq.~\ref{eq:vDEO} the surprising thing is that the velocity depends logarithmically on $c_{\infty}$ and not linearly. 
which is qualitatively similar to the exact results in \eqref{DDO} and predicts  log-sensing for  diffusio-osmosis with electrolytes.

This behavior is confirmed by experimental investigations of water flows under salinity gradients in nanofluidic circuits \cite{lee2014osmotic}, see Fig.~\ref{fig:DEO}. Diffusio-osmotic flow of water under salinity gradients was also evidenced across carbon nanotube membranes\cite{lokesh2018osmotic}, confirming further that diffusio-osmosis was acting against bare osmosis. In an alternative configuration, diffusio-osmosis was also shown to induce very large ionic currents under salinity gradients~\cite{siria2013giant,feng2016single,siria2017new}. We will come back to such cross effects associated with diffusio-osmosis in Sec.~\ref{cross}, as well as in the section dedicated to blue energy harvesting, Sec.~\ref{BlueEnergy}. In a very different field, diffusio-osmotic flows were also shown to strongly impact and shape the reactive fluid flows occurring in the solid Earth~\cite{plumper2017fluid}.
%
%A number of recent experiments
%This is well verified experimentally (see Fig.~\ref{fig:DEO}-b).
{Log-sensing has also many counter-intuitive consequences and a variety of applications~\cite{palacci2012osmotic,palacci2010colloidal}, which we will discuss more specifically in the context of diffusio-phoresis in Sec.~\ref{sec:DP}. 
%\LB{ Article geoscience (Nat. Geo.) sur DO dans les roches \cite{plumper2017fluid} ``evidence that a lot of the reactive fluid flow in the solid Earth occurs through nanoscale fluid pathways.''}

%Note that in the same geometry, we may expect net ionic currents driven by the flow within the charged electrical double layer.~\cite{siria2013giant} We come back to this effect in the section on blue energy harvesting. 
%The phenomenon by which -- reversely -- particles (with charged surfaces) move in a salt concentration gradient is sometimes called chemiphoresis~\cite{anderson1989colloid}. We will stick with diffusio-phoresis. 
}

%\LB{Discuss slippage ????} 

{\it Solvo-osmosis and diffusio-osmosis with mixtures --}
Up to now we considered merely dilute solute solutions, but all previous results can be generalized to mixtures of liquids with any molar fraction
of its constituents. The key ingredient remains that the two constituents interact differently with the solid substrate. As shown in \citefull{marbach2017osmotic}, the diffusio-osmotic velocity now takes the expression
\begin{equation}
v_\infty= \mu_{DO} (-\nabla_x \Pi[X_\infty(x)]),
\label{vDO2}
\end{equation}
where $\Pi$ is the generalized osmotic pressure defined in \eqref{PiGeneral}, calculated for  the molar fraction $X_\infty$, hence generalizing the expression in \eqref{vDO}.
The diffusio-osmotic mobility $\mu_{DO}$ is still given by the initial expression 
%
%
%\begin{equation}
$\mu_{DO}=  \frac{1}{\eta}\int_0^\infty dz^\prime\, z^\prime\, \left({c_s(x,z^\prime)\over c_\infty(x)} -1\right)$. 
%\label{KDO2}
%\end{equation}
However, for a solute-substrate interaction potential ${\cal U}$, 
%However in this equation, $c(x,z)$ denotes the local concentration of the specie interacing specifically with the substrate via a surface potential ${\cal U}$; 
the concentration profile $c_s(x,z)$ is now implicitly related to the value in the bulk $c_\infty(x)$ via the local equilibrium condition
%
%\begin{equation}
$\mu[c_s(x,z)]+{\cal U}(z) \simeq \mu[c_\infty(x)]$.
%\label{LocalEq}
%\end{equation}
%

Diffusio-osmosis with ethanol-water mixtures was investigated recently in \citefull{lee2017nanoscale}. But the majority of existing 
experimental investigations merely explored the reverse configuration of phoretic transport of particles under gradients of liquid composition, denoted as \qote{solvo-phoresis}~\cite{kosmulski1992solvophoresis,paustian2015direct}. Interestingly in \citefull{paustian2015direct}, the phoretic transport of colloidal (polystyrene) particles in ethanol-water mixtures resulted in a \qote{log-sensing} behavior of the particle diffusio-phoretic velocity, obeying $V=D_{SP}\nabla \log X$, with here $X$ the ethanol mole fraction.

\subsubsection{{... And electro-chemical equivalence.}}
\label{ECequ}
In the case of electrolytes, the two previous transport phenomena, electro- and diffusio- osmosis, 
are fundamentally intertwined. Indeed,
from the thermodynamic point of view, the chemical potential and the electric potential contributions merge 
into the electrochemical potential: $\mu_{\rm el} = \mu + q V$ (with $q$ the ion charge and  $V$ the electric potential).
There is accordingly a deep analogy when driving the system under gradients of chemical potential (diffusio-osmosis) or driving
under gradients of electric potential (electro-osmosis).
An illuminating discussion on this point and the corresponding force balance is provided by T. Squires in \citefull{squires2016particles,shi2016diffusiophoretic}
and we reproduce the essentials of the argumentation here.

Let us consider in full generality that a gradient of the electrochemical potential is applied in the bulk far from the boundary, $\nabla \mu_{\rm el,i}^B$
(along the direction of the solid surface, say $x$); the index $i$ runs over  the various ion species in the solution. 
As we discussed above for both electro- and diffusio- osmosis, this will generate 
net thermodynamic forces on individual ion specie $i$, which may be written as $f_i(x,z) = -\nabla \mu_{\rm el,i}(x,z)$. 
A key remark is that the electrochemical potential is approximately constant across the EDL, {\it i.e.} $\mu_{\rm el,i}(x,z)\simeq  \mu_{\rm el,i}^B(x)$, so that the individual force rewrites $f_i(z) \simeq -\nabla \mu_{\rm el,i}^B$. 
The interfacial motion results from the forces in excess to the bulk, so that the corresponding total force acting on the fluid rewrites
\begin{equation}
f_T=\sum_{i=1,n} \Delta c_i \times f_i(z)  =\sum_{i=1,n}  \Delta c_i \times (-\nabla \mu_{\rm el,i}^B)
\end{equation}
where the sum runs over $n$ ion species and $\Delta c_i=c_i(x,z)-c_i^B(x)$ is the excess ion concentration in the boundary layer, 
as compared to the bulk.
This driving force will generate a flow according to the Stokes equation $\eta \partial^2_{z} v_x + f_T=0$ and
following the same steps as above, one obtains the far field slip velocity as 
\begin{equation} v_\infty= -\sum_{i=1,n} {1\over \eta} \nabla \mu_{el,i}^B \int_0^\infty dz\,z\,\Delta c_i(z)\equiv \sum_i M_i (-\nabla \mu_{el,i}^B), \end{equation}
where the mobility $M_i$ takes the expression $M_i={1\over \eta} \int_0^\infty dz\, z  \Delta c_i(z)$.
For  symmetric and monovalent ions,  these mobilities can be exactly calculated using  Poisson-Boltzmann framework, leading to
\begin{equation}
M_\pm= {\epsilon \over e\eta} \left[\mp {\zeta\over 2} + {k_BT\over e} \times \log \left( \cosh^2 {\Phi_0\over 4}\right) \right]
\end{equation}
%$$D_{DO}= { k_BT\over 2\pi \eta \ell_B} {\epsilon \over \eta}  \times \log\left[\left(1-\tanh^2{\Phi_0\over 4}\right)^{-1} \right]$$
with $\Phi_0=eV_0/k_BT$ the dimensionless surface potential and here $\zeta \equiv V_0$ the zeta potential.

Under a constant electric field $\nabla \mu_\pm^B = \mp eE_0 $ and the electro-osmotic mobility is predicted as $\mu_{EO}={1\over e} (M_+-M_-)$, in full agreement with the previous result in \eqref{eq:vEO}-(\ref{eq:zetaPotential}). Under an imposed ionic strength gradient in the bulk, then
$\nabla \mu^B_\pm=k_BT \nabla \log c_\pm$ are identical and  $M_{DO}=M_++M_-\equiv D_{DO}$, again in full agreement with the previous result in \eqref{DDO}. 
%This general formalism reduces to the previous formulas in the previous sections.

%\cite{marshall1977osmotic} consequence on the reflection coefficient through a pore; 
%\LB{verif signes}

%\textcolor{blue}{Jusque la je trouve que le papier manque aussi d'une discussion pratique sur la façon de mesurer les choses à la fois en manip et en dynamique moléculaire}. 

%\textcolor{blue}{et quid du bilan entropique ? A calculer dans un cas simple pour montrer qu'il n'y a pas de probleme avec la diffusio-osmose}.

%The results for diffusio-osmosis in Eqs.~\eqref{vDO} and \eqref{KDO} are analogous to electro-osmosis and other 
%We can also see an analogy of Eq.~\eqref{vDO} with other 
%surface-driven flows, with the mobility defined in terms of the first spatial moment of a density profile
%(solute concentration profile for diffusio-osmosis and charge density profile in the case of electro-osmosis).
%The diffusio-osmotic mobility plays the role of the zeta potential
%in the electro-osmotic flow, which is driven by the external electric field, the {\it zeta potential}
%plays a similar role as the surface adsorption $\Gamma$.

\vskip0.5cm
\subsection{\SM{Transport matrix and symmetry considerations}}

\subsubsection{\SM{Transport matrix and cross fluxes}}
\label{cross}

%Let us consider in this section a more specific geometry with a channel in the form of a slit of width $w$ and height $h$ (with $w \gg h$ to simplify).
As introduced in Sec.\ref{fluxforces}, the framework of irreversible processes allows one to write a linear relation between thermodynamic forces and fluxes \cite{degroot2013non}. Adding the electric forces to the set of forces, one may generalize the results in \eqref{Larray} to obtain linear transport equations now
relating the solvent flux $Q$, excess solute flux $J_s-c_s Q$ and electric current $I_e$ to the pressure gradient $- \nabla p$, chemical potential gradient $- \nabla \mu$ and the applied electric field
$- \nabla V_e$, and summarized as
\begin{equation}
\label{Larray2}
\left(\begin{array}{c} Q \\ {J_s-c_sQ} \\ I_e \end{array}\right)= 
{\mathbb L}
\times \left(\begin{array}{c} - \nabla p \\ - \nabla \mu\\ - \nabla V_e\end{array}  \right),
\end{equation}

\begin{figure}[h!]
\centering
  \includegraphics[width=0.49\textwidth]{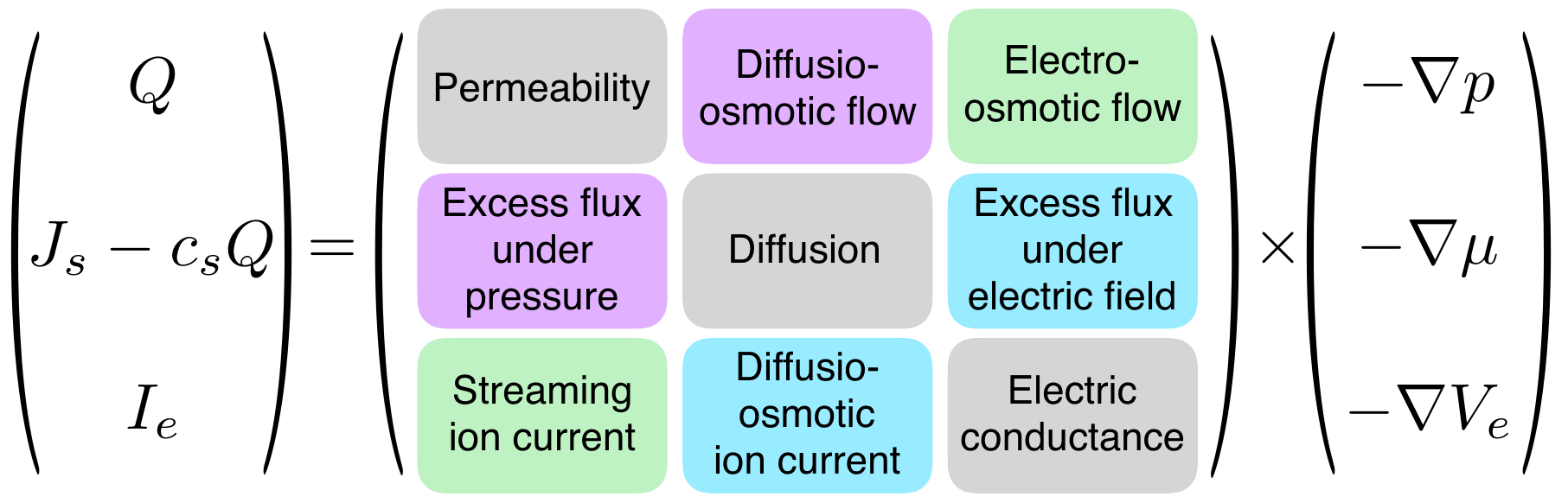}
  \caption{\SM{\textbf{: Transport Matrix.} Explicit transport matrix ${\mathbb L}$ as presented in \eqref{Larray2}, with colors indicating symmetric terms.}}
  \label{fig:TransportMat}
\end{figure}

Due to Onsager principle, this matrix is symetric and positive definite \cite{degroot2013non}.
Each term of this matrix corresponds to a specific transport phenomenon. Diagonal terms are associated respectively with permeability (characterizing solvent flux under a pressure drop), diffusion (characterizing solute flux under an applied solute gradient) and electrical conductance (characterizing ionic current under an applied electric field).
The off-diagonal terms correspond to cross effects. We detail below the cross effects that are all recapitulated in Fig.~\ref{fig:TransportMat}.

In the first row of the matrix, electro-osmosis and diffusio-osmosis -- explored so far -- correspond to the terms relating the solvent flux $Q$ to a chemical gradient $-\nabla \mu$ and an electric field $-\nabla V_e$ respectively. 
A key consequence of the symmetry of the matrix is that the same mobilities characterize symmetric transport phenomena. For example consider the first column of the matrix ${\mathbb L}$, one finds that the electro-osmotic mobility and diffusio-osmotic mobility also describe respectively the electric current and excess solute flux generated under a pressure drop, as 
\begin{eqnarray}
&I_e = {\cal A}\, \mu_{EO}\times (-\nabla p)\nonumber \\
&J_s-c_s Q = {\cal A}\, \mu_{DO}\times (-\nabla p)
\end{eqnarray}
 where ${\cal A}$ is the channel cross section. 
 %similarly the diffusio-osmotic mobility relates the excess solute flux under pressure drop $
The first corresponds to the so-called streaming current and takes its origin in the motion of mobile ions in the EDL which are carried by the pressure-driven flow; the pressure-driven excess solute flux has a similar physical origin.

Streaming currents are commonly measured in experiments \cite{schoch2008transport,bocquet2010nanofluidics}, even down to single carbon and boron-nitride nanotubes \cite{siria2013giant}. To our knowledge, no experimental measurement of pressure-driven excess solute flux has been performed up to now. However this is not the case in molecular dynamics simulations where it is far easier to measure the diffusio-osmotic mobility via the pressure-driven excess flux \cite{ajdari2006giant,lee2017nanoscale}-- see details in Sec.~\ref{Sec:numerics}.

Now, the transport matrix suggests that an electric current can be generated under an osmotic gradient, \SM{which we term here the diffusio-osmotic ion current,} following
\begin{equation}
I_{DO} = K_{\rm osm} \times (-\nabla \log c_{\infty}).
\label{IDO}
\end{equation}
Let us consider a channel in the form of a slit of width $w$ and height $h$ (with $w \gg h$ to simplify).
Using Poisson-Boltzmann to describe the EDL, one can calculate the corresponding osmotic electric current 
\cite{fair1971reverse,siria2013giant,Mouterde2019}
and the mobility takes the 
form 
%$K_{osm} = K^{(1)}_{\rm osm}+ K^{(2)}_{\rm osm}$ with:
\begin{equation}
K_{\rm osm}=  \alpha \, \times (-\Sigma)  {k_BT\over 2\pi\eta\,\ell_B}\left(1 - { \sinh^{-1} \chi \over \chi}\right) 
\label{K1}
\end{equation}
where $\alpha\simeq 2w$ is the perimeter of the channel cross section.
%and
%\begin{equation}
%K^{(2)}_{\rm osm}= {2w }\times (-\Sigma) { k_B T \over 2\pi \eta\, \ell_B } 
%  {b \over \lambda_D } \,  \left[ \sqrt{1+\chi^2}-1\right] 
  %&K_{surf}= & - {2w\over L}  \Sigma   \times
% {4 b_{\rm eff} \alpha_s \lambda_D  \over \eta} \,  \left[ \sqrt{1+\left(2\pi \ell_B\lambda_D \Sigma_n\right)^2}-1\right] \nonumber \\
%&& \times 2 k_BT   C_\infty
%\end{equation}
%\LB{slit, define $w$}
In this expression we introduced $\chi=\sinh {\vert \Phi_0 \vert\over 2}$ with $\Phi_0=eV_0/k_BT$  the dimensionless surface potential $V_0$. In the Poisson-Boltzmann framework $\chi$ is related to the surface charge $\Sigma$ according to $\chi=2\pi \ell_B \lambda_D \vert\Sigma\vert/e$ with $\lambda_D$ the Debye length, so that $\chi \propto \vert\Sigma\vert$. This formula can be extended to take slippage on the surface into account, as well as mobile surface charges~\cite{Mouterde2019}. More precisely the diffusio-osmotic ion current takes its origin in the motion of ions in the EDL which are carried by the diffusio-osmotic flow. 
As a simple estimate we may write that $I_{DO} \approx (-\Sigma) \times v_{DO}$, where 
$v_{DO}$ is the diffusio-osmotic velocity: using the expression \eqref{vDO} for $v_{DO}$, one indeed recovers \eqref{K1}.
However the prediction of \eqref{K1} reports a more complex dependence, since the linear dependence in $\Sigma$ is only valid for large enough $\Sigma$, while  for low $\Sigma$, one finds that
$K_{\rm osm} \propto \Sigma^3$, {\it i.e.} vanishingly small. 
%The corresponding
%%Discuss consequences. Interesting that $K_{\rm osm} \propto \Sigma$, can be understood qualitatively
%%as $ I_{DO} \approx \Sigma \times v_{DO}$, ions carried by the flow. 
%But not trivial: for example in low charge regime, $K_{\rm osm} \propto \Sigma^3$: very small ! Only systems with large charge may
%yield osmotic currents. Like BN, and other oxides.
Such osmotically driven currents have been measured experimentally in various systems, nanochannels, single nanotubes, single nanopores -- see  \citefulls{kim2010power,siria2013giant,feng2016single,zhang2017ultrathin} -- to cite a few. This effect finds important applications in the context of blue energy harvesting\cite{siria2017new}, that we will explore in detail in Sec.~\ref{BlueEnergy}.

\subsubsection{Entropy production with diffusio-osmosis}
\label{transportmatrix}

%\subsubsection{Symmetry considerations}
%Important to consider transport as a whole: an example here.
%
We pointed out above that the sign of the diffusio-osmostic mobility, $\mu_{DO}$, can be either positive or negative, so that the corresponding flux can be along or against the concentration gradient. A negative $\mu_{DO}$ may appear at first sight striking since the direction of the solvent flow corresponds to that of an increase in salt concentration, thus leading to an apparent violation of the second principle. This is however not the case, as it can be verified from a calculation taking into account all relevant fluxes. To highlight this situation, let us consider a membrane separating two reservoirs with fixed volumes; the concentration on the left/right reservoir is $c_s(t)=c_0 \mp {\Delta c_s(t)\over 2}$. The pore size is assumed here to be larger than the solute diameter so that the membrane is permeable  to the solute and there is no bare osmotic pressure. A salinity gradient however generates a diffusio-osmotic flow on the pore surface. Based on the transport matrix formulation, \eqref{Larray}, one may write the solvent and (excess) solute fluxes as a function of the solute concentration and pressure gradients according to:
\begin{eqnarray}
&Q&={{\cal A}_p\over L} \left[ \kappa (-\Delta p)+ \mu_{DO}  (-k_B T \Delta  c_s)\right] \nonumber \\
&J_s-c_0 Q&= {{\cal A}_p\over L} \left[\mu_{DO}\, c_0 (-\Delta p) + \lambda_{s} (- k_B T\Delta c_s)\right]
\label{Larray3}
\end{eqnarray}
with ${\cal A}_p$ the total (open) pore area of the membrane, $L$ its thickness and $\lambda_s=D_s/k_BT$ the diffusive mobility of the solute across the membrane, defined in terms of the solute diffusion coefficient;  $\kappa$ is defined in terms of the permeance as $\mathcal{L}_{\rm hyd}\equiv \kappa {{\cal A}_p/ L}$ (note that $\kappa=\kappa_{\rm hyd}/\eta$ where $\kappa_{\rm hyd}$ is  the permeability introduced above). The second principle 
imposes that the transport matrix in \eqrefs{Larray} \& (\ref{Larray3}) should be definite positive. Accordingly, \SM{the determinant $\det (\mathbb{L}) \propto \kappa\lambda_s - \mu_{DO}^2 c_0$ must be strictly positive.}

On the other hand, since the volume is fixed, the flux vanishes, $Q=0$, and the solute flux writes
\begin{equation}
J_s= {{\cal A}_p\over L} {1\over \kappa} \left[\lambda_{s}\, \kappa-{\mu_{DO}^2\, c_0 }    \right]  (- k_B T\Delta c_s)
\end{equation}
\SM{We find that the term in brackets is proportional to the determinant $\det (\mathbb{L})$, and therefore is constrained by the second principle to be positive.} Accordingly, whatever the sign
of the diffusio-osmotic mobility $\mu_{DO}$ and the corresponding diffusio-osmotic solvent flux, the total solute flux will go down the solute gradient, as expected from the second principle. 

\subsection{The peculiarity of diffusio-osmosis across an orifice}

In the previous sections, we implicitly considered (diffusio-osmotic) transport across long channels, so that fluid flow is translationally invariant along the channel's length. However transport across thin membrane pores \cite{wanunu2010electrostatic,siwy2006ion,bell2011dna} raises the question of the specificity of these geometries in which the channel length $L$ may decrease down to molecular lengths, in particular with the advent of 2D materials such as graphene, h-BN and MoS$_2$ as membranes for fluidic transport~\cite{feng2016single,walker2017extrinsic,wang2017fundamental}. For example, recent measurements across nanopores in MoS$_2$ membranes reported huge diffusio-osmotic ion currents under salinity gradients~\cite{feng2016single}. 
In another experiment, gradients of salts were shown to strongly increase the capture rate of DNA molecules across solid-state nanopores~\cite{wanunu2010electrostatic}.

\begin{figure}[h!]
\centering
  \includegraphics[width=0.45\textwidth]{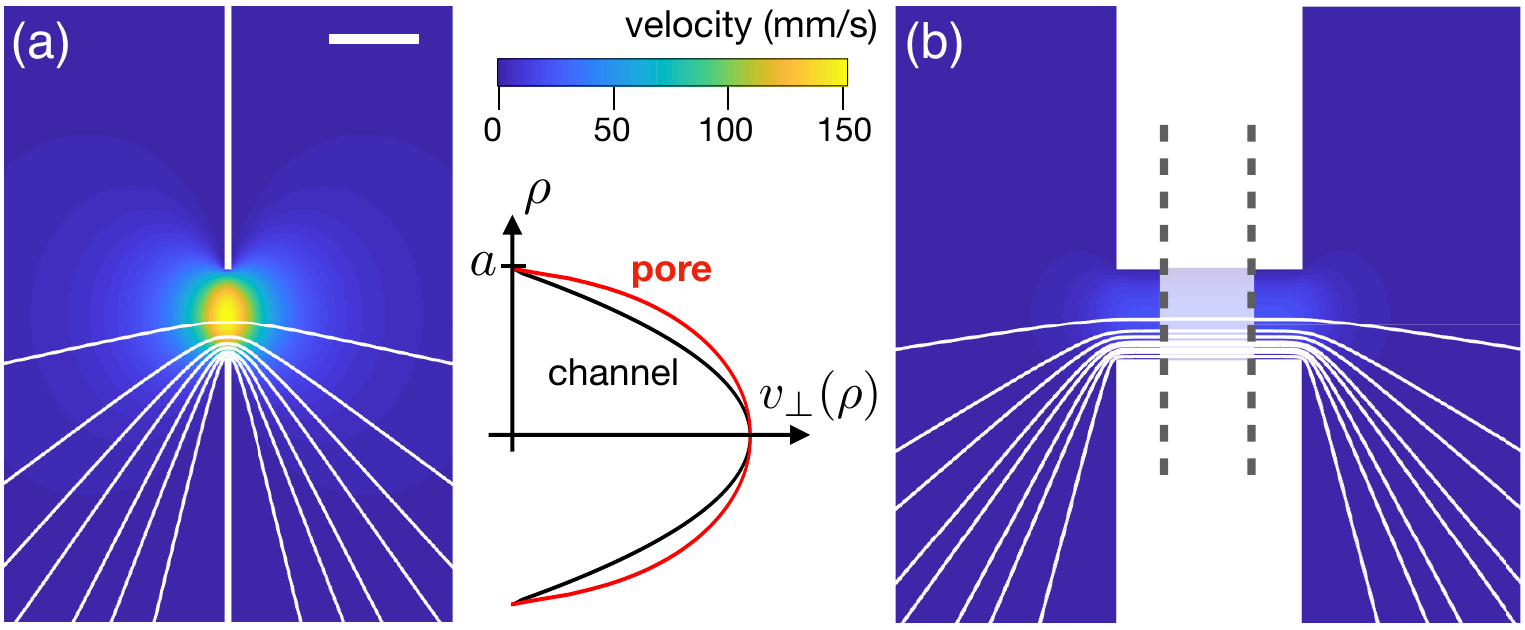}
  \caption{\SM{\textbf{: Peculiarity of pressure flow across an orifice.} (a) Simulated flow velocity and streamlines across a pore with, say, radius $a=10\mathrm{nm}$ and thickness $L = a/10$ under a pressure drop $\Delta p = 1 \mathrm{bar}$. The scale bar is $10\mathrm{nm}$. The streamlines are spaced equally in magnitude at the center of the pore. (b) Simulated flow velocity and streamlines across a channel with same radius $a$ and thickness $L = 10 a$ under the same pressure drop. The scales (velocity and geometry)  are the same as for (a). For readability the whole channel length is not plotted. (Center) Normalized velocity profile (perpendicular to the membrane) at the center of the membrane, comparing the channel and pore cases. }}
  \label{fig:Peculiarity}
\end{figure}

For long channels the driving force for fluid transport, {\it e.g.} the gradient of the chemical potential, is expected to scale as its inverse length, $\nabla \mu = {\Delta \mu/L}$. This would suggest that the driving force diverges as $1/L$ in the limit of nanopores where $L\rightarrow 0$. However entrance effects level off this diverging behavior to a value typically fixed by the lateral size of the pore, say $a$ its radius -- see Fig.~\ref{fig:Peculiarity}. As a rule of thumb, one may expect that $\nabla \mu \approx {\Delta \mu/a}$ (see for example \citefull{hall1975access} for the conductance of ion channels). However the flow in and out of the pore is expected to be strongly disturbed, as shown for example for electro-osmosis across nanopores in thin membranes~ \cite{mao2014electro,sherwood2014electroosmosis,melnikov2017electro}. Similar effects are accordingly expected to apply to diffusio-osmotic transport.

%Above translational invariance, interesting to take the opposite geometry: a nanopore drilled in a infinitely thin membrane. While this may first appear as ideal, this is the case of graphene, and MoS2, explored recently by Radenovic, \cite{feng2016single}.
%Naively take the gradient over a vanishing length $L$ of the channel, so that $L\rightarrow 0$ and the driving force $\nabla c_\infty \sim \Delta c/L$ would become infinite.
%Not possible, and usually one may expect the pore size, say $a$ to act as a limiting bound so that $\nabla c_\infty \sim \Delta c/a$

The diffusio-osmotic flow across a nanopore with vanishing thickness was recently calculated analytically by Rankin {\it et al.}~\cite{Rankin2019diffusio}. The calculation is best performed in oblate-spheroidal coordinates, in line with a similar calculation for electro-osmosis in  \citefull{mao2014electro}. \LBNEW{The averaged diffusio-osmotic velocity $v_{DO}$ across the pore, which is defined in terms of the diffusio-osmotic flux $Q=\pi a^2 v_{DO}$, % where $v_{DO}$ is the averaged velocity across the pore. Accordingly 
is proportional to $\Delta c_s$  the {\it difference} (and not the gradient as in \eqref{vDO}) of the solute concentration between the two sides of the membrane:
$v_{DO} = \mu_{DO}^{\rm pore} (-k_B T \Delta c_s)$. 
%similarly to \eqref{vDO}, one finds $v_{DO} = \mu_{DO}^{\rm pore} (-k_B T \Delta c_s)$, with $\Delta c_s$  the difference of solute concentration between the two sides of the membrane.
The general expression for the mobility derived in \citefull{Rankin2019diffusio} takes the form
\begin{equation}
\mu^{\rm pore}_{DO} = -\frac{2 a}{\pi^2\eta} \int_0^1 d\xi\,\xi^2\int_0^\infty d\nu\,\frac{e^{-{\cal U}/k_B T}-1}{1+\nu^2} 
\label{eq:Qzetanu}
\end{equation}
where ($\nu$,$\xi$) are the oblate spheroidal coordinates (iso-$\nu$ and iso-$\xi$ curves are respectively oblate spheroids and hyperboloids of revolution). This  expression involves a complex spatial average of the Boltzmann weight $e^{-{\cal U}/k_B T}-1$, which should be compared to the corresponding simple expression in \eqref{KDO} for the planar case.}

\LBNEW{The above result can be simplified for certain functional forms of the potential ${\cal U}$. For example, assuming that the interaction potential ${\cal U}$ depends only on variable $\xi$ allows the mobility to be rewritten in terms of the two-dimensional interaction within the pore only as 
$\mu^{\rm pore}_{DO}~=~{1\over\pi a^2 \eta} \int_0^a d\rho \rho \sqrt{a^2-\rho^2} \big[ e^{-{{\cal U} (\rho) / k_BT}} -1  \big] $ with $\rho$ the axi-symmetric distance to the center of the pore~\cite{Rankin2019diffusio}. This expression for the mobility can be recovered thanks to the symmetry of the transport matrix \eqref{Larray2}. Indeed $\mu^{\rm pore}_{DO}$ can also be calculated in terms of the excess solute flux under a pressure driven flow: in this case the velocity profile was shown to be semicircular (and not parabolic) \cite{sampson1891stokes,dagan1982infinite,happel2012low}, see Fig.~\ref{fig:Peculiarity}, and the excess solute flux conveyed by the circular flow reduces to the above expression. }

\LBNEW{
The complexity associated with diffusio-osmosis across an orifice is also highlighted by the predicted dependence of the mobility on the pore size $a$ and the range $\lambda$ of the interaction $\cal U$. Let us focus the discussion for the thin diffuse layer case, where  $\lambda \ll a$ (we refer to  \citefull{Rankin2019diffusio} for a full discusson). As a reference, the diffusio-osmotic mobility across a long channel with length $L$  was shown previoulsy to scale as $\mu_{DO} \sim \frac{\lambda^2}{\eta L}$, see {\it e.g.} Eq.(\ref{lambda2}).
%showing a diverging behavior as $L$ approaches a thin limit. 
However, for an orifice in a thin membrane, Rankin {\it et al.} showed on the basis of Eq.(\ref{eq:Qzetanu}) that 
%
%a variety of scalings emerges in this case, depending on the geometric details of the interaction potential $\cal U$. %, the mobility exhibits various scaling regimes as a function of $\lambda$ and pore size $a$.
%For thick interacting layers $\lambda \gg a$ the diffusio-osmotic mobility scales as $\mu^{\rm pore}_{DO}\sim {a\over \eta}$, up to numerical factors depending on the potential strength.
%If the potential $\cal U$ is long-range, {\it i.e.} longer than the pore size, then the diffusio-osmotic mobility derives immediatly from the Sampson results and $Q_{DO} = {k_BT a^3\over 3 \eta} (\exp[-{\cal U}_0/k_BT]-1) \,  (-\Delta c_s)$. 
% This result amounts indeed to replacing in a phenomenological way the pore length $L$ by its radius $a$, as one usually guesses for entrance effects \sam{Actually not really, also $\lambda$ would have to be replaced, this doesn't work}.
%Now for thin diffuse layers, $\lambda \ll a$ 
the mobility exhibits a variety of non-trivial scalings, with $\mu^{\rm pore}_{DO}\sim \lambda^{\gamma}a^{1-\gamma}$,  and
an exponent $\gamma$ that depends on the details of the interaction potential $\cal U$. For example, for the potential discussed above, which assumes a dependence as ${\cal U}(\xi)$, one finds $\gamma=3/2$; but for a potential depending on the distance to the membrane or to the edge of the pore, then $\gamma=2$~\cite{Rankin2019diffusio}. In the latter case, $\gamma=2$, the diffusio-osmotic mobility scales as $\mu^{\rm pore}_{DO}\sim \lambda^{2}/a$, which corresponds to the long channel result with the length $L$ replaced by $a$. But for other values of the exponent $\gamma$ this simple rule of thumb does not apply, making the diffusio-osmotic transport across the orifice quite peculiar.}

As a last comment, it is possible to extend qualitatively these results to electrolyte solutions, by assuming that the potential range $\lambda$ identifies with the Debye length. \LBNEW{This suggests an anomalous salinity dependence for the diffusio-osmotic mobility $D_{DO} =v_{DO}/[-k_BT \Delta \log c_s] \propto c_s^{1-\gamma/2}$, in contrast to long channels where $D_{DO} \propto c_s^0$. The nanopore geometry may thus depart from the log-sensing behavior of diffusio-osmotic transport under salinity gradients.}
%For a potential range $\lambda$ which is small as compared to the pore radius, the mobility is found to scale as
%This pecularity originates in the square-root dependence of the velocity profile close to the edge, indeed an edge effect.
%As a rule  of thumb, if $\lambda$ is the Debye length, then $\lambda\sim c_s^{-1/2}$, and $D_{DO}\sim c_s^{1/4}$, while usually for long channels $ D_{DO}\sim c_s^0$ is independent of salt. Strange scaling and not log-sensing !
These results remain however to be fully assessed experimentally.

\subsection{Alternative interfacial transport: thermo-osmosis}
%\LB{why do we discuss TO: important to have an integrated views of all interfacial transport and combination interesting ! cf Sano}

\begin{figure}[h!]
\centering
  \includegraphics[width=0.49\textwidth]{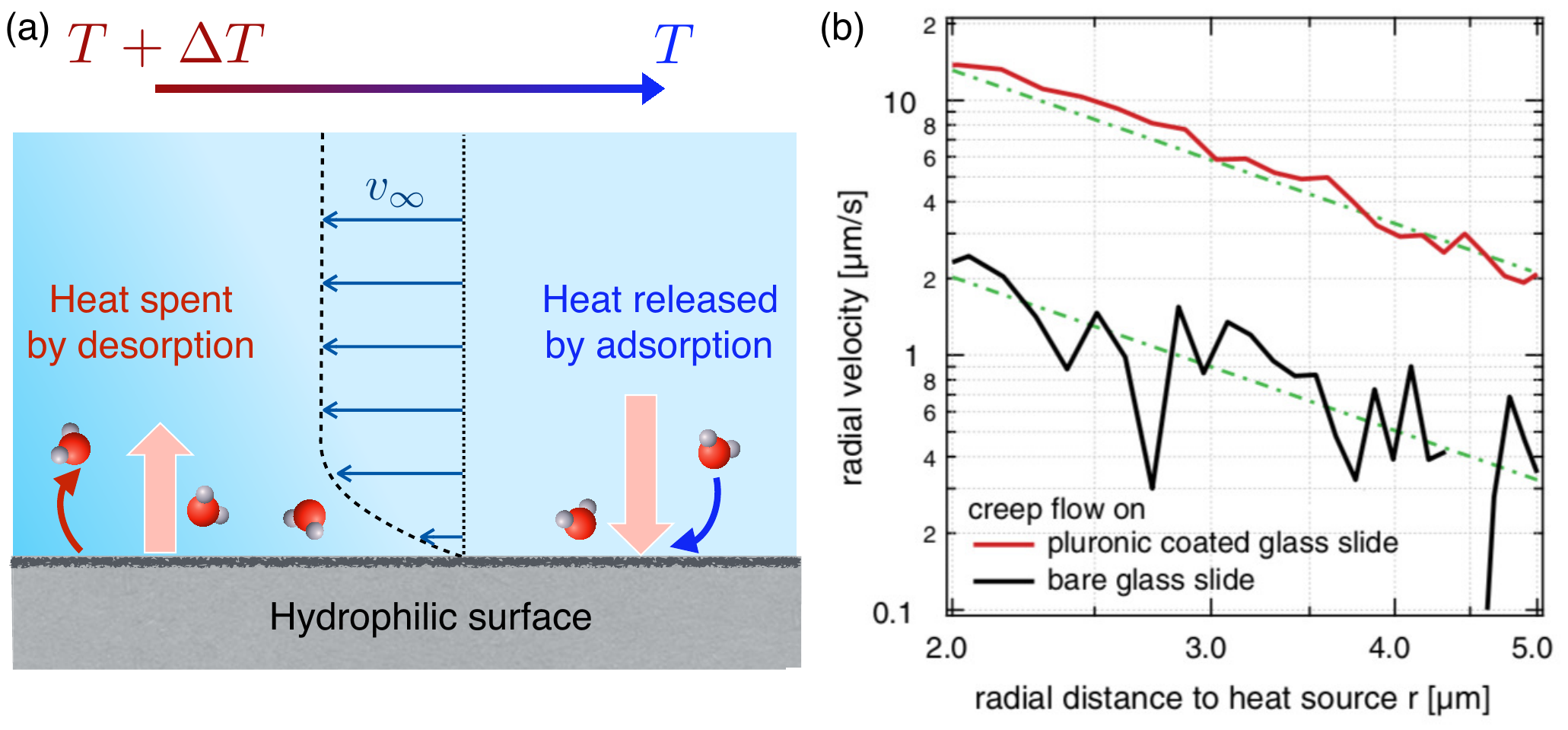}
  \caption{\textbf{: Thermo-osmosis near an interface} (a) Geometry; a temperature gradient is imposed far from a hydrophilic surface. A thermal flux from the hot to the cold region is therefore installed. The interaction of water with the surface induces a force (light red arrows) that varies along the surface due to the thermal gradient, inducing a net flow. (b) Thermo-osmotic flow measured on two different surfaces (going towards the higher temperatures) as a function of the distance to the heat source. Reproduced from \citefull{bregulla2016thermo} with permission from the APS, copyright 2016.}
  \label{fig:TO}
\end{figure}

Extending on electro- and diffusio- osmosis, thermo-osmosis corresponds to fluid motion under gradients of temperature; see Fig.~\ref{fig:TO}-a.
Such effects were reported %It has been demonstrated that osmotic flows may arise under thermal gradients 
as early as in the 1900s~\cite{lippmann1907endosmose,aubert1912thermo}. Thermo-osmosis was first rationalized in terms of thermodynamic forces by Derjaguin et al.~\cite{derjaguin1987some,anderson1989colloid,derjaguin1987thermo}. Similarly as for diffusio-osmosis in \eqref{KDO} and electro-osmosis in \eqref{eq:zetaPotential}, the net velocity generated far from the surface is predicted as \cite{anderson1989colloid}
\begin{equation}
v_{\infty} = \left( \frac{-2}{\eta} \int_0^{\infty} z \delta h(z) dz \right) \nabla \log T_{\infty}
\end{equation}
where $T_{\infty}$ is the temperature far from the surface and $\delta h(z)$ is the excess specific enthalpy in the interfacial layer as compared to the bulk liquid. If the solid surface is \textit{e.g.} hydrophilic, then $\delta h(z) \lesssim 0$ and the flow of water is directed toward higher temperatures, see Fig.~\ref{fig:TO} and~\citefulls{bregulla2016thermo,barragan2017thermo}. 
%Though we do not give full details here, similarly the thermal gradient may also be interpreted as a force acting on the fluid and therefore thermo-osmosis is trully an osmotic process. 
An interpretation of thermo-osmosis (and -phoresis) in terms of interfacial surface tension modification, and therefore Marangoni-like flow generation, has also been suggested~\cite{ruckenstein1981can} and formalized~\cite{piazza2004thermal,parola2004particle}. 
The transport of fluids or particles under thermal forces led to strong debates between the  interfacial approach discussed above and  an \qote{energetic} approach~\cite{duhr2006molecules,piazza2008thermophoresis,piazza2008thermophoresisJP}, which 
%The energetic approaches 
attempts to write the net driving force acting on a particle as the gradient of a thermodynamic quantity \cite{duhr2006molecules}. The resulting Soret coefficient -- defined as the ratio between the thermophoretic mobility and particle diffusion coefficient -- highlights a different dependence on the particle size as compared to the interfacial framework discussed above. Although attractive, the energetic approach was then extensively criticized  
%with sometimes confusing results from the experiments, in particular concerning the dependence of the Soret coefficient on the colloid radius 
~\cite{piazza2008thermophoresis,piazza2008thermophoresisJP}. 
%existing models of thermophoresis
%follow either an ??energetic?? or a ??hydrodynamic??
%route. The starting point of the
%former group of theories is trying and
%write the net force f as the gradient of
%a suitable thermodynamic quantity,
%The pertinence of the various theoretical framework to rationalize experiments was extensively debated in the literature
 %the hydrodynamic approach or the thermodynamic approach is properly justified has been extensively debated
% ~\cite{duhr2006molecules,piazza2008thermophoresis,piazza2008thermophoresisJP}.

%&encoding=UTF-8 Unicodemove large particles such as DNA~\cite{duhr2006molecules}  moller2017steep

%allowing in particular to show that the different approaches to thermo-osmotic velocity measurements would yield fairly similar results. 
As for electro- and diffusio- osmosis, the details of the interfacial dynamics, for example slippage at the interface, is expected to strongly affect thermo-osmotic flows. This has been evidenced for example in molecular dynamics  where a huge enhancement of thermo-osmosis was measured with slip~\cite{morthomas2008thermophoresis,fu2017controls}, although the exact dependence of thermo-osmosis on the  interfacial properties was measured to be  substantially complex.~\cite{fu2017controls}
We refer to the \citefulls{piazza2008thermophoresis,barragan2017thermo} for more in-depth discussion on thermo-osmosis and -phoresis. 

Lately thermo-osmosis has gained growing attention in terms of applications and we briefly comment here on this aspect.
%As we will not come back to the applications of thermo-osmosis later we cover most applications here. 
Many applications of the phenomenon are done in the context of thermo-phoresis, or displacement of colloidal particles under thermal gradients (in a similar way to diffusio-phoresis, see Fig.~\ref{fig:OsmoToPhoresis}-a). This phenomenon was harvested to manipulate colloids and build structures~\cite{golestanian2007designing,piazza2008thermophoresis,di2009colloidal,wurger2010thermal} with  advanced applications in microfluidics~\cite{bregulla2016thermo} or towards DNA detection~\cite{yu2015concentration}. 
 Among other phenomena, it was shown that couplings between thermo- and diffusio- phoretic drivings  allow to finely manipulate colloidal structures \cite{jiang2009manipulation}.  Also, thermophoresis of molecules can provide detailed information about particles and molecules (size, charge and hydration shell) and this provide very efficient analytical tools to probe protein in biological liquids \cite{wienken2010protein,dau2016quantitative}. 
%and led to  was developed for example by the company Nanotemper.
In a different context, applications of thermo-osmosis were suggested for the recovery of water from organic waste-water~\cite{al2016thermal},
as well as for 
%In the context of energy harvesting, 
%applications of thermo-osmosis was suggested with the objective of 
energy harvesting from thermal differences (and waste heat).~\cite{kuipers2015simultaneous,straub2016harvesting,fu2017controls}

%\LB{cite papers by Frenkel: force balance is delicate ! \cite{ganti2018hamiltonian} + \cite{ganti2017molecular}}
%\LB{quote coupling between thermo osmosis and diffusio osmosis by Sano et al. \cite{jiang2009manipulation}}

%Influence of interfacial dynamics on thermo-osmosis; and lots of references; shows that the sign of the direction of the flow is determined by interfacial dynamics~\cite{fu2017controls} (and their slippage/hydrophobic/philic properties) but also says Marangoni can only give a qualitative interpretation of what is going on; the enthalpic formulation is actually really nice; 

%Karnik paper on transport at hydrophobic/hydrophilic interface; he claims he has evaporation at some point but there seems to be no measurement proving so, so I am rather sceptical~\cite{lee2014nanofluidic}. 

%Thermo-osmosis setup with measurement of water flow versus temperature difference; water goes from cold to hot (what ?) \cite{kim2009investigation}

%Another excellent review; though more on thermophoresis~\cite{piazza2008thermophoresis}
%It seems that thermo-osmosis can be explained in two ways (change in enthalpy <-> change in entropy) or a sort of Marangoni effect with a change in surface tension: qui se recoupe mais cependant avec quelques descriptions plus claires il y a ~\cite{piazza2004thermal}

%Sinon j'aime vraiment bien cette manip là ~\cite{bregulla2016thermo}; vraiment très propre 

%and large dependence on slip~\cite{morthomas2008thermophoresis}

\subsection{Numerical simulations of (diffusio-)osmotic transport: methodologies and results}
\label{Sec:numerics}

Molecular simulations have now become a highly efficient tool to explore the fundamental properties of fluids and materials.
%Interfacial transport has been extensively explored using in particular 
Molecular dynamics simulate the many-body dynamics of particles and molecules, either at equilibrium or far from equilibrium, submitted to various thermodynamic forces.
They provide detailed information on the molecular processes
at play. \review{In the present context of studying osmotic forces and related fluxes, this represents a key opportunity to understand the fundamental and subtle origins underlying interfacial transport 
%subtle phenomena 
and how these can be affected by the microscopic details of the interface. }

%In this context, 
Simulating electro-osmosis is relatively straightforward in the sense that the effect of the electric field converts
directly into an electric force acting on the suspended ions. This has led to numerous molecular dynamics studies of electro-osmosis, as well as of streaming currents, allowing to decipher a wealth of phenomena associated with transport within the electric double layer \cite{rotenberg2013electrokinetics,bocquet2010nanofluidics}.
Now, simulating diffusio- and thermo- osmosis is by far more difficult and subtle. Indeed such transport occurs under thermodynamic forces associated with the gradient of concentration or temperature, and these can not obviously be represented in terms of mechanical forces acting on the simulated particles.
We discuss in this section recent developments in the numerical methodologies allowing to perform simulations of osmotic transport. 

%\SM{In this context where transport at the interface may be harvested for a number of applications, it is of utmost importance to be able to test different materials, different compounds (solute and solvent) and different geometries, and to find how they respond to such interfacial driving. Numerical simulations such as molecular dynamics can, in particular, probe new settings and include some insight on specific transport properties of the infinitesimal scales. Electro-osmotic driven transport is easily studied with molecular dynamics. This is done making use of explicit coulombian interactions and an external electric field. Such a simulation with a driving field is typically referred to as \textit{Non-equilibrium molecular dynamics simulation} or NEMD. However, diffusio-osmotic or thermo-osmotic flows are much harder to simulate because it is difficult 
%-- or at least it requires a lot of computational power -- 
%to have a mechanical representation of the thermodynamic forces associated with the corresponding gradients.}

For bare osmosis, direct simulations can be performed using two explicit reservoirs with difference of solute concentration. 
%is  proposed. 
%{Even simulation of osmosis -- simple osmosis occurring between two reservoirs at different concentrations -- requires some cunning implementation. 
For example such implementation was used  by Kalra {\it et al.} in the study of osmosis across carbon nanotubes~\cite{kalra2003osmotic}. 
%make use of two explicit reservoirs containing different salt concentrations -- which is not the least expensive. 
This configuration has the drawback that osmosis occurs in the transient regime since the reservoirs empty/fill during the osmotic process and this limits statistics. 
%no stationary state can be 
%Investigating the osmotic pressure
Osmosis was later rationalized in more simple terms by simplifying the explicit membrane description to reduce it to a confining potential acting on the solute only~\cite{luo2009simulation,lion2012osmosis,yoshida2017osmotic}. This is the numerical pendant to the mechanical views of osmosis described in Sec.~\ref{mech}. 
%Such approaches require simpler implementation the osmotic pressure is the quantity under scrutinity, it may proove  simpler to quantify  osmosis described in Sec.~\ref{mech}.
%This was explored
%use the simplifying approach where 

%\LB{ICICICICI}
The numerical implementation of diffusio-osmosis in molecular dynamics is far more complex since  one should be able to represent the chemical gradient in terms of a microscopic force acting on the particles. Various methods to investigate diffusio-osmotic transport were  proposed in the recent literature and we discuss them now.

\vskip0.2cm
%\subsubsection{Using symmetry relations to infer transport coefficients}
\noindent \textit{Using symmetry relations to infer transport coefficients --}\\
 It turns out that it is far easier to calculate the diffusio-osmotic mobility by exploiting the symmetry of the transport matrix. 
%In the symmetric phenomenon to diffusio-osmosis  an excess solute flux is generated under a pressure gradient. 
Recalling the general relation between fluxes and forces,
%\noindent \textit{Using symmetry relations to infer transport coefficients --}\SM{To compute diffusio-osmosis is even more complex. The strategy followed by several authors is rather to exploit the symmetry of the transport processes. In fact writing explicitly \eqref{Larray} as
\begin{equation}
%\label{Larray}
\left(\begin{array}{c} Q \\ J_s-c_s Q\end{array}\right)= 
\left(\begin{array}{cc} L_{11} & L_{12} \\ L_{21} & L_{22} \end{array}\right)
\times \left(\begin{array}{c} - \Delta p \\ - k_B T \Delta \log c_s \end{array}\right).
\end{equation}
the  Onsager symmetry for the transport matrix implies that $L_{21} = L_{12}$. Accordingly, 
%so that the diffusio-osmotic mobility associated with $L_{21}$ is equivalent, due to Onsager symmetry, to searching for the mobility characterizing the excess solute flux under a pressure driven flow $L_{21} = L_{12}$ 
calculating the diffusio-osmotic mobility as a water flux under a concentration gradient, here $L_{12}=Q/(- k_B T \Delta \log c_s )$, is therefore equivalent to calculating the excess solute flux under a pressure gradient, here $L_{21}=(J_s-c_s Q)/(- \Delta p)$ -- see Fig.~\ref{fig:Numerical}-a. The latter is far easier to implement numerically in non-equilibrium molecular dynamics (NEMD) since it requires only to generate a pressure-driven flow and measure the intergrated solute flux (or locally the velocity  and solute concentration profile). This can be performed with periodic boundary conditions along the flow, so that the resulting diffusio-osmotic mobility is indeed characteristic of the liquid-solid interface under scrunity, and does not depend on {\it e.g.} entrance effects into the pore. This methodology was successfully applied to quantify the diffusio-osmotic mobility on a variety of interfaces, including superhydrophobic surfaces, graphene, and with various liquids~\cite{ajdari2006giant,huang2008massive,lee2017nanoscale,fu2017controls}.  We discuss below some results of the simulations.
%cite Ajdari-LB = excess flux, cf aslo Joly et al. ethanol\\

%\LB{
%EO = easy, less easy to DO and TO because it is difficult to have a mechanical representation of the thermodynamic forces associated with the 
%chemical potential gradients.  For example simulation of osmosis by Kalra {\it et al.} across carbon nanotubes \cite{kalra2003osmotic} make use
%of two explicit reservoirs containing different salt concentration. 
% Also Rosalind Allen.
% To compute diffusio-osmosis is even more complex. The strategy folllowed by different authors \cite{ajdari2006giant,fu2017controls} is to rather measure the excess solute flux under a pressure driven flow, from which the diffusio-osmotic mobility is deduced. Due to the Onsager symmetry, this coefficient is equal to the mobility quantifying the water flux under a solute gradient. 
%%cite Ajdari-LB = excess flux, cf aslo Joly et al. ethanol\\
%FORMULA}

\vskip0.2cm
%\subsubsection{Equilibrium fluctuations for linear response of coefficients}
\noindent \textit{Equilibrium fluctuations for linear response coefficients --} \\
{Transport coefficients may also be inferred from equilibrium fluctuations by making use of Green-Kubo (GK) relations for the various mobilities. 
%If we look for a general relation between specific flows ($Q_i$) and forces ($F_i$) in the linear response regime, we look for the matrix $\mathbb{L}$ such that
%\begin{equation}
%%\label{Larray}
%\left(\begin{array}{c} Q_{\alpha} \\ Q_{\beta} \end{array}\right)= 
%\left(\begin{array}{cc} L_{\alpha\alpha} & L_{\alpha\beta} \\ L_{\beta\alpha} & L_{\beta\beta} \end{array}\right)
%\times \left(\begin{array}{c} F_{\alpha} \\ F_{\beta} \end{array}\right).
%\end{equation}
The transport coefficients introduced in the transport matrix ${ \mathbb L}$ can indeed be written in terms of a time-correlation function of the fluctuating fluxes $Q_i$ {\it at thermal equilibrium}. Such formal relations are obtained thanks to 
linear-response theory and the fluctuation-dissipation theorem~\cite{hansen1986ir,marry2003equilibrium,bocquet1994hydrodynamic}. 
They  provide generic expressions for the non-equilibrium mobilities in terms of {\it equilibrium} correlation functions in the form
\begin{equation}
L_{ij} = \frac{\mathcal{V}}{k_B T} \int_0^{\infty} dt \langle Q_{i}(t) Q_j(0) \rangle 
\end{equation}
where $\mathcal{V}$ is the system volume and $\{Q_i\}$ are the fluxes under scrutiny. 
%\SMM{Il y a une confusion a clarifier ici. $Q$ et $J_s$ n'ont pas la meme unite, et du coup je dirai que $Q_1 = Q$ et $Q_2 = \mathcal{V} (J_s- c_{\infty} Q)$? Je te proposerai donc d'enlever $\mathcal{V}$ dans l'equation ci dessous et de mettre plus bas "The linear-response formalism then leads to (using $Q_1 = Q$ and $Q_2 = \mathcal{V} (J_s- c_{\infty} Q)$, where $\mathcal{V}$ is the volume of the system):"} 
The symmetry of the transport matrix originates in the time-symmetry of the underlying microscopic dynamics
\cite{barrat2003basic}. 
%\LB{explicit formula for DO cf Hiroaki + JLB ?}.
The simplest route to obtain the Green-Kubo formula for the diffusio-osmotic mobility is to consider the solute excess flux generated under a pressure drop since the latter is equivalent to a body force applied to all system particles. The linear-response formalism then immediately leads to\cite{yoshida2017osmotic}
\begin{equation}
L_{21}=L_{21}=\dfrac{\mathcal{V}}{k_\mathrm{B}T}\int_0^{\infty}\langle
 (J_s-c_{\infty}Q)(t)Q(0)\rangle dt.
\label{GK}
\end{equation}
%\SMM{Pourquoi $c^*_{\infty}$ ? je mettrai juste $c_{\infty}$.} 
In the case of a channel with length $L$ and cross area ${\cal A}$, one has $L_{21}=L_{21}\equiv {{\cal A}\over L}\mu_{DO}$.
We refer to Refs\citenum{yoshida2014generic,yoshida2017osmotic} for detailed derivations of these GK equations.

These GK formula allow to calculate numerically the diffusio-osmotic mobility,
as well as any 
%Transport coefficients corresponding to all 
off-diagonal terms of \eqref{Larray2}, by estimating the  correlation functions  in Eq.(\ref{GK}) in {\it equilbrium} simulations.
%~\cite{yoshida2014generic,yoshida2017osmotic}. 
This approach was followed in Refs~\citenum{yoshida2014generic,yoshida2014molecular,yoshida2017osmotic} and the resulting mobilities were  successfully compared with results of NEMD simulations, as discussed below. 
%\citefull{yoshida2014generic} shows a good agreement between both methods at small drivings. NEMD is however crucial to capture non-linear deviations under large driving fields.}

\begin{figure}[h!]
\centering
  \includegraphics[width=0.49\textwidth]{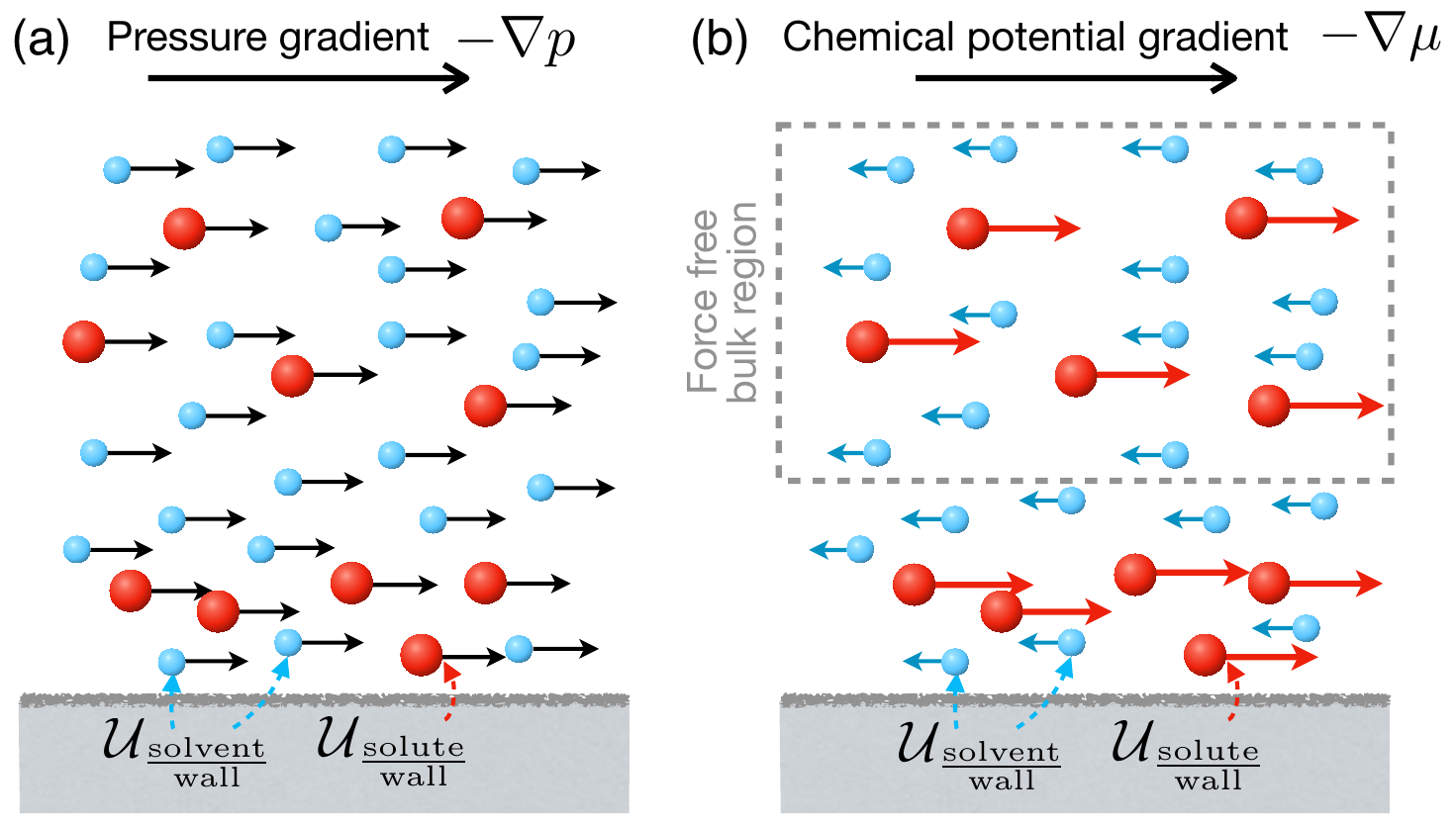}
  \caption{\SM{\textbf{: Non-equilibrium molecular dynamics of osmotic interfacial transport.} Inspired from \citefull{yoshida2017osmotic}. (a) Excess flux under pressure gradient. The pressure gradient is obtained by applying a force acting on each particle and the  solute flux in excess to the bulk $J_s-c_s Q$ is measured. (b)  NEMD method to simulate interfacial transport under chemical potential gradients. The chemical potential gradient is modeled as a forward force per solute particle (red) and a properly defined counter force per solvent particle (blue) such that the total force on the fluid in the bulk is zero. The local diffusio-osmotic velocity profile is directly measured.}}
  \label{fig:Numerical}
\end{figure}

\vskip0.2cm
%\subsubsection{Non-equilibrium molecular dynamics and mechanical representation of chemical gradients}
\noindent \textit{Non-equilibrium molecular dynamics and mechanical representation of chemical gradients--} \\
While the equilibrium approach provides proper foundations to calculate diffusio-osmosis, non-equilibrium simulations  proves usually more practical to calculate transport coefficients, {\it e.g.} in terms of statistics. %to run  to estimate the diffusio-osmotic flows under solute gradients. 
However, as we emphasized above,  this requires to build a proper numerical scheme to implement a mechanical equivalent of the chemical potential gradient.
%suitable to be implemented as a numerical scheme. 
One interesting route was suggested by Yoshida et al. in \citefull{yoshida2014generic} and then applied to electro- and diffusio- osmotic transport of electrolytes: the authors ran different simulations where forces $f_j$ are applied separately to each individual specie, here {\it \{solvent, anions, cations\}}, 
allowing to calculate the corresponding {\it individual} fluxes $Q_i$ and deduce the mobilities for the individual species $M_{i,j}=Q_i/f_j$; the electro- and diffusio- osmotic mobilities are then calculated by proper linear combinations of the mobilities of
 individual species, in order to deduce the electro- and diffusio- osmosis. This approach echoes directly the discussion in Sec.\ref{ECequ}, in which the electro-osmotic and diffusio-osmotic mobilities are deduced from the individual ion mobilities,  defined above as $M_\pm$ \cite{yoshida2014generic}.

It is however relevant to develop numerical methods to simulate explicitly the diffusio-osmotic flows.
%beyond how to simulate really DO ??
Such a numerical scheme was recently proposed in Ref.\citenum{yoshida2017osmotic}, in which a proper mechanical set of driving
forces is applied to the system to mimick the  chemical potential gradient of the solute.
%{Quite recently, a new numerical scheme was proposed to measure interfacial transport generated by diffusio-osmosis using a specific mechanical representation \cite{yoshida2017osmotic}.
%The method
%Inspired from GK relation: look at Eq.(\ref{GK}) and extend to the reverse situation to suggest equivalent of chemical gradient. } \SM{As opposed to standard NEMD applying an identical force on each particle (solute or solvent) to model for instance the impact of pressure -- see Fig.~\ref{fig:Numerical}-a, 
To do so, the scheme applies differential forces on the solute and on the solvent, see Fig.~\ref{fig:Numerical}-b:
%\begin{itemize}
%\item 
 (i) an external force $F_{\mu}$ on each {\it solute} particle in the whole system; (ii)
%\item 
a counter force $-[N^{B}_{s}/(N^{B}-N^{B}_s)]\times F_{\mu}$, acting on each {\it solvent} particle. Here $N^{B}_{s}$ and $N^{B}$ are respectively the number of solute particles and the total number of particles in a properly defined {\it bulk region} (\qote{sufficiently} far from the surface). 
%\end{itemize}
%This set of forces is inspired by the Green-Kubo form
The counter force is set to ensure a force free balance in the bulk volume. %, so that no force act on the system. 
Most important, it can be  verified that applying linear-response theory to the system with this set of forces allows one to 
show that the resulting diffusio-osmotic mobility 
%recover a linear dependence of the (far field) diffusio-osmotic velocity versus $F_{\mu}$, but furthermore that the deduced mobility 
 does  identify with the GK relation in Eq.(\ref{GK}): this therefore fully validates the theoretical foundations of the proposed numerical scheme.
%This set of forces is actually fully justified
The corresponding effective chemical potential is then related to the applied external force $F_{\mu}$  via
\begin{equation}
\label{eqMu}
-\nabla_x\mu=F_{\mu} \frac{N^B}{N^B-N^B_s}.
\end{equation}}
%\LB{\eqref{eqMu} is actually fully confirmed using linear-response theory and identifying the corresponding diffusio-osmotic mobilities via their Green--Kubo expressions, see \citefull{yoshida2017osmotic} for full details.}

%\SM{This non-equilibrium approach was fully validated in the case of diffusio-osmotic driving and compared to theoretical predictions~\cite{yoshida2017osmotic,liu2018pressure}. 

%\SM{COMMENT FROM HIROAKI ON SIMON'S NOTES :The issue about the factor is related to ?low concentration approximation? used on
%the matrix (14), according to Simon. I have not yet gone through all the detailed
%calculations that he did. Daan seems to agree with Simon?s calculations, then the
%equation for the general cases (all range of the concentration) should be slightly
%corrected, which will be reflected in the draft that we are now preparing with Daan
%and Lyderic.}
This approach leads as expected to a velocity profile exhibiting a strong gradient within the interfacial layer, and then a plug flow far from the surface. The deduced diffusio-osmotic mobility obtained from the NEMD scheme was checked to be identical to both the equilibrium GK results and those obtained from the excess flux under pressure-driven flow introduced above \cite{yoshida2017osmotic}. 

%in Ref.\cite{yoshida2017osmotic}, \SM{Further non-equilibrium systems remain to be explored from more complex systems experiencing diffusio-phoresis (see Sec.~\ref{sec:DP} )to other sources of driving such as thermo-osmotic driving. With the growing numerical methods at hand, we may expect further insight to be quickly gained on these systems.}

\vskip0.2cm
%\subsubsection{Difficulties with microscopic stress tensor }
\noindent\textit{Some difficulties with the microscopic stress tensor-- }\\
The continuum approach, as described above, allows one to predict diffusio-osmotic transport in terms of a surface pressure gradient. 
%Following the theoretical understanding of diffusio-osmosis, that is mediated by a pressure driven effect, 
In a different approach, it is accordingly tempting to obtain the diffusio-osmotic flow by a direct numerical calculation of the \textit{local microscopic pressure} in the fluid. 
However, as was demonstrated by Frenkel and collaborators in a series of papers~\cite{liu2017microscopic,liu2018pressure,ganti2017molecular,ganti2018hamiltonian}, a major difficulty in this approach is that
there is no unique expression for the local microscopic pressure tensor ({\it e.g.} in terms of the position and velocities of individual particles
and the microscopic forces acting on them). Accordingly various microscopic definitions of the pressure tensor lead to different numerical results.
Such difficulty was evidenced for diffusio-osmotic flows\cite{liu2018pressure}, but also for thermo-osmotic  flows \cite{ganti2017molecular,ganti2018hamiltonian}.
%It was recently shown in \citefull{liu2018pressure} that due to an ambiguity of the definition of pressure in anisotropic media (\textit{e.g.} close to a surface), local pressure gradients fail to predict diffusio-osmotic flows.
%\LB{ICICICICI:\\
%At microscopic scales, thermo-osmotic flows however raises subtle questions in terms of the microscopic force balance at stake, 
%Beyond the continuum limit. 
%There is no unique definition of the microscopic stress tensor, and various definitions lead to different results.
%As demonstrated recently using molecular dynamics simulations
%%Thermo-osmotic flow measurements have recently been investigated by molecular dynamics tools
%~\cite{ganti2017molecular,ganti2018hamiltonian}. The simulations show  that  near an
%interface, it is not possible to express the microscopic
%forces as the gradient of the stress tensor \cite{ganti2017molecular}, a result which was also highlighted for interfacial transport under solute gradients \cite{liu2017microscopic,liu2018pressure}.
%Problem similar for diffusio-osmosis, cf \cite{liu2018pressure}. A rappeler plus bas. Developper plus ?? car pb def micro pour osmose.\\
%CIter aussi methode par deux gradients, cf ref 38-39 Liu JPCM}

\vskip0.2cm
%\subsubsection{Some results of simulations}
\noindent\textit{Some results of simulations--}\\
Simulations have allowed to gain much insights into diffusio-osmotic transport. Various fluids, {\it e.g.} Lennard-Jones fluids, but also with electrolytes and water-ethanol mixtures, and various interfaces were considered, hydrophilic or hydrophobic surfaces, graphene, superhydrophobic surfaces, \textit{etc}. Among highlighted effects one may quote the impact of hydrodynamic slippage of the fluid at the surface, which does boost considerably the diffusio-osmotic mobility on hydrophobic \cite{ajdari2006giant} and graphene surfaces \cite{yoshida2017osmotic}, and even more on super-hydrophobic surfaces \cite{huang2008massive}. The enhancement of the diffusio-osmotic mobility scales typically like the ratio between the (effective) slip length and the interfacial length, as mentionned in the previous sections. 

Simulations also give some insights on the local diffusio-osmotic velocity profile and its relation to the concentration profiles. Within the continuum framework discussed in the previous section, the velocity profile is obtained simply by integration of the Stokes equation of motion in Eq.(\ref{eq:StDO}), with the pressure expression given in Eq.(\ref{pressure}). 
 Simulations actually show usually a very good agreement between the continuum prediction and the velocity profiles measured in the NEMD simulations, giving strong support to the continuum description. 
 Such an agreement may be considered as surprising in view of the strong velocity gradients occurring on length scales in the range of a few  molecular size. However the Stokes equation is known to be surprisingly robust down to molecular lengthscales \cite{bocquet2010nanofluidics} and this explains its success in predicting diffusio-osmotic flows. 
 
 Finally, the continuum framework allows one to relate the diffusio-osmotic mobility to the concentration profile, and more particularly to its first spatial moment, see Eq.(\ref{KDO}). It is accordingly tempting to relate -- as for Marangoni effects -- the amplitude of diffusio-osmotic transport to the adsorbed quantity, defined as $\Gamma=\int dz\, (c_s(z)-c_\infty)$. The latter is directly connected to the surface tension via the Gibbs-Duhem relation. The adsorbed quantity $\Gamma$ provides in most cases a good estimate for the diffusio-osmotic mobility and its sign. However -- as mentioned earlier -- for complex concentration profiles the relation was found to be more complex than this simple rule of thumb (for example for water-ethanol mixture at interfaces~\cite{lee2017nanoscale}).

\vskip0.5cm
%\section{Nanofluidics and advanced osmotic transport}
\section{Osmosis beyond van\,'t Hoff}
\label{beyondVO}

%\LB{ICICICICI}

\subsection{Advanced osmosis and nanofluidics}

The previous section highlighted molecular insights into osmotic phenomena, unveiling the underlying driving forces at play.
%The interpretation in terms of forces  their link to interactions.
%A wealth of behaviors emerge
However such perspectives also suggest possible extensions to obtain more advanced osmotic transport beyond the linear framework
presented before. In this section we discuss osmotic transport across channels with more complex geometries involving symmetry breaking, or active parts. Our objective in this section is to show that it is possible to extend  simple osmosis  beyond the van\,'t Hoff paradigm and design advanced functionalities resulting in non-linear and active transport. %which offers new routes to go beyond the van\,'t Hoff paradigm. % possible to go beyond.
%such as nanofluidic diodes or rectifiers~\cite{picallo2013nanofluidic,jubin2018dramatic}. 
%It is also of great help to investigate, beyond, nonlinear dependencies of the osmotic reflection coefficient, for instance on salt or solute concentration.~\cite{yamauchi2000membrane,marbach2017osmotic,yoshida2017osmotic}

In this context it is interesting to observe that   in biological species (bacteria, archaea, fungi, ...) many membrane channels do achieve advanced functionalities in order to regulate osmosis: for example rectified osmosis~\cite{kiyosawa1973rectification} \SM{--\textit{e.g.} an osmotic flow with a non-linear dependence on the concentration gradient--}, or gated osmosis to prevent lysis and survive osmotic shocks in mechanosensitive channels~\cite{kung2010mechanosensitive} (with diffusio-osmosis identified as a potential mechanism for the gating mechanism in deformable structures \cite{bonthuis2014mechanosensitive}).
These few examples highlight the possibility of going far beyond the van\,'t Hoff paradigm, thanks to properly designed \SM{(active)} nanochannels.
%many bacteria, archaea, and fungi avert

\begin{figure}[h!]
\centering
  \includegraphics[width=0.49\textwidth]{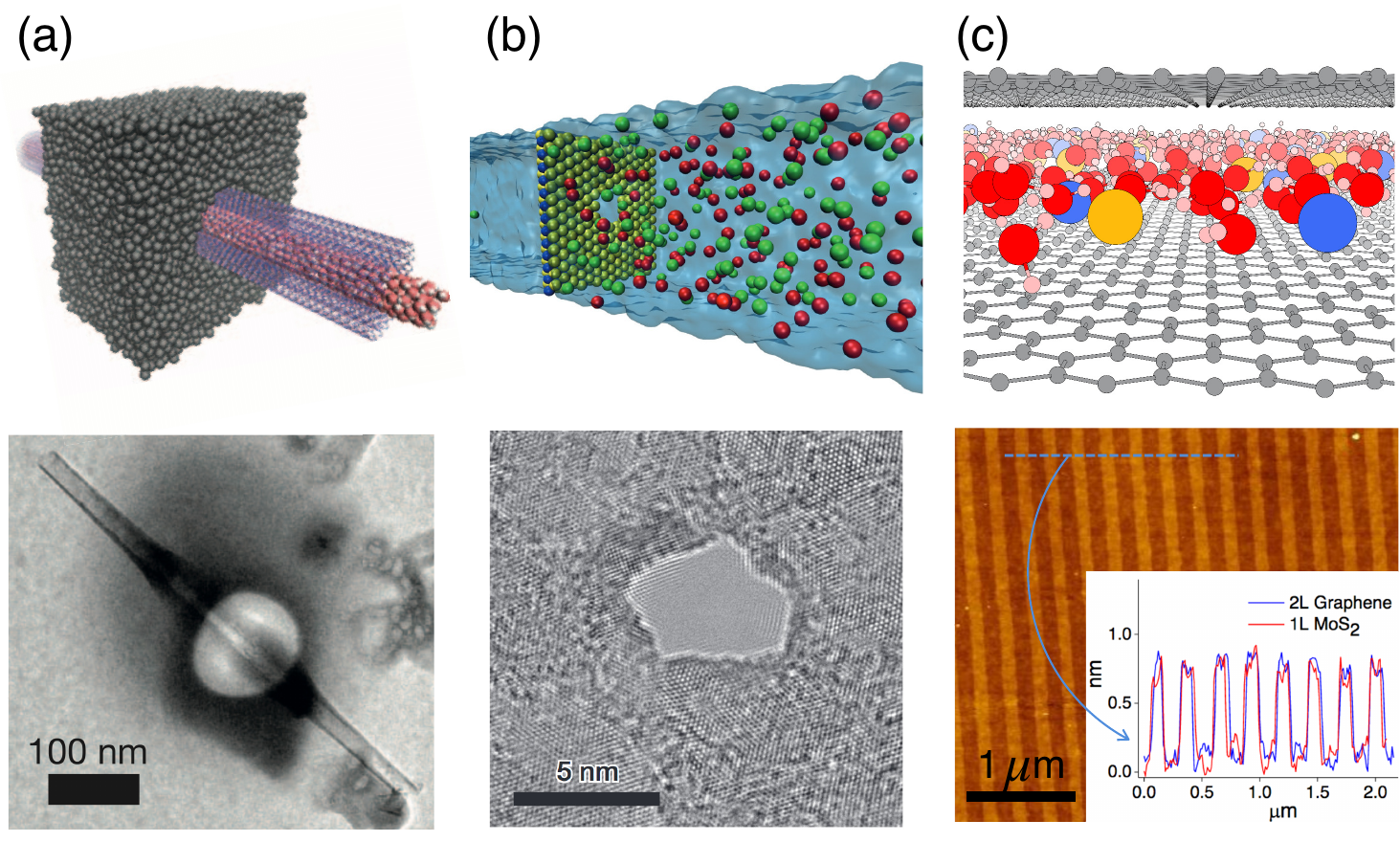}
  \caption{\textbf{: From nanoscale to \.{a}ngstr\"{o}m scale pores.} (a) Reproduced from \citefull{siria2013giant}. (Top) Molecular dynamics representation of water flowing through a transmembrane multi-wall boron-nitride nanotube and (bottom) Transmission electron microscope (TEM) picture of its experimental counterpart. (b) (top) Molecular dynamics representation of a nanopore in a mono-layer MoS$_2$ membrane (in blue and yellow) and the salt (green and red) in solution and (bottom) TEM picture of its experimental counterpart, a 5 nm pore. Reproduced from \citefull{feng2016single} with permission from Springer Nature, copyright 2016. (c) (top) \SM{Molecular dynamics representation of water and ions (orange and blue) flowing through a graphene slit with $7$  \.{a}ngstr\"{o}m spacing. (Courtesy from Timoth\'{e}e Mouterde)} (bottom) AFM image  of bilayer graphene spacers on top of the bottom graphite layer. Inset: Height profiles yield an estimate of 7\AA  
  for the thickness of spacers made from 2 layers of graphene or 1 layer of MoS$_2$ (the blue line shows the scan position for the corresponding trace in the inset). Reproduced from \citefull{esfandiar2017size} with permission from AAAS, copyright 2017.} 
  \label{fig:NanoTech}
\end{figure}

We believe that the advent of nanofluidics has a key role to play in this regard, in order to identify new types of behaviors which could be scaled-up to macroscopic membranes. The new opportunities brought by nanofluidics in terms of the variety of nanoscale geometries and materials, combined with state-of-the-art experimental instrumentation, allows one to fabricate and investigate fundamentally the transport in ever smaller channels, with ever more complex and rich behaviors.
%\LB{The advent of nanofluidics: allows to explore fundamentals; new properties at nanoscales, new transport phenomena, new opportunities, example here with osmotic diodes and active transport; also new nanomaterials -- cf also part on filtration --}
Carbon nanotubes, down to nanometric sizes~\cite{siria2013giant,choi2013diameter,secchi2016massive,tunuguntla2017enhanced} can now be manipulated and inserted in devices were water is flown through -- see Fig. ~\ref{fig:NanoTech}-a. %Carbon nanotubes can also be functionalized~\cite{nednoor2007carbon}. 
Single nanopores can  be carved or etched in membranes that are only an atomic layer thick~\cite{feng2016single} and may be accordingly functionalized~\cite{mahmoud2015functional}, see Fig.~\ref{fig:NanoTech}-b. It is also now possible to fabricate \.{a}ngstr\"om scale slits using graphene sheets as spacers, reaching confinement thicknesses down to $\sim 3$ \AA~\cite{nair2012unimpeded,radha2016molecular} -- see Fig.~\ref{fig:NanoTech}-c. 

Such nanofluidic technologies offer new possibilities in the context of osmotic transport.
%promisse to bring significant potential improvement in many aspects. 
They allow nearly molecular scale designs, leading to various nanofluidic-specific effects which may be key assets for new separation techniques and water filtration: from specific ion exclusion effects~\cite{fornasiero2008ion,choi2013diameter,radha2016molecular,tunuguntla2017enhanced, walker2017extrinsic} with a number of anomalous ionic effects to be investigated~\cite{feng2016observation}, to extremely fast permeation of water, in particular through carbon nanotubes~\cite{majumder2005nanoscale,holt2006fast,whitby2008enhanced,secchi2016massive,tunuguntla2017enhanced}.
Also new types of nanoscale membranes have also emerged recently, offering new designs as compared to traditional membranes: for example,  with dedicated patterns of hydrophilic and hydrophobic regions \cite{lee2014nanofluidic}; or tailor-designed DNA origami channels~\cite{bell2011dna,langecker2012synthetic}, and -- last but not least --  the multilayer membranes of graphene (so-called graphene oxide  membranes)~\cite{joshi2014precise,majumder2017flows,siria2017new}.

This constitutes a new and exciting playground, in which osmotic phenomena may (and should) flourish in various forms. 
We discuss in the next paragraphs two such examples: the development of osmotic diodes, and an active counterpart of osmosis, 
which both lead to tunable osmotic driving beyond van\,'t Hoff.

%\LB{specific treatments, cf hydrophobic nanopores: \cite{lee2014nanofluidic}}

%A number of other materials are triggering broad interest such as graphene oxide based membranes~\cite{majumder2017flows,siria2017new} or tailor-designed DNA origami channels~\cite{bell2011dna,langecker2012synthetic}.

%\LB{To be cited \cite{bonthuis2014mechanosensitive}
%Osmotic shock presents a fatal risk to
%unicellular organisms. A sudden increase of the environmental
%solute concentration, known as hypertonic shock,
%leads to water loss and cell volume decline, whereas a
%sudden decrease, referred to as hypotonic shock, causes
%water to enter the cell rapidly, inducing cytolysis. As a final
%resort in case of severe hypotonic shock, many bacteria,
%archaea, and fungi avert cell lysis by activating nonselective
%membrane channels to release solutes from the cytoplasm [C. Kung, B. Martinac, and S. Sukharev, Annu. Rev.
%Microbiol. 64, 313 (2010).] \cite{kung2010mechanosensitive}
%}

\subsection{Osmotic diodes and osmotic pressure rectification}

One of the successes of nanofluidics was to demonstrate the possibility to design diodes for ionic transport, in full analogy with their electronic
counterpart~\cite{siwy2006ion,karnik2007rectification,bocquet2010nanofluidics}. This takes the form of a non-linear and rectified response for the ionic current versus the applied voltage. Typically an ionic diode behavior manifests itself in channels with an asymmetric design, {\it e.g.} an asymmetric surface charge or an asymmetric geometry. Such behavior is expected to occur
%Overall a diode like behavior in the ionic current versus applied voltage is obtained 
 in the regime where the Dukhin number
is of order one and asymmetric along the channel~\cite{Poggioli2019}: the Dukhin number is defined here as $Du=\Sigma/c_s h$, where $\Sigma$ is the surface charge density, $c_s$ the bulk salt concentration and $h$ a characteristic channel dimension. It quantifies the importance of surface versus bulk electric conduction. As such ionic diodes may find interesting applications to boost osmotic power generation under salinity gradients, see \citefull{siria2017new} and Sec.~\ref{BlueEnergy}.

\begin{figure}[h!]
\centering
  \includegraphics[width=0.49\textwidth]{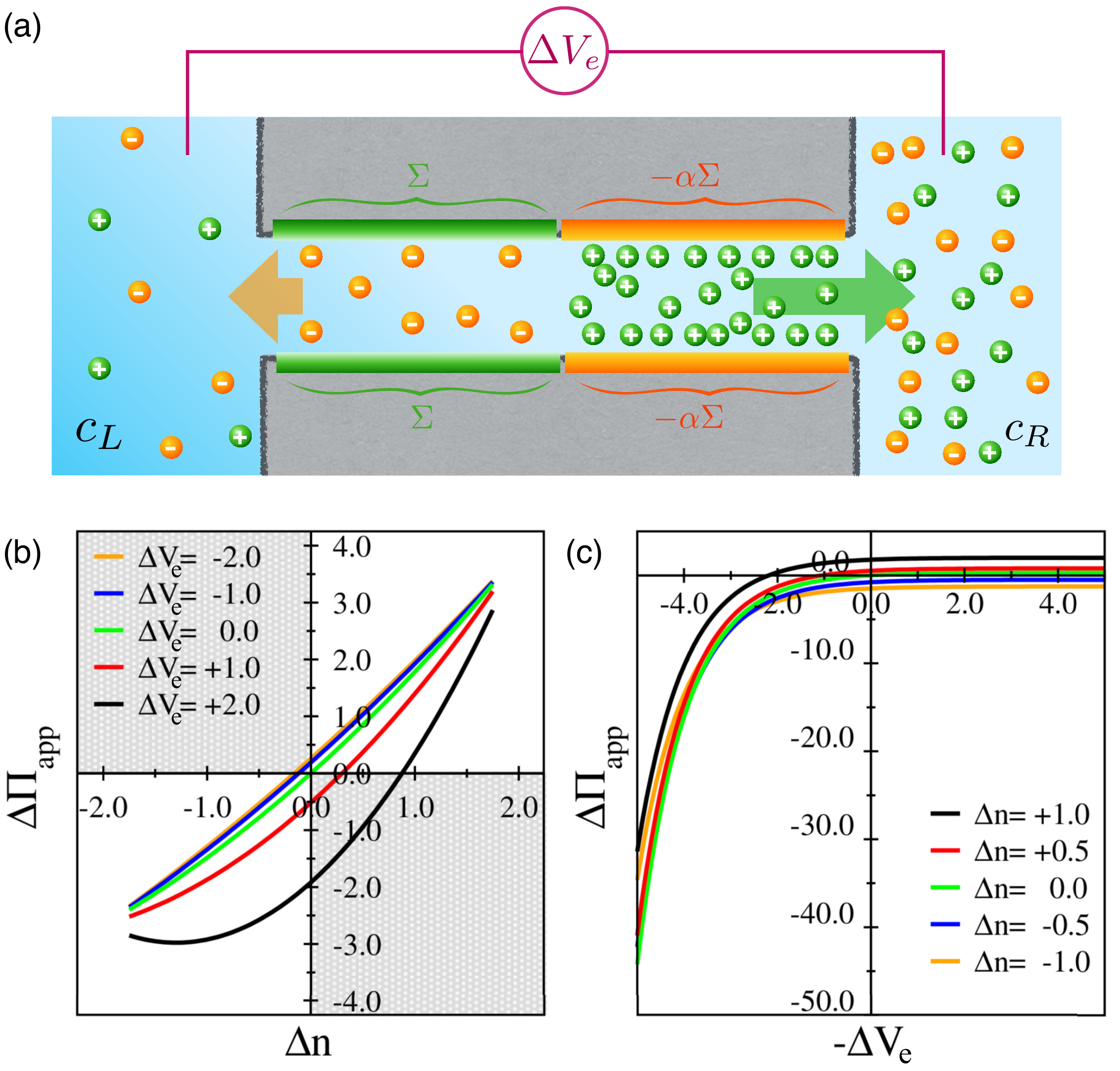}
  \caption{\textbf{: Osmotic rectification in an osmotic diode.} (a) A nanochannel presents an asymmetric surface charge with $\Sigma >0$ and $-\alpha\Sigma$ on the other side with $\alpha \neq 1$. The ions are therefore submitted to an asymmetric force (between one side and the other, in colored arrows) that drives the osmotic flow. Inspired from \citefull{picallo2013nanofluidic}. Apparent osmotic pressure $\Delta \Pi_{\rm app}$ versus (b) salinity gradient $\Delta n = n_R - n_L$ (where $n_i = c_i/c_0$ is a normalized concentration)  and versus (c) applied voltage drop $\Delta V_e = V_R - V_L$ (normalized by $k_BT/e$) as obtained from an analytical solution for the flows of all species in the nanochannel. (b) and (c) are reproduced from \citefull{picallo2013nanofluidic} with permission from the APS, copyright 2013. More information can be found in \citefull{picallo2013nanofluidic}. 
  %\textcolor{blue}{I would expect the contribution at $\Delta V = 0$ to still be diode-like but here it's not so obvious, is it just a zooming question ?}.
  }
  \label{fig:OsmoDiodeR}
\end{figure}

Now, coming back to osmosis, the asymmetry of the design may be expected to yield an asymmetric interaction of the membrane with the electrolyte, hence an {\it asymmetric push} on the liquid.
It was shown in \citefull{picallo2013nanofluidic} that  such asymmetric geometry -- depicted in Fig.~\ref{fig:OsmoDiodeR}-a -- results in an osmotic diode, with a rectified osmotic pressure versus the concentration gradient (non linear dependence), furthermore tunable by the applied electric field. 
%A special case of the osmotic diode applicable to electrolytes (such as salt, very interesting for desalination) is presented in Fig.~\ref{fig:OsmoDiode} and was thoroughly studied~\cite{picallo2013nanofluidic}. In that case the asymmetry comes from an asymmetric surface charge. Hope to realize such a refined patterned structure is given by advanced nano-structured materials~\cite{radha2016molecular,perez2015polydopamine} and refined gating patterning~\cite{karnik2007rectification,kalman2009control}. 

%The geometry considered in  ~\citenum{picallo2013nanofluidic} is that of Fig.~\ref{fig:OsmoDiode}. 
The description builds on the previous mechanical views of osmosis, in Sec.~\ref{mech}.
%One writes the coupled  Poisson-Nernst-Planck equations for the electrolyte densities and 
The Stokes equation for  fluid motion writes  %writing similarly as in Eq.~\ref{St},
\begin{equation}
0 = -\bm{\nabla} p + \eta \nabla^2 \bm{u} + \rho_e (-\bm{\nabla} V_e),
\label{StOsmo}
\end{equation}
with $\rho_e=e (c_+ - c_-)$ the charge density,  $c_\pm$ is the concentration of positive and negative ions (assumed here as monovalent for simplicity) and $V_e$ is the electric potential.
%Here the valence of the electrolytes were taken equal to $1$ for simplicity. 
Following the same steps as in Sec.~\ref{mech} to integrate the fluid equations of motion in the channel, 
%Upon integration with the proper boundary conditions on the channels, 
the general relation between flow and pressure takes the expression
\begin{equation}
Q=\mathcal{ L}_{\rm hyd} \times {-\Delta [p-\Pi_{\rm app} ]}
\end{equation}
%$\mathcal{A}$ the cross section of the pore and $L$ its length; 
where $\mathcal{ L}_{\rm hyd}$ is the channel permeance introduced above.
%One can accordingly extract 
%From Eq.~\ref{StOsmo} it is clear that the electrolyte concentration imbalance yields an 
The apparent osmotic pressure between the two sides of the channel is accordingly defined as
%drives the fluid flow
\begin{equation}
\Delta \Pi_{\rm app} =  \frac{1}{\mathcal{A}} \int\int  {d\mathcal{A}} dx\,  \rho_e\times (-\bm{\nabla} V_e) 
\label{Piapp}
\end{equation}
with $\mathcal{A}$ the cross section of the pore, $L$ its length. % and $\mathcal{V}=\mathcal{A}\times e$ its volume. 
The ion concentration profiles obey the Poisson-Nernst-Planck equations, coupling the diffusive dynamics to the applied electric forces. 
In spite of the expected non-linear dependence of the osmotic pressure in terms of driving forces, the symmetry in the force balance and ionic transport equations, which was highlighted  in Sec.~\ref{mech} and  \eqref{sym} for the simplest geometry,  still holds. There is accordingly a {\it linear} relation between the apparent osmotic pressure in \eqref{Piapp} and the total surface ion flux $j_s$: 
\begin{equation} 
\Delta \Pi_{\rm app}=k_BT \Delta c_s + j_s\times{L\over D}. 
\end{equation} 
It is therefore expected that the rectifying behavior in the ion flux, akin to the current diode, thus translates into a rectifying osmotic pressure. 
% in the osmotic forces  in Sec.\ref{mech}, an intrinsic symmetry 
%\LB{Revenir sur la symetrie entre flux solute et force dans l'osmose, car meme si $Q$ est non lineaire en $\Delta V$, $Q$ est lineaire en $J_s$}

Solving the full equations in the geometry presented in Fig.~\ref{fig:OsmoDiodeR}-a yields a fluid flux:
%(in the absence of a hydrostatic pressure difference) :
\cite{picallo2013nanofluidic}
\begin{equation}
Q = \mathcal{L}_{\rm hyd} [\sigma k_B T \Delta c_s -\Delta p]+ Q_S[\Delta c_s] \left( \exp\left(\frac{e \Delta V_e}{k_B T}\right) - 1 \right)
\end{equation}
where the reflection coefficient $\sigma$ is now a non-linear function of the concentrations in both reservoirs and $Q_S$ plays the role of a \qote{limiting fluid flux}~\cite{picallo2013nanofluidic}. %and corresponds to the diode-like contribution. 
The apparent osmotic pressure $\Delta \Pi_{\rm app} = Q/\mathcal{L}_{\rm hyd}$ is plotted in Fig.~\ref{fig:OsmoDiodeR}. The rectification and diode behavior versus concentration is weak for zero voltage but strongly enhanced for higher applied voltage bias. 
%of the diode behavior is visible, and highly non-intuitive, as it linked to non-linear effects. 

Examples of permeability rectification are actually numerous in the biological world. They are harnessed \textit{e.g.} in plant cells~\cite{kiyosawa1973rectification} or in animal cells~\cite{farmer1970perturbation,toupin1989permeability,peckys2011rectification}. Surprisingly in all the studies that we are aware of, entering flows are notably larger than outer flows, and up to 10 times higher in some mammalian fibroplasts~\cite{peckys2011rectification}. It is fascinating to see how most cells are therefore adapted to \textit{fill in} faster than they would \textit{swell} under the same conditions, probably with a link to survival strategies. We highlight that rectified osmotic flows could be used in a variety of fields. In fact, Fig.~\ref{fig:OsmoDiodeR}-b and the results reported in \citefull{picallo2013nanofluidic}, demonstrate that water may be flown against the natural osmotic gradient, with water flowing to the high salinity reservoir, depending on the voltage applied. Furthermore, under an oscillating electric field, with proper conditions, this induced water flow against the natural osmotic gradient will be maintained. This opens new perspectives \textit{e.g.} for advanced water purification strategies and active filtration with oscillating fields, as we discuss later.

\subsection{Towards active osmosis}

\begin{figure}[h!]
\centering
  \includegraphics[width=0.49\textwidth]{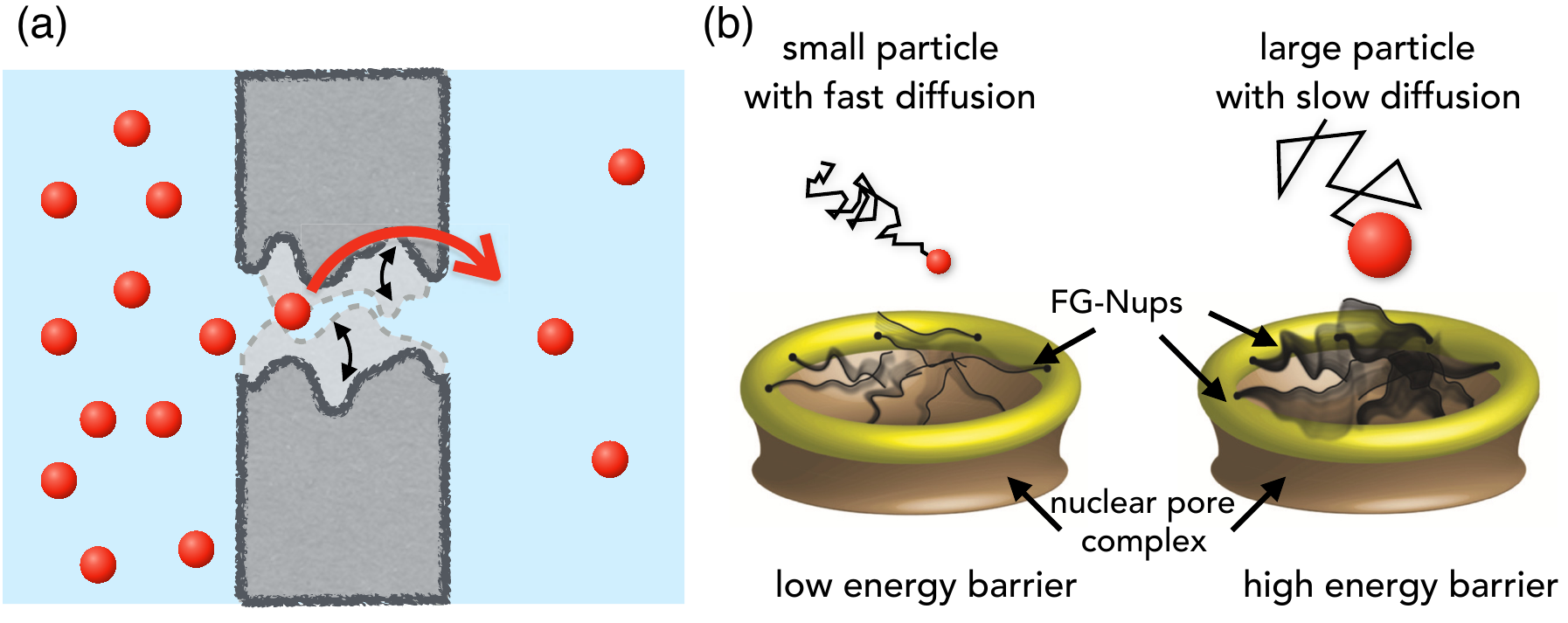}
  \caption{\textbf{: Osmosis through out-of-equilibrium pores.} (a) Illustration of a pore with a time-dependent shape, where the inner pore size may be either smaller either larger than the typical solute size. (b) Reproduced and adapted from \citefull{sakiyama2016spatiotemporal} with permission from Springer Nature, copyright 2016. Representation of the spatiotemporal motion of grafted phenylalanine-glycine nucleoporins (FG-Nups) inside the selectivity filter of the nuclear pore complex. Small (resp. large) particles that diffuse fast (resp. slow) see effectively the FG-Nups as static leaving small openings (resp. moving and everywhere) and therefore encounter only a "small" energy barrier for translocation (resp. large).} 
  \label{fig:activeFiltration}
\end{figure}

We discuss now a second class of examples of osmotic phenomena that goes beyond the van\,'t Hoff paradigm. As we exhaustively discussed, the idea of osmosis is closely related to semi-permeability and sieving -- with the membrane playing the role of a simple colander. However one may consider how the osmotic pressure builds up in membranes with time-dependent pores: a pore which opens and closes over time -- see Fig.~\ref{fig:activeFiltration} -- will exhibit a time-dependent size exclusion and sieving  is thus expected to generate an intermittent osmotic push 
on the fluid. The resulting osmotic pressure is expected to be some time-average of the push, which remains to be properly defined. But injecting energy at the pore scale -- here via the time-dependent opening of the pore -- may also lead to far-from-equilibrium behaviors, allowing possibly  to bypass the entropy bottleneck.
Osmosis across dynamically stimulated pores 
%How these dynamic properties modify osmotic pressure and osmotic transport 
is therefore a subtle problem, which requires proper microscopic foundations.
%as it goes beyond the equilibrium or weakly non-equilibrium framework. 
Beyond the fundamental question, adding the sieving frequency as a new tuning parameter may improve separation and selectivity properties of the membranes
%The underlying question is whether an active version of osmosis, where the pore shape is dynamically stimulated, may enhance the efficiency of osmotic phenomena, brings new perpsective to enhanced selectivity 
~\cite{marbach2017active,marbach2018transport}. 
%in``second principle''
%bypass a lead to  %bypass some of the limitations of 
%advanced osmotic phenomena beyond
%osmotic transport is not defined in equilibrium. In fact -- and as may happen through activated bio-channels -- imagine a situation where a pore is dynamically broadened and narrowed, and where at some times the solute of interest may pass or not because it is not always excluded by size in time -- see Fig.~\ref{fig:activeFiltration}. Then the question of how to define the osmotic pressure and also the rejection coefficient $\sigma$ is completely open. 

The question of dynamic osmosis actually arises naturally in biological pores, {\it e.g.} ion channels, since their shape is affected by thermal fluctuations or demonstrates out-of-equilibrium motion~\cite{noskov2004control}. The question of out-of-equilibrium osmosis was discussed in the literature in the 1980s~\cite{eisenman1983ionic} and  a few molecular dynamics studies were pursued~\cite{lauger1980fluctuations,schroder1983rate,noskov2004control}, usually with a focus on the specificities of the biological channels under scrutinity.
%Understanding out-of-equilibrium osmosis is of utmost interest. 
In fact, temporal dynamics of biochannels are strongly believed to be connected to the selectivity properties of the channel. For example, the fluctuations of the refined structure of the selectivity filter of the KcsA channel is believed to be a key factor for extremely refined passage of the potassium ion~\cite{noskov2004control}. Further, such temporal dynamics of the structure may provide an efficient alternative to simple steric sieving for selectivity. 
%be an alternative sieving way to at the origin of selectivity targeted upon alternative features than steric exclusion, 
This was noticed in the nuclear pore complex~\cite{sakiyama2016spatiotemporal}, where particles see an effective translocation barrier which is dependent on their diffusion properties (see Fig.~\ref{fig:activeFiltration}-b). 

To our knowledge, there is no general framework discussing the concept of \qote{dynamic osmosis}. Several simple models were considered recently by the authors in ~\citefulls{marbach2017active,marbach2018transport}. Here
we illustrate a few basic concepts underlying this active osmosis process and the opportunities that it offers. 

%\LB{introduire d'abord ici l'idee du JCP active sieving qu avec des pores actifs, on peut aller differentier la permeabilite des pores de leur taille: cf figure dans les applications }

In line with our previous discussion of osmotic phenomena, it is particularly fruitful to address the question under the perspective of the mechanical insights, where the pore with fluctuating shape is modeled as a time-dependent energy barrier, say ${\cal U}(x,t)$, using similar notations as previously.  %here with active pores with fluctuating shapes.  
The average osmotic force acting on the fluid is again obtained in terms of the force acting on the fluid integrated over the channel size and averaged over time. It writes within this framework
\begin{equation}
\Delta \Pi_{\rm app}=\langle \int dx\, c_s(x,t)\times (-\nabla_x{\cal U}(x,t))\rangle_t.
\end{equation}
where here $\langle \cdot\rangle_t$ denotes a time average. 
As for the static (passive) case, the Smoluchowski equation for the solute allows one to rewrite the apparent osmotic pressure in terms of the solute flux across the fluctuating barrier: 
\begin{equation}
\Delta \Pi_{\rm app}=k_BT \Delta c_s + \langle j_s\rangle_t \times{L\over D_s}.
\label{Piapp_act}
\end{equation}
It is interesting to note that the concept of osmotic force $\Delta \Pi_{\rm app}$ connects directly to the question of translocation of solute molecules across a fluctuating barrier -- via the solute surface flux $j_s$. 
That specific question was actually the topic of an exhaustive literature in the context of ratchets, molecular motors, or stochastic resonance~\cite{gammaitoni1998stochastic,reimann2002introduction}. 
Numerous counter-intuitive consequences were highlighted, both theoretically and experimentally, like directed motion, \qote{uphill} transport against gradients, enhanced translocation, \textit{etc}. Accordingly the previous symmetry relation \eqref{Piapp_act} shows that the existence of a finite flux $ \langle j_s\rangle_t$, with possibly unconventional dependencies on the concentrations in the reservoirs, will have consequences on osmotic transport, {\it i.e.} leading to flow of the suspending fluid itself and not only solute motion.

\begin{figure}[h!]
\centering
  \includegraphics[width=0.49\textwidth]{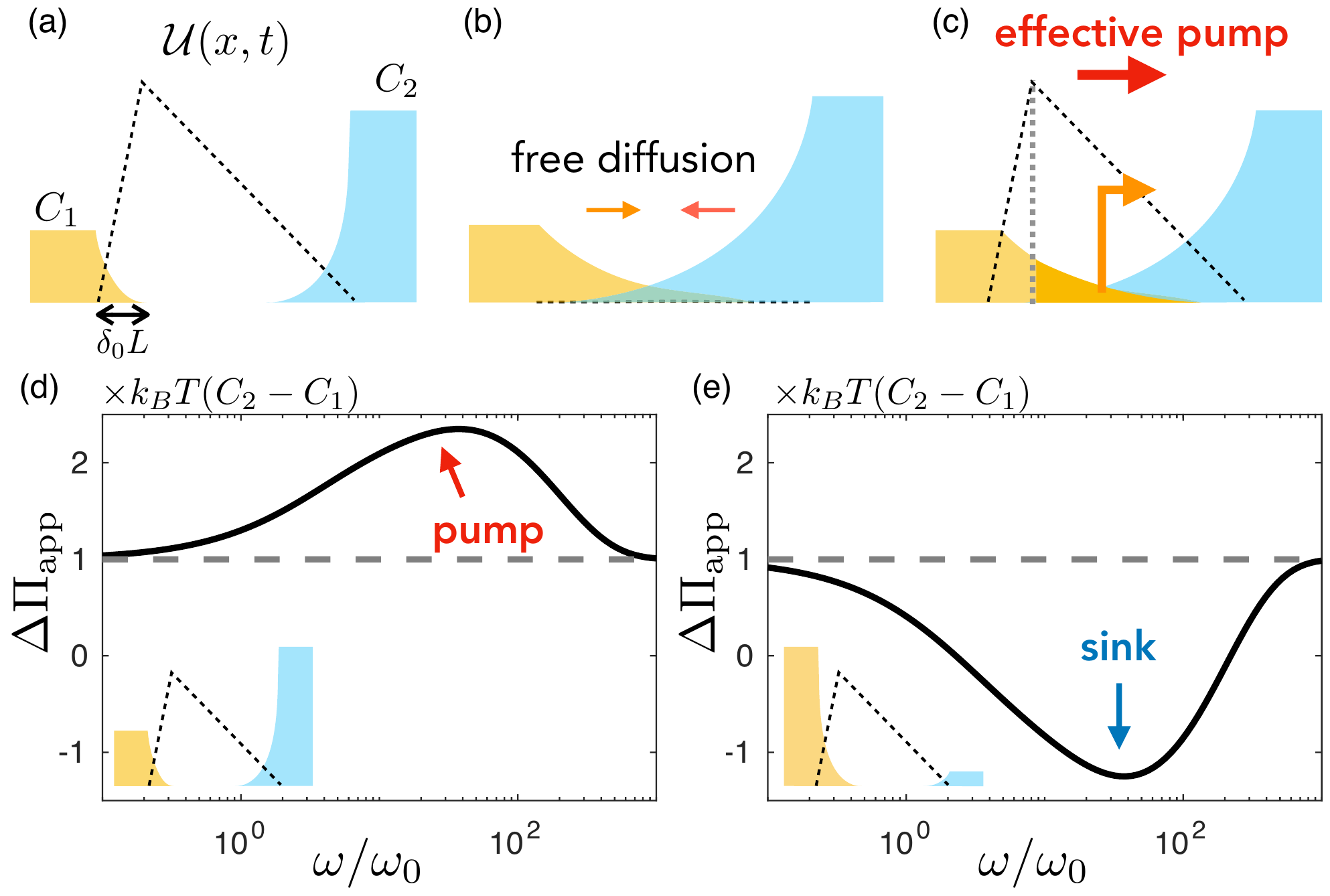}
  \caption{\textbf{: Active membrane as a pump (or sink).} (a) An asymmetric potential barrier (representing the membrane acting on the solute) separates two solute reservoirs with different concentrations. (b) As the barrier is temporarily lowered, the solute may diffuse inwards from both sides. (c) If the barrier is risen back, the solute that has diffused beyond the maximum point of the energy barrier will be carried to the other side. In the example shown here, more solute from the lower concentration side has traversed beyond the maximal point. This solute is then transported  to the highly concentrated side. The active membrane therefore acts as a pump. (d) and (e) Apparent rejection coefficient as obtained from the on/off energy barrier model described in the text~\cite{Marbach2019}.  %by Eq.~\ref{eq:sigmaApp} 
  %with an on/off barrier at frequency 
The normalizing frequency is  $\omega_0=D_s/L^2$, and the asymmetry parameter is here $\delta_0 = 0.1$. Insets indicate the solute concentration on both sides; $\Delta c_s = C_2 - C_1$; for (d), $C_1 = 0.4 C_0$ and $C_2 = C_0$ where $C_0$ is some arbitrary concentration and for (e) $C_1 = C_0$ and $C_2 = 0.1 C_0$.  } 
  \label{fig:activeOsmosis}
\end{figure}

To illustrate this behavior, it is instructive to consider a simple example, made of an asymmetric membrane as in Fig.~\ref{fig:activeOsmosis}, which oscillates in time as an \qote{on/off} process over a time interval $\tau/2 = \pi/\omega$. When \qote{on}, the barrier height is considered as much larger than the thermal energy.
%the barrier is \textit{on}, and for the same time the barrier is \textit{off} meaning $\mathcal{U}(x) \equiv 0$. 
This process bares similarities with the ratchet process in \citefull{rousselet1994directional} and subsequent references, where solute pumping was demonstrated. 
%Considering a simple model of active osmosis allows to highlight its potential for separation and filtration. We take a simple asymmetric membrane represented by an energy barrier $\mathcal{U}(x)$. This profile is typically increasing to a maximum over a length $\delta 0 \times L$ then decreasing over $(1-\delta_0) \times L$ (see Fig.~\ref{fig:activeOsmosis}-a), with $L$ the thickness of the membrane. 
%We now oscillate the barrier in the following way. For a time $T/2 = \pi/\omega$ the barrier is \textit{on}, and for the same time the barrier is \textit{off} meaning $\mathcal{U}(x) \equiv 0$. 
In elementary terms, when the barrier is \qote{off}, solute molecules from both sides diffuse freely (see Fig.~\ref{fig:activeOsmosis}-b). Now, when the barrier is back \qote{on}, solute that has crossed the maximum point of the barrier will slide down to the opposite side (see Fig.~\ref{fig:activeOsmosis}-c).
This process leads to a finite flux of solute averaged over a period,  $\langle j_s\rangle_t$, which can be exactly calculated in the simple model considered, see \citefull{Marbach2019}. For example, in the quasi-static (low frequency) regime, 
the averaged flux reduces in this simple system to 
\begin{equation}
\langle j_s\rangle_t \underset{\omega\rightarrow 0}{\sim} L \omega \times \left[C_2 \times \delta_0-C_1 \times(1-\delta_0)\right],
\end{equation}
with $C_1$ and $C_2$ the solute concentrations in both reservoirs and $\delta_0$ the potential asymmetry, see Fig.\ref{fig:activeOsmosis}. 
%prefactor ${2\over \sqrt{\pi}}$ a verifier
Beyond the specific expression (restricted to this specific regime and model), this result highlights the possibility of uphill solute transport (pumping against concentration gradients) or enhanced solute flux, depending on the direction of the concentration gradient versus the pore asymmetry. 
%the averaged particle velocity is expected to be non-vanishing and scale simply as $V_p\sim -\epsilon L/T$, where $\epsilon=1/2-\delta_0$ characterizes the barrier asymmetry. The corresponding convective flux behaves as $J_c \sim \bar{C} \times V_p$ with $\bar{C}=(C_1+C_2)/2$ the averaged concentration. %On the other hand, the diffusive contribution to the flux behaves as $J_{\rm diff} \sim -D/L (C_2-C_1) \times (T_0/T)$ where the last factor accounts for the fraction of time where the pore is permeable.
This behavior is summarized in Fig.~\ref{fig:activeOsmosis}-d-e.

Now inserting this result for the flux in \eqref{Piapp_act}, one therefore predicts highly counter-intuitive behaviors for the osmotic pressure, {\it i.e.} the driving force acting on the fluid. In Fig.~\ref{fig:activeOsmosis}-d-e, we have introduced and plotted the apparent osmotic pressure
\begin{equation}
\Delta \Pi_{\rm app} = k_BT(C_2-C_1) + {k_B T L \over D_s}\times\langle j_s\rangle_t (C_1,C_2, \omega) 
\end{equation}
based on the full solution of the simplistic previous model. Notably this plot highlights the possibility of \qote{resonant osmosis} for a characteristic frequency (in the form of an extremum of $\sigma_{\rm app}$), but this simple model also suggests that -- depending on the direction of the potential asymmetry against the concentration gradient -- the rejection may be larger than unity (pumping regime) or even decrease towards negative values~\cite{Marbach2019}. 
% Simplifying and assuming that solute from both sides do not interact, we can write the apparent rejection coefficient as
% \begin{align}
%\label{eq:sigmaApp}
% \sigma_{\rm app}[\omega,C_1,C_2]=&1+{1\over C_2-C_1}\sqrt{{  \omega \over 2 \omega_0}}\times \biggl[ ...
%\nonumber \\
%&C_1 \left(\Psi\left({1\over\sqrt{2}}\sqrt{{\omega\over \omega_0}}\right)-\Psi\left({\delta_0\over\sqrt{ 2 }}\sqrt{{\omega\over \omega_0}}\right)\right)
% \nonumber \\
%&
%-C_2 \left(\Psi\left({1\over\sqrt{2}}\sqrt{{\omega\over \omega_0}}\right)-\Psi\left({1 - \delta_0\over\sqrt{ 2}}\sqrt{{\omega\over \omega_0}}\right)\right)\biggr] 
%\end{align}
%with
%\begin{equation}
%\Psi(x)=\left[1-{1\over 2}\left(1+{\rm Erf} (x) \right)
%+{1\over 2}\left(1+{\rm Erf} (-x)\right)
% \right] 
%\end{equation}
%and $\omega_0 = 2\pi/\tau_0$ with $\tau_0 =  L^2/D$ and $D$ diffusion coefficient of solute. DISCUSS
%\LB{Voir si approche simplifi\'ee possible, cf book by Nelson p392}

 % was investigated recently ~\cite{marbach2017active,marbach2018transport}
%The concept of sieving through a dynamical pore was recently investigated theoretically by the authors~\cite{marbach2017active,marbach2018transport}. A dynamical pore introduces a new characteristic timescale in the system (describing the dynamics of the pore). Now the timescale of translocation of a solute particle through the pore may be compared to this new timescale, and determine transport -- connecting to stochastic resonance, although the processes may be different.

This points to a wealth of intriguing behaviors for osmotic phenomena, which were barely explored up to now. 
As emphasized above, 
the recent development of nanofluidics  %~\cite{siria2013giant,feng2016single,radha2016molecular} now 
suggests many routes to develop such active pores in artificial channels, {\it e.g} using voltage-gated nanochannels~\cite{karnik2005electrostatic,kalman2009control,guan2011field,zhou2018electrically} or UV light~\cite{moorthy2001active} or stimulated surface chemical reactivity~\cite{zhang2013bioinspired,liu2018universal}. %In Fig.~\ref{fig:activeFiltration2}-b we present a micro to nano-door, a mechanical gate, whose opening and closing may be imposed by an external forcing oscillator such as a piezo-electric. 
Other externally controlled existing devices include thermally responsive nanochannels~\cite{liu2017bioinspired,hu2015photothermal}. The challenge awaiting is to achieve such active control in yet smaller devices to significantly impact water or ion transport. 

The foundations of active osmosis remain therefore to be properly investigated. The present discussion is merely an appetizer to illustrate the abundance of \qote{exotic} behaviors which could be unveiled in this context. 
%tbut they constitute promising routes for the design of artificial pumps. 

\vskip0.5cm
\section{From diffusio-phoresis of particles to active matter}
\label{sec:DP}

The previous sections showed how gradients of solutes induce fluid motion in the presence of an interface via the diffusio-osmotic phenomenon. Symmetrically when a (solid) particle is suspended in a quiescent fluid, gradients of solute will induce motion of the particle. This motion, called \qote{diffusio-phoresis}, relies on the very same osmotic driving forces, occurring within the interfacial layer at the particle boundary. 
The idea to transport large particles harnessing osmotic forces appeared first in the Russian literature with the works of Derjaguin and Dukhin 
\cite{derjaguin1993kinetic,derjaguin1993diffusiophoresis,derjaguin1987some} and was more thoroughly investigated in the 1990s~\cite{prieve1987diffusiophoresis,ebel1988diffusiophoresis,kosmulski1992solvophoresis}. We refer to the review by Anderson in \citefull{anderson1989colloid} for a dedicated discussion of the underlying transport mechanisms and some of its subtle features. % ({\it e.g.} their force-free nature).

Diffusio-phoresis and its consequences have gained renewed interest for the last decade, highlighting an increasing number of situations where this phenomenon is shown to play a role, as well as dedicated applications in various domains. Basically diffusio-phoresis occurs whenever there is a gradient of solute or of a mixture of solutes and such situations are ubiquitous \cite{velegol2016origins,moller2017steep}. Here
we summarize the main elements of the phenomenon and focus on a number of elementary implications.
% several manifestation of the phenomena  
More explicit applications will be discussed in the next section, Sec.~\ref{applications}.

%\subsection{diffusio-phoretic transport harnessing osmotic forces}
\begin{figure}[h!]
\centering
  \includegraphics[width=0.49\textwidth]{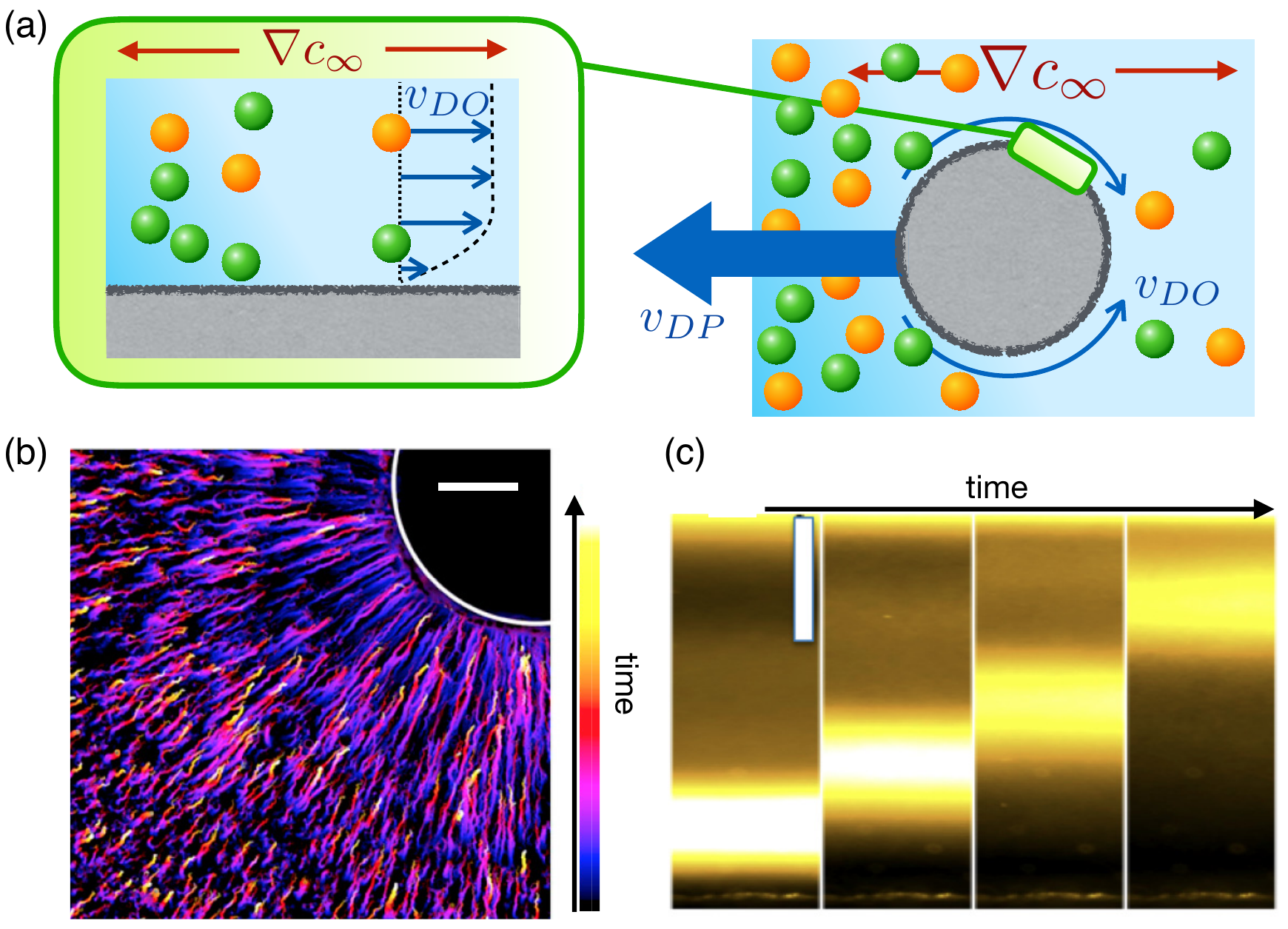}
  \caption{\textbf{: From osmosis to phoresis.} (a) Under a concentration gradient, a particle is put into motion via 
  diffusio-osmosis occurring at its surface. 
  (b) \SM{Time-stamped stream lines of decane droplet migration towards a hydrogel beacon initially loaded with sodium dodecyl sulfate (SDS), acting as a long-range solute source. Adapted from \citefull{banerjee2016soluto} with permission from the United States National Academy of Science, copyright 2016. The scale bar is 100 $\mathrm{\mu}m$. (c) Diffusio-phoretic transport of fluorescent  $\lambda$-DNA under a LiCl gradient ($\Delta c_s = 100 mM$ over a range of $800 \mathrm{\mu}m$, highest concentration being up), scale bar is 100 $\mathrm{\mu m}$.  Images at 100, 150, 200 and 300 s. Adapted from \citefull{palacci2012osmotic}.}
}
  \label{fig:OsmoToPhoresis}
\end{figure}

\subsection{{\it E pur si muove}: from diffusio-osmosis to diffusio-phoretic motion}
%\subsection{{\it E pur si muove}: Osmotic Force balance and diffusio-phoretic motion}

The diffusio-phoretic velocity of a particle under a (dilute) solute gradient writes as \cite{anderson1989colloid}
\begin{equation}
\bm{v}_{DP} = \mu_{DP}  \times (-k_BT\bm{\nabla} c_\infty)
\label{eq:PhoresisDOneutral}
\end{equation}
Physically the phenomenon at stake is sketched in Fig.~\ref{fig:OsmoToPhoresis}: the solute gradient at the solute surface induces a diffusio-osmotic slip velocity of the fluid (relative to the solid particle) beyond the interfacial diffuse layer; the particle is put in motion in order to precisely negate the corresponding velocity.  %As for all interfacially driven transport, diffusio-phoretic motion is {\it force free}.
For spherical particles, the value of the mobility $\mu_{DP}$ defined in \eqref{eq:PhoresisDOneutral} identifies with {\it minus}  the corresponding diffusio-osmotic mobility, as given previously in Sec.~\ref{EO-DO}: 
\begin{equation}
\mu_{DP}=-\mu_{DO}.
\end{equation}
For example, for a solute interacting via a potential $\cal U$ with the particle, 
the diffusio-phoretic mobility is minus the mobility in \eqref{KDO}:
\begin{equation}
\mu_{DP}= -  \mu_{DO} = - \frac{1}{\eta}\int_0^\infty  \, z \, \left(\exp\left(\frac{-\UU(z)}{k_B T}\right) - 1\right) dz.
\label{KDP}
\end{equation}
%where $\UU$ is the potential of interaction between the solute and the particle, with typical range $\lambda$
Interestingly, provided the value for the diffusio-osmotic mobility is constant over the particle's surface, it was shown by Morrison that this result holds for any particle shape (the argument is valid for any interfacially driven transport~\cite{morrison1970electrophoresis,anderson1989colloid}).

\subsubsection{Phoresis in the thin layer limit} 
Summarizing briefly the derivation, \eqref{KDP} is obtained by separating the diffusio-osmotic driving, which occurs at the particle surface within the diffuse layer of thickness $\lambda$, and the far-field flow occurring beyond the diffuse layer. Following the prediction in Sec.~\ref{EO-DO}, 
the near-field diffusio-osmotic flow results in a  diffusio-osmotic \qote{slip} velocity of the fluid relative to the particle, with amplitude $\mu_{DO}  \times (-k_BT\bm{\nabla} c_s)_t$,  where the index $t$ refers to the tangential component along the particle surface.
The concentration gradient in the vicinity of the surface $(\bm{\nabla} c_s)_t$ is related to the far concentration field $c_s$. It obeys Ficks' equation $\nabla^2 c_s=0$ together with the boundary condition at infinity fixing the concentration gradient, $c_s(r\rightarrow \infty)\simeq z \nabla c_\infty$, with $z$ the coordinate along the direction of the gradient. This gives (in spherical coordinates):
\begin{equation}
c_s(r,\theta)= R\, \nabla c_\infty \left( {r\over R}+ {1\over 2} \left(R\over r\right)^2\right) \times \cos\theta
\label{conc}
\end{equation}
outside of the diffuse layer; with $r,\theta$ the spherical coordinates.
Back to the flow, the velocity field outside the diffuse layer obeys the Stokes equation 
%with the proper boundary conditions on the surface and at infinity
\begin{equation}
\eta \bm{\nabla} \bm{v} - \bm{\nabla} p =0, 
\end{equation}
with the boundary conditions on the particle given by the tangential slip velocity and at infinity given by the uniform flow field
%\begin{eqnarray}
%&\mathbf{v} (r=R^+)=v_{slip}\, \mathbf{t}\, \nonumber \\
%&\mathbf{v}(r\rightarrow \infty)= -\bm{v}_{DP}
%\end{eqnarray}
\begin{equation}
\bm{v} (r=R^+)= - \sopp{\frac{3 \sin \theta}{2}} v_{slip}  \, \bm{t}, \,\,\;\&\,\,\; \bm{v}(r\rightarrow \infty)= -\bm{v}_{DP}
\end{equation}
in the particle frame of reference; \sopp{$v_{slip} = \mu_{DO} \times (- k_B T \nabla c_{\infty})$}, $R^+\approx R+\lambda$ denotes the position on the particle surface but located beyond the diffuse layer (here considered as infinitesimal); $\bm{t}$ is the tangential vector on the particle's surface. 
%with the boundary condition on the particle given by the tangential slip velocity $v_{slip}$ and at infinity $\mathbf{v}(r\rightarrow \infty)= -\bm{v}_{DP}$. 
To lowest order, the solution to the previous equations results in a flow dominated by a Stokeslet
\begin{equation}
\begin{split}
v_r &= -v_{DP}\cos\theta + {F\over 4\pi \eta} {\cos \theta \over r} + {\cal O} \left( {1\over r^3}\right) \\
v_\theta &= v_{DP}\sin\theta-{F\over 8\pi \eta}{\sin \theta \over r} + {\cal O} \left( {1\over r^3}\right) % {\sin \theta \over r} + {\cal O} \left( {1\over r^3}\right)
\end{split}
\label{Stokes}
\end{equation}
where  $F=6\pi \eta R ({v}_{DP}+v_{slip})$. As can be easily verified, $F$ identifies with (minus) the force acting on the particle. \SM{At steady-state, the particle is moving with a constant velocity $v_{DP}$ and no force acts on it.} Accordingly, 
the diffusio-phoretic velocity $\bm{v}_{DP}$ is fixed by imposing $F=0$ and this results in \eqref{eq:PhoresisDOneutral}.
%that 
%the total force on the particle is zero: $F=0$, {\it i.e.} the prefactor of the $1/r$ term of the velocity profile -- which identifies with the force on the particle, as can be verified from the force calculated in the far-field -- does vanish. 
This shows implicitly that the far velocity profile scales like $1/r^3$ and can be rewritten as in \citefull{anderson1989colloid}
\begin{equation}
\bm{v}_l({\bm r})={1\over 2} \left({R\over r}\right)^3 \left[3 {{\bm r}{\bm r}\over r^2}-\bm{I}\right]\cdot \bm{v}_{DP}.
\end{equation}
where we came back to the lab frame of reference, $\bm{v}_l({\bm r})=\bm{v}({\bm r})-\bm{v}_{DP}$. 
Accordingly, the hydrodynamic interaction between particles undergoing diffusio-phoretic transport is weak, in contrast to {\it e.g.} gravity driven transport where the fluid velocity scales like $1/r$ far from the particle. This has important consequences for the phoretic transport in confinement \cite{anderson1989colloid,rallabandi2019motion,chamolly2017active}.

{
\subsubsection{ Osmotic force balance on particles } 
\label{discussionforce}
Let us come back to the force balance underlying diffusio-phoresis. We emphasized above that diffusio-phoresis, like any interfacially driven transport, is a force-free motion: the particle moves without any force acting on it, {\it i.e.} \SM{the global resulting force acting on the particle vanishes}~\cite{anderson1989colloid}. 
This has counter-intuitive consequences and led to various mis-interpretations and debates concerning osmotically-driven transport of particles~\cite{cordova2008osmotic,julicher2009comment,fischer2009comment,PhysRevLett.103.079802,PhysRevLett.102.159802,brady2011particle}, in particular in the context of phoretic self-propulsion (see Sec.~\ref{active} and \citefull{moran2017phoretic}). We thus take the proper space here to discuss the osmotic force balance.

A naive interpretation of diffusio-phoresis is that the particle velocity $v_{DP}$ under a solute gradient results from the balance of Stokes' viscous force $F_v=6\pi\eta R v_{DP}$ and the osmotic force resulting from the gradient of the osmotic pressure integrated over the particle surface, hypothetically scaling as $F_{osm} \sim R^2\times R\nabla \Pi$. Balancing the two forces one finds a phoretic velocity behaving as $v_{DP} \sim R^2 {k_BT\over \eta} \nabla c_{s}$. Looking at the expression for the diffusio-phoretic mobility in the thin layer limit, \eqrefs{eq:PhoresisDOneutral} and~\noeqref{KDP}, the latter argument does not match the previous estimate  by a factor of order $(R/\lambda)^2$, where $\lambda$ is the range of the potential of interaction between the solute and the particle. 
 
 The difference between the two scalings originates %origin of this discrepancy actually takes its root 
 in the fact that for interfacially driven motion, the velocity gradients occur mostly over the thickness $\lambda$ of the diffuse layer, and not on  the particle size $R$, as {\it e.g.} for the Stokes flow. More fundamentally, this raises the question of how osmotic pressure is expressed: the existence of a difference of solute concentration between the two sides of the colloidal particle does not obviously imply the existence of a corresponding osmotic pressure and this belief led to much confusion. The argument above based on the global force balance is globally flawed and needs to be properly clarified. In his exhaustive work in \citefull{brady2011particle} following the debate of \citefulls{cordova2008osmotic,julicher2009comment,fischer2009comment,PhysRevLett.103.079802,PhysRevLett.102.159802}, Brady  tackled the question based on a \qote{micromechanical} analysis of the solute and solvent transport in the presence of the colloidal particles. 
 %\LB{However a proper }
 
\LB{However}, in order to properly solve the riddle and reconcile the various approaches, one needs to go into the details of the force distribution and write properly the force balance on the particle undergoing diffusio-phoretic transport. % and identify the mechanical origin of the osmotic pressure.
It is accordingly interesting to relax the hypothesis of a thin diffuse layer, and consider more explicitly the transport inside the diffuse layer, as was explored by various authors, using {\it e.g.} controlled asymptoptic expansions \cite{sabass2012dynamics,sharifi2013diffusiophoretic,cordova2013osmotic}. 

%As we now discuss,  
General results for the hydrodynamic flow and mobility can be obtained {\it without assuming a thin diffuse layer}.
%\LB{extending on similar work performed for electrophoresis \cite{ohshima1983approximate}}.
% was done for electrophoresis in the case of 
%\LB{citer Ohshima et al. \cite{ohshima1983approximate}: electrophoresis, weak potential but no thin EDL}
%
We consider that the interaction between the solute and the particle occurs via a radially symmetric potential ${\cal U}(r)$, so that the Stokes equations now writes
 \begin{equation}
\eta \bm{\nabla^2} \bm{v} - \bm{\nabla} p +c_s(\bm{r}) (-\bm{\nabla} {\cal U})=0.
\label{Stokesplus}
\end{equation}
The boundary conditions are now replaced by the no-slip boundary condition {\it on the particle's surface}, as well as the prescribed velocity at infinity (in the frame of reference of the particle):
%\begin{eqnarray}
%&\mathbf{v} (r=R)=0, \nonumber \\
%&\mathbf{v}(r\rightarrow \infty)= -\bm{v}_{DP}
%\end{eqnarray}
\begin{equation}
\bm{v} (r=R)= \bm{0} \,\,\;\mathrm{and}\,\,\; \bm{v}(r\rightarrow \infty)= -\bm{v}_{DP}
\end{equation}
The concentration profile obeys a Smoluchowski equation in the presence of the external potential ${\cal U}(r)$, in the form
\begin{equation}
0=-\bm{\nabla}\cdot \left[-D_s\bm{\nabla} c_s + \lambda_s c_s (-\bm{\nabla} {\cal U})\right]
\end{equation}
with the boundary condition at infinity accounting for a constant solute gradient $c_s(r\rightarrow \infty)\simeq r\cos\theta \nabla c_\infty$ (convective transport is neglected here). Given the symmetry of the problem, the solution takes the general form $c_s(r,\theta)=c_0(r)\cos\theta$, with $c_0$ scaling with the gradient at infinity as $c_0\propto R\nabla c_\infty$. Altogether this is a self-consistent equation for the solute concentration field. It should therefore be considered as a source term 
%an input data 
for the fluid transport \eqref{Stokesplus}. For large distances to the particle ($r\gg \lambda$), 
 it reduces to the previous result in \eqref{conc}. %for distances to the particle beyond the potential range. 

Interestingly, the solution of \eqref{Stokesplus} for the velocity profile can be calculated {\it exactly}  for any radially symmetric potential ${\cal U}(r)$, by extending textbook techniques for the Stokes problem in \citefull{happel2012low}; see also \citenum{ohshima1983approximate} for a related calculation in the context of electro-phoresis. It can be demonstrated that the solution for $\bm{v}(\bm{r})$ still takes the same form as in \eqref{Stokes}, but the force along the axis of the gradient appearing in the Stokeslet term ($v\sim F/r$) term now takes the expression
\begin{equation}
F= 6 \pi R \eta v_{DP} 
-  \pi R^2  \int_R^{\infty} c_0(r) (-\partial_r{\cal U})(r) \times \varphi(r) dr 
%&&+\frac{2}{3} \pi R^3    \int_R^{\infty} \frac{c_0(r) (-\partial_r{\cal U}(r)}{ r} dr + \frac{4}{3} \pi   \int_R^{\infty}r^2 c_0(r) (-\partial_r{\cal U}(r) dr 
%&F= &6 \pi R \eta v_{DO} - 2 \pi R  \int_R^{\infty} c_0(r) (-\partial_r{\cal U}(r) r dr \nonumber\\
%&&+\frac{2}{3} \pi R^3    \int_R^{\infty} \frac{c_0(r) (-\partial_r{\cal U}(r)}{ r} dr + \frac{4}{3} \pi   \int_R^{\infty}r^2 c_0(r) (-\partial_r{\cal U}(r) dr 
\label{FDO}
\end{equation}
with $\varphi(r)={2 \over 3}\left(3{r\over R} -2 \left({r\over R}\right)^2 - {R\over r}\right)$ a dimensionless function, the factor ${2 \over 3}$ originating from the angular average. 
The diffusio-phoretic velocity results from the force-free condition, $F=0$, and therefore it writes
%If we require that the total force on the sphere vanish, $F_x = 0$ then we have
\begin{equation}
v_{DP} = \frac{\pi R^2}{6\pi\eta R}   \int_R^{\infty} c_0(r) (-\partial_r{\cal U})(r) \times \varphi(r) dr 
\label{VDPexact}
%\int_R^{R+\lambda} \frac{c(r,\theta)}{\cos\theta} \partial_r(-\mathcal{U}) r dr -  \frac{R}{9\eta} \int_R^{R+\lambda} \frac{c(r,\theta)}{r \cos\theta} \partial_r(-\mathcal{U}) dr -   \frac{2}{9\eta} \int_R^{R+\lambda} \frac{c(r,\theta)}{ \cos\theta}  \partial_r(-\mathcal{U}) r^2 dr}
\end{equation} 
Remembering that $c_0(r) \propto R\nabla c_\infty$, this equation generalizes the previous result obtained in the thin layer limit. 

At first sight, \eqrefs{FDO} and \eqref{VDPexact} appear as a force balance between the Stokes friction $6 \pi R \eta v_{DP} $ and the osmotic force, here written in terms of the local force $c_0(r) (-\partial_r{\cal U})(r)$ integrated over the particle surface (and potential range). The latter represents the push of the solute molecules on the particle.
Actually, \eqref{VDPexact} is very similar to Eq.~(2.7) in \citefull{brady2011particle}, with the $r$-dependent term $\pi R^2\times \varphi(r)$ replaced in \citefull{brady2011particle} by the prefactor $L(R)$. 
However the integrated \qote{osmotic push} is weighted here by the local factor $\varphi(r)$ (in contrast to \citefull{brady2011particle}) and this detail actually changes the whole scaling for the mobility. 

Indeed in the thin diffuse layer limit, with $r-R \sim\lambda \ll R$, then one may expand $\varphi(r) \simeq -2 (r-R)^2/R^2$, while the concentration profile $c_0(r)$ can be approximated as 
\begin{equation}
c_0(r)\simeq R \nabla c_\infty \times \left[{r\over R} + {1\over 2} \left({R\over r}\right)^2\right] \exp[-{\cal U}(r)/k_BT].
\end{equation}
One may then verify that the above \eqref{VDPexact} indeed reduces to the results in \eqrefs{eq:PhoresisDOneutral} and \noeqref{KDP} predicted by the thin layer approach. In other words, the weight $\varphi(r)\sim \lambda^2/R^2$ is a signature of the fact that the velocity gradients occur on the potential width $\lambda$ and not on the particle size $R$. An osmotic pressure is indeed expressed at the particle's surface and yields diffusio-phoretic transport, but in a very subtle way which does not reduce to considering only the direct solute force.
 %to the subtle force balance underlying the diffusio-phoretic transport, which 
 This corrects the naive argument suggested at the beginning of the section.

The exact calculation above also allows one to gain key insight into the {\it local} force acting on the particle. The latter is the sum of the hydrodynamic shear force, normal pressure and direct interaction with the solute. Using the exact results for the velocity profile in  the thin layer regime, $\lambda\ll R$, one predicts
%Now expanding and simplifying yields the force over the solid angle
\begin{equation}
\begin{split}
f_r &=  3 L_s R^2  k_B T \nabla c_{\infty} \cos \theta \\
f_{\theta} &= \frac{3}{2}L_s  R^2 k_B T \nabla c_{\infty} \sin \theta 
\end{split}
\label{eqForceLocale}
\end{equation}
where 
$L_s = \int_R^{\infty} \left( e^{-\beta \mathcal{U}(x)} - 1\right) dx$ has the dimension of a length and quantifies the excess adsorption on the interface. {\eqref{eqForceLocale} can be recovered easily with a simplistic argument: one expects this osmotic force to scale as $\mathcal{V}_{\rm int} \nabla \Pi = \mathcal{V}_{\rm int}  \nabla (k_B T c_{\infty})$ where $\mathcal{V}_{\rm int} $ is the interaction volume. Writing $L_s$ the typical interaction lengthscale we have $\mathcal{V}_{\rm int} \approx 4\pi R^2 L_s$, leading accordingly to \eqref{eqForceLocale}}. While the integrated total force does vanish as expected, the osmotic gradients do generate an inhomogeneous local tension on the surface of the particle, as plotted in Fig.~\ref{fig:sphereForce}-a. Accordingly, if one considers that the particle is elastically deformable, such tensions would generate a deformation of the particle following the shape sketched in Fig.~\ref{fig:sphereForce}-b.
\begin{figure}[h!]
\centering
 \includegraphics[width=0.45\textwidth]{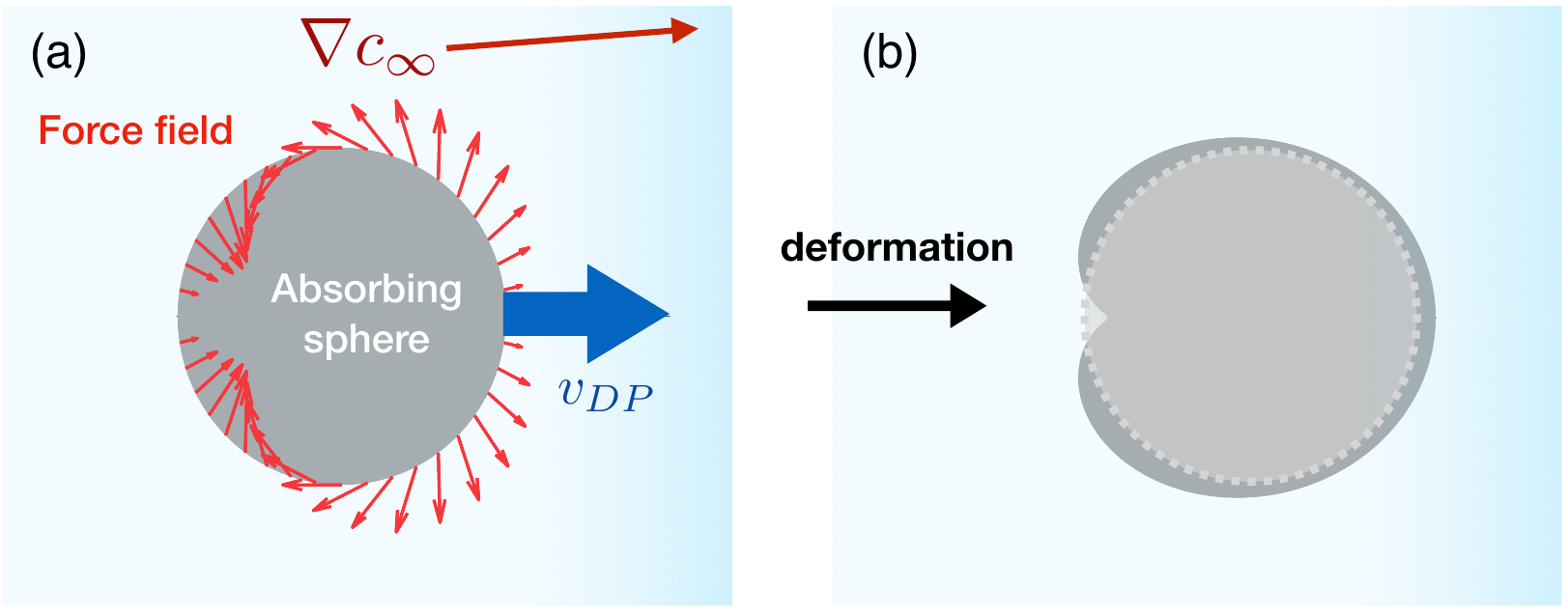}
 \caption{\textbf{: Local force acting on a diffusio-phoretic sphere.} (a) Local force field acting on a sphere experiencing diffusio-phoresis with absorption at its surface in a solute gradient. The local force is plotted with an arbitrary factor amplitude factor (the same for each vector) and projected in the 2D plane; (b) Potential resulting deformation of the sphere, axisymmetric view, when the deformation is assumed to be proportional to the local force. }
 \label{fig:sphereForce}
\end{figure}
%Note that $\Delta \Pi = L_s k_B T \nabla c_{\infty}$

The situation is very different for  electro-phoretic transport. As was first discussed in \citefull{long1996simultaneous}, for electro-phoresis there is a {\it local} force balance between the direct electric force acting on the particle and the hydrodynamic shear acting on its surface:
accordingly the local force simply vanishes identically. In physical terms, this is due to the fact that the electric force acting on the colloid particle exactly balances the electrical force acting on the electric double layer because of local electroneutrality (the charge in the electric double layer being exactly opposite to the surface charge). 
This can be actually verified explicitly by extending the previous calculations to electro-phoresis. This can be performed for weak electrostatic potential along the lines in Ref.\cite{ohshima1983approximate}, and it predicts indeed a vanishing local force.

Accordingly,  particles undergoing electro-phoresis are not expected to deform, in contrast to diffusio-phoresis which leads to local deformations. Such results remain to be experimentally studied in order to observe the modification of a particle conformation undergoing diffusio-phoretic drift. We note however that in the context of thermo-phoresis, 
DNA was reported to stretch under a temperature gradient ~\cite{jiang2007stretching}.
Such effects could have interesting applications in the context of separation of particles, since their shape will differ depending on their size. 

{As a last comment, the previous discussion neglected surface transport at the surface of the particle: this involves convection of solute in the interfacial region, but also fluid slippage at the particle surface, which will both affect the steady-state concentration field of the solute around the particle. This leads to corrections to the mobility as a function of a (properly defined) P\'{e}clet number, as introduced in \citefulls{anderson1991diffusiophoresis, ajdari2006giant,sabass2012dynamics,michelin2014phoretic}.}

\subsubsection{The diffusio-phoretic mobility} 
Let us now focus on the mobility. 
As for diffusio-osmosis, the diffusio-phoretic mobility scales as $\mu_{DP} \approx {\pm\, \lambda^2  / \eta}$ where $\lambda$ is the thickness of the interfacial layer. For electrolytes, the latter identifies with the Debye layer thickness and one expects accordingly that $\mu_{DP} \sim 1/c_s$ so that one usually writes the diffusio-phoretic velocity under salt gradients as:
\begin{equation}
\bm{v}_{DP} = D_{DP}  \bm{\nabla}\log c_s
\label{eq:PhoresisDO}
\end{equation}
The diffusio-phoretic mobility $D_{DP}$ has now the dimension of a diffusion coefficient. According to the previous estimates, one expects for electrolytes that $D_{DP} \approx k_BT/(8\pi\eta \ell_B)$ with $\ell_B$ the Bjerrum length, so that the value for $D_{DP}$ is expected to be in the range -- though slightly smaller -- of   diffusion coefficients of molecules (thus far larger than any colloid diffusion coefficient): experimentally typical values for $D_{DP}$ are in the range $D_{DP} \sim 0.1-1$~10$^{-10}$m$^2/$s \cite{abecassis2008boosting,palacci2010colloidal}.
Note that, as for diffusio-osmosis, diffusio-phoresis under electrolyte gradients with unequal diffusion coefficients for the anions and cations ($D_+\ne D_-$) has an electro-phoretic contribution similar to \eqref{DOdiff} which can become quantitatively predominant; see also \citefull{anderson1989colloid}.
The expression in \eqref{eq:PhoresisDO} highlights a \qote{log-sensing} behavior, similar to that observed in bacteria, {\it e.g.} in \textit{E. coli}~\cite{kalinin2009logarithmic}. It is at the basis of various unconventional transport phenomena which we discuss below.
%bares similarities with log sensing

Aside the case of electrolytes, other classes of relevant interactions involve steric exclusions -- {\it e.g.} for neutral polymers -- for which the mobility is expected to scale as $\mu_{DP} =  R_p^2/\eta$ with $R_p$ the excluded particle diameter of the solute \cite{anderson1989colloid,jiang2009manipulation,sear2017diffusiophoresis}.

On the experimental side, diffusio-phoresis has been investigated in numerous studies. First measurements were performed by the Russian school~\cite{derjaguin1993diffusiophoresis}, and later by Prieve, Anderson and collaborators in the 90's   \cite{prieve1987diffusiophoresis,ebel1988diffusiophoresis}. However the development of microfluidics over the last two decades has allowed to develop dedicated systems in which concentration gradients are perfectly controlled and tunable. It was then possible to measure diffusio-phoretic motion and obtain further insights into the phenomenon and its consequences \cite{abecassis2008boosting,palacci2012osmotic,palacci2010colloidal,paustian2013microfluidic,nery2017diffusiophoresis,paustian2015direct,kosmulski1992solvophoresis,shin2017membraneless,guha2017chemotaxis,florea2014long}.

While the above discussion assumed implicitly a dilute solute, the phenomenon is expected to occur under gradients of mixtures, and is denoted as solvo-phoresis \cite{kosmulski1992solvophoresis}. This was for example investigated in a recent study by 
Paustian et al. \cite{paustian2015direct}, who showed that polystyrene colloids undergo motion in gradients of water-ethanol mixtures. 
The velocity of the particles was shown experimentally to obey
\begin{equation}
\bm{v}_{DP}= D_{DP} \bm{\nabla} \log X
\end{equation}
where $X$ is the ethanol {\it molar fraction}, thus pointing to non-ideality effects. It would be interesting to disentangle the contribution of the dependence of the interfacial thickness with the molar fraction.
%Points rather to non-ideality. Q: what is the width of the interfacial structure, not obvious.}
As a final remark, a slightly different phenomenon is the so-called osmo-phoresis, which is obtained for permeable particles in which their interface plays the role of a semi-permeable membrane and reported in \citefull{nardi1999vesicles}.
%being imaginative:\\
%\LB{Osmophoresis:} particle = permeable membrane, cf Anderson review + PRL paper by Sackmann \cite{nardi1999vesicles}

%Thereby, it is possible to induce phoretic motion of colloids in solute gradients, mimicking to some extent chemotaxis. This is called diffusio-phoresis. Similarly, it is possible to drive colloids with electric fields by electrophoresis and also by thermal gradients via electrophoresis~\cite{piazza2008thermophoresis}. 
%

%\LB{applications: use dedicated solute gradients to tune the driving forces; cf also cargoing and log-sensing cf below 4.4}
%Another class of phoretic transport is dielectro-phoresis. Dielectrophoresis harnesses the difference in dielectric response between the colloidal particle and the solvent. The dielectric mismatch creates an electric dipole on the colloid, and if a gradient of electric field is applied, it will induce a force on the dipole, that will displace the colloid~\cite{squires2016particles}. Dielectrophoresis is harnessed for transport but is very different from the other phoretic phenomena discussed above because it is not an interfacial process. Also it is not possible to induce dielectro-osmosis ? Is it ? if the particle can't move, then the water does move around it no ? 

\subsection{Harnessing diffusio-phoresis: membrane less separation, log-sensing and localization}
\label{DPharness}
%Instead of discussing further the underlying subtelties of the phenomenon, for which we refer to \cite{andersion1989colloid}, we rather 

In this section we highlight a number of chosen examples to illustrate the impact and the applications of diffusio-phoresis in diverse physical situations. 
An interesting feature of diffusio-phoresis is that complex patterns of solute gradients can be rather easily achieved -- in relative contrast to electric fields as driving forces -- so that this phenomenon can induce particle motion in quite subtle ways leading to a wealth of counter-intuitive behaviors.  
Such solute patterns may occur naturally, for example due to evaporation leading then to drying film stratification \cite{sear2017diffusiophoresis},  in membrane fouling \cite{florea2014long}, or at dead-end pores, allowing for boosted extraction of particles in porous media \cite{kar2015enhanced,shin2016size}, as well as in hydrothermal pores with steep pH gradients \cite{moller2017steep}. 
Alternatively static or dynamic patterns of solutes can be designed thanks to dedicated microfluidic devices \cite{palacci2010colloidal,paustian2013microfluidic}.
An illuminating example was reported recently in \citefull{banerjee2016soluto}, showing that \qote{chemical} beacons emitting solutes may allow to engineer ultra-long range nonequilibrium interactions between particles, up to millimeters -- see Fig.~\ref{fig:OsmoToPhoresis}-b.

%A  first  example, evaporation is expected to induce concentration of solute and diffusio-phoresis was shown recently to be the driving force of the formation of stratification in drying films \cite{sear2017diffusio-phoresis}. 
%\LB{discuss here the effective diffusion, rather than above ???}
%More amazingly, it is also possible to harness these osmotic forces to move large particles such as DNA~\cite{duhr2006molecules} and more~\cite{squires2005microfluidics}. Notably, unlike van der Waals or electrostatic forces, such osmotic forces are intrinsically long range: as long as the driving gradient is present, it generates the force. 

%For example, diffusio-phoresis was shown to be harnessed up to half a millimeter~\cite{banerjee2016soluto} [ARTICLE TODD A REVOIR CITATION]. 

Instead of being exhaustive, we discuss here several \qote{elementary mechanisms}, which
serve the purpose of highlighting the versatile manipulation of particle assemblies via diffusio-phoretic motion. %whose signatures are  observed in various  experiments.
%manipulation of particle assemblies by diffusio-phoretic

\begin{figure}[h!]
\centering
  \includegraphics[width=0.49\textwidth]{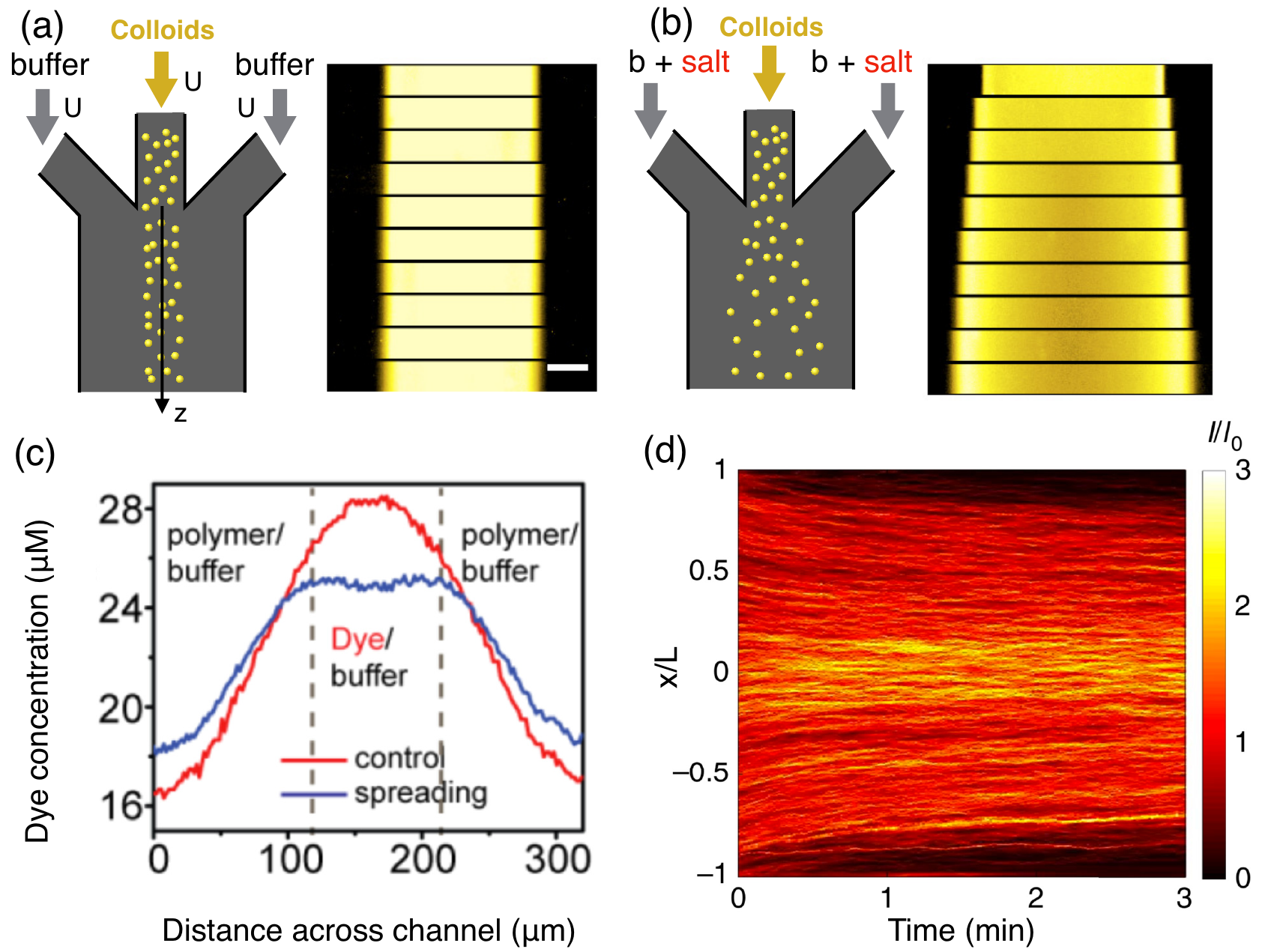}
  \caption{\textbf{: Harnessing diffusio-phoresis to transport particles.} 
  %(a) adapted from \citefull{abecassis2008boosting}. Fluorescent colloids are injected in the central branch of a three branch merging microfluidic channel, all branches having the same inlet velocity. They are imaged at different positions of the channel (side). In case (a) all channels have the same buffer composition, whereas in (b) salt (typically 10 mM of NaCl, LiCl or KCl) is added to the side channels and in (c) is added to the central channel. Although little dispersion is seen in case (a) due to the low diffusivity of colloids, it is very remarkable in (b). The horizontal scale bar is $50 \mu$m.\LB{CAPTION (c)+(d)}
  (a-b) Adapted from \citefull{abecassis2008boosting}. Fluorescent colloids are injected in the central branch of a microfluidic channel with three branches,  with the same inlet velocity. They are imaged at different positions along the channel (side). In case (a) all channels have the same buffer composition (control), whereas in (b) salt (typically 10 mM of NaCl, LiCl or KCl) is added to the side channels. Although little dispersion is seen in case (a) due to the low diffusivity of colloids, a strong migration towards the high salinity is observed in (b). The horizontal scale bar is 50 $\mathrm{\mu}$m. (c) Dispersion of a dye (Rh6G, 50 $\mathrm{\mu}$M) at a cross-section of a microfluidic system with three branches similar to (a). The dye enters the central channel and the side channels are filled with polymer (5 wt \% Ficoll 400K) for the spreading experiment (control, without polymer). Adapted from \citefull{guha2017chemotaxis} with permission from Springer Nature, copyright 2017.  (d) Spatio-temporal evolution of particles upon exposure to CO$_2$ gradients (CO$_2$ is flown above and below, in $x = \pm L$). The particles are polystyrene, diameter $0.5 \mathrm{\mu}$m, dispersed in a liquid buffer, and $L = 400 \mathrm{\mu}m$. Adapted from \citefull{shin2017membraneless}, image under Creative Commons Attribution 4.0 International License.}
  \label{fig:diffusiophoresis}
\end{figure}

{\it Boosting migration --} As a first example, we highlight how diffusio-phoretic transport leads to strongly enhanced migration of particles, 
with the fast solute \qote{towing} the large, slow, particles.
This effect was demonstrated in particular   %we discuss the $\Psi$ 
in coflow geometries, such as in Fig.~\ref{fig:diffusiophoresis}, which was considered in various papers \cite{abecassis2008boosting,guha2017chemotaxis,shin2017membraneless,shin2017accumulation}. This geometry is quite generic in microfluidics in the context of mixing and serves here the purpose of highlighting consequences of diffusio-phoretic motion.
Colloids have a low diffusion constant and therefore barely mix in such a geometry, see Fig.~\ref{fig:diffusiophoresis}-a. Adding tiny amounts of salt, typically millimolars, drastically boosts colloid dispersion -- see Fig.~\ref{fig:diffusiophoresis}-b-c -- with an observed effective diffusion coefficient of the colloids which is 10 to 100 times larger than the equilibrium diffusion coefficient. As mentioned earlier, this is a consequence of diffusio-phoretic motion of the colloids
under the salinity gradients present across the various parts of the channel. 
%Another notable feature of diffusio-phoresis is that it allows to transport particles over much faster timescales than they would have with equilibrium diffusion~\ref{fig:diffusio-phoresis}. We illustrate this in Fig.~\ref{fig:diffusio-phoresis}, where colloidal particles are driven by salt gradients. In this experiment, the effective diffusion coefficient of colloids is 10 to 100 times larger than the  equilibrium diffusion coefficient. 
{This can be rationalized on the basis of simple arguments. The growth rate for the width of the colloid suspension writes as 
$dw/dt=2v_{DP}=2D_{DP}
\nabla \log c_{\rm s}$, with $c_s$ the inhomogeneous salt concentration.  The latter relaxes via Fick's diffusion and  $ \nabla \log
c_{s}\approx \pm 1/\sqrt{D_{s} t}$ (the sign
depending on the salt gradient direction), so that 
\begin{equation}
{dw\over dt}\approx \pm {2D_{DP}\over \sqrt{D_{s}\, t}}\,\,.
\end{equation}
Note that in the experiments of Fig.\ref{fig:diffusiophoresis}, the effective time is related to the position $z$ along the channel as $t=z/U$ ($U$ the average flow velocity). Integrating this
equation yields immediately the observed diffusive like behavior, 
\begin{equation}
w(t)-w_{0}=\pm \sqrt{2 D_{\rm eff} t}.
\label{diffusion}
\end{equation}
 with a diffusion coefficient 
%
%\begin{equation}
$D_{\rm eff} \simeq {D_{\rm DP}^2/ D_{\rm s}}$. Quantitatively  $D_{\rm eff}$ is of the order of a (fraction of) salt coefficient $D_s$ so that 
 $D_{\rm eff}\gg D_{0}$ (the colloid diffusion coefficient) and colloids \qote{diffuse} much faster in the presence of (even minute) concentration gradients. 
%\end{equation}
%
Similar behaviors in equivalent geometries have been reported under CO$_2$ gradients \cite{shin2017membraneless,shin2017accumulation} or polymer gradients \cite{guha2017chemotaxis}.
}

{\it Localization  --} As a second example, diffusio-phoretic motion can be harnessed to manipulate and {\it localize} particle assemblies. 
%of manipulation of particle assemblies, 
%If it is possible to harness diffusio-osmotic forces to transport particles, one may wonder if it is then possible to \textit{localize} particles at a specific location. Surprisingly 
Interestingly in the biological world, bacteria are capable of using solute contrasts to localize proteins~\cite{shapiro2009and}. Localization is then used as an information for further vital processes, for instance localization of the ring of the FtsZ protein at midcell is used for cellular division~\cite{osawa2008reconstitution,loose2008spatial}. 
Similar features can be obtained on the basis of diffusio-phoresis under salt gradients, harnessing the {\it log-sensing} feature discussed above
in \eqref{eq:PhoresisDO}, and leading to particle accumulation~\cite{palacci2012osmotic,palacci2010colloidal}.
%
%Diffusio-osmotic transport may also be used to localize particles. As we have seen in Eq.~\ref{eq:vDEO}, the diffusio-osmotic velocity scales logarithmically with the salinity concentration gradient
%\begin{equation}
%v_{\infty} = - \mu_{DO} \partial_x \log c_{\infty}.
%\end{equation}
Indeed, under a linear salt concentration gradient, %(as is often the case at steady state between two reservoirs) 
the diffusio-osmotic velocity  is not uniform and will be larger in regions of lower salt concentration -- see Fig.~\ref{fig:LogSensing}-a. 
Alternating the gradient over time leads to rectification of diffusio-phoretic motion and accumulation of the colloids towards \textit{e.g.}the center of the cell, as highlighted in Fig.~\ref{fig:LogSensing}-b.
%Therefore, it can be harnessed \textit{e.g.} for particle focusing~\cite{palacci2012osmotic,palacci2010colloidal}. In fact, upon alternating concentration gradients, see Fig.~\ref{fig:LogSensing}, this \textit{log-sensing} may be used to induce larger forces towards the center of the setup. One can thus create osmotic traps. Furthermore, log-sensing transport was shown to be extremely sensitive to concentration gradients, \textit{e.g.} extremely robust upon small concentration gradients and with time~\cite{palacci2012osmotic}. 
%\LB{Introduce osmotic shock here, very counterintuitive}
%\LB{discuss two aspects which are consequence of log-sensing: \\
%1. rectification\\
%2. osmotic shock \\ 
%cf soft matter Palacci 2011
%
It is interesting on this elementary example to formalize the rectification process. The colloid and salt equations of transport obey the coupled Smoluchowski and diffusion equations
\begin{eqnarray}
&\partial_t \rho &= -\bm{\nabla}\cdot \left( - D_{0} \bm{\nabla}\rho + D_{DP} \bm{\nabla}[\log c_s] \times \rho \right) \nonumber \\
&\partial_t c_s&= D_s \bm{\nabla}^2 c_s
\label{Smolu2}
\end{eqnarray}
with $\rho$ the colloid density and $c_s$ the salt concentration; 
 $D_{0}$ and $D_s$ are the particle and salt diffusion constants and $D_{DP}$ the particle diffusio-phoretic mobility.
%
%Leads to complex interplay. Two examples here.\\
Let us simplify the geometry to fix ideas and consider a one-dimensional channel. We consider an oscillating salt concentration profile, 
%Oscillating  concentration profile, say in 1D, 
$\nabla c_s(x,t)= f(t) \Delta c_s/\ell$, with $\ell$ a characteristic length scale and
$f(t)$ an oscillating function of time with zero average.
Averaging over the rapid salt concentration oscillations, the mean diffusio-phoretic velocity which enters the Smoluchowski equation simplifies to $ \bar v_{DP}=D_{DP} \langle \nabla[\log c_s] \rangle_t \approx  -D_{DP} (\langle f^2\rangle_t/\ell^2) \times x$, where $x$ is the distance to the center of the cell and $\langle\cdot\rangle_t$ an average over time. 
%Back in the Smoluchowski equation for the colloids, Eq.(\ref{Smolu2}), the averaged diffusio-phoretic term can thus
It can be rewritten in terms of an effective potential via 
\begin{equation} \bar v_{DP}  \equiv \mu_0 \times -\partial_x{\cal U}_{\rm eff},
\end{equation} with $\mu_0=D_{0}/k_BT$ the colloid mobility
%takes the form of an effective trapping
%Results in trapping potential with spring like restoring force 
%and the effective  potential ${\cal U}_{\rm eff} $ takes the form
and
\begin{equation}
{\cal U}_{\rm eff} (x)= {k_B T \over 2} \langle f^2\rangle_t  {x^2 \over \sigma_{\ell}^2} ,
\end{equation} 
with $\sigma_{\ell}=\frac{\ell}{2 }\sqrt{{D_{0}\over D_{DP}}}\ll \ell$. % {\it i.e.} a harmonic trapping potential
This illustrates that the rectified diffusio-phoresis of the colloids can be interpreted in terms of a harmonic trapping potential towards the central node ($x=0$) of the solute concentration oscillation pattern. % Cite Palacci PRL + SoftMatter. [voir Richard Sear ?]\\
This allows one to manipulate the colloidal population via time-dependent solute gradients. 

{Alternative routes for focusing colloidal populations were proposed using diffusio-phoretic transport without the requirement of
time-dependent fields~\cite{shi2016diffusiophoretic} -- see Fig.~\ref{fig:LogSensing}-c. These make use of combined steady gradients of salt and pH which are shown to yield localization of
the particles.  
The interpretation of this subtle phenomenon incidentally
highlights  that the effects of gradients cannot be simply superimposed for diffusio-phoresis and a new formulation of coupled diffusio-phoretic transport was required, rewriting the driving solute gradients in terms of the corresponding ion fluxes. Such combination of gradients of pH and salts was suggested to occur in hydrothermal pores, with potential consequences on the emergence of an ion-gradient-driven early
protometabolism and the origin of life \cite{moller2017steep}. Finally focusing of colloidal particles was demonstrated in dead-end pore geometries \cite{shin2016size}, with potential applications to preconcentration, separation, and
sorting of particles. 
}

\begin{figure}[h!]
\centering
  \includegraphics[width=0.49\textwidth]{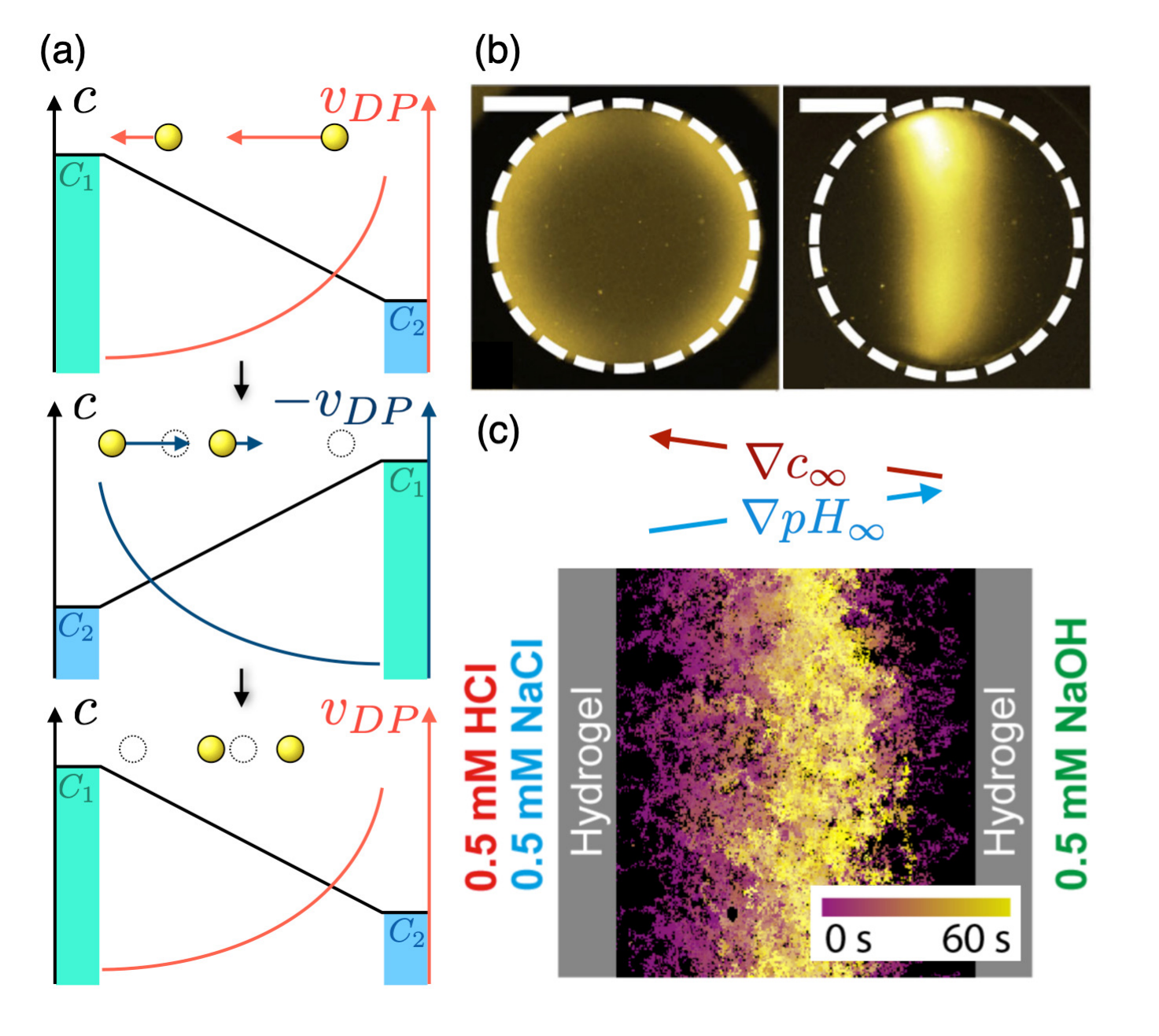}
  \caption{\textbf{: Focusing particles with diffusio-phoretic transport.} (a) Trajectories of two (yellow) particles starting at different lateral positions upon an alternating concentration gradient. As the diffusio-phoretic velocity scales logarithmically with solute concentration, the particles closer to lower concentrations move faster, resulting in localization of the particles at the center. (b) Alternating fluids are flushed on both sides of a circular microfluidic well containing fluorescent colloids. The concentration of LiCl in the two side channels alternates (left/right) with period 480s, between buffer alone and 100 mM. The scale bar is 200 $\mu$m. (Left) Initial particle distribution in the well and (Right) stationary colloid distribution under the alternating concentration gradient.  Reproduced from \citefull{palacci2012osmotic}. \SM{(c) Diffusio-phoresis under combined steady pH and salt gradients. Flowing NaOH and HCl solutions separately in two reservoir channels establishes a gradient in pH from 3.3 (left) to 10.7 (right), within which diffusio-phoretic particles proceed monotonically to the right. A NaCl gradient is superimposed on the pH gradient inducing diffusio-phoresis to the left. Streamline images reveal unexpected focusing at a location within the channel. Adapted from \citefull{shi2016diffusiophoretic} with permission from the APS copyright 2016.}} 
  \label{fig:LogSensing}
\end{figure}

{\it Osmotic shock --} As a last example, we discuss a very striking and counter-intuitive behavior steming from log-sensing, coined as osmotic shock, which was discussed in \citefull{palacci2012osmotic}. It illustrates that diffusio-phoresis keeps a long-lasting \qote{memory} of solute gradients, even when they would be expected to be already homogenized. 
%works even though diffusion processes have long disappeared. memory
Consider a situation where the colloids are spread in a reservoir with lateral size $\ell$, with initially a uniform solute concentration $c_0$. Then at time $t=0$,
solute is {\it flushed at the boundaries}, $c_s(x=\pm\ell/2,t) =0$ for $t>0$ (simplifying to 1D, and $x$ is the coordinate from the center of the reservoir). 
%For a 1D geometry, 
After a short transient, the solute concentration profile will decay to zero according to
$c_s(x,t) \simeq c_0 \exp[-t/\tau] \, \cos(\pi x/\ell)$, with $\tau=\ell^2/D_s$ the diffusion timescale of the solute. The diffusio-phoretic velocity of the colloids then writes $v_{DP} = D_{DP} \times \nabla \log c(x,t)$, so that
\begin{equation}
v_{DP} =  D_{DP} \times {\pi\over \ell}\tan\left({\pi x\over\ell}\right) \approx D_{DP} \times {\pi^2\over \ell^2} \times x,
\label{eqVdpInd}
\end{equation}
pointing towards the center of the reservoir, $x=0$, hence gathering the colloid population towards this position.
From \eqref{eqVdpInd} it is clear that the DP velocity is therefore {\it independent of time} ! This leads to the counter-intuitive result that  diffusio-phoretic motion occurs on far longer timescales than the solute diffusion timescale. %[cite Palacci 2011].
This behavior was highlighted experimentally in \citefull{palacci2012osmotic}, where diffusio-phoretic motion of colloid particles was observed on timescales ten times longer than the naive diffusive timescale for the salt.
Log-sensing is an efficient approach to localize colloids, but its application for trapping of other types of particles, \textit{e.g.} polymers, remains to be explored. This could  possibly be harnessed to improve sensors or traps for high throughput chemical reactions~\cite{krishnan2010geometry}. Applications to information storage and retrieval could also be explored~\cite{myers2015information}. Log-sensing also helps remove particles or fluids trapped in dead-end pores, as we will discuss in Sec.\ref{deadendpores}.

As a final word on this section, the role of diffusio-phoretic transport in biological context has yet to be readily explored and quantified.
In the toolbox of living systems, concentration gradients play a versatile role, readily exploited in many aspects of the biological machinery, such as energy reservoirs, but also serving more sophisticated functionalities, associated with spatial signaling, localization and pattern formation at the various scales involved in the biological processes. One may cite \textit{e.g.}  enzyme transport \cite{agudo2018enhanced}, protein localization in bacteria \cite{shapiro2009and} or more generally in spatial cell biology the use of concentration patterns for positional information~\cite{lutkenhaus2008tinkering,rothfield2005spatial,eldar2002robustness}, to quote a few. Obviously chemotaxis in biological organisms under solute gradients is a highly complex phenomenon steming from the interplay of complex signalling pathways, quite far from the simple diffusio-phoretic transport discussed here. But one may reversely remark that the consequences of diffusio-phoresis as a physical phenomenon cannot be overlooked in biological materials, especially in the presence of ubiquitous gradients. This has been barely explored \cite{Sear2019}.

\subsection{From self-propulsion to self-assembly}
\label{active}
%\LB{cf Rev. \cite{agudo2018enhanced} for a recent review for enzymes}

%\LB{cite Brady, osmotic motor and comments: force free or not; comments and answers \cite{cordova2008osmotic,fischer2009comment,PhysRevLett.102.159802,julicher2009comment,PhysRevLett.103.079802}}

%\SM{CITER ICI GINOT 2018 AGGREGATION \cite{ginot2018aggregation}}
%
%\SM{CITER AUSSI PAPIER RECENT GOLESTANIAN \cite{ibrahim2018shape} Shape dependent phoretic propulsion}
%
%\SM{ATTENTION ! veut-on inclure que le developpement "exact" peut etre utilise pour decrire aussi des systemes qui genere les gradients a la surface?  je penche plutot pour en rester la. }

Beyond the idea of passive diffusio-phoresis, where particles move under externally imposed solute gradients, arose the idea that the solute concentration gradients could be generated {\it on the particle themselves}, {\it e.g.} via chemical reactions occurring at their surfaces.
For an asymmetric chemical reactivity, this self-diffusio-phoresis process thus generates self-propulsion of the particles, fueled by the chemical reactions \cite{aubret2017eppur,moran2017phoretic,illien2017fuelled}.
Together with other phenomena leading to self-propulsion, this triggered the emergence of the field of active matter, which has exploded
over the last decade.
It is not our purpose to review this field, as this goes beyond the scope of our focus on osmotic forces and we refer to some recent reviews
on this topic~\cite{bechinger2016active,illien2017fuelled,colberg2014chemistry}. We however highlight here a few phenomena where osmosis, via diffusio-phoresis and related mechanisms, is explicitly at play.

{\it Self-diffusio-phoresis --} On the experimental side, the phenomenon of self-diffusio-phoresis  was pioneered by Paxton and coworkers~\cite{paxton2004catalytic}, who showed self propulsion of Platinum/Gold nanorods in hydrogen peroxide. Hydrogen peroxide is chemically transformed differentially on both metals, either forming or being depleted on each side of the rod, and this creates a gradient of the reacting specie (here hydrogen peroxide) driving diffusio-phoresis, see Fig.~\ref{fig:activeParticles}. For such bimetallic particles with redox reactions on both sides of the particle, self-electro-phoresis may actually contribute to the driving force, via motion of charges (electrons and ions) within and outside the particle. 
Self-diffusio-phoresis was further demonstrated in colloidal janus particles of various materials, see \citefulls{mano2005bioelectrochemical,howse2007self,palacci2010sedimentation,theurkauff2012dynamic,palacci2013living,buttinoni2013dynamical,mino2011enhanced,ginot2015nonequilibrium,aubret2018targeted,brown2014ionic}, and \citefulls{bechinger2016active,brown2017ionic} for a more exhaustive literature on this aspect.

\begin{figure}[h!]
\centering
  \includegraphics[width=0.49\textwidth]{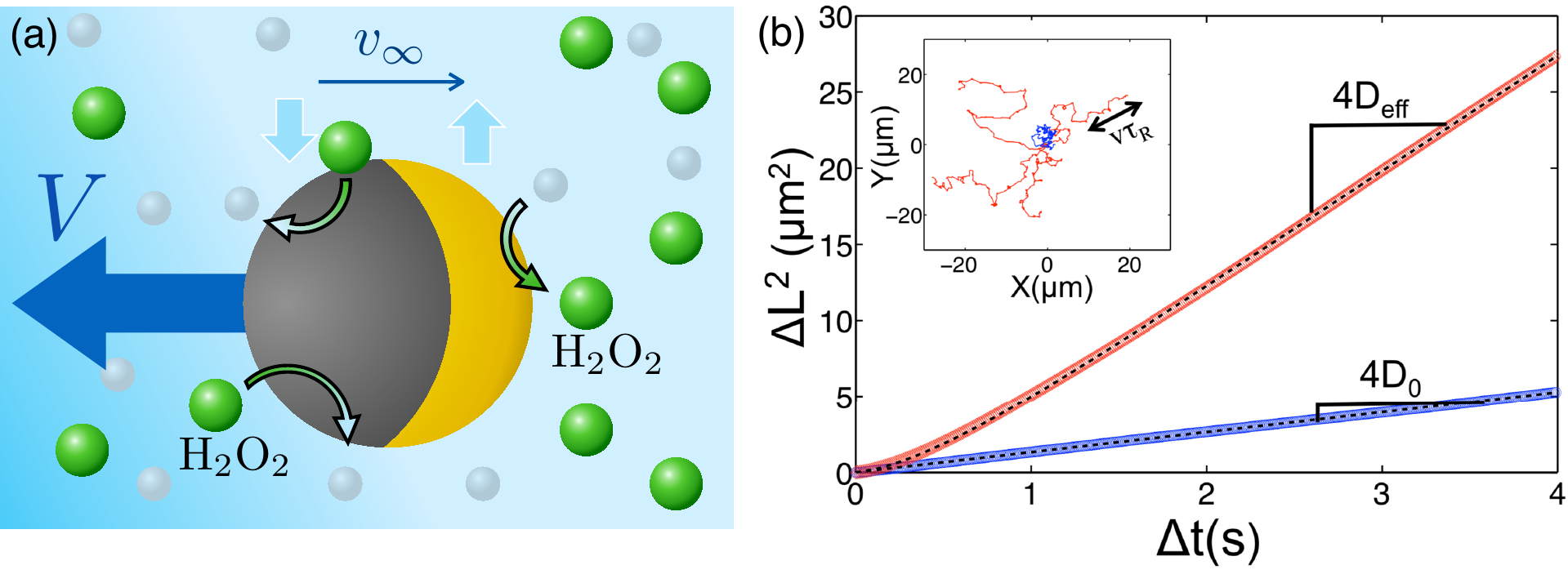}
  \caption{\textbf{: Self-propelled particles.} (a) Chemical reactions occur differentially at the front and at the rear of a reactive colloidal particle, thereby inducing a chemical concentration gradient. This leads to  diffusio-osmotic driving at the surface, hereby displacing the particle. (b) Experimental mean squared displacements $\Delta L^2(\Delta t)$ and 2D trajectories (inset) for bare (blue) and active colloids (red) in 7.5$\%$ H$_2$O$_2$ solution. Bare colloids (bottom) show standard diffusion ($\Delta L^2$ linear in time), while the mean squared displacement of active colloids shows a ballistic motion at small timescales and a diffusive motion at longer timescales. The measured diffusion coefficients are $D_0 = 0.33\mu$m$^2/$s for bare and $D_{\rm eff} = 1.9 \mu$m$^2/$s for active colloids. Reproduced from \citefull{palacci2010sedimentation} with permission from the APS, copyright 2010.}
  \label{fig:activeParticles}
\end{figure}

On the theoretical side, the mechanisms by which the creation or removal of species on the particle's surface generate an osmotic pressure gradient and motion are not obvious and there has been some initial debate on this question, see \citefulls{cordova2008osmotic,fischer2009comment,PhysRevLett.102.159802,julicher2009comment,PhysRevLett.103.079802}. This  echoes directly our discussion about the diffusio-phoretic force balance in Sec.~\ref{discussionforce}. 
Actually the question was pioneered by Lammert {\it et al.} on the putative self-electro-phoresis of biological cells or vesicles driven by 
non-uniform ion pumping across the bounding membrane \cite{lammert1996ion}. 
%Non!uniform ion pumping across the bounding membrane of a biological cell or arti_cial vesicle can generate electro!osmotic effects
%In fact, with an asymmetric colloid, the properties of adsorption/interaction with the solute may be different at the rear, thereby producing a gradient driving the colloid in that direction. This was achieved for the first time by Paxton and coworkers~\cite{paxton2004catalytic}, with Platinum/Gold nanorods in hydrogen peroxide. Since the Nernst Potential on Platinum and Gold is not the same, hydrogen peroxide is either forming or being depleted on each side of the rod. This creates a gradient of reacting specie (here hydrogen peroxide) driving diffusio-phoresis, see Fig.~\ref{fig:activeParticles}. 
Echoing this situation, an illuminating model for self-propulsion via asymmetric osmotic driving force was introduced by 
Golestanian {\it et al.} \cite{golestanian2005propulsion,golestanian2007designing}.
They considered a particle exhibiting a non-uniform chemical reactivity on its surface, as defined by the corresponding solute flux on its surface
$\alpha(\bm{r})=-D_r \nabla_\perp C_{\rm react}$ (corresponding to the generation or consumption of the solute by the chemical reaction), 
$D_r$ being the diffusion constant of the reactant. This boundary term is coupled to the diffusive dynamics of the solute concentration
in the bulk. The resulting concentration gradient induces a diffusio-osmotic slip velocity $v_{DO}$ at the surface, depending on the position, and accordingly particle motion. In the case of a janus sphere, 
%The self-phoretic process was first considered by Lammert {\it et al.} for self-electropheresis. Later the self-diffusio-osmotic was proposed as a source of propulsion 
%by Golestanian {\it et al.} \cite{golestanian2005propulsion,golestanian2007designing}. %[REF NJP2007] + Brady + comments. 
exhibiting a contrasting chemical reactivity on its two moieties, the self-diffusio-phoretic velocity $V$ takes the simple expression 
\begin{equation}
V=\langle v_{DO} \rangle_{sphere} ={1\over 8D_r} (\alpha_--\alpha_+)(\mu_++\mu_-) 
\end{equation}
where $\alpha_\pm$ is the chemical reactivity on the two sides and $\mu_\pm$ the local surface phoretic mobility. 
We emphasize though that on the experimental side, other mechanisms also contribute to the motion of catalytically self-propelled particles, like self-electro-phoresis \cite{moran2017phoretic}.

To some extent, such "active particles" mimick  self-propelled biological organisms.
%nanoscale directed machinery, consuming energy right at the place where transport happens, making them really interesting.  \textcolor{blue}{what happens that doesn't break the second principle here ? H2O2 is consumed but also released...}. These type of particles started the field of "active matter", and 
The possibility to fabricate artificially these systems  constitutes a playground to study far-from-equilibrium behaviors \cite{bechinger2016active}.
%for a thorough review. The interesting thing with these particles is that they 
%Such particles : 
Because active particles consume energy at a local scale, their collective behavior is {\it a priori } not constrained -- at least to some extent -- by thermodynamics and may possibly allow to break the bottleneck of the second principle; {\it cf.} the beautiful example of the rotating Feynman ratchet with active materials in \citefull{di2010bacterial}.
%being transported and creating structures in a far more complex way since consumption of energy is performed on a local scale.

{\it Active suspensions --} Such active particles have fascinating behaviors, and we focus on a few examples. 
First, self propulsion leads to ballistic motion on short timescales, but orientational random motion leads to diffusive behavior on long timescales, in a way similar to the so-called \qote{run and tumble} motion of bacteria. The effective diffusion coefficient is however far larger than the bare diffusion coefficient based on the Stokes-Einstein estimates $D_0$ \cite{howse2007self,palacci2010sedimentation}, see Fig.~\ref{fig:activeParticles}-b.
As a rule of thumb the effective diffusion coefficient is typically $D_{\rm eff} \approx V^2\times \tau_R$, where $V$ is the self-propelling velocity and $\tau_R$ is the timescale for rotational brownian motion: $\tau_R \sim D_{\rm rot}^{-1} \sim R^2/D_0$, where $D_{\rm rot}$ is the rotational diffusion coefficient, $R$ the particle size.
As a consequence the particles behave as a \qote{hot} bath, with a high effective temperature defined from a \qote{fluctuation-dissipation}-like relation as
${k_B T_{\rm eff}}= D_{\rm eff}/\mu_0$, where $\mu_0=D_0/k_BT$ the bare particle mobility, so that
\begin{equation}
{k_B T_{\rm eff}} \approx k_BT\times{V^2\, \tau_R\over D_0}
\end{equation}
(up to numerical prefactors).
Altogether this predicts that
$T_{\rm eff}/T \sim Pe^2$ where the P\'{e}clet number is defined in terms of the self-phoretic velocity as $Pe=V\,R/D_0$. This prediction was further confirmed experimentally\cite{palacci2010sedimentation}.

\SM{\it Osmotic pressure of active suspensions --} The question of the osmotic pressure created by active particles was raised in a number of theoretical and experimental works~\cite{lion2014osmosis,solon2015pressure,rodenburg2017van,ginot2015nonequilibrium,takatori2016acoustic}.
As we introduced above, the osmotic pressure acting on the fluid can be defined mechanically via the average force exerted by the active particles on a 
semi-permeable wall. In the case where the active particles exhibit a Boltzmann-like equilibrium in the presence of the wall (say, represented by an 
external potential),  as shown {\it e.g.} for sedimentation profiles \cite{palacci2010sedimentation}, then the osmotic pressure
reduces to the van\,'t Hoff law, except that the temperature is replaced by the effective temperature of the suspension:
\begin{equation}
\Delta \Pi \simeq k_BT_{\rm eff} \times \Delta \rho
\end{equation}
where  $\rho$ is here the concentration of active particles, and the effective temperature $T_{\rm eff}$ was introduced above. 
This matches the equation of state measured using sedimentation profiles~\cite{ginot2015nonequilibrium}. 
%\SMM{In orders of magnitude, for the active colloids used in \citefull{palacci2010sedimentation} the osmotic pressure is of the order of $10^{-5} Pa$. BUT THERE's probably better in Ginot ? Can you add an order of magnitude !?}

However  Boltzmann-like equilibrium is expected to fail in some limiting situations for active particles. In particular it is commonly observed that self-propelled particles do hit \qote{compulsively} hard surfaces, similarly to a fly on a window \cite{di2010bacterial}: in such cases, the particle remains stuck at the membrane's surface until it reorients, and there is a non-boltzmannian accumulation of particles at the membrane. This is nicely exemplified in the run-and-tumble model under an external field, where strong deviations from the Boltzmann profile is predicted when  
the typical drift velocity $V_d=\mu_0 {\cal F}_{\rm ext}$ (with ${\cal F}_{\rm ext}$  the maximum external force, say, due to the separating membrane),
is larger than the particle self-propulsion speed $V$ \cite{tailleur2008statistical}. This situation occurs for steep potentials. %In such case, 
%where $V < V_d$, 
When such accumulation occurs,  
the corresponding osmotic push will differ from the simple van\,'t Hoff law, see~\citefulls{rodenburg2017van,takatori2016acoustic}. In particular, the osmotic pressure  depends on the properties of the membrane itself and its interaction with the particles -- typically via the ratio between the typical
membrane characteristic steepness and the particle mean-free path \cite{takatori2016acoustic} --, 
 in strong contrast to the van\,'t Hoff \qote{universal} relation.
%the ballistic mean-free path is larger than the potential characteristic length scale. 
%colloids persistently propelling ?into? the repulsive membrane.
Similar deviations from the van\,'t Hoff law also occurs when the interaction between the active particles and the membrane involves wall-induced rotational torques~\cite{solon2015pressure}. 
%except different for gravity, no wall \cite{ginot2015nonequilibrium}
%osmotic pressure, $\Delta \Pi \sim k_BT_{\rm eff} \times \Delta c$ but equilibrium on the surface not obvious 
%but leads to difficulties \cite{solon2015pressure}.
%Soft confinement then ok van 't Hoff like. But strong confinement leads to non-Boltzmann equilibrium (=
%the density is peaked along the perimeter), then deviations from van 't Hoff law \cie{takatori2016acoustic}
%The swim pressure depends only on the parameter An external file that holds a picture, illustration, etc. Object name is ncomms10694-m37.jpg, a ratio of the swimmers' run length An external file that holds a picture, illustration, etc. Object name is ncomms10694-m38.jpg to the size of the trap Rc=?U0/k.

%which ias they are self-propelled at short timescales they are demonstrate ballistic motion and at longer timescales diffusion is much larger than for passive colloids~\cite{howse2007self,palacci2010sedimentation}, see Fig.~\ref{fig:activeParticles}-b. This induces a hot effective temperature in the system~\cite{palacci2010sedimentation} 
%\LB{\begin{equation}
%{k_B T_{\rm eff}}=  k_BT\times \left( 1+ {2\over 9} Pe^2\right)
%\end{equation}
%where the Peclet number is defined in terms of the self-phoretic velocity $Pe=V.R/D_0$ ($D_0$ the bare colloid diffusion constant).}
%The equation of state is far more complex, highlighting specific properties of active fluids~\cite{ginot2015nonequilibrium,solon2015pressure} and allowing to set the grounds for basic understanding of far-from-equilibrium physics. 

{\it Towards self-organization and self-assembly --} 
Another interesting feature of particles propelling via self-diffusio-phoresis is that they interact via {\it chemical signaling}. Propelled particles act as a beacon -- similarly to 
the situation considered in \citefull{banerjee2016soluto} -- and leave 
%Another fascinating feature of active particles is that they may self-assemble at relatively fast kinetic rates, driven by diffusio-phoresis, see Fig.~\ref{fig:activeAssembly}-a. In fact, each particle 
 a \qote{trace} of their passage in the form of a diffusing cloud of chemicals which will be felt by other paticles, see Fig.~\ref{fig:activeAssembly}. 
 Accordingly other active particles will be reoriented towards or against~\cite{moerman2017solute,reigh2018diffusiophoretically} the active particle via diffusio-phoretic motion (on top of their self-driving motion). 
Indeed the surface creation or consumption of solutes generates long-distance distortion of the solute concentration
profile, typically relaxing spatially as a monopole $\delta c_s \sim 1/r$ (\SM{the scaling deriving from Fick's equation with a sink}). This long-range interaction is for example highlighted in Fig.~\ref{fig:activeAssembly}-b, showing the diffusio-phoretic attractive motion induced by a single beacon, from \citefull{palacci2013living}. \review{At shorter range,} the interaction may become more complex and requires detailed investigation of the chemical drivings \cite{reigh2018diffusiophoretically}.  % Depending on the sign of the diffusio-phoretic mobility for the solute under consideration, this  diffusio-phoretic interaction can be attractive or repulsive~\cite{moerman2017solute}.
 
 This osmotic-induced chemical interaction is at the origin of many advanced collective properties of active particles, such as
clustering \cite{theurkauff2012dynamic,buttinoni2013dynamical,bechinger2016active,ginot2018aggregation}, or self-assembly \cite{palacci2013living,aubret2018targeted},
see Fig.~\ref{fig:activeAssembly}-c-d.
Out-of-equilibrium self-assembly has raised enormous interest, since activity leads to unexpected structures, with the hope of designing novel and smart materials~\cite{bechinger2016active}.

%Reorientation was mostly harvested to drive attraction and thus assembly of colloids -- see Fig.~\ref{fig:activeAssembly}-b, though under some conditions it is also possible to harness this added diffusio-phoretic contribution for repulsion~\cite{moerman2017solute}. 

It is interesting to  formalize the basics of the phenomenon at stake by writing the coupled diffusion-reaction equation for the colloid population and solute concentration. For the purpose of illustration, we only consider here a single neutral chemically generated specie which acts as a chemo-attractant to the colloids. These dynamical equations actually identify with the so-called Keller-Segel equation, which were
written to describe the chemotactic aggregation of a slime mold (amoebae) under the perspective of a dynamical instability \cite{keller1970initiation,brenner1998physical,theurkauff2012dynamic}:
\begin{equation}
\begin{split}
\partial_t \rho & =  - \bm{\nabla} . \left( - D_{\rm eff} \bm{\nabla} \rho + \mu_{DP}  \bm{\nabla} c_s \times \rho) \right)  \\
\partial_t c_s & = D_s \bm{\nabla}^2 c_s + \alpha \rho \simeq 0
\label{KS}
\end{split}
\end{equation}
 with  $D_{\rm eff}$  the effective diffusion coefficient of the active colloids,  $D_s$ the diffusion coefficient of the \qote{chemo-attractant} specie
 %$\mu$ the diffusio-phoretic mobility
  and $\alpha$ the chemical rate of the powering chemical reaction occurring at the surface of each colloid; we assume here that the solute dynamics are fast. By analogy to electrostatics, the second equation for the solute allows one to obtain the solute concentration as a function of the colloid density, as $c_s (\bm{r}) \simeq {\alpha/D_s} 
  \int d\bm{r^\prime}\, \rho(\bm{r^\prime})/4\pi \vert \bm{r}-\bm{r^\prime}\vert$.
  When introduced in the first equation in \eqref{KS}, this shows that the diffusio-phoretic attraction acts as a self-consistent effective potential
  such that 
  \begin{equation} 
  {\cal U}_{\rm eff}(\bm{r},\{\rho\}) =-{\mu_{DP}\over \mu_0}\, {\alpha\over D_s} 
  \int d\bm{r^\prime}\, {\rho(\bm{r^\prime})\over 4\pi \vert \bm{r}-\bm{r^\prime}\vert}.
  \end{equation}
  with $\mu_0$ the bare colloid mobility. %$=D_{\rm eff}/k_BT_{\rm eff} $  
  Equivalently, the dynamics of the colloid population can be formally derived from an effective free energy functional of the colloid system which takes the simple form
  \begin{eqnarray}
 & {\cal F}_{\rm eff} = &k_BT_{\rm eff} \int d\bm{r} \left[\rho(\bm{r})\log \rho(\bm{r}) - \rho(\bm{r})\right]\nonumber \\
 && -{1\over 2} \left({\mu_{DP}\over \mu_0}\, {\alpha\over D_s} \right)  \int d\bm{r^\prime}d\bm{r^{\prime\prime}}\, 
 {\rho(\bm{r^\prime})  \, \rho(\bm{r^{\prime\prime}}) \over 4\pi \vert \bm{r^{\prime}}-\bm{r^{\prime\prime}}\vert}
 \end{eqnarray}
 This highlights that the \qote{osmotic interaction} via the diffusio-phoretic motion induced by the solute traces leads  to {\it long range non-equilibrium interactions}. This is expected to lead to  strong collective effects, such as the clusterization observed experimentally. For attractive systems, $\mu_{DP}>0$, these equations are formally analogous to a (non-inertial) gravitational system.    Keller-Segel and subsequent works have shown that the above equations predicts an aggregation mechanism, similar to a Chandrasekhar gravitational collapse \cite{keller1970initiation,brenner1998physical}. 
 Similar conclusions were predicted for thermally active colloids \cite{golestanian2012collective} where the threshold for collapse was rederived.

\begin{figure}[h!]
\centering
  \includegraphics[width=0.49\textwidth]{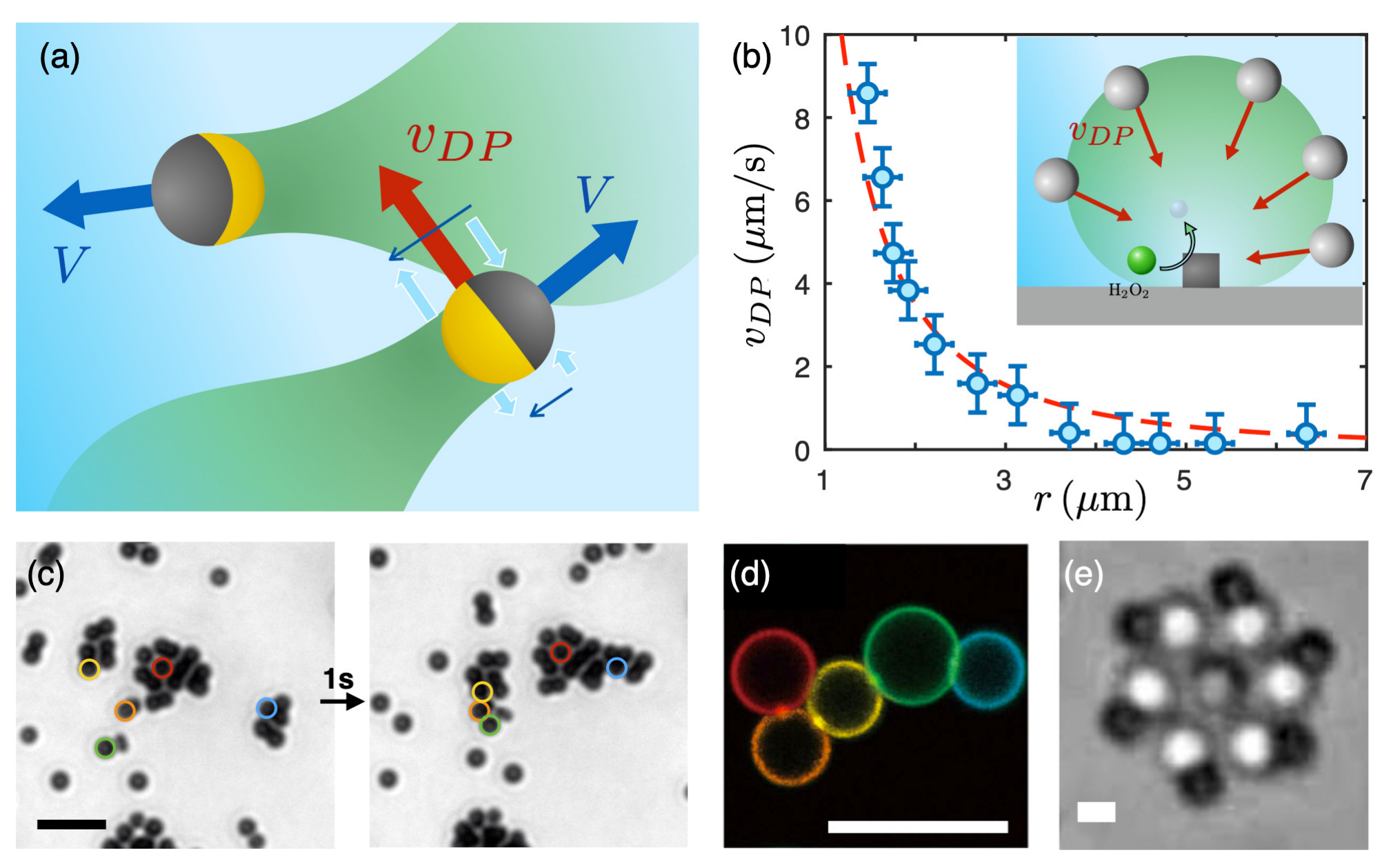}
  \caption{\textbf{: Out-of-equilibrium self-assembly.} (a) Sketch of the diffusing chemical trace left behind active particles, and which modifies the local chemical gradient. Another active particle approaching this chemical gradient will therefore sense a different driving velocity along its edge, changing its trajectory (here in an attractive configuration) (b) Courtesy from Jeremie Palacci, data from \citefull{palacci2013living}. Passive colloid diffusio-phoretic speed as a function of the distance to a hematite cube that can be used -- with blue light -- to catalyse the dissociation of H$_2$O$_2$. (c) Spontaneous self-assembly of active particles. The particles form various cluster sizes and shapes. The scale bar is 10$\mu$m. Reproduced from \citefull{theurkauff2012dynamic} with permission from the APS, copyright 2012. (d) Sequential self-assembly of DNA-grafted droplets, the different colors represent different functionalizations. The scale bar is 10$\mu$m. Reproduced from \citefull{zhang2017sequential}, image under Creative Commons Attribution 4.0 International License. (e) Targeted assembly of phototactic swimmers into nanogears. The scale bar is 1$\mu$m. Reproduced from \citefull{aubret2018targeted} with permission from Springer Nature, copyright 2018.} 
  \label{fig:activeAssembly}
\end{figure}  
  
  However  the  clusters observed in the experiments, {\it e.g.} in \citefulls{theurkauff2012dynamic,palacci2013living,ginot2018aggregation}, do not correspond to full collapse and  are rather dynamic, with clusters reaching a finite size and continuously rearranging over time, see Fig.~\ref{fig:activeAssembly}-c. As shown in \citefulls{pohl2014dynamic,stark2018artificial}, this behavior can be reproduced by  Keller-Segel-like dynamics provided both translational and rotational 
  phoretic conditions are properly taken into account in the kinetics. Using this framework, it is furthermore possible to predict the condition in which dynamic clusterization occurs \cite{pohl2014dynamic,stark2018artificial}, in good agreement with the experiments. 
%  dynamic clustering
%only when these two phoretic contributions give rise to competing attractive and repulsive interactions,
%  More in detail the condition for 
%  In particular it suggests to introduce a threshold 'Chandrasekhar' number, $N_c$, 
% 
%%\begin{equation}
%$N_c= {4 D_\rho D_c / \mu \alpha}$ (in 2D) \cite{brenner1998physical}. Cite Brenner

%%\bibitem{Brenner1998}
%M.P. Brenner, L.S. Levitov, and E.O. Budrene.
%%\newblock Physical mechanisms for chemotactic pattern formation by bacteria.
%\newblock {\em Biophys J}, {\bf 74} 1677 (1998).
%Whether this is at play is not obvious, but show patterns

%Reorientation was mostly harvested to drive attraction and thus assembly of colloids -- see Fig.~\ref{fig:activeAssembly}-b, though under some conditions it is also possible to harness this added diffusio-phoretic contribution for repulsion~\cite{moerman2017solute}. Out-of-equilibrium self-assembly has risen enormous interest, in particular because the study of the structures provided insight in new out-of-equilibrium physics, but also were hoped to lead to novel and smart materials~\cite{bechinger2016active}. 

Beyond clusterization, the self-diffusio-phoretic motion of particles and their osmotic interactions were shown lately to 
%obtain higher level of structuration leading 
lead to the self-assembly of active particles into higher levels of  structure organization. This is highlighted in Fig.~\ref{fig:activeAssembly}-e, from \citefull{aubret2018targeted}, where self-spinning microgears are built on the basis of these non-equilibrium interactions.  
%Note that the variety of assembly pathways has outgrown simple diffusio-phoresis and now 
Beyond, more "on demand" structures are possible, like structures assembled through DNA-grafted interfaces~\cite{feng2013dna,zhang2017sequential,mcmullen2018freely} (see Fig.~\ref{fig:activeAssembly}-d).

%\textcolor{blue}{Is there something to be said in relation to this Golestanian paper~\cite{golestanian2005propulsion} ? with Ajdari with a colloid which has a small reactive site; }

%\LB{ajouter discussion Keller-Segel, collective, aussi  rotors Jeremie car bel exemple d'organisation liee a DP entre colloides}

%%% TRANSITION %%%

%Now coming back to osmosis, the osmotic pressure generated by active fluids in the vicinity of passive semi-permeable membranes has been explored with \textit{e.g.} the kinetic approach~\cite{rodenburg2017van,lion2014osmosis}. Simplistically, active fluids increase the effective osmotic pressure because they contain active particles hitting on the membrane and therefore increasing the effective force on the membrane.
%In some conditions, activity was even shown to induce reverse osmosis \cite{lion2014osmosis}.
%However, conversely \textit{how an active membrane (with a passive fluid) influences osmotic transport} is still a broad question, also with a connection to biological understanding. We give some ideas below. 

%Optically driven micropump~\cite{leach2006optically}, why not go to nanoscales ? 

\vskip0.5cm
\section{Osmosis, towards applications}
\label{applications}

From food processing in biological organisms~\cite{sheeler1987cell,jensen2009osmotically,comtet2017passive}, to reverse osmosis for desalination and energy generation from salinity differences~\cite{kedem1961physical,logan2012membrane,siria2013giant,werber2016materials}, osmotic forces are harvested in a considerable number of applications in very different domains. In this section we review more specifically a variety of such applications based on (recently) elucidated transport mechanisms relying on osmotic forces. 

%\LB{ Article geoscience (Nat. Geo.) sur DO dans les roches \cite{plumper2017fluid} ``evidence that a lot of the reactive fluid flow in the solid Earth occurs through nanoscale fluid pathways.''}

\subsection{Water treatment and membrane separation}

\subsubsection{Reverse and forward osmosis and their limitations}
%Water filtration, membrane separation and new nanomaterials}

%\LB{insister key is selectivity, not permeability; choose what you want to leave}

Access to clean water and cleaning water from industrial waste is a great challenge\cite{jones2018state}: in 2015 still 663 million people worldwide lacked access to drinkable water~\cite{world2015progress}, and cleaning waste water is becoming a major challenge in oil and gas industries~\cite{vidic2013impact,gregory2011water}; going further, some new regulations may appear to enforce a zero liquid discharge for industrial waste, thus requiring complete recycling of water ressources~\cite{onishi2017desalination}. On a day-to-day basis, humanity consumes the equivalent of 10-100 cubic kilometers of fresh water~\cite{hoekstra2012water} for all purposes (agriculture, industry, domestic).
%and from the following sources: fresh water sources, from rain, and in terms of pollution). 
Because fresh water is not directly accessible everywhere, and in order to cover the growing need for freshwater, desalination of sea water and cleaning of waste water have become essential. 

\begin{figure}[h!]
\centering
  \includegraphics[width=0.49\textwidth]{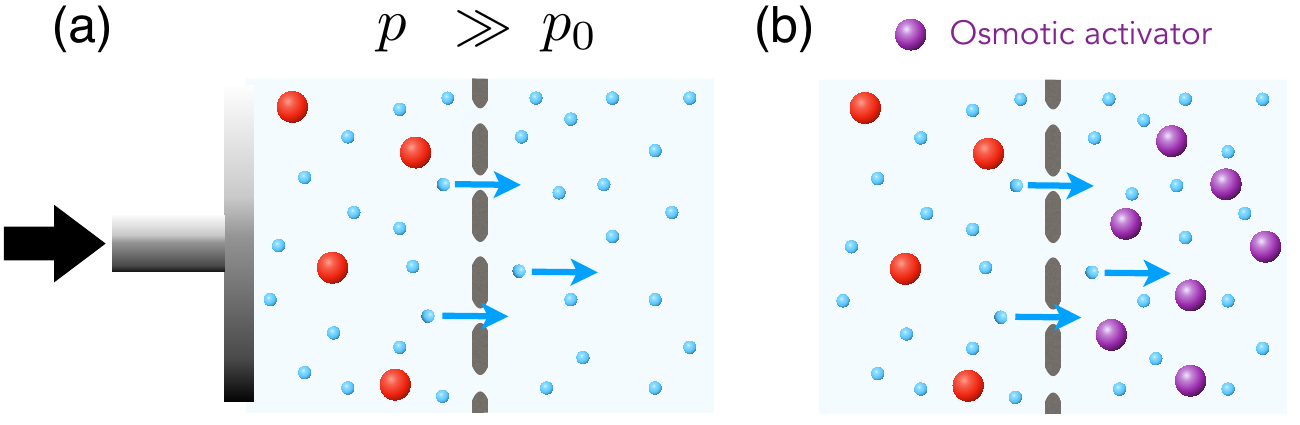}
  \caption{\textbf{: Reverse Osmosis and Forward Osmosis.} (a) Schematic explaining reverse osmosis, occurring via a piston applying a large hydrostatic pressure $p$ such that the difference to the atmospheric pressure (pressure of the other compartment) is larger than the osmotic pressure $p - p_0 \geq \Delta \Pi$. (b) Schematic explaining forward osmosis, occurring via the addition of some soluble species (here the purple solute) that increases the osmotic pressure on the drought side and therefore "attracts" water from the brine side.} 
  \label{fig:RODO}
\end{figure}

Lately, membrane based technology has established itself for water purification. %\LB{citer la logique des passoires}
Reverse osmosis is the most broadly used technique \review{(representing 62-65$\%$ of the installed capacity in 2015 for desalination~\cite{alkaisi2017review,amy2017membrane}, 24$\%$ being thermal based technologies)}. Reverse osmosis relies on the very simple principle of applying an external large hydrostatic pressure to counterbalance the osmotic pressure difference and induce a flow of water towards the low concentration side -- see Fig.~\ref{fig:RODO}-a. In particular, one can therefore extract water from seawater by \textit{concentrating} seawater even more, or extract water from waste water (in a simplistic view). For desalination the pressures involved are typically of 30-50 bars in order to exceed the osmotic pressure.

In a different approach, forward osmosis (combined with thermal methods for desalination) makes use of draw solutions to counterbalance the salinity induced osmotic pressure~\cite{mccutcheon2005novel,linares2016life,chen2019functionalized} -- see Fig.~\ref{fig:RODO}-b. 
\review{Generating a high osmotic pressure, typically above the 30 bars of pressure between sea and fresh water, requires draw solutes which are highly soluble in water, and also with a sufficiently small size (hence low molecular weight).}
%crucially have a low molecular weight 
\review{Indeed, as a rule of thumb,} for a solute with elementary volume ${v}_0\sim r^3$ with $r$ the solute size, the maximum
osmotic pressure which can be achieved is typically $\Delta \Pi \sim {k_BT\over v_0}$. This would suggest that solutes with size above 
$1$~nm are not able to achieve a sufficiently high osmotic pressure for desalination (note that this argument forgets departures from ideality, which could increase more strongly the osmotic pressure). In a more subtle way, a solute with a strong affinity towards water may also decrease the water chemical potential and modify accordingly the pressure. This echoes the huge pressure drops measured with hydrogel structures~\cite{wheeler2008transpiration}.

A number of other membrane based techniques with similar geometry are used or being developed, from electrodialysis (based on electric potential driving of salts)~\cite{semiat2008energy,chao2008feasibility} to capacitive deionization~\cite{oren2008capacitive,suss2012capacitive,kim2017low}, but also shock electrodialysis based on the idea of combining salt recovery with a porous charged material~\cite{deng2013overlimiting}, concentration polarization ~\cite{kim2010direct}, and other techniques harnessing chemical phenomena like adsorption desalination~\cite{shahzad2018adsorption} or biodesalination~\cite{aramburo2017desalination,minas2015biodesalination} and bio-water treatment~\cite{liu2004production}.

%\LB{A word on {\bf FO+draw solutes \cite{mccutcheon2005novel,chen2019functionalized}, Elimelech: ``An
%effective draw solution solute must have very
%specific characteristics. It must have a high osmotic
%efficiency, meaning that it has to be highly
%soluble in water and have a low molecular weight
%in order to generate a high osmotic pressure.
%Higher osmotic pressure leads to higher water flux
%and feed water recovery.''
%Discuss the fact that gelly systems reduce the chemical potential of water, cf Stroock $\Delta P_{gel}$.}  \cite{wheeler2008transpiration}
%}

Membrane-based technologies suffer from a number of limitations. First, they have a high-propensity to fouling by molecules which are larger than the critical molecule size allowed to pass~\cite{elimelech2011future}; also due to the large pressures applied during reverse osmosis.  Second, because they are passive membranes -- essentially discriminating particles upon their size -- they can not be at the same time \textit{very selective and highly permeable}. This was formalized for ultrafiltration membranes in \citefulls{mehta2005permeability,park2017maximizing}. In fact increasing the permeability of a membrane (and therefore the energy required to recover a given amount of cleaned water) requires essentially to broaden the size of the pores, as water flow within pores is limited by friction on the pore walls. However this leads inevitably to a decrease in the selectivity or separation properties of the membrane; and reciprocally. This is called the selectivity-permeability trade-off. For nanofiltration membranes (used for reverse osmosis and so on) the same trade-off exists although the proper establishment of a limiting regime is still empirical~\cite{werber2016materials} -- see Fig.~\ref{fig:Selectivity}. 
Finally, one challenging progress route for membrane separation is the ability to perform molecular scale design~\cite{werber2016critical} and therefore to ensure the best selectivity properties to eliminate \textit{e.g.} micropollutants, some of which are of great concern for health~\cite{stephens2016pcbs}. Overall, it should be realized that the main current challenge in desalination and water purification is not really the permeability of the membrane, but rather achieving a well-controlled selectivity to retain/reject specific species.

\begin{figure}[h!]
\centering
  \includegraphics[width=0.49\textwidth]{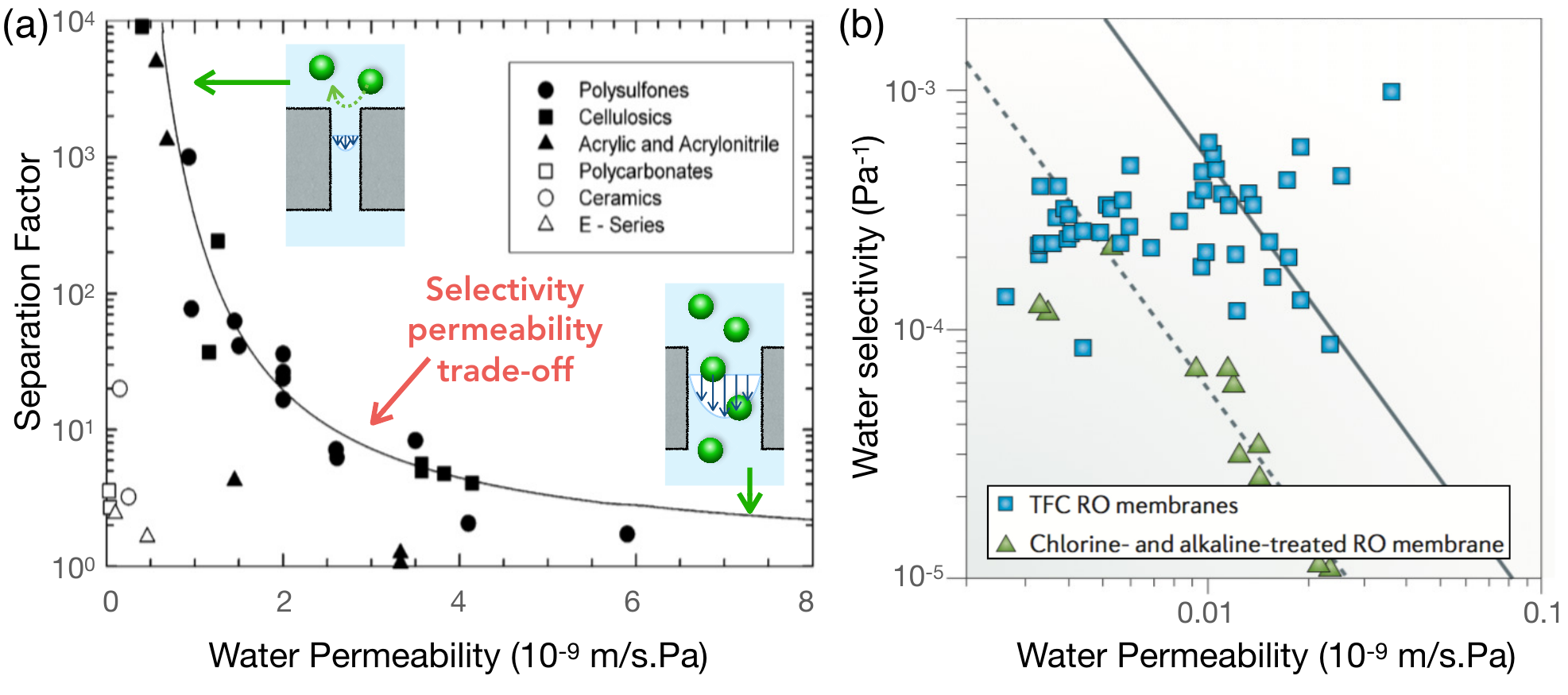}
  \caption{\textbf{: Selectivity Permeability Trade-off.} (a) Adapted from \citefull{mehta2005permeability}. Selectivity versus permeability values for ultrafiltration membranes (used for separation of larger molecules than salt ions). Bovine serum albumine is the model molecule for selectivity. The line indicates the standard selectivity permeability model (with a log normal distribution of pores, Poiseuille flow and selectivity given by Zeman and Wales exclusion rules~\cite{zeman1981polymer}).  (b) Adapted from \citefull{werber2016materials} with permission from Springer Nature, copyright 2016.  Selectivity versus permeability values for reverse osmosis membranes using salt as the model specie for selectivity. The lines correspond to the empirical models inspired from \citefull{geise2011water} to relate maximal selectivity and permeability in reverse osmosis membranes.} 
  \label{fig:Selectivity}
\end{figure}

%Alternatively some people try to perform ressource recovery from waste water [59–61], both to generate fresh water but also to find added value to the (solid) waste. Re-use and reclamation of waste water from industrial or municipal sources to restore it to potable quality seems like a good idea to achieve on-site sustainable wastewater treatment in cities – especially since pumping water over long distances is also very expensive [62,63]. The technical challenge is great !
%A side yet critical issue for drinkable water is micropollutants, because of their variety [57], and also because they may be found in a great range of concentrations (from ng/L to $\mu$g/L) [4]. Some of these micropollutants are of great concern for health [58].

\subsubsection{What can we expect from new nanomaterials and nanofluidic devices ?}

It may appear that developing the \qote{ideal membrane}, which is both highly selective and highly permeable, is like squaring the circle. However, Nature has achieved this \textit{tour-de-force}, with water porins like aquaporins exhibiting unrivalled performances in terms of selectivity and permeability \cite{barboiu2016artificial}.
This requires to develop new artificial materials with properly decorated nanopores allowing for such exquisite design, for example self-assembled artificial water channels \cite{barboiu2016artificial,shen2018achieving,sun2018imidazole}, or tailor-made DNA origami channels~\cite{bell2011dna,langecker2012synthetic}. 
This is actually a challenge that
nanoscale science may be up to \cite{majumder2017flows,wang2017fundamental,barboiu2016artificial,mauter2018role}.
%A number of other materials are triggering broad interest such as graphene oxide based membranes~\cite{majumder2017flows,siria2017new} or tailor-designed DNA origami channels~\cite{bell2011dna,langecker2012synthetic}.

The development of new nanomaterials has indeed allowed the emergence of new avenues for  membrane separation.
Graphitic materials of various forms and geometries, such as carbon nanotubes, graphene and lately graphene oxides membranes, have raised
considerable promises, see \citefulls{majumder2017flows,wang2017fundamental} for reviews on this topic. Carbon materials were consistently shown to exhibit ultra-low water friction 
and high permeability, and this represents a key asset to minimize the viscous loss in separation processes. 
Furthermore advanced functionalization allows one to decorate the nanotubes improving selectivity \cite{lokesh2018osmotic,nednoor2007carbon,chan2016novel}. 
Membranes made of nanopores in \textit{e.g.} 2D graphene sheets have a molecular thickness, while keeping high mechanical strength: this accordingly increases the driving forces for transport (which scale like the inverse thickness) by orders of magnitude, hence all transport coefficients and the overall efficiency of the process \cite{wang2017fundamental}.

Still, these graphitic systems -- carbon nanotubes, graphene slits and more generally 2D materials -- remain difficult to fabricate as large-scale membranes. Upscaling towards industrial applications is a considerable challenge. The advent of graphene oxide membranes and their derivatives may change the story. These are constituted of graphene flakes, which organize into parallel stacks of  graphene layers, having nanoslits in a staggered alignment and an interlayer distance which is typically below the nanometer.  In spite of the complex labyrinthine flow 
across multiple graphene layers \cite{yoshida2016labyrinthine}, the membranes demonstrate large permeability 
\cite{abraham2017tunable,thebo2018highly}.  Last but not least, they are relatively easy to fabricate at large scale. Such systems therefore appear as ideal membranes for ionic separation 
\cite{mi2014graphene} and may well revolutionize the domain of filtration. 

Now, beyond materials themselves, it should be realized that nanoscales allow for many new \qote{exotic} transport phenomena, the consequences of which  have  -- up to now -- been barely harnessed. One may quote for example the membranes made of hydrophobic nanopores, making use of nanobubbles as a semipermeable sieve for osmotic phenomena \cite{lee2014nanofluidic};
%
%smosis membranes comprising short hydrophobic nanopores that separate two fluid reservoirs.
or the ionic and osmotic diodes, allowing for rectified transport in membranes, or the active osmotic phenomenon, as we discussed above; 
or in a different context,  the specific adsorption properties of graphene oxide membranes allowing for water-ethanol separation in membranes\cite{gravelle2016carbon}, which are far more efficient than standard distillation. 
Such routes would deserve a proper exploration to go beyond the relatively basic sieving principles underlying membrane science.
These should offer alternative routes for filtration and separation which still need to be invented. 

%\LB{LB: figure active pore enlev\'ee pour l'instant, \`a d\'ecider}
%\begin{figure}[h!]
%\centering
%  \includegraphics[width=0.49\textwidth]{FIGURES/SelectivityPermeabilityActive.pdf}
%  \caption{\textbf{Active filtration beyond the sieving paradigm} (a) Selectivity versus permeability at different forcing frequencies of opening of an active pore presented in (b) (nondimensional units). The pore is a circular pore of radius $r(t) = r_0( 1 + \varepsilon \sin (\omega t)) + \delta r$ where $\varepsilon = 0.2$ in this example, $\delta r$ is a gaussian noise with relaxation rate $\lambda$ and amplitude $\langle \delta r^2 \rangle = \theta$. $\lambda$ and $\sqrt{2\theta}$ are used to nondimensionalize respectively time and space variables. The selectivity is defined as the ratio of the permeance of a particle $K_{\infty}^k$ with mobility $k$ to the permeance of a reference particle with mobility $k_0$. In this example $k/k_0 = 10$. More details can be found in Ref.~\citenum{marbach2017active}.}
%  \label{fig:ActiveSP}
%\end{figure}

\subsection{Osmosis in biological systems: aquaporins, ion pumps and the kidney}

%\LB{NPO: in physiology, osmosis is usually the method of choice to quantify the permeability of porins, eg AQP. Use vesicle which inflates under an osmotic stress. Discuss rectification also ?}

Osmotic forces are harvested in the biological world in a considerable variety of phenomena and contexts: to store energy,  induce mechanical motion,  control ejection and absorption of compounds, \textit{etc}. Osmotic pressure was much studied at first in plants~\cite{dutrochet1826agent}, and one may in fact assess that plant life depends on osmotic forces. Indeed plants can not rely on muscle for force generation, yet they are able to achieve tensile and compressive stresses on a much wider range~\cite{dumais2012vegetable}. To produce motion or growth, they rely on an underlying hydraulic machinery driven by osmotic or humidity gradients. For example, phloem~\footnote{Phloem is a living tissue that transports soluble compounds in particular sugar in plants} flow harnesses osmotic driving to transport sugar over long distances~\cite{broyer1947movement,thompson2003scaling,jensen2009osmotically,jensen2011optimality,comtet2017passive}. Osmotic transport is critical to regulate size in \textit{e.g.} conifer leaf~\cite{rademaker2017sugar}. The opening and closing of stomata on leaves~\footnote{Stomata are small pores at the leaves surface that control leaf transpiration}~\cite{mansfield1990some} and the circadian motion of various plants and flowers~\cite{satter1981mechanisms,van2003flower} is regulated by swelling or shrinking driven by water flows. Those flows are generally actuated by active transport of solutes through specialized pumps~\cite{hedrich1989physiology,irving1997phototropic}. Even biofilms harvest osmotic pressure gradients in the extracellular space for surface motility~\cite{seminara2012osmotic}.

In animals and human beings, a number of processes involve osmotic flows for water or volume regulation and transport: from the kidney~\cite{marbach2016active} to the liver~\cite{meyer2017predictive} and the intestine~\cite{thiagarajah2018water}, not forgetting salivary secretion~\cite{turner2002understanding}. Cells harvest osmotic forces in a variety of ways, most obviously to control expansion and regulate size, \textit{e.g.} in cysts~\cite{gin2010model}, and also to regulate absorption~\cite{rauch2000endocytosis} or ejection of genetic material~\cite{evilevitch2003osmotic} via small capsules. A number of processes also harvest more subtle forces in a fascinating way and we cite a few to engage the curious reader. % in curiosity driven research. 
Osmotic pressure changes may affect frequency of miniature end-plate potentials in neuromuscular junctions~\cite{hubbard1968examination,furshpan1956effects}, but also drive oscillatory flows for cell regeneration~\cite{kucken2008osmoregulatory}. Electro-osmosis is harvested for uphill transport of water by insects in draught areas~\cite{kuppers1986uphill} but also more generally for epithelial transport~\cite{fischbarg2017epithelial,dvoriashyna2018osmotic}.

%\LB{ajouter ref Stone Nature Com 2018: `` bacterial cells to establish an osmotic pressure difference between the biofilm and the
%external environment. This pressure difference promotes biofilm expansion on nutritious
%surfaces by physically swelling the colony, which enhances nutrient uptake, and enables
%matrix-producing cells to outcompete non-matrix-producing cheaters via physical exclusion''}
The list of examples of osmotic transport in biological systems is nearly infinite, and occurs at all possible scales from individual molecules to organs and tissues. It is pointless to attempt a thorough review.  Rather, 
we discuss below in more detail three specific biological phenomena related to osmosis. These examples raise in particular the question of  whether such phenomena may be mimicked artificially to achieve advanced osmotic transport in artificial devices. 
%ask the question of how we may inspire and learn lessons from such objects 
%for artificial devices to perform advanced osmotic transport. 

\subsubsection{Aquaporins: the ideal semi-permeable membrane}

A decisive turnpoint in the study of nanoscale systems was triggered notably by the discovery of nanoscale channels in biology. One of the most famous of these channel families is the aquaporin family (the most common being AQP1 or CHIP-28, see Fig.~\ref{fig:Aquaporin}-a)~\cite{preston1992appearance,murata2000structural}. An aquaporin is a water-specific channel; aquaporins are present in many organs in living systems, animals, but also plants~\cite{maurel2008plant}: they play a central role in the human kidney (see below), are also key role players in red blood cells and many other organs~\cite{agre2006aquaporin}, \SM{and regulate water uptake in plants~\cite{javot2002role}}. The striking specificity of aquaporins is that they are both highly selective to water \textit{and} highly permeable. \SMM{The permeability of an aquaporin was measured notably by P. Agre {\it et al.} to be in the range of $p_f=11.7\times 10^{14}$ cm$^3$.s$^{-1}$ at 37$^\circ$ C~\cite{zeidel1992reconstitution,roux2001structure}\footnote{A more recent measurement in \citefull{horner2015mobility} suggests that this value may actually have been underestimated by a factor 5.} (with $p_f$ here defined as $Q=p_f \,v_w {\Delta p}/k_BT$;  $Q$ is the water flux and $v_w$ the bulk water molecular volume). This corresponds to $\approx 3$ million water molecules translocating per second per bar ($p_f$ being related to the particle flux $dN/dt$ according to $dN/dt=({p_f/ k_BT})\Delta p$).
%by allows 3 billion water molecules through per second under a pressure drop of 1 bar~\cite{zeidel1992reconstitution,roux2001structure} -- 
The value of the permeability of the aquaporin is much larger than that for other channels, see \cite{barboiu2016artificial} for a comparison, or what would predicted by continuum dynamics at these scales~\cite{walz1994biologically,bocquet2010nanofluidics}}. 

\begin{figure}[h!]
\centering
  \includegraphics[width=0.49\textwidth]{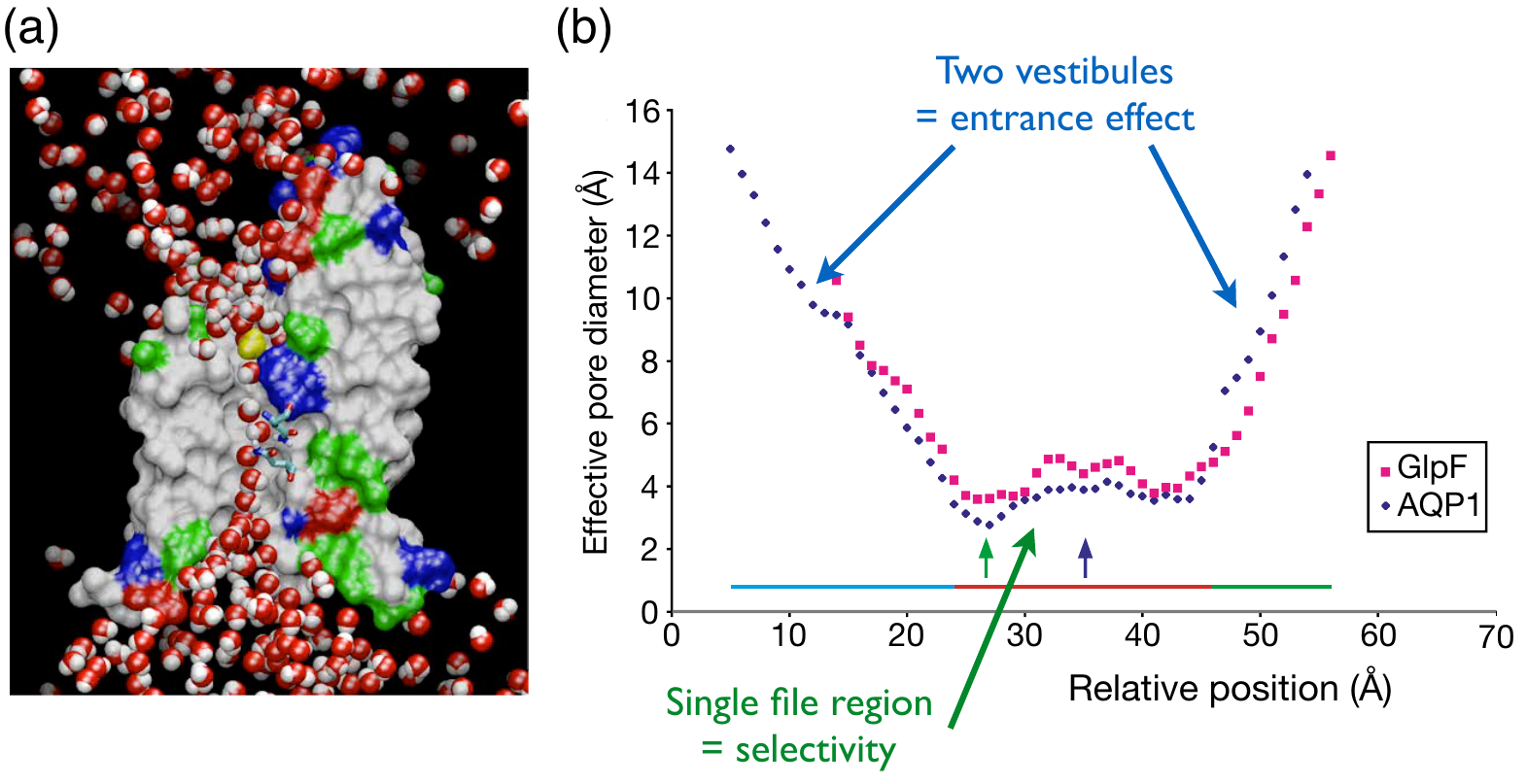}
  \caption{\textbf{: The aquaporin AQP1 water channel.} (a) Molecular dynamics simulation of water transversing an aquaporin channel. Snapshot from movie in \citefull{tajVideo} under Creative Commons license, in complement to \citefull{tajkhorshid2002control}. (b) Effective pore diameter  of the AQP1 and GlpF channels. Pore diameters were determined with AMBER-based van der Waals radii and analysed using the program HOLE38. Reproduced and adapted from \citefull{sui2001structural} with permission from Springer Nature, copyright 2001.} 
  \label{fig:Aquaporin}
\end{figure}

Aquaporins present several intriguing features: surprisingly they are hydrophobic channels~\cite{sui2001structural} and they are extremely constricted~\cite{sui2001structural} --  only 3 \AA~in diameter at the narrowest point that allows for this selectivity. 
An aquaporin-based membrane constitutes therefore a somewhat ideal semipermeable membrane.
All of its exceptional transport properties are intimately connected to its nanoscale (and even \.{A}ngstr\"{o}m-scale) structure -- thus hinting to the striking and appealing properties of fluid flow at the nanoscale. It is thus natural to look for artificial solutions for semi-permeable membranes harvesting properly designed nanoscale structures \cite{shen2018achieving,sun2018imidazole}.  For example, aquaporins present a sophisticated hourglass shape, that is believed to enhance the water permeability~\cite{gravelle2013optimizing}, see Fig.~\ref{fig:Aquaporin}-b. Such a geometry could be readily mimicked using e.g. pore coatings~\cite{perez2015polydopamine} to enhance permeability of membranes.

%\LB{cite slippage combined with hour glass shape}

\subsubsection{Kidney: an ultra-efficient and unconventional osmotic exchanger}

As a second example, we discuss the  separation process occurring in the kidneys. As we highlight, the efficiency of the kidney filtration process   takes its root in a very unconventional osmotic process, 
% as we discuss below, is highligh unconventional and 
and could be inspirational for future separation technologies. 
%\LB{intro: insist on Kidney as a very unconventional sepation process !!}

Per day, the human kidney is capable of recycling about 200 L of water and 1.5 kg of salt, separating urea from water and salt at the low cost of 0.5 kJ/L~\cite{greger1996comprehensive} while readsorbing $\simeq$ 99$\%$ of the water input. The core of the kidney separation process lies in the millions of parallel filtration substructures called nephrons~\cite{greger1996comprehensive}. A striking feature is that the nephrons of all mammals present a precise loop geometry, the so-called Loop of Henle - in the shape of the letter "U" -- see Fig.~\ref{fig:Kidney}-a. This loop plays a key role in the urinary concentrating mechanism and has been extensively studied from a biological and physiological point of view~\cite{greger1996comprehensive,baylis1989water,stephenson1972concentration,layton2010mathematical,edwards2009modeling}. 

\begin{figure}[h!]
\centering
  \includegraphics[width=0.49\textwidth]{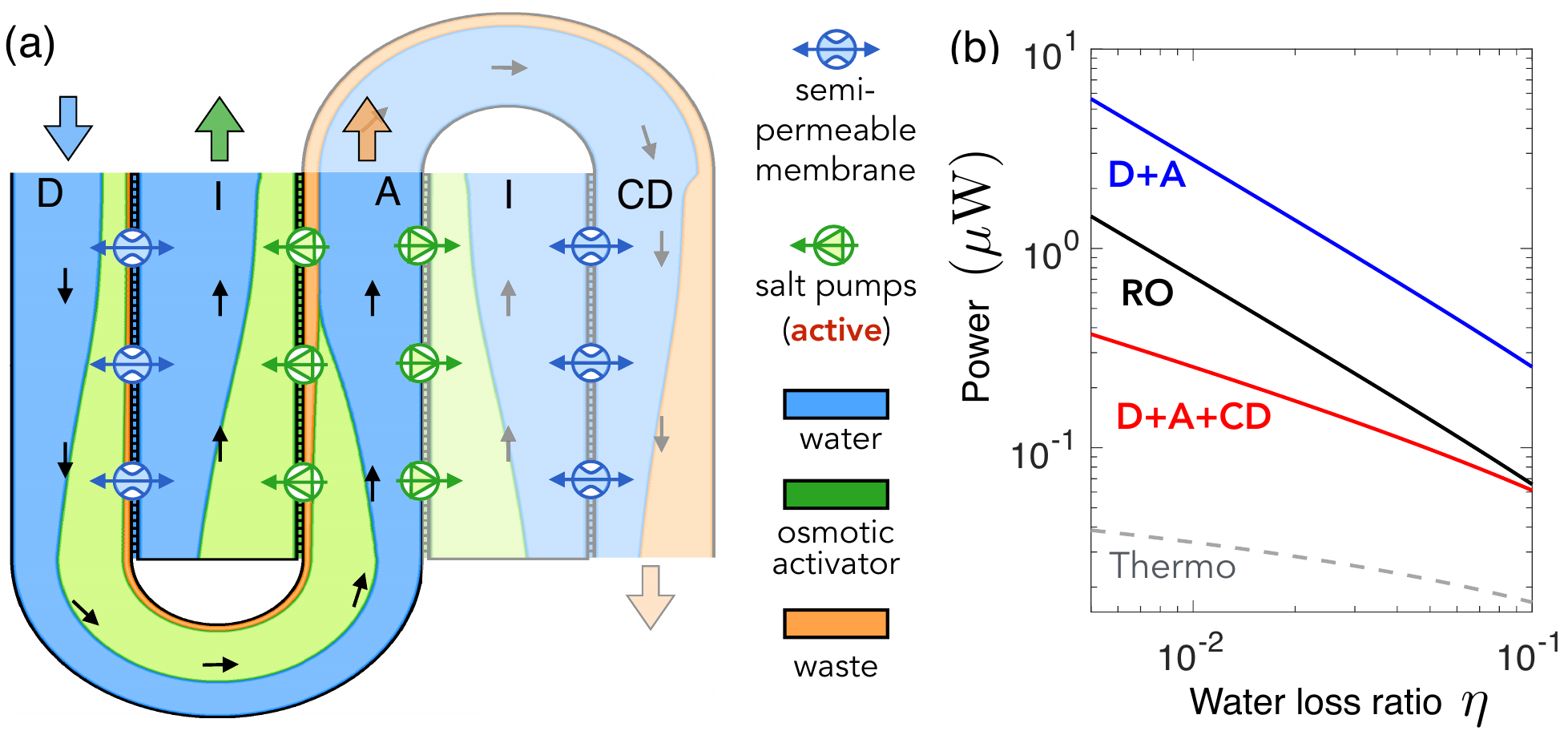}
  \caption{\textbf{: The osmotic exchanger principle of the kidney.} (a) Inspired from \citefull{marbach2016active}. Water, salt and urea molar fractions are represented in various colors along the U tube (descending, D and ascending A) limbs and the interstitium (I). For visibility, the water molar fraction was divided by 100. Black arrows represent the direction of flow.
  %(b) Microfluidic counterpart with (bottom) interstitium as a U shaped tube, (middle) separated by an ion selective or a semipermeable membrane from (top) the primary U tube. Electrodes are inserted in top and bottom to drive salt flow via an electric potential difference. In both schematics 
 A semi-permeable membrane (containing aquaporins) separates the descending limb and the interstitium, while the ascending limb contains salt pumps transporting actively the salt to the interstitium. A third limb, the collecting duct (CD, in lighter collers) also exchanges with the interstitium via a semi-permeable membrane. The latter is crucial for the overall efficiency of the separation process. 
 %The blue arrow represents the primary inlet with water, salt and urea, the orange the outlet with highly concentrated urea, and the green the secondary outlet with water and salt. 
  (b) Adapted from \citefull{marbach2016active}. Power required for the functioning of the separation process as a function of the targeted water loss ratio: for the simple loop geometry (A+D), for the full serpentine (A+D+CD), as compared to the equivalent reverse osmosis process under a pressure gradient (RO). (a) and (b) are under Creative Commons Attribution 3.0 License.  } 
  \label{fig:Kidney}
\end{figure}

%Recently we put forward a physical mechanism based on biological insight, see Ref.~\citenum{marbach2016active}. In fact 
As put forward in \citefull{marbach2016active}, the U loop acts as an osmotic exchanger, similar in concept to a thermal exchanger -- see Fig.~\ref{fig:Kidney}-a. A mixture of water, salt and urea (or any other compound to be separated) enters and flows through the tubular U loop.  
Water and ions may be exchanged through the tube walls with a common interconnecting media, called the interstitium. On the descending side (D), aquaporins allow for water permeation across the walls. On the ascending side (A), salt is actively pumped, using an external source of energy (in the case of the kidneys, the dissociation energy of Adenosine Tri-Phosphate, ATP). This pumped salt results in an increased salt concentration in the interstitium, higher than the concentration of salt and urea in the descending tube (D). The osmotic pressure is therefore inverted and drives water from the U tube to the interstitium across the aquaporin channels. As a result, urea is highly concentrated in the U loop, while salt and water are redirected from the interstitium towards the blood circulatory network. 
%The longitudinal extension of the system allows the full equilibration of the osmotic pressure along the limbs connected by aquaporins (and therefore a semi-permeable channel). 
This U-shape geometrical design is key to the efficient operation of the separation. Note that the third limb following the U-tube plays a crucial role in 
enhancing the separation efficiency~\cite{marbach2016active}. 

{One may actually estimate the working efficiency of this osmotic exchanger in a simple way, providing a lower bound on the separation
ratio. It is quantified in terms of the amount of {\it lost water} $\eta = c_w^{\rm A, top} v_{\rm A,top}/ c_w^{\rm D, top}  v_{\rm D,top}$, where $v$ is the flow velocity calculated at the top of the ascending (A) or descending (D) branch, and $c_w$ is the concentration of water.
  %may be simply rationalized in terms of an osmotic pressure balance. 
  For the system to work, water has to flow from the descending branch towards the interstitium 
  %At the top of the single U loop, we expect water to continue flowing from the descending limb to the interstitium. and this 
  and this requires that chemical activities obey
%\begin{equation}
$a_{\rm D, top}^{\rm Water} \geq a_{\rm I, top}^{\rm Water}$.
%\end{equation}
The latter can be expressed simply (in the low concentration regime) in terms of molar fractions and one obtains 
\begin{equation}
\frac{c_w^{\rm D, top}}{c_w^{\rm D, top} + c_s^{\rm D, top} + c_{\rm waste}^{\rm D, top} }\geq \frac{c_w^{\rm I, top}}{c_w^{\rm I, top} + c_s^{\rm I, top} }
\label{eq:balance}
\end{equation}
where $c_w$, $c_s$ and $c_{\rm waste}$ are respectively the concentrations of water, osmotic activator (salt) and waste. Assuming that all the osmotic activator has been reabsorbed in the upper branch yields $c_s^{\rm I, top} = \frac{v_{\rm D,top}}{v_{\rm I,top}} c_s^{\rm D, top}$. Water flow is conserved and thus $c_w^{\rm D, top} v_{\rm D,top} = c_w^{\rm I, top} v_{\rm I,top} + c_w^{\rm A, top} v_{\rm A,top}$. A lower bound for the fraction of lost water $\eta_{\rm lost}$ %$ =  [{\rm Water}]_{\rm A, top} v_{\rm A,top}/ [{\rm Water}]_{\rm D, top} v_{\rm D,top}$ flow 
can then be simply  deduced from \eqref{eq:balance} as
\begin{equation}
\eta_{\rm lost} \geq \left(\frac{c_{\rm waste}^{\rm D, top}}{c_s^{\rm D, top} + c_{\rm waste}^{\rm D, top}}\right)^n.
\end{equation}
with $n=1$. 
For the geometry including a third reabsorbing branch,  the collecting duct, see Fig.~\ref{fig:Kidney}-a, a similar reasoning yields the same result  with $n=2$. The square exponent thus leads to much smaller lost water fraction $\eta_{\rm lost}$ showing that this third branch is essential in the overall efficiency of the kidney separation. 
%\begin{equation}
%\eta \geq \left(\frac{[{\rm Waste}]_{\rm D, top}}{[{\rm Osm}]_{\rm D, top} + [{\rm Waste}]_{\rm D, top}}\right)^2.
%\end{equation}
}
Using physiological values for the concentration, this estimate provides a prediction for water reabsorption, and thus urea separation, in the range of $\eta_{\rm lost} \sim 1$\%, which is in excellent agreement with every-day life experience; see \citenum{marbach2016active}. %$\eta \sim 1$\%.
%A further interesting feature of the kidney as compared to standard filtration devices is that it operates at a remarkably low pressure -- the arterial blood pressure -- that is two orders of magnitude smaller than the pressure required for reverse osmosis~\cite{bocquet2010nanofluidics}.
\SM{To some extent, note that the osmotic exchanger of the kidney may be compared to a forward osmosis process. However the key difference is the geometry with 3 limbs that allows for a more efficient reabsorption of water.}

In fact, energy wise, this system is also shown to be far more efficient than standard reverse osmosis principles, as can be estimated within the above model, see  Fig.~\ref{fig:Kidney}-b.
In living systems, the nephron operates the separation of urea from water near the thermodynamic limit, $\simeq$ 0.2 kJ/L~\cite{marbach2016active}. Yet, standard dialytic filtration systems, which are based on reverse osmosis and passive equilibration with a dialysate, require more than two orders of magnitude more energy~\cite{gambro}.

Some attemps to build artificial devices mimicking the nephron were reported in the literature, but they rely on biological tissues or cell mediated transport, and cannot be easily scaled up and transferred to other separation devices~\cite{borenstein2007microfabrication,kim2011wearable,armignacco2015wak}. \SM{None of the approaches so far rely on the specific geometry of the U-loop to improve the filtration process.} Mimicking the separation process occurring in the kidney based on the physical perspective described above can now be foreseen using microfluidic elementary building blocks.
 %and we give a general perspective in Fig.~\ref{fig:Kidney}-b. 
 Such an artificial counterpart could have definite impact in improving the early stage of desalination processes consisting in water pre-treatment to suppress large polluting molecules, and, down-the-road, could lead to substantial progress in dialysis for patients suffering from kidney failure.

%in plants, starting investigation for glycerol water mixtures\cite{ray1960theory} 

%also there is this Lyklema studying proteins and interactions electro with interfaces ~\cite{norde1991proteins}

\subsubsection{Proton pumps, chemi-osmosis and advanced ionic machinery}

As a last example, we discuss proton pumps and channels, which are compelling illustrations of how Nature harvests osmotic forces to drive mechanical parts. Biological systems have developed a 
%Among the
fascinating artillery of devices to passively and actively transport ions, namely ionic channels and ion pumps. Among these, proton pumps are canonical examples. We detail a few examples below. 

\noindent  \SM{\textit{Proton pumps to build proton gradients} -- There exists a great variety of ways to actively transport protons in biology, from combined proton-electron transfer in cytochrome oxidase (crucial for respiration~\cite{kaila2010proton}) to proton pumps implying the participation of ATP -- the latter are called H$^+$-ATPases~\cite{pedersen1987ion}.  ATP-ases play a key role in bio-energetics and are ubiquitous in many forms of life and plants~\cite{mulkidjanian2007inventing}. They include three types. The P-type ATP-ases include in particular the plasma membrane H$^+$-ATPase, that uses the dissociation energy of ATP to form gradients of protons. These gradients are crucial for plant movement (from phloem loading, to size regulation in the stomatal aperture, to tip growing systems~\footnote{Tip-growing systems, such as pollen tubes or root hairs, continuously grow in one direction.}~\cite{palmgren2001plant}). The V-type ATP-ases also use the dissociation energy of ATP to form gradients of protons. Interestingly this chemical reaction is accompanied by a rotary motion of the protein. It is central to many processes in animals~\cite{nishi2002vacuolar}, from acid base balance in the kidney, pH maintenance in mechanosensory hair cells, bone resorption, tumour metastasis, sperm motility and maturation \textit{etc.} The last type, the F-type, can work similarly to the V-type~\cite{murata2005structure,meier2005structure} and consumes ATP to form gradients of protons depending on aerobic conditions~\cite{capaldi2002mechanism}. However it most commonly works the reverse way, \textit{e.g.} consuming the proton gradient and synthesizing ATP, and we discuss that below.}

\begin{figure}[h!]
\centering
  \includegraphics[width=0.49\textwidth]{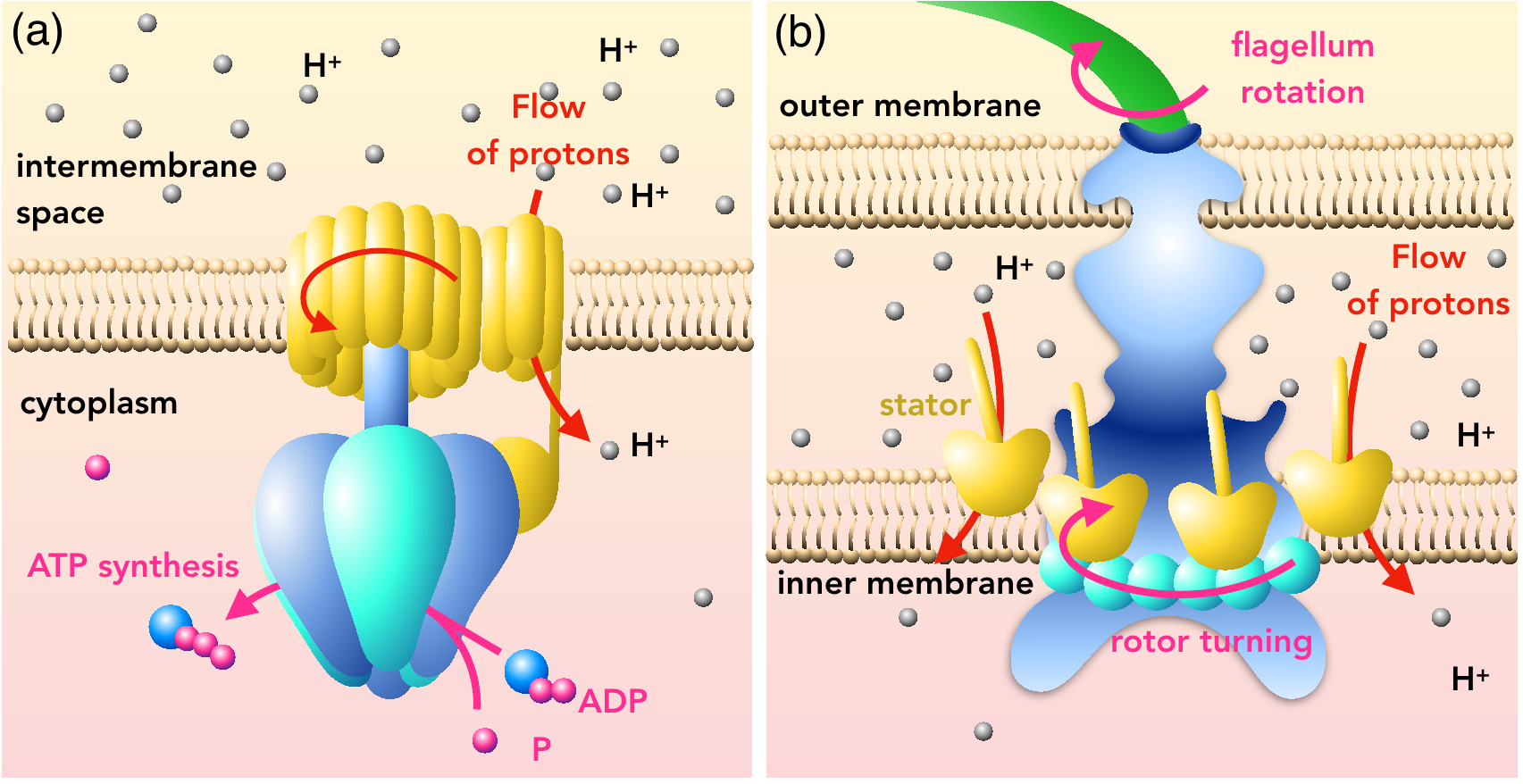}
  \caption{\SM{\textbf{: Harvesting proton gradients: energy vectorization and locomotion} (a) Simplistic view of an F-type H$^{+}$-ATPase, here working as an ATP synthesis enzyme. A proton gradient is maintained between the intermembrane space and the cytoplasm by the respiratory cycle. Protons thus naturally flow inwards through the proton channel of the ATPase (in yellow). This triggers a mechanical rotation of the central element of the ATPase that in turn catalyzes the synthesis of ATP from Adenosine diphosphate (ADP) and Phosphate (P). (b) Simplistic view of the bacterial flagellar motor (in blue). The proton gradient transverses here the stator parts of the motor (in yellow), namely the MotA/MotB complexes. These are responsible for turning the basal rotor of the flagellar motor. As the flagellum is attached to the motor, this induces rotation of the flagellum and allows for bacterial locomotion.}} 
  \label{fig:Proton}
\end{figure}

\noindent \SM{\textit{Proton gradients harvested for energy vectorization and locomotion} -- The idea that osmotic gradients could be harvested for advanced functionalities was introduced as early as in the 1960s, by the seminal work of Mitchell in~\citefull{mitchell1961coupling}. He introduced the concept of \textit{chemi-osmotic coupling}, namely that a chemical reaction may be powered by the directed channeling of a specie. In the case of the F-type ATP-ase, directed motion of protons (note that the full reaction does imply the production of water on one side of the membrane) catalyzes the synthesis of ATP through a rotary motion~\cite{meier2005structure} -- see Fig.~\ref{fig:Proton}-a. As the F-type catalyzes the formation of ATP (that is the vector for energy in all living systems), it is central to all forms of life~\cite{junge2015atp}. The rotary motion occurring during synthesis can be harvested for artificial locomotion of inorganic devices~\cite{soong2000powering} . Harvesting proton gradients for locomotion is more commonly performed not by the F-type ATP-ase but by the bacterial flagellar motor~\cite{sowa2008bacterial} -- see Fig.~\ref{fig:Proton}-b.  The bacterial flagellar motor is an impressive 45 nm~\cite{sowa2008bacterial} ionic machinery at the root of bacterial locomotion via flagellar rotation notably in E. coli~\cite{silverman1974flagellar}. A proton gradient induces spontaneous transport of protons through stator parts (MotA/MotB)~\cite{kojima2004solubilization}. As the transport is gated through these channels, it induces a ratchet-like motion of the rotor part of the motor~\cite{nishihara2015gate}. The flagellum is attached to the rotor and therefore rotates. Around 1200 ions translocating per rotation generate a force at the base of the flagellar motor of about 200 pN~\cite{sowa2008bacterial}. The flagellum rotates at about 100 Hz~\cite{sowa2008bacterial} allowing E. coli to swim at more than 10 body lengths per second ! }

\noindent  \textit{Nanoscale ionic machinery} -- 
The proper function of the F-type enzyme is dependent on a subtle balance of osmotic and chemical potentials for proper function~\cite{cereijo1972osmotic} and the detailed mechanisms involving motion and electric field coupling to the proton flux are still investigated~\cite{miller2013electric}. Further physical insight on the detailed flows in the proton pump but more broadly on ionic channels is required to establish biomimetic principles to construct similar ionic machines with artificial material. Such physical insight is also dependent on better modeling of ion transport at the ultimate scales, with strong charge interactions, breakdown of hydrodynamics, \textit{etc.}

\subsection{Blue energy harvesting: osmotic power and capacitive mixing}
\label{BlueEnergy}

As we have seen, filtration and separation of molecules requires energy input to counteract the entropy of mixing. Reversely, entropic energy harvesting may be possible by \textit{mixing molecules}. The energy harvested from differences in salinity, \textit{e.g.} by mixing sea water and fresh river water, is called \textit{blue energy}. The maximal entropic energy collected by mixing volumes of sea and river water is typically  $0.8$ kWh/m$^3$, see~\citefull{yip2012thermodynamic}. Over the earth, counting the natural potential resources where rivers flow in the ocean such as the Amazonian river, a total of around $1$TW of power could be  harvested, amounting to 8,500 TWh in a year~\cite{logan2012membrane}. This is to be compared with the actual production of other renewable energies: in 2015, hydraulic energy production is $\sim$4,000 TWh, the nuclear energy around 2,600 TWh, and wind and solar 1,100 TWh altogether~\cite{statistics2017key}. In the global energy balance, blue energy, as a renewable and non-intermittent source of energy, has thus a great potential. Here we focus on some energetic and osmotic aspects of blue energy and refer to \citefull{siria2017new} for a more detailed review of the current status of blue energy harvesting.

%entropy of mixing; 
% PV = energy, P = energy/V; ok; so P = Pi = k_B T c avec c = 1mol/L = 1000mol/m^3; R = 8.314; T = 300; energy/V =  8,314*300*1000/3,464e6 = 0.7 kWh/m^3 en fait ça fait c/2 si tu mixes les memes volumes mais bref ARGH en fait ça marche pas ce truc

\begin{figure}[h!]
\centering
  \includegraphics[width=0.49\textwidth]{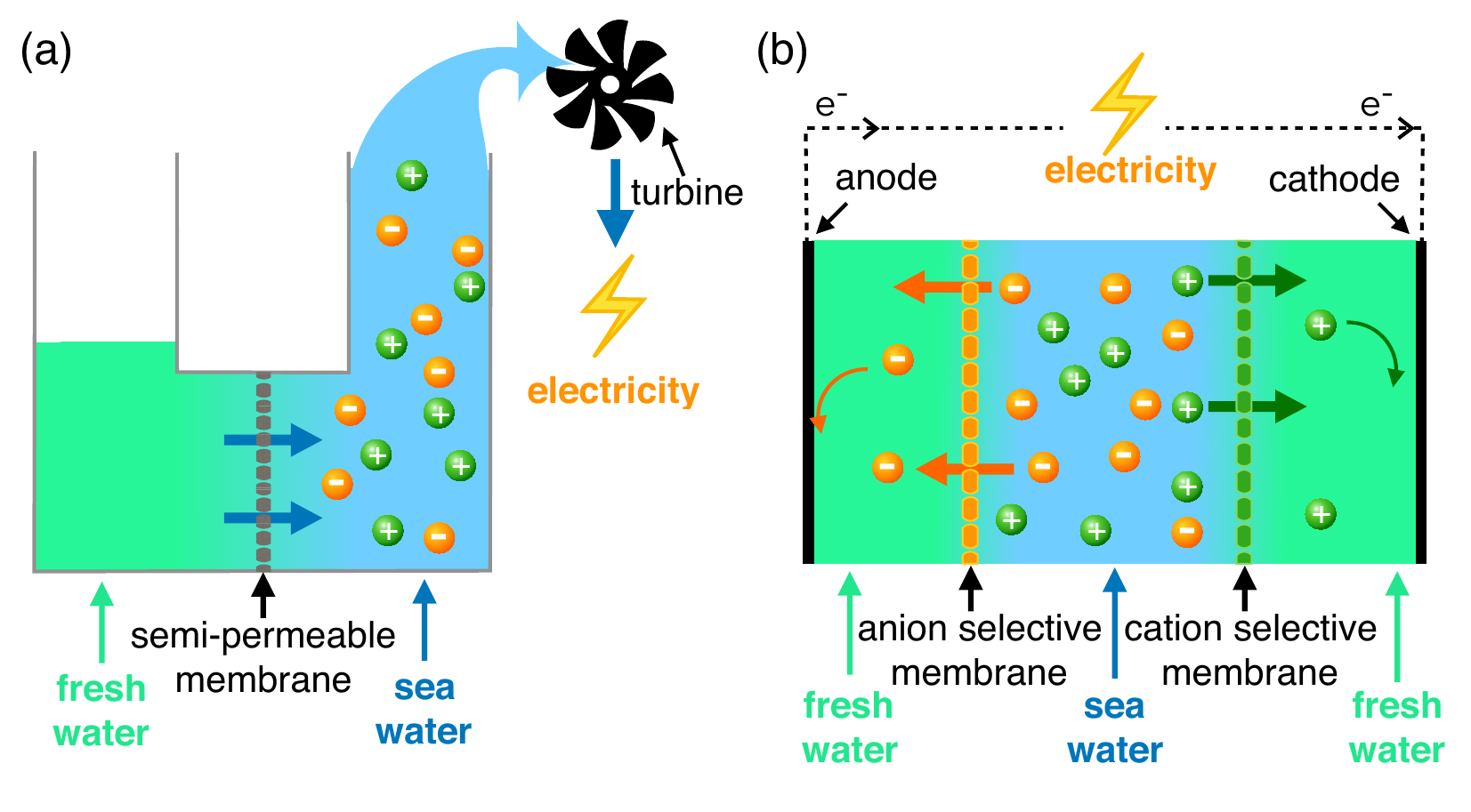}
  \caption{\textbf{: Collecting blue energy.} (a) Pressure retarded osmosis (PRO). The mixing of sea water and fresh water across a semi-permeable membrane drives a water flow that turns a turbine generating energy. (b) Reverse electro-dialysis (RED). Fresh and sea water are separated by stacks of alternating cation and anion selective membranes. Spontaneous diffusion induces fluxes of ions through the selective membranes, which  is captured at the boundaries by reactive electrodes producing an electric current. Usually RED is performed by alternating fresh and sea water a dozen times, although only three layers are represented on the figure.} 
  \label{fig:BlueEnergy}
\end{figure}

The current attempts to harvest blue energy have essentially relied on two techniques, as sketched in Fig.~\ref{fig:BlueEnergy}. Pressure-retarded osmosis (PRO) harvests the natural osmotic force between sea water and river water when they are separated by a semipermeable membrane to activate a turbine to generate electricity. Reverse electro-dialysis (RED) uses diffusion gradients of salts between sea water and river water  to directly generate  (ionic) electric currents by separating the corresponding ion fluxes using a  multi-stack of cation and anion selective membranes \cite{tufa2018progress}. Both strategies rely on separation of water from ions or ions from water, and therefore require subnanoporous structures which impede the water fluxes and diminish energetic efficiency. Current PRO technologies are only able to produce up to 3 W/m$^2$, less than the critical 5 W/m$^2$ for economic viability~\cite{skilhagen2010osmotic}.
The reasons for such a low performance can be readily understood: while the osmotic pressure at the interface between sea water and fresh water is considerable and reaches 30 bars, the permeability of the semi-permeable membrane is extremely small since its pore structure is in the sub-nanometer scale to sieve ions: the power, which is the product of flow rate and pressure drop is accordingly small.

 On the other hand, state-of-the-art RED achieved up to 8 W/m$^2$ in controlled environment~\cite{kim2010power}, and there is an industrial hope for blue energy harvesting which is currently explored with the REDStack project in the Netherlands~\cite{siria2017new,tufa2018progress}. 

Still we note that the above power figures should not be considered as negligible, because membrane systems are quite compact and hundreds of square meters of membranes can be packed over a single ground square meter. Such performances should be compared to the 2.5 W per square meter {\it of ground field} required for a Windmill farm~\cite{mackay2008sustainable}, due to the very large required distance between windmills to prevent flow interactions. This illustrates that blue energy is actually already competitive as such in spite of the poor performances of PRO and RED.

%Be that as it may, 
%\LB{blue energy= no need for semi-permeable membrane; or AEM/CEM: DO makes the job}
%
Beyond PRO and RED, it was shown recently that new nanomaterials and nanofluidic transport constitute key assets that allow to boost considerably these performances \cite{siria2013giant,feng2016single,siria2017new,walker2017extrinsic,rankin2016effect}. 
Experiments across nanotubes of boron-nitride (BN), and subsequently across MoS$_2$ nanoporous membranes, reported huge ionic currents. 
%Osmotic powers reach thousands of Watts per square meter in these devices, and even up to $10^6$W/m$^2$ for the 2D material due to its molecular thickness (leading to huge gradients).
\begin{figure}[h!]
\centering
  \includegraphics[width=0.45\textwidth]{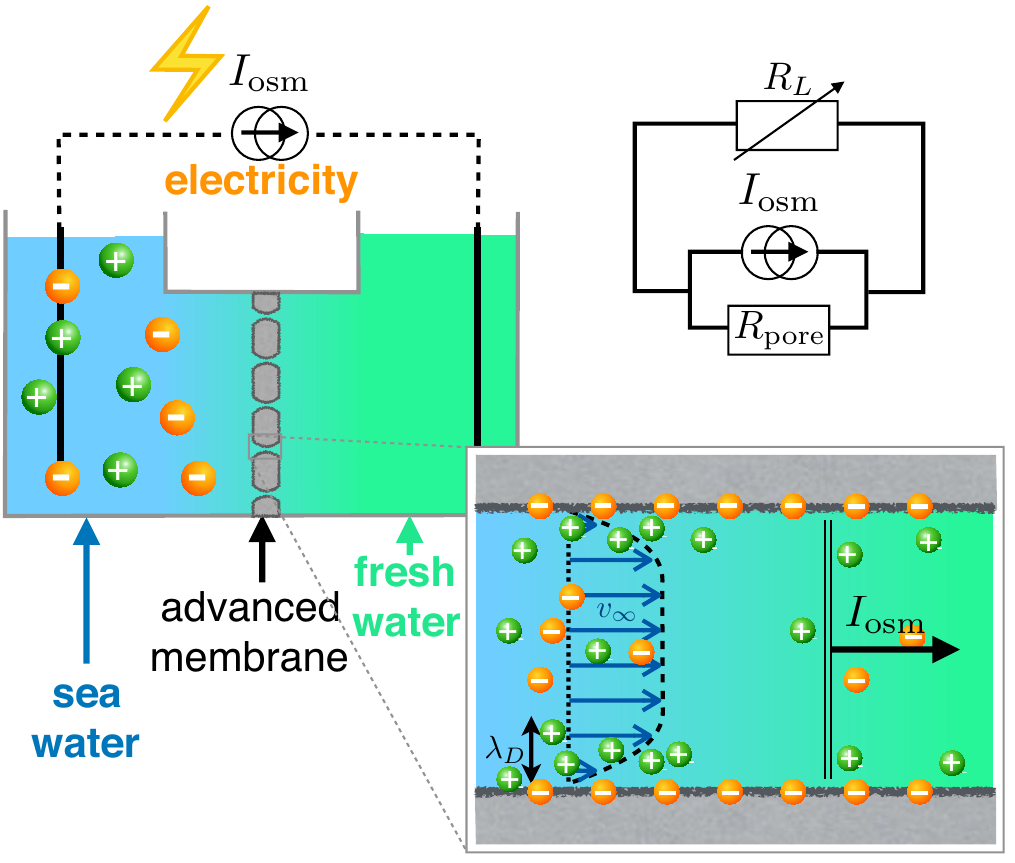}
  \caption{\textbf{: Blue energy with diffusio-osmosis.} A porous membrane with large and charged pores (zoom) induces a diffusio-osmotic plug-like flow with center velocity $v_{\infty}$ upon a salt concentration difference (\textit{e.g.} here between sea and fresh water) as seen in Fig.~\ref{fig:DEO}). This flow drives excess charges in the electric double layer producing a net ionic current $I_{\rm osm}$ that can be harvested in a load resistance $R_L$ -- top right electric schematic.   } 
  \label{fig:BlueDO}
\end{figure}
A puzzling remark is that the BN nanotubes in the experiments of \citefull{siria2013giant} or the MoS$_2$ nanopores of \citefull{feng2016single} are fully permeable to ions, in contrast to the canonical views of RED involving cation and anion selective membranes. 
The origin of the osmotic current was then shown to be the diffusio-osmotic ionic currents taking place at the surface of the materials, 
coupled to the considerable surface charge exhibited by these systems -- see Fig.~\ref{fig:BlueDO}. We reported in the previous section the corresponding ionic current in
Eqs.(\ref{IDO})-(\ref{K1}), and for a membrane constituted of $N$ tubes of radius $R$, length $L$ and surface charge $\Sigma$, the ionic current can be estimated as 
\begin{equation}
I_{\rm osm} \approx  N 2\pi R\, \Sigma\times v_{DO}\approx N {2\pi R\over L} \Sigma\, \, D_{DO} \times  \Delta \log c_{s}
\end{equation}  
where $v_{DO}$ the diffusio-osmotic water flow speed and $D_{DO}$ is the diffusio-osmotic mobility, typically $D_{DO} \sim k_BT/(8\pi\eta \ell_B)$.
This prediction was fully confirmed experimentally in Ref.\citenum{siria2013giant}. %The large current takes its origin in the high surface charge $\Sigma$ of BN
Therefore -- and this is a key asset -- the blue energy does not require full selectivity of the membrane, in contrast to RED standards. 
Connecting the membrane to a load resistance $R_L$ --Fig.~\ref{fig:BlueDO} --, the maximum  osmotic power which can be harvested is easily found to be
% is $\mathcal{P} = U_L^2/ R_L$ where $U_L$ the voltage drop through the membrane, see the electric schematic in Fig.~\ref{fig:BlueDO}. A simple optimization leads to the maximal power harvested as:
\begin{equation}
\mathcal{P} = \frac{1}{4} R_{\rm pore} I_{\rm osm}^2
\end{equation}
%ICICICICICI
where $R_{\rm pore}$ is the pore or membrane resistance (that can be obtained from standard conductance measurements).
\SMM{Osmotic power reaches thousands of Watts per square meter in BN nanotubes, and even up to $10^6$W/m$^2$ for the 2D MoS$_2$ due to its molecular thickness (leading to huge gradients).}
This estimate actually suggests to couple diffusio-osmotic current generation with an asymmetric pore geometry leading to ionic diode behavior\cite{siria2017new}: blocking the ionic backflow thanks to the diode property allows one to boost the output power by reducing Joule losses (see details in \citefull{siria2017new} and Fig.~\ref{fig:BlueDO}). Asymmetric channels were indeed shown to improve energy harvesting~\cite{zhu2018unique,lin2019rectification}.

This methodology can be readily generalized to other materials which are better suited for upscaling as compared to BN nanotubes. Key progress has been made recently in this direction \cite{siria2017new}.
%The key point is to find a nanoporous material with sufficiently high surface charge to allow for a notable diffusio-osmotic flow. h-BN, MoS$_2$ and TiO$_2$ are promissing materials in that regard~\cite{siria2013giant,feng2016single,siria2017new,walker2017extrinsic}. 
%as $R_{\rm pore} = 1/G_{\rm pore}$). 
%The maximal power density through thin and highly charged materials has been measured up to $10^6$ W/m$^2$ (in atomic layer thin MoS$_2$ nanopores~\cite{feng2016single}). 
Using diffusio-osmotic currents thus constitutes a promising route for improved blue energy harvesting, making it possibly relevant to industrial scale.

%Diffusio-osmotic current harvesting has a few limitations, from the difficulty to upscale observations made at the individual pore scale to larger structures, to the fact that power harvested is intrinsically dependent on the load resistance $R_L$, with a maximum when $R_L = R_{\rm pore}$. For all these reasons it is necessary to continue efforts in osmotic energy harvesting, \textit{e.g.} exploiting advanced structures like ionic to osmotic diodes~\cite{zhu2018unique}.

%\LB{ICICICI}
Beyond these membrane-based routes, 
the so-called \qote{capacitive mixing} methodology is an alternative approach to harvest osmotic energy
%Beyond diffusio-osmosis, blue energy may also be harvested by an alternative technique called capacitive mixing
~\cite{brogioli2009extracting}. The principle is to charge and discharge an ionic capacitor by alternating flows of salty and fresh water. 
%For proper function, the capacitor plates, made of highly porous material, are functionalized by specific coating such that one is  positively charged, and the other negatively charged in surface. \textcolor{blue}{Actually for porous carbon you don't seem to need that, but you do need something to ensure that ions separate and I don't get it}. 
Capacitor plates are connected to current collectors. First (step A on Fig.~\ref{fig:CapMix}-a) salty water is flushed in, charging the capacitor plates, resulting in a closed circuit current in the load resistance. Then (step B), salty water is replaced by fresh water. When the circuit is closed again on the load resistance (step C), the capacitor plates discharge into the bulk as fresh water is less salty, resulting in a current in the opposite direction. The circuit is opened and fresh water is replaced by salty water (step D) and the cycle may start again. 

The power generated may be computed from the area of the cycle in the voltage/charge plane -- see Fig.~\ref{fig:CapMix}-b. Typically, over 1 cycle (about 20h~\cite{brogioli2011prototype}), 1 J per gram of carbon electrode may be collected. To compare with previous results, we estimate that 1 carbon plate of $6 \times 6$cm$^2$ is about $1$g, such that one may recover around 0.2 W/m$^2$ with capacitive mixing. Capacitive mixing therefore requires significant progress in optimizing the cell setup and the nanoporous structure to enhance performances~\cite{simoncelli2018blue,marino2016capacitive}. 

% brogioli dit 8.5g de porous carbon electrode, sachant qu'il y a 8 stack, ça fait environ 1g pour le stack, et il recupere environ 1J/g par cycle (un cycle = 20h), donc environ 1 J; sachant que les électrodes font 6x6 cm^2, ça fait (1/20x60) J/(6x6 e-4) m^2 s = 0.2 W/m^2

\begin{figure}[h!]
\centering
  \includegraphics[width=0.49\textwidth]{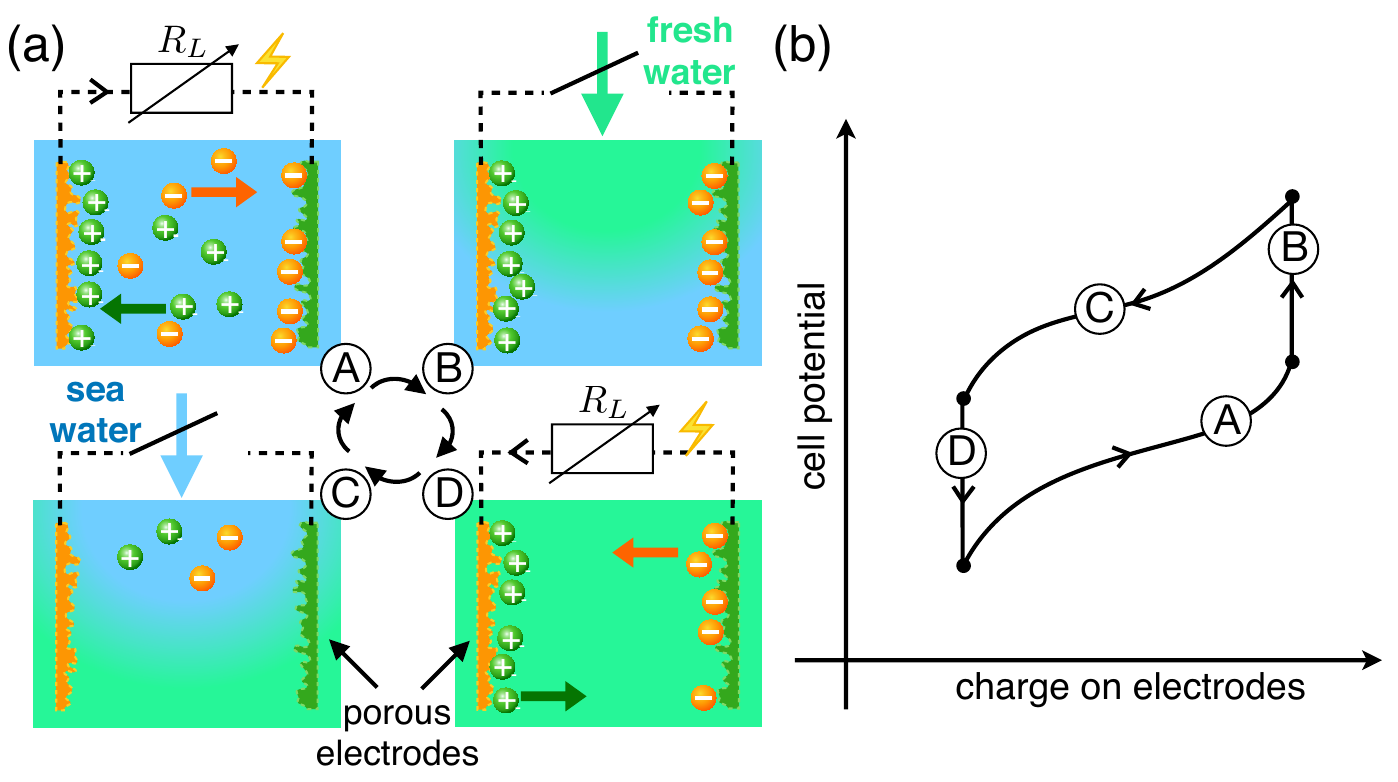}
  \caption{\textbf{: Capacitive Mixing to collect blue energy.} (a) Capacitive mixing cycle. Two electrodes with functionalized surfaces (such that one is positively charged in surface (green) and the other negatively charged in surface (orange)) are embedded in a fluidic device where salty  water and fresh water are alternatively flushed in a cycle. (b) Associated Voltage versus Charge cycle. The cycle is described further in the text.} 
  \label{fig:CapMix}
\end{figure}

%\textcolor{blue}{maybe also say something about the fact that when you measure the current, what you actually measure is the diffusio-osmotic current + the nernst contribution that should be substracted !}

\subsection{Dead-end pores: Detergency, particle and liquid osmotic extraction }
\label{deadendpores}
%Diffusio-osmotic flows may be harvested to transport particles in a variety of contexts. In the following we focus on a few applications and detail their working principle. 

We have demonstrated in the previous sections how efficient diffusio-phoresis is to boost migration of particles. Combined with the ability to
generate gradients of solute (in particular of salts) at small scales, it proves a method of choice in various applications to extract particles or liquids from dead-end pores. We discuss shortly two examples where diffusio-osmotic forces are harnessed.
%diffusio-phoretic transport is actually very efficient 

%{\it Washing out molecules, applications to detergency --}
A nice application of diffusio-osmosis was highlighted recently in the context of cleaning and the significance of rinsing in 
laundry detergency ~\cite{shin2018cleaning}. The question at stake here is how to extract particles which are stuck in dead-end pores in the porous matrix constituting the fabric. A simple flow resulting from mechanical action may not be able to perform this task, \sopp{especially since particles buried in small pores in the interyarn pore space may not be recovered by advection because flow is channelled by larger pores} (see Fig.~\ref{fig:Detergency}-a and~c). Experiments then showed that rinsing with fresh water generates  surfactant gradients at the scale of the fabric fibres and this in turn leads to diffusio-phoretic motion of the particles inside the dead-end pores. This flushing geometry echoes the osmotic shock discussed above.  As highlighted in Fig.~\ref{fig:Detergency}-b and~d, the gradient-induced motion allows one to extract particles from the intertwined network of pores. 
%leaning by Surfactant Gradients: Particulate Removal from Porous Materials and the Significance of Rinsing in Laundry Detergency
%diffusio-phoretic transport may also be harnessed in specific contexts to recover particles from porous materials~\cite{shin2018cleaning}. In fact, consider extracting particles from small pores. The obvious way of doing so is for example to flush water through the structure, therefore entraining particles. Since all pores are not aligned with the direction of flow, particles stuck too deep within those pores may not be recovered simply by advection (see Fig.~\ref{fig:Detergency}-a). However, if first the structure is filled with a highly saline solution (or any other concentrated solution) and then rinsed with a low saline solution (or reversely), a concentration gradient will establish within the pores. This concentration gradient will induce a diffusio-phoretic force on the particles that will entrain them out of the pore whatever their distance to the exit (see Fig.~\ref{fig:Detergency}-b). diffusio-phoresis thus allows to recover particles deep within porous structures. One of the obvious implications is connected to the importance of rinsing in laundry detergency~\cite{shin2018cleaning}. 
This suggests that, after laundering with any kind of detergent, rinsing with fresh water will allow a diffusio-phoretic push to wash out dirt and stains. \sopp{It is worthwhile noting that detergency within this prospect also benefits from the log-sensing and osmotic shock effect discussed in Sec.\ref{DPharness}: particle removal by this mechanism is effective on significantly long time scales, allowing for proper removal of the particles.} 

\begin{figure}[h!]
\centering
  \includegraphics[width=0.45\textwidth]{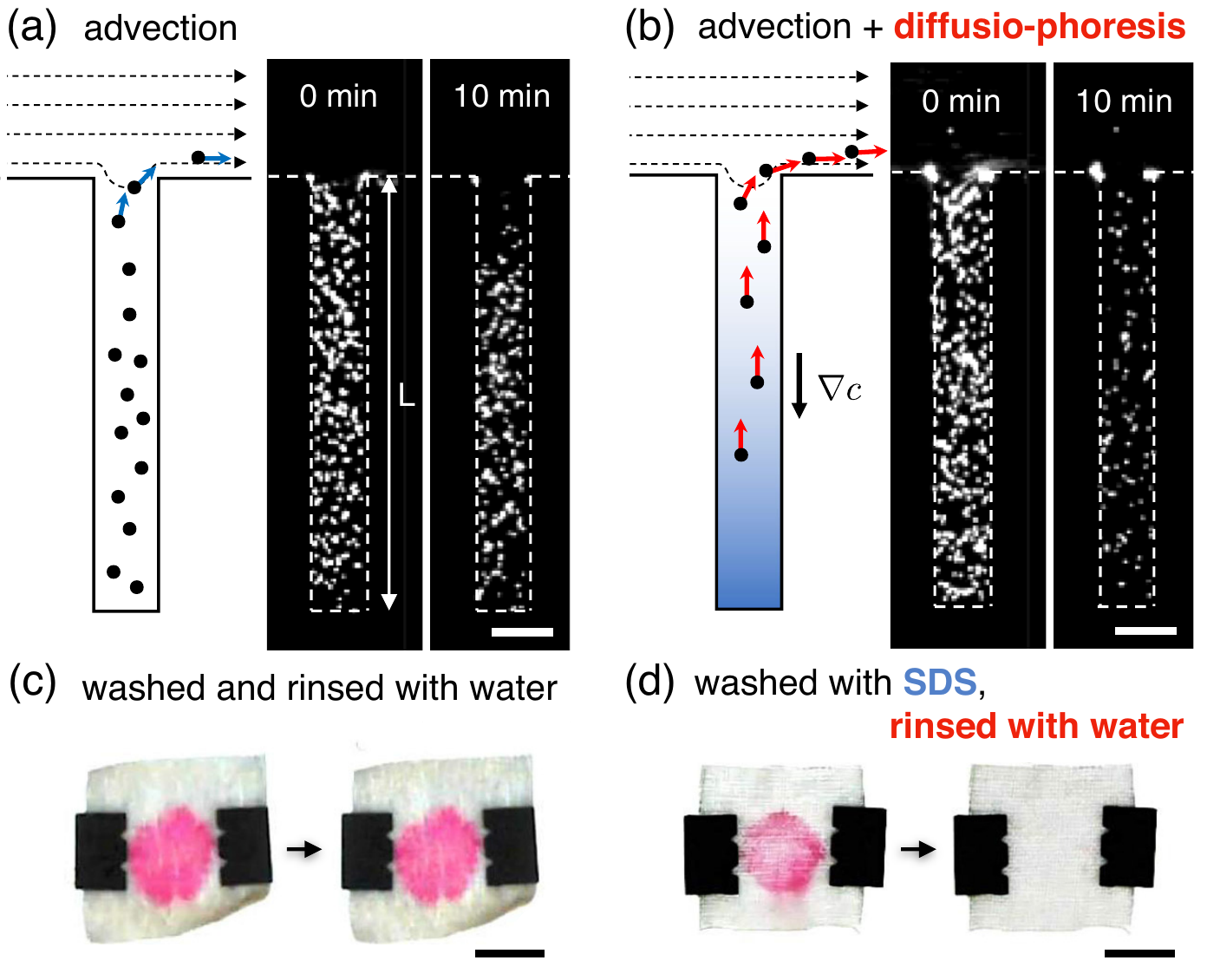}
  \caption{\textbf{: Particle removal with diffusio-phoresis.} Reproduced from \citefull{shin2018cleaning} with permission from the APS, copyright 2018. (a) Fluorescence image sequence showing particles in a dead-end microfluidic pore, upon advection in the main conduct. The solutions are composed of SDS at 10 mM. (b) Same as (a) with a solute gradient, where the inner pore solute concentration is 10 mM and the outer (main channel) is 0.1 mM. All scale bars are 50 $\mu$m. \SM{(c) and (d) A piece of cotton fabric is stained with colored colloidal particles (polystyrene latex). The piece of fabric is washed and rinsed in water (c) or washed in 10 mM SDS and rinsed with water (d) then photographed immediately after rinsing (left) and 120 s afterwards (right). All scale bars are 1 cm.} }
  \label{fig:Detergency}
\end{figure}
This type of mechanism based on diffusio-phoretic migration is versatile and applies to any flushing geometry. Various recent experiments considered
%More generally diffusio-phoretic transport was shown to be very efficient in extracting 
extraction of particles -- colloidal particles and oil emulsions -- from dead-end pores \cite{kar2015enhanced,shin2016size}.
%{\it Washing out fluids? liquid extraction --}
Such results have also obvious applications in a different context, in geology for example, where dissolution and recrystallization at the mineral-fluid interface leads to ubiquituous salt gradients at the root of diffusio-phoretic and -osmotic transport \cite{plumper2017fluid,putnis2014mineral}. In fact, flushing by fresh water was shown to enhance considerably oil recovery, a method coined as \qote{Low salinity enhanced oil recovery}~\cite{lager2008losal}.
%Similar to the detergency discussed above, it was shown that diffusio-phoretic transport may allow 
%%\LB{citer dead end pores, take droplets out of liquid, similar to detergency, cf \cite{kar2015enhanced}, \cite{shin2016size}}
%
%"Transient salt gradients are ubiquitous
%in biological and geological systems. Geological
%examples include flows originating in disturbed mineral
%formations41 and geoengineering involving the
%creation of ion gradients for enhanced oil recovery"
%In a very different field, diffusio-osmotic flows was also shown to strongly impact and shape the reactive fluid flows occurring in the solid Earth \cite{plumper2017fluid}.
%
%low sal EOR \cite{lager2008losal,
%The example of detergency is illuminating because it illustrates how the trivial concept of osmosis may be harvested in a very subtle way to improve specific processes. 
While the very origin of this phenomenon is still debated, it is quite clear that diffusio-osmotic flows will play a key role in recovering
biphasic mixtures using salinity gradients. 
%In that respect, we consider if it is possible to recover fluids from biphasic mixtures using salinity gradients. 
Consider oil in a porous structure with typical pore radius $a$, as sketched in Fig.~\ref{fig:OilRecovery}, where oil is trapped in dead-end pores. 
After a flush with fresh water, a diffusio-osmotic flow may be generated at the surface of the porous material, with velocity $v_{DO} = - D_{DO} \nabla \log c_{s}$.
 %Suppose we flow first highly salty water and then fresh water. This will create a salinity gradient within the pores (see Fig.~\ref{fig:OilRecovery}). A water flow with plug profile will be induced due to diffusio-osmosis. The water velocity in the pore is directed towards the low salinity with velocity $v_{\infty} = - \mu_{DO} \nabla \ln c_{\infty}$. The diffusio-osmotic mobility $\mu_{DO}$ depends weakly on the surface properties (zeta potential). If $c$ is expressed in units/volume, a good estimate is  $\mu_{DO} = k_B T /8 \pi \eta l_B$, with $l_B \simeq 0.7$nm the Bjerrum length (for water at room temperature), and typically $\mu_{DO} \simeq 2 \times 10^{-10}$ m$^2$.s$^{-1}$.
 Assuming first that oil is blocked, this generates a counterbalancing pressure gradient, such that the total flux is vanishing, leading to a pressure drop
%Now, if the oil is blocked, and thus stationary, the diffusio-osmotic flow has to be counterbalanced by a pressure gradient $\nabla P$. Writing that the total flux $Q_T$ (diffusio-osmotic and pressure contributions) should vanish leads to the following balance:
%\begin{equation}
%Q_T=v_{\infty} \pi R^2 - \nabla P \frac{\pi R^4}{8\eta} = 0
%\end{equation}
%with $R$ the characteristic pore radius. This leads to a pressure drop
\begin{equation}
\Delta p_{DO} = - \frac{8 \eta D_{DO}}{a^2} \Delta \left[ \log c_s \right]
\end{equation} 
along the dead-end channel (and independent of the channel length).
%In the case of high salinity close to the oil, the pressure is thus lower at the water-oil interface. 
Putting in numbers, with a strong salinity gradient between salty water at 1 M and fresh water at $0.1$ mM to fix ideas, we find $\Delta p_{DO} = 0.07$ bar for $a = 100$nm and up to $\Delta p_{DO} = 30$ bars for $a = 5$nm. This has to be compared to the oil-water capillary pressure expressed as $\Delta P_{\rm cap} = \frac{\gamma}{a}$ with $\gamma \simeq 10-20$ mN$/$m a typical surface tension at the oil-water interface (possibly decorated with injected surfactants). %(silicon oil~\cite{peters2013interfacial})
While  $\Delta p_{\rm cap} = 2$ bar $> \Delta p_{DO}$ for $a = 100$ nm, it is in the same range  for $a = 5$nm with $\Delta p_{\rm cap}\sim \Delta p_{DO}  \sim  40$ bar. For very small pores, the pressure induced by diffusio-osmosis -- which scales as $1/a^2$ -- is thus able to bypass the capillary pressure, scaling  as $1/a$. %Diffusio-osmotic effect are therefore efficient to unjam  oil inside nanometric dead-end pores. 
These simplistic estimates are made for illustration only and would deserve more detailed experimental investigations. They highlight the efficiency of diffusio-osmotic effects to extract liquids which are deeply confined within nanometric dead-end porosity. 
\begin{figure}[h!]
\centering
  \includegraphics[width=0.45\textwidth]{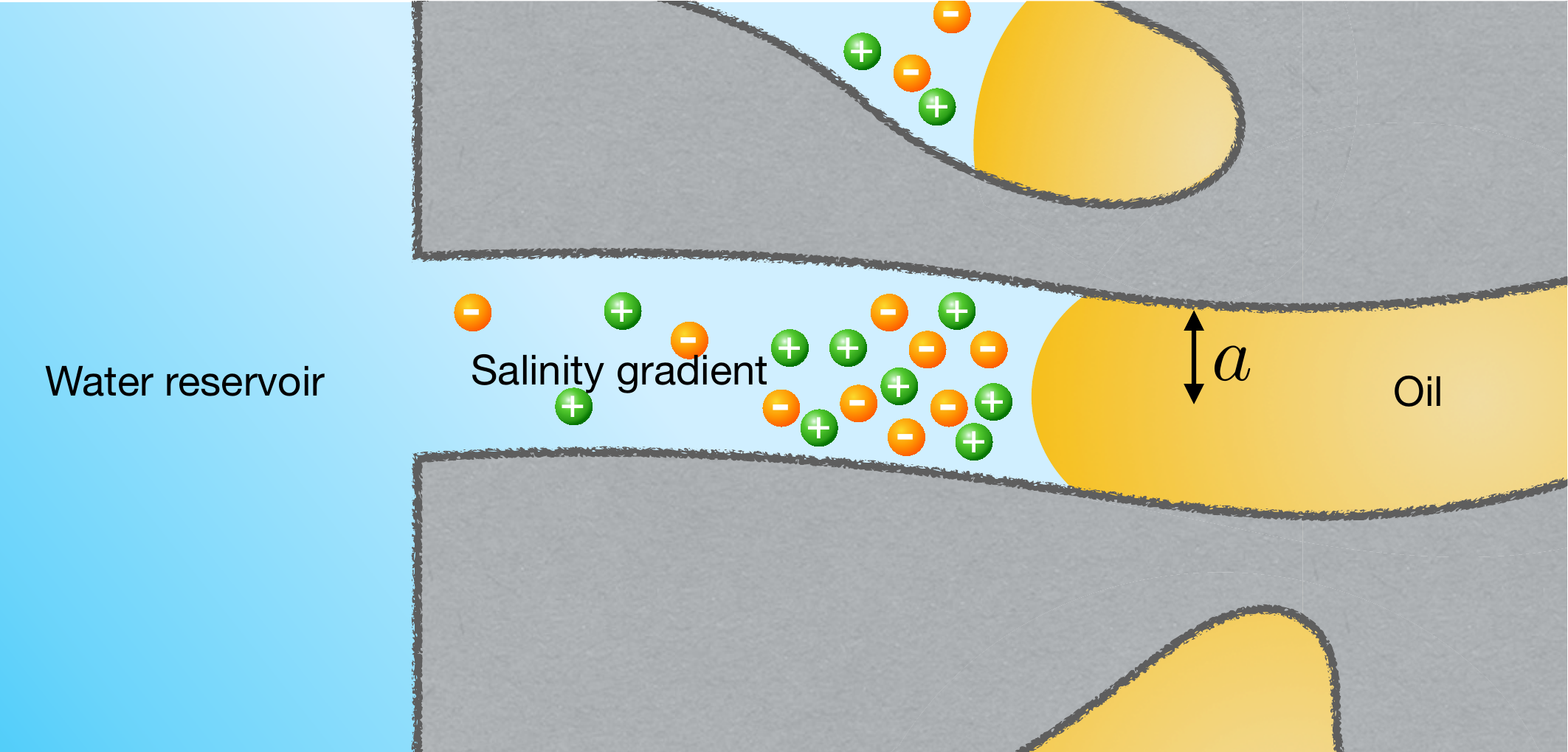}
  \caption{\textbf{: Diffusio-osmotic effects for oil recovery under salinity gradients.} Enhanced oil  recovery is traditionally preformed by injecting sea water in the reservoir to push the oil. However flushing with fresh water slugs is known to boost the process. Gradients of salinity within dead end pores may help bypassing the capillary forces blocking the oil within the porosity. 
   %from porous structures by flushing water at large pressures through the structure. On the schematic saline water was first flushed and then fresh water.
   } 
  \label{fig:OilRecovery}
\end{figure}

%\textcolor{blue}{We could also put a note on water pumping, namely inverting the system; instead of using water to move things, you use things to move water.  There's this nice paper by Todd~\cite{paustian2014induced} where they pump water flows thanks to Janus pillars (do these pillars need to be redone with time ? It seems that AC currents drive the mechanism so no need) and unfortunately he did not seem to do an energetic balance, but I wouldn't be surprised if it were to consume less than standard pumps no? (no clue how standard pumps work but you have to convert electricity and mechanical energy at some point so it's a mess. Preliminary paper driving flow orthogonal to the surfaces~\cite{pascall2010induced}and also \cite{mansuripur2009asymmetric}. This connects also to paper by Roland - to some extent. }
%
%paper Roland \cite{rinne2012nanoscale}

%\begin{figure}[h]
%\centering
%  \includegraphics[height=3cm]{example1}
%  \caption{An example figure caption \textendash\ the image is from the \textit{Physical Chemistry Chemical Physics} cover gallery.}
%  \label{fgr:example}
%\end{figure}
%
%\begin{figure*}
% \centering
% \includegraphics[height=3cm]{example2}
% \caption{An image from the \textit{Physical Chemistry Chemical Physics} cover gallery, set as a two-column figure.}
% \label{fgr:example2col}
%\end{figure*}

\section{Concluding remarks and perspectives}

\SM{As is clear from our discussion in the previous section, osmosis is ubiquitous and crucial to an impressive number of processes, with extremely diverse manifestations. In spite of this diversity, a key and universal aspect of osmosis is that it may be interpreted as a driving force, exerted by the membrane (or a surface, or a particle's surface, and so on) on the solute particles. As we have seen in many situations in detail, we  typically expect the apparent osmotic pressure to write generically as 
$$\Delta \Pi_{\rm app} \simeq \langle c_s (-\nabla \mathcal{U}_{\rm eff}) \rangle$$ 
with $\langle . \rangle$ some specific average and $\mathcal{U}_{\rm eff}$ the effective interaction potential. This mechanical perspective allows one to interpret most osmotic related phenomena (diffusio-osmosis, diffusio-phoresis, active osmosis, \textit{etc}.). 
Beyond this generic description, a proper description of the forces at play is required in more specific examples, as we showed on the subtle example of the force balance in diffusio-phoresis. %(polymer phoresis, specific chemotactic flows in biophysics, self-assembly of advanced structures) is required to break the entropy bottleneck. 
}

\SM{Our understanding of osmotic related phenomena is still blurred by a number of open riddles. Non-equilibrium osmotic flows should be investigated, in particular to harvest non-equilibrium forces for advanced transport of species, which offer a number of promising avenues. Introducing more reliable descriptions and understanding for ionic transport at the smallest scales should also open the way to build advanced ionic detectors and ionic-powered machinery. At micrometric scales, a number of processes could be improved, harvesting the properties of specific geometries -- as in the kidney -- together with a clever mix of osmotic forces -- as diffusio-phoresis for detergency. }

Overall we still have a lot to learn from Nature and how it harvests osmosis in many forms, for separation purposes, energy storage and harvesting, {\it etc}. Today osmosis is usually harnessed in its most basic form, for example as the prototypical example of osmotic pressure across a semi-permeable membrane. Yet Nature has developed far more clever and far more complex examples. Mimicking the natural wonders with artificial systems is a great challenge but it opens new avenues for many outstanding societal questions that are worth the journey.}

\vskip0.5cm

\review{\section{List of Symbols}}

\review{We report below symbols that are used frequently throughout the review.}

%\vspace{1mm}

\review{\begin{tabularx}{0.47\textwidth}{lX}
$a$ & Pore radius \\
$\mathcal{A}$ & Membrane or pore area \\
$b$ & Slip length of the surface \\
$\beta$ & $= 1/ k_B T$ \\
$c_s$ & Solute concentration \\ 
$c_w$ & Water or solvent concentration \\
$c_{+/-}$ & Concentration of positive or negative ions \\
$D_{DO}$ & Diffusio-osmotic "diffusion" coefficient \\
$D_{DP}$ & Diffusio-phoretic "diffusion" coefficient \\
$D_0$ & Colloid diffusion coefficient \\
$D_s$ & Diffusion coefficient of the solute \\
$e$ & Elementary charge \\
$E$ & Electric field \\
$\epsilon$ & Dielectric permittivity of the fluid \\
$\eta$ & Solvent viscosity \\
% $f$ & Helmholtz free energy of the system \\
% $G$ & Gibb's free energy of the system \\
% $\delta h$ & Excess specific enthalpy \\
$I_{DO}$ & Diffusio-osmotic ion current \\
$I_e$ & Electric current \\
$J_e$ & Exchange or Excess solute flow  \\
$J_s$ & Solute flow \\
$j_s$ & $= J_s/\mathcal{A}$ Solute flow per unit area \\
$k_B$ & Boltzmann's constant \\ 
$K_{\rm osm}$ & Osmotic electric mobility \\
$L$ & Thickness of the membrane or length of the pore \\
$\kappa_{\rm hyd}$ & Permeance of the membrane or pore \\
$\ell_B$ & $= e^2 / 4 \pi \epsilon k_B T$ Bjerrum length \\
$\mathbb{L}$ & Transport matrix (or a part of the full matrix) \\
$\mathcal{L}_D$ & $ = \frac{D_s \mathcal{A}}{\eta L}$, Solute permeability of the membrane or pore \\
$\mathcal{L}_{\rm hyd}$ & $ = \frac{\kappa_{\rm hyd} \mathcal{A}}{\eta L}$, Hydrodynamic permeability of the membrane or pore \end{tabularx}}

\clearpage
\review{\begin{tabularx}{0.47\textwidth}{lX}
$\lambda$ & range of potential interactions \\
$\lambda_D$ & $=1/\sqrt{8\pi \ell_B c_s}$ Debye length \\
$\lambda_s$ & $= \frac{D_s}{k_B T}$ mobility of the solute \\
$\mu_{DO}$ & Diffusio-osmotic mobility \\
$\mu_{DP}$ & Diffusio-phoretic mobility \\
$\mu_{EO}$ & Electro-osmotic mobility \\
$\mu^0_i$ & Chemical potential of the pure specie $i$ \\
$\mu_i$ & Chemical potential of specie $i$ \\
$N_i$ & Number of molecules of specie $i$ \\
$\omega_s$ & Solute "mobility" across the membrane \\
$p$ & Pressure \\
$\Pi$ & Osmotic pressure \\
%$\Phi$ & Entropy production rate \\ 
%$\Phi_0$ & $= eV_0/k_B T$ Dimensionless surface potential \\
$Q$ & Volume flow\\
$R$ & Particle size \\
$\rho_e$ & Charge density \\
%$S$ & Entropy \\
$\sigma$ & Reflection or selectivity coefficient \\
$\Sigma$ & Surface charge \\
$T$ & Temperature \\ 
$\mathcal{U}(x)$ & Potential barrier representing the membrane \\
$\bm{v}$ & Velocity field of the fluid \\
$v_{DO}$ & Diffusio-osmotic velocity \\
$v_{DP}$ & Diffusio-phoretic velocity \\
$v_{EO}$ & Electro-osmotic velocity \\
$v_w$ & Molar volume of water \\ 
$V_e$ & Electric potential \\
%$\mathcal{V}$ & System volume \\
$X$ & Solute molar fraction \\
$\zeta$ & Zeta potential 
\end{tabularx}}

\vskip0.5cm
\section*{Conflicts of interest}
%In accordance with our policy on \href{http://www.rsc.org/journals-books-databases/journal-authors-reviewers/author-responsibilities/#code-of-conduct}{Conflicts of interest} please ensure that a conflicts of interest statement is included in your manuscript here.  Please note that this statement is required for all submitted manuscripts.  If no conflicts exist, please state that ``There are no conflicts to declare''.
There are no conflicts to declare

\vskip0.5cm
\section*{Acknowledgements}
\SM{The authors are grateful for the numerous discussions they enjoyed with Marie-Laure Bocquet, Daan Frenkel, David Huang, J\'{e}r\'{e}mie Palacci, Benjamin Rotenberg, Alessandro Siria, Todd Squires, Emmanuel Trizac, Patrick Warren. Authors also acknowledge pertinent feedback from Maarten Biesheuvel and Alan Kay.}
%S.M. from ?  fundings.. 
L.B. acknowledges funding from the European Union's H2020 Framework Programme/ERC Advanced Grant {\it Shadoks} and 
European Union's H2020 Framework Programme/FET {\it NanoPhlow}.
S.M. and L. B. acknowledge funding from ANR project {\it Neptune}.
%%%END OF MAIN TEXT%%%


\balance


%%%REFERENCES%%%
\bibliography{Osmosis} %You need to replace "rsc" on this line with the name of your .bib file

\providecommand*{\mcitethebibliography}{\thebibliography}
\csname @ifundefined\endcsname{endmcitethebibliography}
{\let\endmcitethebibliography\endthebibliography}{}
\begin{mcitethebibliography}{381}
\providecommand*{\natexlab}[1]{#1}
\providecommand*{\mciteSetBstSublistMode}[1]{}
\providecommand*{\mciteSetBstMaxWidthForm}[2]{}
\providecommand*{\mciteBstWouldAddEndPuncttrue}
  {\def\EndOfBibitem{\unskip.}}
\providecommand*{\mciteBstWouldAddEndPunctfalse}
  {\let\EndOfBibitem\relax}
\providecommand*{\mciteSetBstMidEndSepPunct}[3]{}
\providecommand*{\mciteSetBstSublistLabelBeginEnd}[3]{}
\providecommand*{\EndOfBibitem}{}
\mciteSetBstSublistMode{f}
\mciteSetBstMaxWidthForm{subitem}
{(\emph{\alph{mcitesubitemcount}})}
\mciteSetBstSublistLabelBeginEnd{\mcitemaxwidthsubitemform\space}
{\relax}{\relax}

\bibitem[Meyer and van~'t Hoff(1890)]{meyer1890pression}
M.~L. Meyer and J.~van~'t Hoff, \emph{Recueil des Travaux Chimiques des
  Pays-Bas}, 1890, \textbf{9}, 157--161\relax
\mciteBstWouldAddEndPuncttrue
\mciteSetBstMidEndSepPunct{\mcitedefaultmidpunct}
{\mcitedefaultendpunct}{\mcitedefaultseppunct}\relax
\EndOfBibitem
\bibitem[van~'t Hoff(1890)]{van1890wesen}
J.~van~'t Hoff, \emph{Zeitschrift f{\"u}r Physikalische Chemie}, 1890,
  \textbf{5}, 174--176\relax
\mciteBstWouldAddEndPuncttrue
\mciteSetBstMidEndSepPunct{\mcitedefaultmidpunct}
{\mcitedefaultendpunct}{\mcitedefaultseppunct}\relax
\EndOfBibitem
\bibitem[Kung \emph{et~al.}(2010)Kung, Martinac, and
  Sukharev]{kung2010mechanosensitive}
C.~Kung, B.~Martinac and S.~Sukharev, \emph{Annual review of microbiology},
  2010, \textbf{64}, 313--329\relax
\mciteBstWouldAddEndPuncttrue
\mciteSetBstMidEndSepPunct{\mcitedefaultmidpunct}
{\mcitedefaultendpunct}{\mcitedefaultseppunct}\relax
\EndOfBibitem
\bibitem[Guell and Brenner(1996)]{guell1996physical}
D.~C. Guell and H.~Brenner, \emph{Industrial \& engineering chemistry
  research}, 1996, \textbf{35}, 3004--3014\relax
\mciteBstWouldAddEndPuncttrue
\mciteSetBstMidEndSepPunct{\mcitedefaultmidpunct}
{\mcitedefaultendpunct}{\mcitedefaultseppunct}\relax
\EndOfBibitem
\bibitem[Dutrochet(1826)]{dutrochet1826agent}
R.~J.~H. Dutrochet, \emph{L'agent imm{\'e}diat du mouvement vital
  d{\'e}voil{\'e} dans sa nature et dans son mode d'action, chez les
  veg{\'e}taux et chez les animaux}, Bailli{\`e}re, 1826\relax
\mciteBstWouldAddEndPuncttrue
\mciteSetBstMidEndSepPunct{\mcitedefaultmidpunct}
{\mcitedefaultendpunct}{\mcitedefaultseppunct}\relax
\EndOfBibitem
\bibitem[Dutrochet(1995)]{dutrochet1995new}
R.~H. Dutrochet, \emph{Journal of Membrane Science}, 1995, \textbf{100},
  5--7\relax
\mciteBstWouldAddEndPuncttrue
\mciteSetBstMidEndSepPunct{\mcitedefaultmidpunct}
{\mcitedefaultendpunct}{\mcitedefaultseppunct}\relax
\EndOfBibitem
\bibitem[Graham \emph{et~al.}(1854)Graham\emph{et~al.}]{graham1854vii}
T.~Graham \emph{et~al.}, \emph{Philosophical Transactions of the Royal Society
  of London}, 1854, \textbf{144}, 177--228\relax
\mciteBstWouldAddEndPuncttrue
\mciteSetBstMidEndSepPunct{\mcitedefaultmidpunct}
{\mcitedefaultendpunct}{\mcitedefaultseppunct}\relax
\EndOfBibitem
\bibitem[Fick(1855)]{fick1855v}
A.~Fick, \emph{The London, Edinburgh, and Dublin Philosophical Magazine and
  Journal of Science}, 1855, \textbf{10}, 30--39\relax
\mciteBstWouldAddEndPuncttrue
\mciteSetBstMidEndSepPunct{\mcitedefaultmidpunct}
{\mcitedefaultendpunct}{\mcitedefaultseppunct}\relax
\EndOfBibitem
\bibitem[Mauro(1957)]{mauro1957nature}
A.~Mauro, \emph{Science}, 1957, \textbf{126}, 252--253\relax
\mciteBstWouldAddEndPuncttrue
\mciteSetBstMidEndSepPunct{\mcitedefaultmidpunct}
{\mcitedefaultendpunct}{\mcitedefaultseppunct}\relax
\EndOfBibitem
\bibitem[Robbins and Mauro(1960)]{robbins1960experimental}
E.~Robbins and A.~Mauro, \emph{The Journal of general physiology}, 1960,
  \textbf{43}, 523--532\relax
\mciteBstWouldAddEndPuncttrue
\mciteSetBstMidEndSepPunct{\mcitedefaultmidpunct}
{\mcitedefaultendpunct}{\mcitedefaultseppunct}\relax
\EndOfBibitem
\bibitem[Pfeffer(1877)]{pfeffer1877osmotische}
W.~Pfeffer, \emph{Osmotische untersuchungen: studien zur zellmechanik}, W.
  Engelmann, 1877\relax
\mciteBstWouldAddEndPuncttrue
\mciteSetBstMidEndSepPunct{\mcitedefaultmidpunct}
{\mcitedefaultendpunct}{\mcitedefaultseppunct}\relax
\EndOfBibitem
\bibitem[Wald(1986)]{wald1986theory}
G.~Wald, \emph{Journal of Chemical Education}, 1986, \textbf{63}, 658\relax
\mciteBstWouldAddEndPuncttrue
\mciteSetBstMidEndSepPunct{\mcitedefaultmidpunct}
{\mcitedefaultendpunct}{\mcitedefaultseppunct}\relax
\EndOfBibitem
\bibitem[van~'t Hoff(1887)]{van1887rolle}
J.~H. van~'t Hoff, \emph{Zeitschrift f{\"u}r physikalische Chemie}, 1887,
  \textbf{1}, 481--508\relax
\mciteBstWouldAddEndPuncttrue
\mciteSetBstMidEndSepPunct{\mcitedefaultmidpunct}
{\mcitedefaultendpunct}{\mcitedefaultseppunct}\relax
\EndOfBibitem
\bibitem[Borg(2003)]{borg2003osmosis}
F.~G. Borg, \emph{arXiv preprint physics/0305011}, 2003\relax
\mciteBstWouldAddEndPuncttrue
\mciteSetBstMidEndSepPunct{\mcitedefaultmidpunct}
{\mcitedefaultendpunct}{\mcitedefaultseppunct}\relax
\EndOfBibitem
\bibitem[Einstein(1905)]{einstein1905movement}
A.~Einstein, \emph{Ann. d. Phys}, 1905, \textbf{17}, 1\relax
\mciteBstWouldAddEndPuncttrue
\mciteSetBstMidEndSepPunct{\mcitedefaultmidpunct}
{\mcitedefaultendpunct}{\mcitedefaultseppunct}\relax
\EndOfBibitem
\bibitem[McMillan~Jr and Mayer(1945)]{mcmillan1945statistical}
W.~G. McMillan~Jr and J.~E. Mayer, \emph{The Journal of Chemical Physics},
  1945, \textbf{13}, 276--305\relax
\mciteBstWouldAddEndPuncttrue
\mciteSetBstMidEndSepPunct{\mcitedefaultmidpunct}
{\mcitedefaultendpunct}{\mcitedefaultseppunct}\relax
\EndOfBibitem
\bibitem[Gibbs(1897)]{gibbs1897semi}
J.~W. Gibbs, \emph{Nature}, 1897, \textbf{55}, 461\relax
\mciteBstWouldAddEndPuncttrue
\mciteSetBstMidEndSepPunct{\mcitedefaultmidpunct}
{\mcitedefaultendpunct}{\mcitedefaultseppunct}\relax
\EndOfBibitem
\bibitem[Callen(1998)]{callen1998thermodynamics}
H.~B. Callen, \emph{Thermodynamics and an Introduction to Thermostatistics},
  1998\relax
\mciteBstWouldAddEndPuncttrue
\mciteSetBstMidEndSepPunct{\mcitedefaultmidpunct}
{\mcitedefaultendpunct}{\mcitedefaultseppunct}\relax
\EndOfBibitem
\bibitem[Guggenheim(1967)]{guggenheim1985thermodynamics}
E.~A. Guggenheim, \emph{Amsterdam, North-Holland, 1967}, 1967,  Chapter 4\relax
\mciteBstWouldAddEndPuncttrue
\mciteSetBstMidEndSepPunct{\mcitedefaultmidpunct}
{\mcitedefaultendpunct}{\mcitedefaultseppunct}\relax
\EndOfBibitem
\bibitem[Barrat and Hansen(2003)]{barrat2003basic}
J.-L. Barrat and J.-P. Hansen, \emph{Basic concepts for simple and complex
  liquids}, Cambridge University Press, 2003\relax
\mciteBstWouldAddEndPuncttrue
\mciteSetBstMidEndSepPunct{\mcitedefaultmidpunct}
{\mcitedefaultendpunct}{\mcitedefaultseppunct}\relax
\EndOfBibitem
\bibitem[Paustian \emph{et~al.}(2015)Paustian, Angulo, Nery-Azevedo, Shi,
  Abdel-Fattah, and Squires]{paustian2015direct}
J.~S. Paustian, C.~D. Angulo, R.~Nery-Azevedo, N.~Shi, A.~I. Abdel-Fattah and
  T.~M. Squires, \emph{Langmuir}, 2015, \textbf{31}, 4402--4410\relax
\mciteBstWouldAddEndPuncttrue
\mciteSetBstMidEndSepPunct{\mcitedefaultmidpunct}
{\mcitedefaultendpunct}{\mcitedefaultseppunct}\relax
\EndOfBibitem
\bibitem[Talen and Staverman(1965)]{talen1965negative}
J.~Talen and A.~Staverman, \emph{Transactions of the Faraday Society}, 1965,
  \textbf{61}, 2800--2804\relax
\mciteBstWouldAddEndPuncttrue
\mciteSetBstMidEndSepPunct{\mcitedefaultmidpunct}
{\mcitedefaultendpunct}{\mcitedefaultseppunct}\relax
\EndOfBibitem
\bibitem[Talen and Staverman(1965)]{talen1965osmometry}
J.~Talen and A.~Staverman, \emph{Transactions of the Faraday Society}, 1965,
  \textbf{61}, 2794--2799\relax
\mciteBstWouldAddEndPuncttrue
\mciteSetBstMidEndSepPunct{\mcitedefaultmidpunct}
{\mcitedefaultendpunct}{\mcitedefaultseppunct}\relax
\EndOfBibitem
\bibitem[Weinstein and Caplan(1968)]{weinstein1968charge}
J.~N. Weinstein and S.~R. Caplan, \emph{Science}, 1968, \textbf{161},
  70--72\relax
\mciteBstWouldAddEndPuncttrue
\mciteSetBstMidEndSepPunct{\mcitedefaultmidpunct}
{\mcitedefaultendpunct}{\mcitedefaultseppunct}\relax
\EndOfBibitem
\bibitem[Lee \emph{et~al.}(2014)Lee, Cottin-Bizonne, Biance, Joseph, Bocquet,
  and Ybert]{lee2014osmotic}
C.~Lee, C.~Cottin-Bizonne, A.-L. Biance, P.~Joseph, L.~Bocquet and C.~Ybert,
  \emph{Physical Review Letters}, 2014, \textbf{112}, 244501\relax
\mciteBstWouldAddEndPuncttrue
\mciteSetBstMidEndSepPunct{\mcitedefaultmidpunct}
{\mcitedefaultendpunct}{\mcitedefaultseppunct}\relax
\EndOfBibitem
\bibitem[Lee \emph{et~al.}(2017)Lee, Cottin-Bizonne, Fulcrand, Joly, and
  Ybert]{lee2017nanoscale}
C.~Lee, C.~Cottin-Bizonne, R.~Fulcrand, L.~Joly and C.~Ybert, \emph{The journal
  of physical chemistry letters}, 2017, \textbf{8}, 478--483\relax
\mciteBstWouldAddEndPuncttrue
\mciteSetBstMidEndSepPunct{\mcitedefaultmidpunct}
{\mcitedefaultendpunct}{\mcitedefaultseppunct}\relax
\EndOfBibitem
\bibitem[Staverman(1951)]{staverman1951theory}
A.~Staverman, \emph{Recueil des Travaux Chimiques des Pays-Bas}, 1951,
  \textbf{70}, 344--352\relax
\mciteBstWouldAddEndPuncttrue
\mciteSetBstMidEndSepPunct{\mcitedefaultmidpunct}
{\mcitedefaultendpunct}{\mcitedefaultseppunct}\relax
\EndOfBibitem
\bibitem[Kedem and Katchalsky(1958)]{kedem1958thermodynamic}
O.~Kedem and A.~Katchalsky, \emph{Biochimica et biophysica Acta}, 1958,
  \textbf{27}, 229--246\relax
\mciteBstWouldAddEndPuncttrue
\mciteSetBstMidEndSepPunct{\mcitedefaultmidpunct}
{\mcitedefaultendpunct}{\mcitedefaultseppunct}\relax
\EndOfBibitem
\bibitem[Onsager(1931)]{onsager1931reciprocal}
L.~Onsager, \emph{Physical review}, 1931, \textbf{37}, 405\relax
\mciteBstWouldAddEndPuncttrue
\mciteSetBstMidEndSepPunct{\mcitedefaultmidpunct}
{\mcitedefaultendpunct}{\mcitedefaultseppunct}\relax
\EndOfBibitem
\bibitem[De~Groot and Mazur(2013)]{degroot2013non}
S.~R. De~Groot and P.~Mazur, \emph{Non-equilibrium thermodynamics}, Courier
  Corporation, 2013\relax
\mciteBstWouldAddEndPuncttrue
\mciteSetBstMidEndSepPunct{\mcitedefaultmidpunct}
{\mcitedefaultendpunct}{\mcitedefaultseppunct}\relax
\EndOfBibitem
\bibitem[Kedem and Katchalsky(1961)]{kedem1961physical}
O.~Kedem and A.~Katchalsky, \emph{The Journal of general physiology}, 1961,
  \textbf{45}, 143--179\relax
\mciteBstWouldAddEndPuncttrue
\mciteSetBstMidEndSepPunct{\mcitedefaultmidpunct}
{\mcitedefaultendpunct}{\mcitedefaultseppunct}\relax
\EndOfBibitem
\bibitem[Kedem and Katchalsky(1963)]{kedem1963permeabilitya}
O.~Kedem and A.~Katchalsky, \emph{Transactions of the Faraday Society}, 1963,
  \textbf{59}, 1918--1930\relax
\mciteBstWouldAddEndPuncttrue
\mciteSetBstMidEndSepPunct{\mcitedefaultmidpunct}
{\mcitedefaultendpunct}{\mcitedefaultseppunct}\relax
\EndOfBibitem
\bibitem[Kedem and Katchalsky(1963)]{kedem1963permeabilityb}
O.~Kedem and A.~Katchalsky, \emph{Transactions of the Faraday Society}, 1963,
  \textbf{59}, 1931--1940\relax
\mciteBstWouldAddEndPuncttrue
\mciteSetBstMidEndSepPunct{\mcitedefaultmidpunct}
{\mcitedefaultendpunct}{\mcitedefaultseppunct}\relax
\EndOfBibitem
\bibitem[Kedem and Katchalsky(1963)]{kedem1963permeabilityc}
O.~Kedem and A.~Katchalsky, \emph{Transactions of the Faraday Society}, 1963,
  \textbf{59}, 1941--1953\relax
\mciteBstWouldAddEndPuncttrue
\mciteSetBstMidEndSepPunct{\mcitedefaultmidpunct}
{\mcitedefaultendpunct}{\mcitedefaultseppunct}\relax
\EndOfBibitem
\bibitem[Starling(1896)]{starling1896absorption}
E.~H. Starling, \emph{The Journal of physiology}, 1896, \textbf{19},
  312--326\relax
\mciteBstWouldAddEndPuncttrue
\mciteSetBstMidEndSepPunct{\mcitedefaultmidpunct}
{\mcitedefaultendpunct}{\mcitedefaultseppunct}\relax
\EndOfBibitem
\bibitem[Pappenheimer(1953)]{pappenheimer1953passage}
J.~R. Pappenheimer, \emph{Physiological Reviews}, 1953, \textbf{33},
  387--423\relax
\mciteBstWouldAddEndPuncttrue
\mciteSetBstMidEndSepPunct{\mcitedefaultmidpunct}
{\mcitedefaultendpunct}{\mcitedefaultseppunct}\relax
\EndOfBibitem
\bibitem[Adamson \emph{et~al.}(2004)Adamson, Lenz, Zhang, Adamson, Weinbaum,
  and Curry]{adamson2004oncotic}
R.~Adamson, J.~Lenz, X.~Zhang, G.~Adamson, S.~Weinbaum and F.~Curry, \emph{The
  Journal of physiology}, 2004, \textbf{557}, 889--907\relax
\mciteBstWouldAddEndPuncttrue
\mciteSetBstMidEndSepPunct{\mcitedefaultmidpunct}
{\mcitedefaultendpunct}{\mcitedefaultseppunct}\relax
\EndOfBibitem
\bibitem[Anderson and Malone(1974)]{anderson1974mechanism}
J.~L. Anderson and D.~M. Malone, \emph{Biophysical journal}, 1974, \textbf{14},
  957--982\relax
\mciteBstWouldAddEndPuncttrue
\mciteSetBstMidEndSepPunct{\mcitedefaultmidpunct}
{\mcitedefaultendpunct}{\mcitedefaultseppunct}\relax
\EndOfBibitem
\bibitem[Yamauchi \emph{et~al.}(2000)Yamauchi, Shin, Shinozaki, and
  Kawabe]{yamauchi2000membrane}
A.~Yamauchi, Y.~Shin, M.~Shinozaki and M.~Kawabe, \emph{Journal of Membrane
  Science}, 2000, \textbf{170}, 1--7\relax
\mciteBstWouldAddEndPuncttrue
\mciteSetBstMidEndSepPunct{\mcitedefaultmidpunct}
{\mcitedefaultendpunct}{\mcitedefaultseppunct}\relax
\EndOfBibitem
\bibitem[Fujita and Kobatake(1968)]{fujita1968interpretation}
H.~Fujita and Y.~Kobatake, \emph{Journal of Colloid and Interface Science},
  1968, \textbf{27}, 609--615\relax
\mciteBstWouldAddEndPuncttrue
\mciteSetBstMidEndSepPunct{\mcitedefaultmidpunct}
{\mcitedefaultendpunct}{\mcitedefaultseppunct}\relax
\EndOfBibitem
\bibitem[Hill(1979)]{hill1979osmosis}
A.~Hill, \emph{Quarterly reviews of biophysics}, 1979, \textbf{12},
  67--99\relax
\mciteBstWouldAddEndPuncttrue
\mciteSetBstMidEndSepPunct{\mcitedefaultmidpunct}
{\mcitedefaultendpunct}{\mcitedefaultseppunct}\relax
\EndOfBibitem
\bibitem[Ferry(1936)]{ferry1936ultrafilter}
J.~D. Ferry, \emph{Chemical Reviews}, 1936, \textbf{18}, 373--455\relax
\mciteBstWouldAddEndPuncttrue
\mciteSetBstMidEndSepPunct{\mcitedefaultmidpunct}
{\mcitedefaultendpunct}{\mcitedefaultseppunct}\relax
\EndOfBibitem
\bibitem[Ray(1960)]{ray1960theory}
P.~M. Ray, \emph{Plant physiology}, 1960, \textbf{35}, 783\relax
\mciteBstWouldAddEndPuncttrue
\mciteSetBstMidEndSepPunct{\mcitedefaultmidpunct}
{\mcitedefaultendpunct}{\mcitedefaultseppunct}\relax
\EndOfBibitem
\bibitem[Giddings \emph{et~al.}(1968)Giddings, Kucera, Russell, and
  Myers]{giddings1968statistical}
J.~C. Giddings, E.~Kucera, C.~P. Russell and M.~N. Myers, \emph{The Journal of
  Physical Chemistry}, 1968, \textbf{72}, 4397--4408\relax
\mciteBstWouldAddEndPuncttrue
\mciteSetBstMidEndSepPunct{\mcitedefaultmidpunct}
{\mcitedefaultendpunct}{\mcitedefaultseppunct}\relax
\EndOfBibitem
\bibitem[Renkin(1954)]{renkin1954filtration}
E.~M. Renkin, \emph{The Journal of general physiology}, 1954, \textbf{38},
  225--243\relax
\mciteBstWouldAddEndPuncttrue
\mciteSetBstMidEndSepPunct{\mcitedefaultmidpunct}
{\mcitedefaultendpunct}{\mcitedefaultseppunct}\relax
\EndOfBibitem
\bibitem[Anderson and Quinn(1974)]{anderson1974restricted}
J.~L. Anderson and J.~A. Quinn, \emph{Biophysical Journal}, 1974, \textbf{14},
  130\relax
\mciteBstWouldAddEndPuncttrue
\mciteSetBstMidEndSepPunct{\mcitedefaultmidpunct}
{\mcitedefaultendpunct}{\mcitedefaultseppunct}\relax
\EndOfBibitem
\bibitem[Chou(1998)]{chou1998fast}
T.~Chou, \emph{Physical Review Letters}, 1998, \textbf{80}, 85\relax
\mciteBstWouldAddEndPuncttrue
\mciteSetBstMidEndSepPunct{\mcitedefaultmidpunct}
{\mcitedefaultendpunct}{\mcitedefaultseppunct}\relax
\EndOfBibitem
\bibitem[Chou(1999)]{chou1999kinetics}
T.~Chou, \emph{The Journal of Chemical Physics}, 1999, \textbf{110},
  606--615\relax
\mciteBstWouldAddEndPuncttrue
\mciteSetBstMidEndSepPunct{\mcitedefaultmidpunct}
{\mcitedefaultendpunct}{\mcitedefaultseppunct}\relax
\EndOfBibitem
\bibitem[Manning(1968)]{manning1968binary}
G.~S. Manning, \emph{The Journal of Chemical Physics}, 1968, \textbf{49},
  2668--2675\relax
\mciteBstWouldAddEndPuncttrue
\mciteSetBstMidEndSepPunct{\mcitedefaultmidpunct}
{\mcitedefaultendpunct}{\mcitedefaultseppunct}\relax
\EndOfBibitem
\bibitem[Picallo \emph{et~al.}(2013)Picallo, Gravelle, Joly, Charlaix, and
  Bocquet]{picallo2013nanofluidic}
C.~B. Picallo, S.~Gravelle, L.~Joly, E.~Charlaix and L.~Bocquet, \emph{Physical
  Review Letters}, 2013, \textbf{111}, 244501\relax
\mciteBstWouldAddEndPuncttrue
\mciteSetBstMidEndSepPunct{\mcitedefaultmidpunct}
{\mcitedefaultendpunct}{\mcitedefaultseppunct}\relax
\EndOfBibitem
\bibitem[Marbach \emph{et~al.}(2017)Marbach, Yoshida, and
  Bocquet]{marbach2017osmotic}
S.~Marbach, H.~Yoshida and L.~Bocquet, \emph{The Journal of Chemical Physics},
  2017, \textbf{146}, 194701\relax
\mciteBstWouldAddEndPuncttrue
\mciteSetBstMidEndSepPunct{\mcitedefaultmidpunct}
{\mcitedefaultendpunct}{\mcitedefaultseppunct}\relax
\EndOfBibitem
\bibitem[Debye \emph{et~al.}(1954)Debye, Debye, Eckstein, Barber, and
  Arquette]{debye1954equilibrium}
P.~J.~W. Debye, P.~Debye, B.~Eckstein, W.~Barber and G.~Arquette,
  \emph{Equilibrium and sedimentation of uncharged particles in inhomogeneous
  electric fields}, Academic Press, 1954, pp. 273--285\relax
\mciteBstWouldAddEndPuncttrue
\mciteSetBstMidEndSepPunct{\mcitedefaultmidpunct}
{\mcitedefaultendpunct}{\mcitedefaultseppunct}\relax
\EndOfBibitem
\bibitem[Grim and Sollner(1957)]{grim1957contributions}
E.~Grim and K.~Sollner, \emph{The Journal of general physiology}, 1957,
  \textbf{40}, 887--899\relax
\mciteBstWouldAddEndPuncttrue
\mciteSetBstMidEndSepPunct{\mcitedefaultmidpunct}
{\mcitedefaultendpunct}{\mcitedefaultseppunct}\relax
\EndOfBibitem
\bibitem[Anderson(1989)]{anderson1989colloid}
J.~L. Anderson, \emph{Annual review of fluid mechanics}, 1989, \textbf{21},
  61--99\relax
\mciteBstWouldAddEndPuncttrue
\mciteSetBstMidEndSepPunct{\mcitedefaultmidpunct}
{\mcitedefaultendpunct}{\mcitedefaultseppunct}\relax
\EndOfBibitem
\bibitem[Derjaguin \emph{et~al.}(1993)Derjaguin, Sidorenkov, Zubashchenko, and
  Kiseleva]{derjaguin1993kinetic}
B.~Derjaguin, G.~Sidorenkov, E.~Zubashchenko and E.~Kiseleva, \emph{Progress in
  Surface Science}, 1993, \textbf{43}, 138--152\relax
\mciteBstWouldAddEndPuncttrue
\mciteSetBstMidEndSepPunct{\mcitedefaultmidpunct}
{\mcitedefaultendpunct}{\mcitedefaultseppunct}\relax
\EndOfBibitem
\bibitem[Derjaguin \emph{et~al.}(1993)Derjaguin, Dukhin, and
  Korotkova]{derjaguin1993diffusiophoresis}
B.~Derjaguin, S.~Dukhin and A.~Korotkova, \emph{Progress in Surface Science},
  1993, \textbf{43}, 153--158\relax
\mciteBstWouldAddEndPuncttrue
\mciteSetBstMidEndSepPunct{\mcitedefaultmidpunct}
{\mcitedefaultendpunct}{\mcitedefaultseppunct}\relax
\EndOfBibitem
\bibitem[Hunter(2001)]{hunter2001foundations}
R.~J. Hunter, \emph{Foundations of colloid science}, Oxford university press,
  2001\relax
\mciteBstWouldAddEndPuncttrue
\mciteSetBstMidEndSepPunct{\mcitedefaultmidpunct}
{\mcitedefaultendpunct}{\mcitedefaultseppunct}\relax
\EndOfBibitem
\bibitem[Andelman(1995)]{andelman1995electrostatic}
D.~Andelman, \emph{Handbook of biological physics}, Elsevier, 1995, vol.~1, pp.
  603--642\relax
\mciteBstWouldAddEndPuncttrue
\mciteSetBstMidEndSepPunct{\mcitedefaultmidpunct}
{\mcitedefaultendpunct}{\mcitedefaultseppunct}\relax
\EndOfBibitem
\bibitem[Squires(2016)]{squires2016particles}
T.~M. Squires, \emph{Fluids, Colloids and Soft Materials: An Introduction to
  Soft Matter Physics}, 2016,  59--79\relax
\mciteBstWouldAddEndPuncttrue
\mciteSetBstMidEndSepPunct{\mcitedefaultmidpunct}
{\mcitedefaultendpunct}{\mcitedefaultseppunct}\relax
\EndOfBibitem
\bibitem[Schoch \emph{et~al.}(2008)Schoch, Han, and
  Renaud]{schoch2008transport}
R.~B. Schoch, J.~Han and P.~Renaud, \emph{Reviews of modern physics}, 2008,
  \textbf{80}, 839\relax
\mciteBstWouldAddEndPuncttrue
\mciteSetBstMidEndSepPunct{\mcitedefaultmidpunct}
{\mcitedefaultendpunct}{\mcitedefaultseppunct}\relax
\EndOfBibitem
\bibitem[Devasenathipathy and
  Santiago(2005)]{devasenathipathy2005electrokinetic}
S.~Devasenathipathy and J.~Santiago, \emph{Microscale Diagnostic Techniques},
  Springer, 2005, pp. 113--154\relax
\mciteBstWouldAddEndPuncttrue
\mciteSetBstMidEndSepPunct{\mcitedefaultmidpunct}
{\mcitedefaultendpunct}{\mcitedefaultseppunct}\relax
\EndOfBibitem
\bibitem[Bocquet and Charlaix(2010)]{bocquet2010nanofluidics}
L.~Bocquet and E.~Charlaix, \emph{Chemical Society Reviews}, 2010, \textbf{39},
  1073--1095\relax
\mciteBstWouldAddEndPuncttrue
\mciteSetBstMidEndSepPunct{\mcitedefaultmidpunct}
{\mcitedefaultendpunct}{\mcitedefaultseppunct}\relax
\EndOfBibitem
\bibitem[Khair and Squires(2009)]{khair2009influence}
A.~S. Khair and T.~M. Squires, \emph{Physics of Fluids}, 2009, \textbf{21},
  042001\relax
\mciteBstWouldAddEndPuncttrue
\mciteSetBstMidEndSepPunct{\mcitedefaultmidpunct}
{\mcitedefaultendpunct}{\mcitedefaultseppunct}\relax
\EndOfBibitem
\bibitem[Mouterde and Bocquet(2018)]{Mouterde2019}
T.~Mouterde and L.~Bocquet, \emph{The European physical journal E}, 2018,
  \textbf{41}, 148\relax
\mciteBstWouldAddEndPuncttrue
\mciteSetBstMidEndSepPunct{\mcitedefaultmidpunct}
{\mcitedefaultendpunct}{\mcitedefaultseppunct}\relax
\EndOfBibitem
\bibitem[Messinger and Squires(2010)]{messinger2010suppression}
R.~Messinger and T.~Squires, \emph{Physical Review Letters}, 2010,
  \textbf{105}, 144503\relax
\mciteBstWouldAddEndPuncttrue
\mciteSetBstMidEndSepPunct{\mcitedefaultmidpunct}
{\mcitedefaultendpunct}{\mcitedefaultseppunct}\relax
\EndOfBibitem
\bibitem[Pascall and Squires(2010)]{pascall2010induced}
A.~J. Pascall and T.~M. Squires, \emph{Physical Review Letters}, 2010,
  \textbf{104}, 088301\relax
\mciteBstWouldAddEndPuncttrue
\mciteSetBstMidEndSepPunct{\mcitedefaultmidpunct}
{\mcitedefaultendpunct}{\mcitedefaultseppunct}\relax
\EndOfBibitem
\bibitem[Bonthuis and Netz(2012)]{bonthuis2012unraveling}
D.~J. Bonthuis and R.~R. Netz, \emph{Langmuir}, 2012, \textbf{28},
  16049--16059\relax
\mciteBstWouldAddEndPuncttrue
\mciteSetBstMidEndSepPunct{\mcitedefaultmidpunct}
{\mcitedefaultendpunct}{\mcitedefaultseppunct}\relax
\EndOfBibitem
\bibitem[Anderson and Prieve(1991)]{anderson1991diffusiophoresis}
J.~L. Anderson and D.~C. Prieve, \emph{Langmuir}, 1991, \textbf{7},
  403--406\relax
\mciteBstWouldAddEndPuncttrue
\mciteSetBstMidEndSepPunct{\mcitedefaultmidpunct}
{\mcitedefaultendpunct}{\mcitedefaultseppunct}\relax
\EndOfBibitem
\bibitem[Ajdari and Bocquet(2006)]{ajdari2006giant}
A.~Ajdari and L.~Bocquet, \emph{Physical Review Letters}, 2006, \textbf{96},
  186102\relax
\mciteBstWouldAddEndPuncttrue
\mciteSetBstMidEndSepPunct{\mcitedefaultmidpunct}
{\mcitedefaultendpunct}{\mcitedefaultseppunct}\relax
\EndOfBibitem
\bibitem[Huang \emph{et~al.}(2008)Huang, Cottin-Bizonne, Ybert, and
  Bocquet]{huang2008massive}
D.~M. Huang, C.~Cottin-Bizonne, C.~Ybert and L.~Bocquet, \emph{Physical Review
  Letters}, 2008, \textbf{101}, 064503\relax
\mciteBstWouldAddEndPuncttrue
\mciteSetBstMidEndSepPunct{\mcitedefaultmidpunct}
{\mcitedefaultendpunct}{\mcitedefaultseppunct}\relax
\EndOfBibitem
\bibitem[Prieve \emph{et~al.}(1984)Prieve, Anderson, Ebel, and
  Lowell]{prieve1984motion}
D.~Prieve, J.~Anderson, J.~Ebel and M.~Lowell, \emph{Journal of Fluid
  Mechanics}, 1984, \textbf{148}, 247--269\relax
\mciteBstWouldAddEndPuncttrue
\mciteSetBstMidEndSepPunct{\mcitedefaultmidpunct}
{\mcitedefaultendpunct}{\mcitedefaultseppunct}\relax
\EndOfBibitem
\bibitem[Lokesh \emph{et~al.}(2018)Lokesh, Youn, and Park]{lokesh2018osmotic}
M.~Lokesh, S.~K. Youn and H.~G. Park, \emph{Nano letters}, 2018, \textbf{18},
  6679--6685\relax
\mciteBstWouldAddEndPuncttrue
\mciteSetBstMidEndSepPunct{\mcitedefaultmidpunct}
{\mcitedefaultendpunct}{\mcitedefaultseppunct}\relax
\EndOfBibitem
\bibitem[Siria \emph{et~al.}(2013)Siria, Poncharal, Biance, Fulcrand, Blase,
  Purcell, and Bocquet]{siria2013giant}
A.~Siria, P.~Poncharal, A.-L. Biance, R.~Fulcrand, X.~Blase, S.~T. Purcell and
  L.~Bocquet, \emph{Nature}, 2013, \textbf{494}, 455\relax
\mciteBstWouldAddEndPuncttrue
\mciteSetBstMidEndSepPunct{\mcitedefaultmidpunct}
{\mcitedefaultendpunct}{\mcitedefaultseppunct}\relax
\EndOfBibitem
\bibitem[Feng \emph{et~al.}(2016)Feng, Graf, Liu, Ovchinnikov, Dumcenco,
  Heiranian, Nandigana, Aluru, Kis, and Radenovic]{feng2016single}
J.~Feng, M.~Graf, K.~Liu, D.~Ovchinnikov, D.~Dumcenco, M.~Heiranian,
  V.~Nandigana, N.~R. Aluru, A.~Kis and A.~Radenovic, \emph{Nature}, 2016,
  \textbf{536}, 197\relax
\mciteBstWouldAddEndPuncttrue
\mciteSetBstMidEndSepPunct{\mcitedefaultmidpunct}
{\mcitedefaultendpunct}{\mcitedefaultseppunct}\relax
\EndOfBibitem
\bibitem[Siria \emph{et~al.}(2017)Siria, Bocquet, and Bocquet]{siria2017new}
A.~Siria, M.-L. Bocquet and L.~Bocquet, \emph{Nature Reviews Chemistry}, 2017,
  \textbf{1}, 0091\relax
\mciteBstWouldAddEndPuncttrue
\mciteSetBstMidEndSepPunct{\mcitedefaultmidpunct}
{\mcitedefaultendpunct}{\mcitedefaultseppunct}\relax
\EndOfBibitem
\bibitem[Pl{\"u}mper \emph{et~al.}(2017)Pl{\"u}mper, Botan, Los, Liu,
  Malthe-S{\o}renssen, and Jamtveit]{plumper2017fluid}
O.~Pl{\"u}mper, A.~Botan, C.~Los, Y.~Liu, A.~Malthe-S{\o}renssen and
  B.~Jamtveit, \emph{Nature geoscience}, 2017, \textbf{10}, 685\relax
\mciteBstWouldAddEndPuncttrue
\mciteSetBstMidEndSepPunct{\mcitedefaultmidpunct}
{\mcitedefaultendpunct}{\mcitedefaultseppunct}\relax
\EndOfBibitem
\bibitem[Palacci \emph{et~al.}(2012)Palacci, Cottin-Bizonne, Ybert, and
  Bocquet]{palacci2012osmotic}
J.~Palacci, C.~Cottin-Bizonne, C.~Ybert and L.~Bocquet, \emph{Soft Matter},
  2012, \textbf{8}, 980--994\relax
\mciteBstWouldAddEndPuncttrue
\mciteSetBstMidEndSepPunct{\mcitedefaultmidpunct}
{\mcitedefaultendpunct}{\mcitedefaultseppunct}\relax
\EndOfBibitem
\bibitem[Palacci \emph{et~al.}(2010)Palacci, Ab{\'e}cassis, Cottin-Bizonne,
  Ybert, and Bocquet]{palacci2010colloidal}
J.~Palacci, B.~Ab{\'e}cassis, C.~Cottin-Bizonne, C.~Ybert and L.~Bocquet,
  \emph{Physical Review Letters}, 2010, \textbf{104}, 138302\relax
\mciteBstWouldAddEndPuncttrue
\mciteSetBstMidEndSepPunct{\mcitedefaultmidpunct}
{\mcitedefaultendpunct}{\mcitedefaultseppunct}\relax
\EndOfBibitem
\bibitem[Kosmulski and Matuevi(1992)]{kosmulski1992solvophoresis}
M.~Kosmulski and E.~Matuevi, \emph{Journal of colloid and interface science},
  1992, \textbf{150}, 291--294\relax
\mciteBstWouldAddEndPuncttrue
\mciteSetBstMidEndSepPunct{\mcitedefaultmidpunct}
{\mcitedefaultendpunct}{\mcitedefaultseppunct}\relax
\EndOfBibitem
\bibitem[Shi \emph{et~al.}(2016)Shi, Nery-Azevedo, Abdel-Fattah, and
  Squires]{shi2016diffusiophoretic}
N.~Shi, R.~Nery-Azevedo, A.~I. Abdel-Fattah and T.~M. Squires, \emph{Physical
  Review Letters}, 2016, \textbf{117}, 258001\relax
\mciteBstWouldAddEndPuncttrue
\mciteSetBstMidEndSepPunct{\mcitedefaultmidpunct}
{\mcitedefaultendpunct}{\mcitedefaultseppunct}\relax
\EndOfBibitem
\bibitem[Fair and Osterle(1971)]{fair1971reverse}
J.~Fair and J.~Osterle, \emph{The Journal of Chemical Physics}, 1971,
  \textbf{54}, 3307--3316\relax
\mciteBstWouldAddEndPuncttrue
\mciteSetBstMidEndSepPunct{\mcitedefaultmidpunct}
{\mcitedefaultendpunct}{\mcitedefaultseppunct}\relax
\EndOfBibitem
\bibitem[Kim \emph{et~al.}(2010)Kim, Duan, Chen, and Majumdar]{kim2010power}
D.-K. Kim, C.~Duan, Y.-F. Chen and A.~Majumdar, \emph{Microfluidics and
  Nanofluidics}, 2010, \textbf{9}, 1215--1224\relax
\mciteBstWouldAddEndPuncttrue
\mciteSetBstMidEndSepPunct{\mcitedefaultmidpunct}
{\mcitedefaultendpunct}{\mcitedefaultseppunct}\relax
\EndOfBibitem
\bibitem[Zhang \emph{et~al.}(2017)Zhang, Sui, Li, Xie, Kong, Xiao, Gao, Wen,
  and Jiang]{zhang2017ultrathin}
Z.~Zhang, X.~Sui, P.~Li, G.~Xie, X.-Y. Kong, K.~Xiao, L.~Gao, L.~Wen and
  L.~Jiang, \emph{Journal of the American Chemical Society}, 2017,
  \textbf{139}, 8905--8914\relax
\mciteBstWouldAddEndPuncttrue
\mciteSetBstMidEndSepPunct{\mcitedefaultmidpunct}
{\mcitedefaultendpunct}{\mcitedefaultseppunct}\relax
\EndOfBibitem
\bibitem[Wanunu \emph{et~al.}(2010)Wanunu, Morrison, Rabin, Grosberg, and
  Meller]{wanunu2010electrostatic}
M.~Wanunu, W.~Morrison, Y.~Rabin, A.~Y. Grosberg and A.~Meller, \emph{Nature
  nanotechnology}, 2010, \textbf{5}, 160\relax
\mciteBstWouldAddEndPuncttrue
\mciteSetBstMidEndSepPunct{\mcitedefaultmidpunct}
{\mcitedefaultendpunct}{\mcitedefaultseppunct}\relax
\EndOfBibitem
\bibitem[Siwy(2006)]{siwy2006ion}
Z.~S. Siwy, \emph{Advanced Functional Materials}, 2006, \textbf{16},
  735--746\relax
\mciteBstWouldAddEndPuncttrue
\mciteSetBstMidEndSepPunct{\mcitedefaultmidpunct}
{\mcitedefaultendpunct}{\mcitedefaultseppunct}\relax
\EndOfBibitem
\bibitem[Bell \emph{et~al.}(2011)Bell, Engst, Ablay, Divitini, Ducati, Liedl,
  and Keyser]{bell2011dna}
N.~A. Bell, C.~R. Engst, M.~Ablay, G.~Divitini, C.~Ducati, T.~Liedl and U.~F.
  Keyser, \emph{Nano letters}, 2011, \textbf{12}, 512--517\relax
\mciteBstWouldAddEndPuncttrue
\mciteSetBstMidEndSepPunct{\mcitedefaultmidpunct}
{\mcitedefaultendpunct}{\mcitedefaultseppunct}\relax
\EndOfBibitem
\bibitem[Walker \emph{et~al.}(2017)Walker, Ubych, Saraswat, Chalklen,
  Braeuninger-Weimer, Caneva, Weatherup, Hofmann, and
  Keyser]{walker2017extrinsic}
M.~I. Walker, K.~Ubych, V.~Saraswat, E.~A. Chalklen, P.~Braeuninger-Weimer,
  S.~Caneva, R.~S. Weatherup, S.~Hofmann and U.~F. Keyser, \emph{ACS nano},
  2017, \textbf{11}, 1340--1346\relax
\mciteBstWouldAddEndPuncttrue
\mciteSetBstMidEndSepPunct{\mcitedefaultmidpunct}
{\mcitedefaultendpunct}{\mcitedefaultseppunct}\relax
\EndOfBibitem
\bibitem[Wang \emph{et~al.}(2017)Wang, Boutilier, Kidambi, Jang,
  Hadjiconstantinou, and Karnik]{wang2017fundamental}
L.~Wang, M.~S. Boutilier, P.~R. Kidambi, D.~Jang, N.~G. Hadjiconstantinou and
  R.~Karnik, \emph{Nature nanotechnology}, 2017, \textbf{12}, 509\relax
\mciteBstWouldAddEndPuncttrue
\mciteSetBstMidEndSepPunct{\mcitedefaultmidpunct}
{\mcitedefaultendpunct}{\mcitedefaultseppunct}\relax
\EndOfBibitem
\bibitem[Hall(1975)]{hall1975access}
J.~E. Hall, \emph{The Journal of general physiology}, 1975, \textbf{66},
  531--532\relax
\mciteBstWouldAddEndPuncttrue
\mciteSetBstMidEndSepPunct{\mcitedefaultmidpunct}
{\mcitedefaultendpunct}{\mcitedefaultseppunct}\relax
\EndOfBibitem
\bibitem[Mao \emph{et~al.}(2014)Mao, Sherwood, and Ghosal]{mao2014electro}
M.~Mao, J.~Sherwood and S.~Ghosal, \emph{Journal of Fluid Mechanics}, 2014,
  \textbf{749}, 167--183\relax
\mciteBstWouldAddEndPuncttrue
\mciteSetBstMidEndSepPunct{\mcitedefaultmidpunct}
{\mcitedefaultendpunct}{\mcitedefaultseppunct}\relax
\EndOfBibitem
\bibitem[Sherwood \emph{et~al.}(2014)Sherwood, Mao, and
  Ghosal]{sherwood2014electroosmosis}
J.~D. Sherwood, M.~Mao and S.~Ghosal, \emph{Langmuir}, 2014, \textbf{30},
  9261--9272\relax
\mciteBstWouldAddEndPuncttrue
\mciteSetBstMidEndSepPunct{\mcitedefaultmidpunct}
{\mcitedefaultendpunct}{\mcitedefaultseppunct}\relax
\EndOfBibitem
\bibitem[Melnikov \emph{et~al.}(2017)Melnikov, Hulings, and
  Gracheva]{melnikov2017electro}
D.~V. Melnikov, Z.~K. Hulings and M.~E. Gracheva, \emph{Physical Review E},
  2017, \textbf{95}, 063105\relax
\mciteBstWouldAddEndPuncttrue
\mciteSetBstMidEndSepPunct{\mcitedefaultmidpunct}
{\mcitedefaultendpunct}{\mcitedefaultseppunct}\relax
\EndOfBibitem
\bibitem[Rankin \emph{et~al.}(2019)Rankin, Bocquet, and
  Huang]{Rankin2019diffusio}
D.~J. Rankin, L.~Bocquet and D.~M. Huang, \emph{submitted}, 2019,
  arxiv.org/abs/1904.10636\relax
\mciteBstWouldAddEndPuncttrue
\mciteSetBstMidEndSepPunct{\mcitedefaultmidpunct}
{\mcitedefaultendpunct}{\mcitedefaultseppunct}\relax
\EndOfBibitem
\bibitem[Sampson(1891)]{sampson1891stokes}
R.~A. Sampson, \emph{Philosophical Transactions of the Royal Society of London.
  A}, 1891, \textbf{182}, 449--518\relax
\mciteBstWouldAddEndPuncttrue
\mciteSetBstMidEndSepPunct{\mcitedefaultmidpunct}
{\mcitedefaultendpunct}{\mcitedefaultseppunct}\relax
\EndOfBibitem
\bibitem[Dagan \emph{et~al.}(1982)Dagan, Weinbaum, and
  Pfeffer]{dagan1982infinite}
Z.~Dagan, S.~Weinbaum and R.~Pfeffer, \emph{Journal of Fluid Mechanics}, 1982,
  \textbf{115}, 505--523\relax
\mciteBstWouldAddEndPuncttrue
\mciteSetBstMidEndSepPunct{\mcitedefaultmidpunct}
{\mcitedefaultendpunct}{\mcitedefaultseppunct}\relax
\EndOfBibitem
\bibitem[Happel and Brenner(2012)]{happel2012low}
J.~Happel and H.~Brenner, \emph{Low Reynolds number hydrodynamics: with special
  applications to particulate media}, Springer Science \& Business Media, 2012,
  vol.~1\relax
\mciteBstWouldAddEndPuncttrue
\mciteSetBstMidEndSepPunct{\mcitedefaultmidpunct}
{\mcitedefaultendpunct}{\mcitedefaultseppunct}\relax
\EndOfBibitem
\bibitem[Lippmann(1907)]{lippmann1907endosmose}
G.~Lippmann, \emph{Compt. rend}, 1907, \textbf{145}, 104--105\relax
\mciteBstWouldAddEndPuncttrue
\mciteSetBstMidEndSepPunct{\mcitedefaultmidpunct}
{\mcitedefaultendpunct}{\mcitedefaultseppunct}\relax
\EndOfBibitem
\bibitem[Aubert(1912)]{aubert1912thermo}
M.~Aubert, \emph{Ann. Chim. Phys.}, 1912, \textbf{26}, 165\relax
\mciteBstWouldAddEndPuncttrue
\mciteSetBstMidEndSepPunct{\mcitedefaultmidpunct}
{\mcitedefaultendpunct}{\mcitedefaultseppunct}\relax
\EndOfBibitem
\bibitem[Derjaguin(1987)]{derjaguin1987some}
B.~V. Derjaguin, \emph{Surface Forces and Surfactant Systems}, Springer, 1987,
  pp. 17--30\relax
\mciteBstWouldAddEndPuncttrue
\mciteSetBstMidEndSepPunct{\mcitedefaultmidpunct}
{\mcitedefaultendpunct}{\mcitedefaultseppunct}\relax
\EndOfBibitem
\bibitem[Derjaguin \emph{et~al.}(1987)Derjaguin, Churaev, and
  Muller]{derjaguin1987thermo}
B.~Derjaguin, N.~Churaev and V.~Muller, \emph{Surface Forces}, Springer, 1987,
  pp. 390--409\relax
\mciteBstWouldAddEndPuncttrue
\mciteSetBstMidEndSepPunct{\mcitedefaultmidpunct}
{\mcitedefaultendpunct}{\mcitedefaultseppunct}\relax
\EndOfBibitem
\bibitem[Bregulla \emph{et~al.}(2016)Bregulla, W{\"u}rger, G{\"u}nther, Mertig,
  and Cichos]{bregulla2016thermo}
A.~P. Bregulla, A.~W{\"u}rger, K.~G{\"u}nther, M.~Mertig and F.~Cichos,
  \emph{Physical Review Letters}, 2016, \textbf{116}, 188303\relax
\mciteBstWouldAddEndPuncttrue
\mciteSetBstMidEndSepPunct{\mcitedefaultmidpunct}
{\mcitedefaultendpunct}{\mcitedefaultseppunct}\relax
\EndOfBibitem
\bibitem[Barrag{\'a}n and Kjelstrup(2017)]{barragan2017thermo}
V.~M. Barrag{\'a}n and S.~Kjelstrup, \emph{Journal of Non-Equilibrium
  Thermodynamics}, 2017, \textbf{42}, 217--236\relax
\mciteBstWouldAddEndPuncttrue
\mciteSetBstMidEndSepPunct{\mcitedefaultmidpunct}
{\mcitedefaultendpunct}{\mcitedefaultseppunct}\relax
\EndOfBibitem
\bibitem[Ruckenstein(1981)]{ruckenstein1981can}
E.~Ruckenstein, \emph{Journal of Colloid and Interface Science}, 1981,
  \textbf{83}, 77--81\relax
\mciteBstWouldAddEndPuncttrue
\mciteSetBstMidEndSepPunct{\mcitedefaultmidpunct}
{\mcitedefaultendpunct}{\mcitedefaultseppunct}\relax
\EndOfBibitem
\bibitem[Piazza(2004)]{piazza2004thermal}
R.~Piazza, \emph{Journal of Physics: Condensed Matter}, 2004, \textbf{16},
  S4195\relax
\mciteBstWouldAddEndPuncttrue
\mciteSetBstMidEndSepPunct{\mcitedefaultmidpunct}
{\mcitedefaultendpunct}{\mcitedefaultseppunct}\relax
\EndOfBibitem
\bibitem[Parola and Piazza(2004)]{parola2004particle}
A.~Parola and R.~Piazza, \emph{The European Physical Journal E}, 2004,
  \textbf{15}, 255--263\relax
\mciteBstWouldAddEndPuncttrue
\mciteSetBstMidEndSepPunct{\mcitedefaultmidpunct}
{\mcitedefaultendpunct}{\mcitedefaultseppunct}\relax
\EndOfBibitem
\bibitem[Duhr and Braun(2006)]{duhr2006molecules}
S.~Duhr and D.~Braun, \emph{Proceedings of the National Academy of Sciences},
  2006, \textbf{103}, 19678--19682\relax
\mciteBstWouldAddEndPuncttrue
\mciteSetBstMidEndSepPunct{\mcitedefaultmidpunct}
{\mcitedefaultendpunct}{\mcitedefaultseppunct}\relax
\EndOfBibitem
\bibitem[Piazza(2008)]{piazza2008thermophoresis}
R.~Piazza, \emph{Soft Matter}, 2008, \textbf{4}, 1740--1744\relax
\mciteBstWouldAddEndPuncttrue
\mciteSetBstMidEndSepPunct{\mcitedefaultmidpunct}
{\mcitedefaultendpunct}{\mcitedefaultseppunct}\relax
\EndOfBibitem
\bibitem[Piazza and Parola(2008)]{piazza2008thermophoresisJP}
R.~Piazza and A.~Parola, \emph{Journal of Physics: Condensed Matter}, 2008,
  \textbf{20}, 153102\relax
\mciteBstWouldAddEndPuncttrue
\mciteSetBstMidEndSepPunct{\mcitedefaultmidpunct}
{\mcitedefaultendpunct}{\mcitedefaultseppunct}\relax
\EndOfBibitem
\bibitem[Morthomas and W{\"u}rger(2008)]{morthomas2008thermophoresis}
J.~Morthomas and A.~W{\"u}rger, \emph{Journal of Physics: Condensed Matter},
  2008, \textbf{21}, 035103\relax
\mciteBstWouldAddEndPuncttrue
\mciteSetBstMidEndSepPunct{\mcitedefaultmidpunct}
{\mcitedefaultendpunct}{\mcitedefaultseppunct}\relax
\EndOfBibitem
\bibitem[Fu \emph{et~al.}(2017)Fu, Merabia, and Joly]{fu2017controls}
L.~Fu, S.~Merabia and L.~Joly, \emph{Physical Review Letters}, 2017,
  \textbf{119}, 214501\relax
\mciteBstWouldAddEndPuncttrue
\mciteSetBstMidEndSepPunct{\mcitedefaultmidpunct}
{\mcitedefaultendpunct}{\mcitedefaultseppunct}\relax
\EndOfBibitem
\bibitem[Golestanian \emph{et~al.}(2007)Golestanian, Liverpool, and
  Ajdari]{golestanian2007designing}
R.~Golestanian, T.~Liverpool and A.~Ajdari, \emph{New Journal of Physics},
  2007, \textbf{9}, 126\relax
\mciteBstWouldAddEndPuncttrue
\mciteSetBstMidEndSepPunct{\mcitedefaultmidpunct}
{\mcitedefaultendpunct}{\mcitedefaultseppunct}\relax
\EndOfBibitem
\bibitem[Di~Leonardo \emph{et~al.}(2009)Di~Leonardo, Ianni, and
  Ruocco]{di2009colloidal}
R.~Di~Leonardo, F.~Ianni and G.~Ruocco, \emph{Langmuir}, 2009, \textbf{25},
  4247--4250\relax
\mciteBstWouldAddEndPuncttrue
\mciteSetBstMidEndSepPunct{\mcitedefaultmidpunct}
{\mcitedefaultendpunct}{\mcitedefaultseppunct}\relax
\EndOfBibitem
\bibitem[W{\"u}rger(2010)]{wurger2010thermal}
A.~W{\"u}rger, \emph{Reports on Progress in Physics}, 2010, \textbf{73},
  126601\relax
\mciteBstWouldAddEndPuncttrue
\mciteSetBstMidEndSepPunct{\mcitedefaultmidpunct}
{\mcitedefaultendpunct}{\mcitedefaultseppunct}\relax
\EndOfBibitem
\bibitem[Yu and Chen(2015)]{yu2015concentration}
L.-H. Yu and Y.-F. Chen, \emph{Analytical chemistry}, 2015, \textbf{87},
  2845--2851\relax
\mciteBstWouldAddEndPuncttrue
\mciteSetBstMidEndSepPunct{\mcitedefaultmidpunct}
{\mcitedefaultendpunct}{\mcitedefaultseppunct}\relax
\EndOfBibitem
\bibitem[Jiang \emph{et~al.}(2009)Jiang, Wada, Yoshinaga, and
  Sano]{jiang2009manipulation}
H.-R. Jiang, H.~Wada, N.~Yoshinaga and M.~Sano, \emph{Physical Review Letters},
  2009, \textbf{102}, 208301\relax
\mciteBstWouldAddEndPuncttrue
\mciteSetBstMidEndSepPunct{\mcitedefaultmidpunct}
{\mcitedefaultendpunct}{\mcitedefaultseppunct}\relax
\EndOfBibitem
\bibitem[Wienken \emph{et~al.}(2010)Wienken, Baaske, Rothbauer, Braun, and
  Duhr]{wienken2010protein}
C.~J. Wienken, P.~Baaske, U.~Rothbauer, D.~Braun and S.~Duhr, \emph{Nature
  communications}, 2010, \textbf{1}, 1--7\relax
\mciteBstWouldAddEndPuncttrue
\mciteSetBstMidEndSepPunct{\mcitedefaultmidpunct}
{\mcitedefaultendpunct}{\mcitedefaultseppunct}\relax
\EndOfBibitem
\bibitem[Dau \emph{et~al.}(2016)Dau, Edeleva, Seidel, Stockley, Braun, and
  Jenne]{dau2016quantitative}
T.~Dau, E.~Edeleva, S.~Seidel, R.~Stockley, D.~Braun and D.~E. Jenne,
  \emph{Scientific reports}, 2016, \textbf{6}, 35413\relax
\mciteBstWouldAddEndPuncttrue
\mciteSetBstMidEndSepPunct{\mcitedefaultmidpunct}
{\mcitedefaultendpunct}{\mcitedefaultseppunct}\relax
\EndOfBibitem
\bibitem[Al-Alawy and Al-Alawy(2016)]{al2016thermal}
A.~F. Al-Alawy and R.~M. Al-Alawy, \emph{Iraqi Journal of Chemical and
  Petroleum Engineering}, 2016, \textbf{17}, 53--68\relax
\mciteBstWouldAddEndPuncttrue
\mciteSetBstMidEndSepPunct{\mcitedefaultmidpunct}
{\mcitedefaultendpunct}{\mcitedefaultseppunct}\relax
\EndOfBibitem
\bibitem[Kuipers \emph{et~al.}(2015)Kuipers, Hanemaaijer, Brouwer, van
  Medevoort, Jansen, Altena, van~der Vleuten, and Bak]{kuipers2015simultaneous}
N.~Kuipers, J.~H. Hanemaaijer, H.~Brouwer, J.~van Medevoort, A.~Jansen,
  F.~Altena, P.~van~der Vleuten and H.~Bak, \emph{Desalination and Water
  Treatment}, 2015, \textbf{55}, 2766--2776\relax
\mciteBstWouldAddEndPuncttrue
\mciteSetBstMidEndSepPunct{\mcitedefaultmidpunct}
{\mcitedefaultendpunct}{\mcitedefaultseppunct}\relax
\EndOfBibitem
\bibitem[Straub \emph{et~al.}(2016)Straub, Yip, Lin, Lee, and
  Elimelech]{straub2016harvesting}
A.~P. Straub, N.~Y. Yip, S.~Lin, J.~Lee and M.~Elimelech, \emph{Nature Energy},
  2016, \textbf{1}, 16090\relax
\mciteBstWouldAddEndPuncttrue
\mciteSetBstMidEndSepPunct{\mcitedefaultmidpunct}
{\mcitedefaultendpunct}{\mcitedefaultseppunct}\relax
\EndOfBibitem
\bibitem[Rotenberg and Pagonabarraga(2013)]{rotenberg2013electrokinetics}
B.~Rotenberg and I.~Pagonabarraga, \emph{Molecular Physics}, 2013,
  \textbf{111}, 827--842\relax
\mciteBstWouldAddEndPuncttrue
\mciteSetBstMidEndSepPunct{\mcitedefaultmidpunct}
{\mcitedefaultendpunct}{\mcitedefaultseppunct}\relax
\EndOfBibitem
\bibitem[Kalra \emph{et~al.}(2003)Kalra, Garde, and Hummer]{kalra2003osmotic}
A.~Kalra, S.~Garde and G.~Hummer, \emph{Proceedings of the National Academy of
  Sciences}, 2003, \textbf{100}, 10175--10180\relax
\mciteBstWouldAddEndPuncttrue
\mciteSetBstMidEndSepPunct{\mcitedefaultmidpunct}
{\mcitedefaultendpunct}{\mcitedefaultseppunct}\relax
\EndOfBibitem
\bibitem[Luo and Roux(2009)]{luo2009simulation}
Y.~Luo and B.~Roux, \emph{The Journal of Physical Chemistry Letters}, 2009,
  \textbf{1}, 183--189\relax
\mciteBstWouldAddEndPuncttrue
\mciteSetBstMidEndSepPunct{\mcitedefaultmidpunct}
{\mcitedefaultendpunct}{\mcitedefaultseppunct}\relax
\EndOfBibitem
\bibitem[Lion and Allen(2012)]{lion2012osmosis}
T.~W. Lion and R.~J. Allen, \emph{The Journal of Chemical Physics}, 2012,
  \textbf{137}, 244911\relax
\mciteBstWouldAddEndPuncttrue
\mciteSetBstMidEndSepPunct{\mcitedefaultmidpunct}
{\mcitedefaultendpunct}{\mcitedefaultseppunct}\relax
\EndOfBibitem
\bibitem[Yoshida \emph{et~al.}(2017)Yoshida, Marbach, and
  Bocquet]{yoshida2017osmotic}
H.~Yoshida, S.~Marbach and L.~Bocquet, \emph{The Journal of Chemical Physics},
  2017, \textbf{146}, 194702\relax
\mciteBstWouldAddEndPuncttrue
\mciteSetBstMidEndSepPunct{\mcitedefaultmidpunct}
{\mcitedefaultendpunct}{\mcitedefaultseppunct}\relax
\EndOfBibitem
\bibitem[McDonald and Hansen(1986)]{hansen1986ir}
I.~McDonald and J.~Hansen, \emph{Academic Press, London, ed}, 1986, \textbf{2},
  179\relax
\mciteBstWouldAddEndPuncttrue
\mciteSetBstMidEndSepPunct{\mcitedefaultmidpunct}
{\mcitedefaultendpunct}{\mcitedefaultseppunct}\relax
\EndOfBibitem
\bibitem[Marry \emph{et~al.}(2003)Marry, Dufr{\^e}che, Jardat, and
  Turq]{marry2003equilibrium}
V.~Marry, J.-F. Dufr{\^e}che, M.~Jardat and P.~Turq, \emph{Molecular Physics},
  2003, \textbf{101}, 3111--3119\relax
\mciteBstWouldAddEndPuncttrue
\mciteSetBstMidEndSepPunct{\mcitedefaultmidpunct}
{\mcitedefaultendpunct}{\mcitedefaultseppunct}\relax
\EndOfBibitem
\bibitem[Bocquet and Barrat(1994)]{bocquet1994hydrodynamic}
L.~Bocquet and J.-L. Barrat, \emph{Physical review E}, 1994, \textbf{49},
  3079\relax
\mciteBstWouldAddEndPuncttrue
\mciteSetBstMidEndSepPunct{\mcitedefaultmidpunct}
{\mcitedefaultendpunct}{\mcitedefaultseppunct}\relax
\EndOfBibitem
\bibitem[Yoshida \emph{et~al.}(2014)Yoshida, Mizuno, Kinjo, Washizu, and
  Barrat]{yoshida2014generic}
H.~Yoshida, H.~Mizuno, T.~Kinjo, H.~Washizu and J.-L. Barrat, \emph{Physical
  Review E}, 2014, \textbf{90}, 052113\relax
\mciteBstWouldAddEndPuncttrue
\mciteSetBstMidEndSepPunct{\mcitedefaultmidpunct}
{\mcitedefaultendpunct}{\mcitedefaultseppunct}\relax
\EndOfBibitem
\bibitem[Yoshida \emph{et~al.}(2014)Yoshida, Mizuno, Kinjo, Washizu, and
  Barrat]{yoshida2014molecular}
H.~Yoshida, H.~Mizuno, T.~Kinjo, H.~Washizu and J.-L. Barrat, \emph{The Journal
  of chemical physics}, 2014, \textbf{140}, 214701\relax
\mciteBstWouldAddEndPuncttrue
\mciteSetBstMidEndSepPunct{\mcitedefaultmidpunct}
{\mcitedefaultendpunct}{\mcitedefaultseppunct}\relax
\EndOfBibitem
\bibitem[Liu \emph{et~al.}(2017)Liu, Ganti, Burton, Zhang, Wang, and
  Frenkel]{liu2017microscopic}
Y.~Liu, R.~Ganti, H.~G. Burton, X.~Zhang, W.~Wang and D.~Frenkel,
  \emph{Physical review letters}, 2017, \textbf{119}, 224502\relax
\mciteBstWouldAddEndPuncttrue
\mciteSetBstMidEndSepPunct{\mcitedefaultmidpunct}
{\mcitedefaultendpunct}{\mcitedefaultseppunct}\relax
\EndOfBibitem
\bibitem[Liu \emph{et~al.}(2018)Liu, Ganti, and Frenkel]{liu2018pressure}
Y.~Liu, R.~Ganti and D.~Frenkel, \emph{Journal of Physics: Condensed Matter},
  2018, \textbf{30}, 205002\relax
\mciteBstWouldAddEndPuncttrue
\mciteSetBstMidEndSepPunct{\mcitedefaultmidpunct}
{\mcitedefaultendpunct}{\mcitedefaultseppunct}\relax
\EndOfBibitem
\bibitem[Ganti \emph{et~al.}(2017)Ganti, Liu, and Frenkel]{ganti2017molecular}
R.~Ganti, Y.~Liu and D.~Frenkel, \emph{Physical Review Letters}, 2017,
  \textbf{119}, 038002\relax
\mciteBstWouldAddEndPuncttrue
\mciteSetBstMidEndSepPunct{\mcitedefaultmidpunct}
{\mcitedefaultendpunct}{\mcitedefaultseppunct}\relax
\EndOfBibitem
\bibitem[Ganti \emph{et~al.}(2018)Ganti, Liu, and
  Frenkel]{ganti2018hamiltonian}
R.~Ganti, Y.~Liu and D.~Frenkel, \emph{Physical Review Letters}, 2018,
  \textbf{121}, 068002\relax
\mciteBstWouldAddEndPuncttrue
\mciteSetBstMidEndSepPunct{\mcitedefaultmidpunct}
{\mcitedefaultendpunct}{\mcitedefaultseppunct}\relax
\EndOfBibitem
\bibitem[Kiyosawa and Tazawa(1973)]{kiyosawa1973rectification}
K.~Kiyosawa and M.~Tazawa, \emph{Protoplasma}, 1973, \textbf{78},
  203--214\relax
\mciteBstWouldAddEndPuncttrue
\mciteSetBstMidEndSepPunct{\mcitedefaultmidpunct}
{\mcitedefaultendpunct}{\mcitedefaultseppunct}\relax
\EndOfBibitem
\bibitem[Bonthuis and Golestanian(2014)]{bonthuis2014mechanosensitive}
D.~J. Bonthuis and R.~Golestanian, \emph{Physical Review Letters}, 2014,
  \textbf{113}, 148101\relax
\mciteBstWouldAddEndPuncttrue
\mciteSetBstMidEndSepPunct{\mcitedefaultmidpunct}
{\mcitedefaultendpunct}{\mcitedefaultseppunct}\relax
\EndOfBibitem
\bibitem[Esfandiar \emph{et~al.}(2017)Esfandiar, Radha, Wang, Yang, Hu, Garaj,
  Nair, Geim, and Gopinadhan]{esfandiar2017size}
A.~Esfandiar, B.~Radha, F.~Wang, Q.~Yang, S.~Hu, S.~Garaj, R.~Nair, A.~Geim and
  K.~Gopinadhan, \emph{Science}, 2017, \textbf{358}, 511--513\relax
\mciteBstWouldAddEndPuncttrue
\mciteSetBstMidEndSepPunct{\mcitedefaultmidpunct}
{\mcitedefaultendpunct}{\mcitedefaultseppunct}\relax
\EndOfBibitem
\bibitem[Choi \emph{et~al.}(2013)Choi, Ulissi, Shimizu, Bellisario, Ellison,
  and Strano]{choi2013diameter}
W.~Choi, Z.~W. Ulissi, S.~F. Shimizu, D.~O. Bellisario, M.~D. Ellison and M.~S.
  Strano, \emph{Nature communications}, 2013, \textbf{4}, 2397\relax
\mciteBstWouldAddEndPuncttrue
\mciteSetBstMidEndSepPunct{\mcitedefaultmidpunct}
{\mcitedefaultendpunct}{\mcitedefaultseppunct}\relax
\EndOfBibitem
\bibitem[Secchi \emph{et~al.}(2016)Secchi, Marbach, Nigu{\`e}s, Stein, Siria,
  and Bocquet]{secchi2016massive}
E.~Secchi, S.~Marbach, A.~Nigu{\`e}s, D.~Stein, A.~Siria and L.~Bocquet,
  \emph{Nature}, 2016, \textbf{537}, 210\relax
\mciteBstWouldAddEndPuncttrue
\mciteSetBstMidEndSepPunct{\mcitedefaultmidpunct}
{\mcitedefaultendpunct}{\mcitedefaultseppunct}\relax
\EndOfBibitem
\bibitem[Tunuguntla \emph{et~al.}(2017)Tunuguntla, Henley, Yao, Pham, Wanunu,
  and Noy]{tunuguntla2017enhanced}
R.~H. Tunuguntla, R.~Y. Henley, Y.-C. Yao, T.~A. Pham, M.~Wanunu and A.~Noy,
  \emph{Science}, 2017, \textbf{357}, 792--796\relax
\mciteBstWouldAddEndPuncttrue
\mciteSetBstMidEndSepPunct{\mcitedefaultmidpunct}
{\mcitedefaultendpunct}{\mcitedefaultseppunct}\relax
\EndOfBibitem
\bibitem[Mahmoud \emph{et~al.}(2015)Mahmoud, Mansoor, Mansour, and
  Khraisheh]{mahmoud2015functional}
K.~A. Mahmoud, B.~Mansoor, A.~Mansour and M.~Khraisheh, \emph{Desalination},
  2015, \textbf{356}, 208--225\relax
\mciteBstWouldAddEndPuncttrue
\mciteSetBstMidEndSepPunct{\mcitedefaultmidpunct}
{\mcitedefaultendpunct}{\mcitedefaultseppunct}\relax
\EndOfBibitem
\bibitem[Nair \emph{et~al.}(2012)Nair, Wu, Jayaram, Grigorieva, and
  Geim]{nair2012unimpeded}
R.~Nair, H.~Wu, P.~Jayaram, I.~Grigorieva and A.~Geim, \emph{Science}, 2012,
  \textbf{335}, 442--444\relax
\mciteBstWouldAddEndPuncttrue
\mciteSetBstMidEndSepPunct{\mcitedefaultmidpunct}
{\mcitedefaultendpunct}{\mcitedefaultseppunct}\relax
\EndOfBibitem
\bibitem[Radha \emph{et~al.}(2016)Radha, Esfandiar, Wang, Rooney, Gopinadhan,
  Keerthi, Mishchenko, Janardanan, Blake,
  Fumagalli,\emph{et~al.}]{radha2016molecular}
B.~Radha, A.~Esfandiar, F.~Wang, A.~Rooney, K.~Gopinadhan, A.~Keerthi,
  A.~Mishchenko, A.~Janardanan, P.~Blake, L.~Fumagalli \emph{et~al.},
  \emph{Nature}, 2016, \textbf{538}, 222\relax
\mciteBstWouldAddEndPuncttrue
\mciteSetBstMidEndSepPunct{\mcitedefaultmidpunct}
{\mcitedefaultendpunct}{\mcitedefaultseppunct}\relax
\EndOfBibitem
\bibitem[Fornasiero \emph{et~al.}(2008)Fornasiero, Park, Holt, Stadermann,
  Grigoropoulos, Noy, and Bakajin]{fornasiero2008ion}
F.~Fornasiero, H.~G. Park, J.~K. Holt, M.~Stadermann, C.~P. Grigoropoulos,
  A.~Noy and O.~Bakajin, \emph{Proceedings of the National Academy of
  Sciences}, 2008, \textbf{105}, 17250--17255\relax
\mciteBstWouldAddEndPuncttrue
\mciteSetBstMidEndSepPunct{\mcitedefaultmidpunct}
{\mcitedefaultendpunct}{\mcitedefaultseppunct}\relax
\EndOfBibitem
\bibitem[Feng \emph{et~al.}(2016)Feng, Liu, Graf, Dumcenco, Kis, Di~Ventra, and
  Radenovic]{feng2016observation}
J.~Feng, K.~Liu, M.~Graf, D.~Dumcenco, A.~Kis, M.~Di~Ventra and A.~Radenovic,
  \emph{Nature materials}, 2016, \textbf{15}, 850\relax
\mciteBstWouldAddEndPuncttrue
\mciteSetBstMidEndSepPunct{\mcitedefaultmidpunct}
{\mcitedefaultendpunct}{\mcitedefaultseppunct}\relax
\EndOfBibitem
\bibitem[Majumder \emph{et~al.}(2005)Majumder, Chopra, Andrews, and
  Hinds]{majumder2005nanoscale}
M.~Majumder, N.~Chopra, R.~Andrews and B.~J. Hinds, \emph{Nature}, 2005,
  \textbf{438}, 44\relax
\mciteBstWouldAddEndPuncttrue
\mciteSetBstMidEndSepPunct{\mcitedefaultmidpunct}
{\mcitedefaultendpunct}{\mcitedefaultseppunct}\relax
\EndOfBibitem
\bibitem[Holt \emph{et~al.}(2006)Holt, Park, Wang, Stadermann, Artyukhin,
  Grigoropoulos, Noy, and Bakajin]{holt2006fast}
J.~K. Holt, H.~G. Park, Y.~Wang, M.~Stadermann, A.~B. Artyukhin, C.~P.
  Grigoropoulos, A.~Noy and O.~Bakajin, \emph{Science}, 2006, \textbf{312},
  1034--1037\relax
\mciteBstWouldAddEndPuncttrue
\mciteSetBstMidEndSepPunct{\mcitedefaultmidpunct}
{\mcitedefaultendpunct}{\mcitedefaultseppunct}\relax
\EndOfBibitem
\bibitem[Whitby \emph{et~al.}(2008)Whitby, Cagnon, Thanou, and
  Quirke]{whitby2008enhanced}
M.~Whitby, L.~Cagnon, M.~Thanou and N.~Quirke, \emph{Nano letters}, 2008,
  \textbf{8}, 2632--2637\relax
\mciteBstWouldAddEndPuncttrue
\mciteSetBstMidEndSepPunct{\mcitedefaultmidpunct}
{\mcitedefaultendpunct}{\mcitedefaultseppunct}\relax
\EndOfBibitem
\bibitem[Lee \emph{et~al.}(2014)Lee, Laoui, and Karnik]{lee2014nanofluidic}
J.~Lee, T.~Laoui and R.~Karnik, \emph{Nature nanotechnology}, 2014, \textbf{9},
  317\relax
\mciteBstWouldAddEndPuncttrue
\mciteSetBstMidEndSepPunct{\mcitedefaultmidpunct}
{\mcitedefaultendpunct}{\mcitedefaultseppunct}\relax
\EndOfBibitem
\bibitem[Langecker \emph{et~al.}(2012)Langecker, Arnaut, Martin, List, Renner,
  Mayer, Dietz, and Simmel]{langecker2012synthetic}
M.~Langecker, V.~Arnaut, T.~G. Martin, J.~List, S.~Renner, M.~Mayer, H.~Dietz
  and F.~C. Simmel, \emph{Science}, 2012, \textbf{338}, 932--936\relax
\mciteBstWouldAddEndPuncttrue
\mciteSetBstMidEndSepPunct{\mcitedefaultmidpunct}
{\mcitedefaultendpunct}{\mcitedefaultseppunct}\relax
\EndOfBibitem
\bibitem[Joshi \emph{et~al.}(2014)Joshi, Carbone, Wang, Kravets, Su,
  Grigorieva, Wu, Geim, and Nair]{joshi2014precise}
R.~Joshi, P.~Carbone, F.~C. Wang, V.~G. Kravets, Y.~Su, I.~V. Grigorieva,
  H.~Wu, A.~K. Geim and R.~R. Nair, \emph{science}, 2014, \textbf{343},
  752--754\relax
\mciteBstWouldAddEndPuncttrue
\mciteSetBstMidEndSepPunct{\mcitedefaultmidpunct}
{\mcitedefaultendpunct}{\mcitedefaultseppunct}\relax
\EndOfBibitem
\bibitem[Majumder \emph{et~al.}(2017)Majumder, Siria, and
  Bocquet]{majumder2017flows}
M.~Majumder, A.~Siria and L.~Bocquet, \emph{MRS Bulletin}, 2017, \textbf{42},
  278--282\relax
\mciteBstWouldAddEndPuncttrue
\mciteSetBstMidEndSepPunct{\mcitedefaultmidpunct}
{\mcitedefaultendpunct}{\mcitedefaultseppunct}\relax
\EndOfBibitem
\bibitem[Karnik \emph{et~al.}(2007)Karnik, Duan, Castelino, Daiguji, and
  Majumdar]{karnik2007rectification}
R.~Karnik, C.~Duan, K.~Castelino, H.~Daiguji and A.~Majumdar, \emph{Nano
  letters}, 2007, \textbf{7}, 547--551\relax
\mciteBstWouldAddEndPuncttrue
\mciteSetBstMidEndSepPunct{\mcitedefaultmidpunct}
{\mcitedefaultendpunct}{\mcitedefaultseppunct}\relax
\EndOfBibitem
\bibitem[Poggioli \emph{et~al.}(2019)Poggioli, Siria, and
  Bocquet]{Poggioli2019}
A.~Poggioli, A.~Siria and L.~Bocquet, \emph{The journal of physical chemistry
  B}, 2019, \textbf{123}, 1171--1185\relax
\mciteBstWouldAddEndPuncttrue
\mciteSetBstMidEndSepPunct{\mcitedefaultmidpunct}
{\mcitedefaultendpunct}{\mcitedefaultseppunct}\relax
\EndOfBibitem
\bibitem[Farmer and Macey(1970)]{farmer1970perturbation}
R.~E. Farmer and R.~I. Macey, \emph{Biochimica et Biophysica Acta
  (BBA)-Biomembranes}, 1970, \textbf{196}, 53--65\relax
\mciteBstWouldAddEndPuncttrue
\mciteSetBstMidEndSepPunct{\mcitedefaultmidpunct}
{\mcitedefaultendpunct}{\mcitedefaultseppunct}\relax
\EndOfBibitem
\bibitem[Toupin \emph{et~al.}(1989)Toupin, Le~Maguer, and
  McGann]{toupin1989permeability}
C.~Toupin, M.~Le~Maguer and L.~McGann, \emph{Cryobiology}, 1989, \textbf{26},
  431--444\relax
\mciteBstWouldAddEndPuncttrue
\mciteSetBstMidEndSepPunct{\mcitedefaultmidpunct}
{\mcitedefaultendpunct}{\mcitedefaultseppunct}\relax
\EndOfBibitem
\bibitem[Peckys \emph{et~al.}(2011)Peckys, Kleinhans, and
  Mazur]{peckys2011rectification}
D.~B. Peckys, F.~Kleinhans and P.~Mazur, \emph{PloS one}, 2011, \textbf{6},
  e23643\relax
\mciteBstWouldAddEndPuncttrue
\mciteSetBstMidEndSepPunct{\mcitedefaultmidpunct}
{\mcitedefaultendpunct}{\mcitedefaultseppunct}\relax
\EndOfBibitem
\bibitem[Marbach and Bocquet(2017)]{marbach2017active}
S.~Marbach and L.~Bocquet, \emph{The Journal of Chemical Physics}, 2017,
  \textbf{147}, 154701\relax
\mciteBstWouldAddEndPuncttrue
\mciteSetBstMidEndSepPunct{\mcitedefaultmidpunct}
{\mcitedefaultendpunct}{\mcitedefaultseppunct}\relax
\EndOfBibitem
\bibitem[Marbach \emph{et~al.}(2018)Marbach, Dean, and
  Bocquet]{marbach2018transport}
S.~Marbach, D.~S. Dean and L.~Bocquet, \emph{Nature Physics}, 2018,
  \textbf{14}, 1108--1113\relax
\mciteBstWouldAddEndPuncttrue
\mciteSetBstMidEndSepPunct{\mcitedefaultmidpunct}
{\mcitedefaultendpunct}{\mcitedefaultseppunct}\relax
\EndOfBibitem
\bibitem[Sakiyama \emph{et~al.}(2016)Sakiyama, Mazur, Kapinos, and
  Lim]{sakiyama2016spatiotemporal}
Y.~Sakiyama, A.~Mazur, L.~E. Kapinos and R.~Y. Lim, \emph{Nature
  nanotechnology}, 2016, \textbf{11}, 719\relax
\mciteBstWouldAddEndPuncttrue
\mciteSetBstMidEndSepPunct{\mcitedefaultmidpunct}
{\mcitedefaultendpunct}{\mcitedefaultseppunct}\relax
\EndOfBibitem
\bibitem[Noskov \emph{et~al.}(2004)Noskov, Berneche, and
  Roux]{noskov2004control}
S.~Y. Noskov, S.~Berneche and B.~Roux, \emph{Nature}, 2004, \textbf{431},
  830\relax
\mciteBstWouldAddEndPuncttrue
\mciteSetBstMidEndSepPunct{\mcitedefaultmidpunct}
{\mcitedefaultendpunct}{\mcitedefaultseppunct}\relax
\EndOfBibitem
\bibitem[Eisenman and Horn(1983)]{eisenman1983ionic}
G.~Eisenman and R.~Horn, \emph{The Journal of membrane biology}, 1983,
  \textbf{76}, 197--225\relax
\mciteBstWouldAddEndPuncttrue
\mciteSetBstMidEndSepPunct{\mcitedefaultmidpunct}
{\mcitedefaultendpunct}{\mcitedefaultseppunct}\relax
\EndOfBibitem
\bibitem[L{\"a}uger \emph{et~al.}(1980)L{\"a}uger, Stephan, and
  Frehland]{lauger1980fluctuations}
P.~L{\"a}uger, W.~Stephan and E.~Frehland, \emph{Biochimica et Biophysica Acta
  (BBA)-Biomembranes}, 1980, \textbf{602}, 167--180\relax
\mciteBstWouldAddEndPuncttrue
\mciteSetBstMidEndSepPunct{\mcitedefaultmidpunct}
{\mcitedefaultendpunct}{\mcitedefaultseppunct}\relax
\EndOfBibitem
\bibitem[Schr{\"o}der(1983)]{schroder1983rate}
H.~Schr{\"o}der, \emph{The Journal of chemical physics}, 1983, \textbf{79},
  1997--2005\relax
\mciteBstWouldAddEndPuncttrue
\mciteSetBstMidEndSepPunct{\mcitedefaultmidpunct}
{\mcitedefaultendpunct}{\mcitedefaultseppunct}\relax
\EndOfBibitem
\bibitem[Gammaitoni \emph{et~al.}(1998)Gammaitoni, H{\"a}nggi, Jung, and
  Marchesoni]{gammaitoni1998stochastic}
L.~Gammaitoni, P.~H{\"a}nggi, P.~Jung and F.~Marchesoni, \emph{Reviews of
  modern physics}, 1998, \textbf{70}, 223\relax
\mciteBstWouldAddEndPuncttrue
\mciteSetBstMidEndSepPunct{\mcitedefaultmidpunct}
{\mcitedefaultendpunct}{\mcitedefaultseppunct}\relax
\EndOfBibitem
\bibitem[Reimann and H{\"a}nggi(2002)]{reimann2002introduction}
P.~Reimann and P.~H{\"a}nggi, \emph{Applied Physics A}, 2002, \textbf{75},
  169--178\relax
\mciteBstWouldAddEndPuncttrue
\mciteSetBstMidEndSepPunct{\mcitedefaultmidpunct}
{\mcitedefaultendpunct}{\mcitedefaultseppunct}\relax
\EndOfBibitem
\bibitem[Rousselet \emph{et~al.}(1994)Rousselet, Salome, Ajdari, and
  Prostt]{rousselet1994directional}
J.~Rousselet, L.~Salome, A.~Ajdari and J.~Prostt, \emph{Nature}, 1994,
  \textbf{370}, 446\relax
\mciteBstWouldAddEndPuncttrue
\mciteSetBstMidEndSepPunct{\mcitedefaultmidpunct}
{\mcitedefaultendpunct}{\mcitedefaultseppunct}\relax
\EndOfBibitem
\bibitem[Marbach \emph{et~al.}(2019)Marbach, Kavokine, and
  Bocquet]{Marbach2019}
S.~Marbach, N.~Kavokine and L.~Bocquet, \emph{submitted to J. Chem. Phys.},
  2019\relax
\mciteBstWouldAddEndPuncttrue
\mciteSetBstMidEndSepPunct{\mcitedefaultmidpunct}
{\mcitedefaultendpunct}{\mcitedefaultseppunct}\relax
\EndOfBibitem
\bibitem[Karnik \emph{et~al.}(2005)Karnik, Fan, Yue, Li, Yang, and
  Majumdar]{karnik2005electrostatic}
R.~Karnik, R.~Fan, M.~Yue, D.~Li, P.~Yang and A.~Majumdar, \emph{Nano letters},
  2005, \textbf{5}, 943--948\relax
\mciteBstWouldAddEndPuncttrue
\mciteSetBstMidEndSepPunct{\mcitedefaultmidpunct}
{\mcitedefaultendpunct}{\mcitedefaultseppunct}\relax
\EndOfBibitem
\bibitem[Kalman \emph{et~al.}(2009)Kalman, Sudre, Vlassiouk, and
  Siwy]{kalman2009control}
E.~B. Kalman, O.~Sudre, I.~Vlassiouk and Z.~S. Siwy, \emph{Analytical and
  bioanalytical chemistry}, 2009, \textbf{394}, 413--419\relax
\mciteBstWouldAddEndPuncttrue
\mciteSetBstMidEndSepPunct{\mcitedefaultmidpunct}
{\mcitedefaultendpunct}{\mcitedefaultseppunct}\relax
\EndOfBibitem
\bibitem[Guan \emph{et~al.}(2011)Guan, Fan, and Reed]{guan2011field}
W.~Guan, R.~Fan and M.~A. Reed, \emph{Nature communications}, 2011, \textbf{2},
  506\relax
\mciteBstWouldAddEndPuncttrue
\mciteSetBstMidEndSepPunct{\mcitedefaultmidpunct}
{\mcitedefaultendpunct}{\mcitedefaultseppunct}\relax
\EndOfBibitem
\bibitem[Zhou \emph{et~al.}(2018)Zhou, Vasu, Cherian, Neek-Amal, Zhang,
  Ghorbanfekr-Kalashami, Huang, Marshall, Kravets,
  Abraham,\emph{et~al.}]{zhou2018electrically}
K.-G. Zhou, K.~Vasu, C.~Cherian, M.~Neek-Amal, J.~C. Zhang,
  H.~Ghorbanfekr-Kalashami, K.~Huang, O.~Marshall, V.~Kravets, J.~Abraham
  \emph{et~al.}, \emph{arXiv preprint arXiv:1805.06390}, 2018\relax
\mciteBstWouldAddEndPuncttrue
\mciteSetBstMidEndSepPunct{\mcitedefaultmidpunct}
{\mcitedefaultendpunct}{\mcitedefaultseppunct}\relax
\EndOfBibitem
\bibitem[Moorthy \emph{et~al.}(2001)Moorthy, Khoury, Moore, and
  Beebe]{moorthy2001active}
J.~Moorthy, C.~Khoury, J.~S. Moore and D.~J. Beebe, \emph{Sensors and Actuators
  B: Chemical}, 2001, \textbf{75}, 223--229\relax
\mciteBstWouldAddEndPuncttrue
\mciteSetBstMidEndSepPunct{\mcitedefaultmidpunct}
{\mcitedefaultendpunct}{\mcitedefaultseppunct}\relax
\EndOfBibitem
\bibitem[Zhang \emph{et~al.}(2013)Zhang, Hou, Zeng, Yang, Li, Yan, Tian, and
  Jiang]{zhang2013bioinspired}
H.~Zhang, X.~Hou, L.~Zeng, F.~Yang, L.~Li, D.~Yan, Y.~Tian and L.~Jiang,
  \emph{Journal of the American Chemical Society}, 2013, \textbf{135},
  16102--16110\relax
\mciteBstWouldAddEndPuncttrue
\mciteSetBstMidEndSepPunct{\mcitedefaultmidpunct}
{\mcitedefaultendpunct}{\mcitedefaultseppunct}\relax
\EndOfBibitem
\bibitem[Liu \emph{et~al.}(2018)Liu, Xie, Li, Zhang, Yang, Zhao, Zhu, Kong,
  Jiang, and Wen]{liu2018universal}
P.~Liu, G.~Xie, P.~Li, Z.~Zhang, L.~Yang, Y.~Zhao, C.~Zhu, X.-Y. Kong, L.~Jiang
  and L.~Wen, \emph{NPG Asia Materials}, 2018, \textbf{10}, 849--857\relax
\mciteBstWouldAddEndPuncttrue
\mciteSetBstMidEndSepPunct{\mcitedefaultmidpunct}
{\mcitedefaultendpunct}{\mcitedefaultseppunct}\relax
\EndOfBibitem
\bibitem[Liu \emph{et~al.}(2017)Liu, Wang, Yu, Karton, Li, Zhang, Guo, Hou,
  Cheng, Jiang,\emph{et~al.}]{liu2017bioinspired}
J.~Liu, N.~Wang, L.-J. Yu, A.~Karton, W.~Li, W.~Zhang, F.~Guo, L.~Hou,
  Q.~Cheng, L.~Jiang \emph{et~al.}, \emph{Nature communications}, 2017,
  \textbf{8}, 2011\relax
\mciteBstWouldAddEndPuncttrue
\mciteSetBstMidEndSepPunct{\mcitedefaultmidpunct}
{\mcitedefaultendpunct}{\mcitedefaultseppunct}\relax
\EndOfBibitem
\bibitem[Hu \emph{et~al.}(2015)Hu, Gao, Ding, Wang, Jiang, Jin, and
  Jiang]{hu2015photothermal}
L.~Hu, S.~Gao, X.~Ding, D.~Wang, J.~Jiang, J.~Jin and L.~Jiang, \emph{ACS
  nano}, 2015, \textbf{9}, 4835--4842\relax
\mciteBstWouldAddEndPuncttrue
\mciteSetBstMidEndSepPunct{\mcitedefaultmidpunct}
{\mcitedefaultendpunct}{\mcitedefaultseppunct}\relax
\EndOfBibitem
\bibitem[Prieve and Roman(1987)]{prieve1987diffusiophoresis}
D.~C. Prieve and R.~Roman, \emph{Journal of the Chemical Society, Faraday
  Transactions 2: Molecular and Chemical Physics}, 1987, \textbf{83},
  1287--1306\relax
\mciteBstWouldAddEndPuncttrue
\mciteSetBstMidEndSepPunct{\mcitedefaultmidpunct}
{\mcitedefaultendpunct}{\mcitedefaultseppunct}\relax
\EndOfBibitem
\bibitem[Ebel \emph{et~al.}(1988)Ebel, Anderson, and
  Prieve]{ebel1988diffusiophoresis}
J.~Ebel, J.~L. Anderson and D.~Prieve, \emph{Langmuir}, 1988, \textbf{4},
  396--406\relax
\mciteBstWouldAddEndPuncttrue
\mciteSetBstMidEndSepPunct{\mcitedefaultmidpunct}
{\mcitedefaultendpunct}{\mcitedefaultseppunct}\relax
\EndOfBibitem
\bibitem[Velegol \emph{et~al.}(2016)Velegol, Garg, Guha, Kar, and
  Kumar]{velegol2016origins}
D.~Velegol, A.~Garg, R.~Guha, A.~Kar and M.~Kumar, \emph{Soft Matter}, 2016,
  \textbf{12}, 4686--4703\relax
\mciteBstWouldAddEndPuncttrue
\mciteSetBstMidEndSepPunct{\mcitedefaultmidpunct}
{\mcitedefaultendpunct}{\mcitedefaultseppunct}\relax
\EndOfBibitem
\bibitem[M{\"o}ller \emph{et~al.}(2017)M{\"o}ller, Kriegel, Kie{\ss}, Sojo, and
  Braun]{moller2017steep}
F.~M. M{\"o}ller, F.~Kriegel, M.~Kie{\ss}, V.~Sojo and D.~Braun,
  \emph{Angewandte Chemie International Edition}, 2017, \textbf{56},
  2340--2344\relax
\mciteBstWouldAddEndPuncttrue
\mciteSetBstMidEndSepPunct{\mcitedefaultmidpunct}
{\mcitedefaultendpunct}{\mcitedefaultseppunct}\relax
\EndOfBibitem
\bibitem[Banerjee \emph{et~al.}(2016)Banerjee, Williams, Azevedo, Helgeson, and
  Squires]{banerjee2016soluto}
A.~Banerjee, I.~Williams, R.~N. Azevedo, M.~E. Helgeson and T.~M. Squires,
  \emph{Proceedings of the National Academy of Sciences}, 2016, \textbf{113},
  8612--8617\relax
\mciteBstWouldAddEndPuncttrue
\mciteSetBstMidEndSepPunct{\mcitedefaultmidpunct}
{\mcitedefaultendpunct}{\mcitedefaultseppunct}\relax
\EndOfBibitem
\bibitem[Morrison~Jr(1970)]{morrison1970electrophoresis}
F.~Morrison~Jr, \emph{Journal of Colloid and Interface Science}, 1970,
  \textbf{34}, 210--214\relax
\mciteBstWouldAddEndPuncttrue
\mciteSetBstMidEndSepPunct{\mcitedefaultmidpunct}
{\mcitedefaultendpunct}{\mcitedefaultseppunct}\relax
\EndOfBibitem
\bibitem[Rallabandi \emph{et~al.}(2019)Rallabandi, Yang, and
  Stone]{rallabandi2019motion}
B.~Rallabandi, F.~Yang and H.~A. Stone, \emph{arXiv preprint arXiv:1901.04311},
  2019\relax
\mciteBstWouldAddEndPuncttrue
\mciteSetBstMidEndSepPunct{\mcitedefaultmidpunct}
{\mcitedefaultendpunct}{\mcitedefaultseppunct}\relax
\EndOfBibitem
\bibitem[Chamolly \emph{et~al.}(2017)Chamolly, Ishikawa, and
  Lauga]{chamolly2017active}
A.~Chamolly, T.~Ishikawa and E.~Lauga, \emph{New Journal of Physics}, 2017,
  \textbf{19}, 115001\relax
\mciteBstWouldAddEndPuncttrue
\mciteSetBstMidEndSepPunct{\mcitedefaultmidpunct}
{\mcitedefaultendpunct}{\mcitedefaultseppunct}\relax
\EndOfBibitem
\bibitem[C{\'o}rdova-Figueroa and Brady(2008)]{cordova2008osmotic}
U.~M. C{\'o}rdova-Figueroa and J.~F. Brady, \emph{Physical Review Letters},
  2008, \textbf{100}, 158303\relax
\mciteBstWouldAddEndPuncttrue
\mciteSetBstMidEndSepPunct{\mcitedefaultmidpunct}
{\mcitedefaultendpunct}{\mcitedefaultseppunct}\relax
\EndOfBibitem
\bibitem[J{\"u}licher and Prost(2009)]{julicher2009comment}
F.~J{\"u}licher and J.~Prost, \emph{Physical Review Letters}, 2009,
  \textbf{103}, 079801\relax
\mciteBstWouldAddEndPuncttrue
\mciteSetBstMidEndSepPunct{\mcitedefaultmidpunct}
{\mcitedefaultendpunct}{\mcitedefaultseppunct}\relax
\EndOfBibitem
\bibitem[Fischer and Dhar(2009)]{fischer2009comment}
T.~M. Fischer and P.~Dhar, \emph{Physical Review Letters}, 2009, \textbf{102},
  159801\relax
\mciteBstWouldAddEndPuncttrue
\mciteSetBstMidEndSepPunct{\mcitedefaultmidpunct}
{\mcitedefaultendpunct}{\mcitedefaultseppunct}\relax
\EndOfBibitem
\bibitem[C\'ordova-Figueroa and Brady(2009)]{PhysRevLett.103.079802}
U.~M. C\'ordova-Figueroa and J.~F. Brady, \emph{Phys. Rev. Lett.}, 2009,
  \textbf{103}, 079802\relax
\mciteBstWouldAddEndPuncttrue
\mciteSetBstMidEndSepPunct{\mcitedefaultmidpunct}
{\mcitedefaultendpunct}{\mcitedefaultseppunct}\relax
\EndOfBibitem
\bibitem[C\'ordova-Figueroa and Brady(2009)]{PhysRevLett.102.159802}
U.~M. C\'ordova-Figueroa and J.~F. Brady, \emph{Phys. Rev. Lett.}, 2009,
  \textbf{102}, 159802\relax
\mciteBstWouldAddEndPuncttrue
\mciteSetBstMidEndSepPunct{\mcitedefaultmidpunct}
{\mcitedefaultendpunct}{\mcitedefaultseppunct}\relax
\EndOfBibitem
\bibitem[Brady(2011)]{brady2011particle}
J.~F. Brady, \emph{Journal of Fluid Mechanics}, 2011, \textbf{667},
  216--259\relax
\mciteBstWouldAddEndPuncttrue
\mciteSetBstMidEndSepPunct{\mcitedefaultmidpunct}
{\mcitedefaultendpunct}{\mcitedefaultseppunct}\relax
\EndOfBibitem
\bibitem[Moran and Posner(2017)]{moran2017phoretic}
J.~L. Moran and J.~D. Posner, \emph{Annual Review of Fluid Mechanics}, 2017,
  \textbf{49}, 511--540\relax
\mciteBstWouldAddEndPuncttrue
\mciteSetBstMidEndSepPunct{\mcitedefaultmidpunct}
{\mcitedefaultendpunct}{\mcitedefaultseppunct}\relax
\EndOfBibitem
\bibitem[Sabass and Seifert(2012)]{sabass2012dynamics}
B.~Sabass and U.~Seifert, \emph{The Journal of chemical physics}, 2012,
  \textbf{136}, 064508\relax
\mciteBstWouldAddEndPuncttrue
\mciteSetBstMidEndSepPunct{\mcitedefaultmidpunct}
{\mcitedefaultendpunct}{\mcitedefaultseppunct}\relax
\EndOfBibitem
\bibitem[Sharifi-Mood \emph{et~al.}(2013)Sharifi-Mood, Koplik, and
  Maldarelli]{sharifi2013diffusiophoretic}
N.~Sharifi-Mood, J.~Koplik and C.~Maldarelli, \emph{Physics of Fluids}, 2013,
  \textbf{25}, 012001\relax
\mciteBstWouldAddEndPuncttrue
\mciteSetBstMidEndSepPunct{\mcitedefaultmidpunct}
{\mcitedefaultendpunct}{\mcitedefaultseppunct}\relax
\EndOfBibitem
\bibitem[C{\'o}rdova-Figueroa \emph{et~al.}(2013)C{\'o}rdova-Figueroa, Brady,
  and Shklyaev]{cordova2013osmotic}
U.~C{\'o}rdova-Figueroa, J.~Brady and S.~Shklyaev, \emph{Soft Matter}, 2013,
  \textbf{9}, 6382--6390\relax
\mciteBstWouldAddEndPuncttrue
\mciteSetBstMidEndSepPunct{\mcitedefaultmidpunct}
{\mcitedefaultendpunct}{\mcitedefaultseppunct}\relax
\EndOfBibitem
\bibitem[Ohshima \emph{et~al.}(1983)Ohshima, Healy, and
  White]{ohshima1983approximate}
H.~Ohshima, T.~W. Healy and L.~R. White, \emph{Journal of the Chemical Society,
  Faraday Transactions 2: Molecular and Chemical Physics}, 1983, \textbf{79},
  1613--1628\relax
\mciteBstWouldAddEndPuncttrue
\mciteSetBstMidEndSepPunct{\mcitedefaultmidpunct}
{\mcitedefaultendpunct}{\mcitedefaultseppunct}\relax
\EndOfBibitem
\bibitem[Long \emph{et~al.}(1996)Long, Viovy, and Ajdari]{long1996simultaneous}
D.~Long, J.-L. Viovy and A.~Ajdari, \emph{Physical Review Letters}, 1996,
  \textbf{76}, 3858\relax
\mciteBstWouldAddEndPuncttrue
\mciteSetBstMidEndSepPunct{\mcitedefaultmidpunct}
{\mcitedefaultendpunct}{\mcitedefaultseppunct}\relax
\EndOfBibitem
\bibitem[Jiang and Sano(2007)]{jiang2007stretching}
H.-R. Jiang and M.~Sano, \emph{Applied Physics Letters}, 2007, \textbf{91},
  154104\relax
\mciteBstWouldAddEndPuncttrue
\mciteSetBstMidEndSepPunct{\mcitedefaultmidpunct}
{\mcitedefaultendpunct}{\mcitedefaultseppunct}\relax
\EndOfBibitem
\bibitem[Michelin and Lauga(2014)]{michelin2014phoretic}
S.~Michelin and E.~Lauga, \emph{Journal of Fluid Mechanics}, 2014,
  \textbf{747}, 572--604\relax
\mciteBstWouldAddEndPuncttrue
\mciteSetBstMidEndSepPunct{\mcitedefaultmidpunct}
{\mcitedefaultendpunct}{\mcitedefaultseppunct}\relax
\EndOfBibitem
\bibitem[Ab{\'e}cassis \emph{et~al.}(2008)Ab{\'e}cassis, Cottin-Bizonne, Ybert,
  Ajdari, and Bocquet]{abecassis2008boosting}
B.~Ab{\'e}cassis, C.~Cottin-Bizonne, C.~Ybert, A.~Ajdari and L.~Bocquet,
  \emph{Nature materials}, 2008, \textbf{7}, 785\relax
\mciteBstWouldAddEndPuncttrue
\mciteSetBstMidEndSepPunct{\mcitedefaultmidpunct}
{\mcitedefaultendpunct}{\mcitedefaultseppunct}\relax
\EndOfBibitem
\bibitem[Kalinin \emph{et~al.}(2009)Kalinin, Jiang, Tu, and
  Wu]{kalinin2009logarithmic}
Y.~V. Kalinin, L.~Jiang, Y.~Tu and M.~Wu, \emph{Biophysical journal}, 2009,
  \textbf{96}, 2439--2448\relax
\mciteBstWouldAddEndPuncttrue
\mciteSetBstMidEndSepPunct{\mcitedefaultmidpunct}
{\mcitedefaultendpunct}{\mcitedefaultseppunct}\relax
\EndOfBibitem
\bibitem[Sear and Warren(2017)]{sear2017diffusiophoresis}
R.~P. Sear and P.~B. Warren, \emph{Physical Review E}, 2017, \textbf{96},
  062602\relax
\mciteBstWouldAddEndPuncttrue
\mciteSetBstMidEndSepPunct{\mcitedefaultmidpunct}
{\mcitedefaultendpunct}{\mcitedefaultseppunct}\relax
\EndOfBibitem
\bibitem[Paustian \emph{et~al.}(2013)Paustian, Azevedo, Lundin, Gilkey, and
  Squires]{paustian2013microfluidic}
J.~S. Paustian, R.~N. Azevedo, S.-T.~B. Lundin, M.~J. Gilkey and T.~M. Squires,
  \emph{Physical Review X}, 2013, \textbf{3}, 041010\relax
\mciteBstWouldAddEndPuncttrue
\mciteSetBstMidEndSepPunct{\mcitedefaultmidpunct}
{\mcitedefaultendpunct}{\mcitedefaultseppunct}\relax
\EndOfBibitem
\bibitem[Nery-Azevedo \emph{et~al.}(2017)Nery-Azevedo, Banerjee, and
  Squires]{nery2017diffusiophoresis}
R.~Nery-Azevedo, A.~Banerjee and T.~M. Squires, \emph{Langmuir}, 2017,
  \textbf{33}, 9694--9702\relax
\mciteBstWouldAddEndPuncttrue
\mciteSetBstMidEndSepPunct{\mcitedefaultmidpunct}
{\mcitedefaultendpunct}{\mcitedefaultseppunct}\relax
\EndOfBibitem
\bibitem[Shin \emph{et~al.}(2017)Shin, Shardt, Warren, and
  Stone]{shin2017membraneless}
S.~Shin, O.~Shardt, P.~B. Warren and H.~A. Stone, \emph{Nature Communications},
  2017, \textbf{8}, 15181\relax
\mciteBstWouldAddEndPuncttrue
\mciteSetBstMidEndSepPunct{\mcitedefaultmidpunct}
{\mcitedefaultendpunct}{\mcitedefaultseppunct}\relax
\EndOfBibitem
\bibitem[Guha \emph{et~al.}(2017)Guha, Mohajerani, Collins, Ghosh, Sen, and
  Velegol]{guha2017chemotaxis}
R.~Guha, F.~Mohajerani, M.~Collins, S.~Ghosh, A.~Sen and D.~Velegol,
  \emph{Journal of the American Chemical Society}, 2017, \textbf{139},
  15588--15591\relax
\mciteBstWouldAddEndPuncttrue
\mciteSetBstMidEndSepPunct{\mcitedefaultmidpunct}
{\mcitedefaultendpunct}{\mcitedefaultseppunct}\relax
\EndOfBibitem
\bibitem[Florea \emph{et~al.}(2014)Florea, Musa, Huyghe, and
  Wyss]{florea2014long}
D.~Florea, S.~Musa, J.~M. Huyghe and H.~M. Wyss, \emph{Proceedings of the
  National Academy of Sciences}, 2014,  201322857\relax
\mciteBstWouldAddEndPuncttrue
\mciteSetBstMidEndSepPunct{\mcitedefaultmidpunct}
{\mcitedefaultendpunct}{\mcitedefaultseppunct}\relax
\EndOfBibitem
\bibitem[Nardi \emph{et~al.}(1999)Nardi, Bruinsma, and
  Sackmann]{nardi1999vesicles}
J.~Nardi, R.~Bruinsma and E.~Sackmann, \emph{Physical Review Letters}, 1999,
  \textbf{82}, 5168\relax
\mciteBstWouldAddEndPuncttrue
\mciteSetBstMidEndSepPunct{\mcitedefaultmidpunct}
{\mcitedefaultendpunct}{\mcitedefaultseppunct}\relax
\EndOfBibitem
\bibitem[Kar \emph{et~al.}(2015)Kar, Chiang, Ortiz~Rivera, Sen, and
  Velegol]{kar2015enhanced}
A.~Kar, T.-Y. Chiang, I.~Ortiz~Rivera, A.~Sen and D.~Velegol, \emph{ACS nano},
  2015, \textbf{9}, 746--753\relax
\mciteBstWouldAddEndPuncttrue
\mciteSetBstMidEndSepPunct{\mcitedefaultmidpunct}
{\mcitedefaultendpunct}{\mcitedefaultseppunct}\relax
\EndOfBibitem
\bibitem[Shin \emph{et~al.}(2016)Shin, Um, Sabass, Ault, Rahimi, Warren, and
  Stone]{shin2016size}
S.~Shin, E.~Um, B.~Sabass, J.~T. Ault, M.~Rahimi, P.~B. Warren and H.~A. Stone,
  \emph{Proceedings of the National Academy of Sciences}, 2016, \textbf{113},
  257--261\relax
\mciteBstWouldAddEndPuncttrue
\mciteSetBstMidEndSepPunct{\mcitedefaultmidpunct}
{\mcitedefaultendpunct}{\mcitedefaultseppunct}\relax
\EndOfBibitem
\bibitem[Shin \emph{et~al.}(2017)Shin, Ault, Warren, and
  Stone]{shin2017accumulation}
S.~Shin, J.~T. Ault, P.~B. Warren and H.~A. Stone, \emph{Physical Review X},
  2017, \textbf{7}, 041038\relax
\mciteBstWouldAddEndPuncttrue
\mciteSetBstMidEndSepPunct{\mcitedefaultmidpunct}
{\mcitedefaultendpunct}{\mcitedefaultseppunct}\relax
\EndOfBibitem
\bibitem[Shapiro \emph{et~al.}(2009)Shapiro, McAdams, and
  Losick]{shapiro2009and}
L.~Shapiro, H.~H. McAdams and R.~Losick, \emph{Science}, 2009, \textbf{326},
  1225--1228\relax
\mciteBstWouldAddEndPuncttrue
\mciteSetBstMidEndSepPunct{\mcitedefaultmidpunct}
{\mcitedefaultendpunct}{\mcitedefaultseppunct}\relax
\EndOfBibitem
\bibitem[Osawa \emph{et~al.}(2008)Osawa, Anderson, and
  Erickson]{osawa2008reconstitution}
M.~Osawa, D.~E. Anderson and H.~P. Erickson, \emph{Science}, 2008,
  \textbf{320}, 792--794\relax
\mciteBstWouldAddEndPuncttrue
\mciteSetBstMidEndSepPunct{\mcitedefaultmidpunct}
{\mcitedefaultendpunct}{\mcitedefaultseppunct}\relax
\EndOfBibitem
\bibitem[Loose \emph{et~al.}(2008)Loose, Fischer-Friedrich, Ries, Kruse, and
  Schwille]{loose2008spatial}
M.~Loose, E.~Fischer-Friedrich, J.~Ries, K.~Kruse and P.~Schwille,
  \emph{Science}, 2008, \textbf{320}, 789--792\relax
\mciteBstWouldAddEndPuncttrue
\mciteSetBstMidEndSepPunct{\mcitedefaultmidpunct}
{\mcitedefaultendpunct}{\mcitedefaultseppunct}\relax
\EndOfBibitem
\bibitem[Krishnan \emph{et~al.}(2010)Krishnan, Mojarad, Kukura, and
  Sandoghdar]{krishnan2010geometry}
M.~Krishnan, N.~Mojarad, P.~Kukura and V.~Sandoghdar, \emph{Nature}, 2010,
  \textbf{467}, 692\relax
\mciteBstWouldAddEndPuncttrue
\mciteSetBstMidEndSepPunct{\mcitedefaultmidpunct}
{\mcitedefaultendpunct}{\mcitedefaultseppunct}\relax
\EndOfBibitem
\bibitem[Myers \emph{et~al.}(2015)Myers, Celebrano, and
  Krishnan]{myers2015information}
C.~J. Myers, M.~Celebrano and M.~Krishnan, \emph{Nature nanotechnology}, 2015,
  \textbf{10}, 886\relax
\mciteBstWouldAddEndPuncttrue
\mciteSetBstMidEndSepPunct{\mcitedefaultmidpunct}
{\mcitedefaultendpunct}{\mcitedefaultseppunct}\relax
\EndOfBibitem
\bibitem[Agudo-Canalejo \emph{et~al.}(2018)Agudo-Canalejo, Adeleke-Larodo,
  Illien, and Golestanian]{agudo2018enhanced}
J.~Agudo-Canalejo, T.~Adeleke-Larodo, P.~Illien and R.~Golestanian,
  \emph{Accounts of chemical research}, 2018, \textbf{51}, 2365--2372\relax
\mciteBstWouldAddEndPuncttrue
\mciteSetBstMidEndSepPunct{\mcitedefaultmidpunct}
{\mcitedefaultendpunct}{\mcitedefaultseppunct}\relax
\EndOfBibitem
\bibitem[Lutkenhaus(2008)]{lutkenhaus2008tinkering}
J.~Lutkenhaus, \emph{Science}, 2008, \textbf{320}, 755--756\relax
\mciteBstWouldAddEndPuncttrue
\mciteSetBstMidEndSepPunct{\mcitedefaultmidpunct}
{\mcitedefaultendpunct}{\mcitedefaultseppunct}\relax
\EndOfBibitem
\bibitem[Rothfield \emph{et~al.}(2005)Rothfield, Taghbalout, and
  Shih]{rothfield2005spatial}
L.~Rothfield, A.~Taghbalout and Y.-L. Shih, \emph{Nature Reviews Microbiology},
  2005, \textbf{3}, 959\relax
\mciteBstWouldAddEndPuncttrue
\mciteSetBstMidEndSepPunct{\mcitedefaultmidpunct}
{\mcitedefaultendpunct}{\mcitedefaultseppunct}\relax
\EndOfBibitem
\bibitem[Eldar \emph{et~al.}(2002)Eldar, Dorfman, Weiss, Ashe, Shilo, and
  Barkai]{eldar2002robustness}
A.~Eldar, R.~Dorfman, D.~Weiss, H.~Ashe, B.-Z. Shilo and N.~Barkai,
  \emph{Nature}, 2002, \textbf{419}, 304\relax
\mciteBstWouldAddEndPuncttrue
\mciteSetBstMidEndSepPunct{\mcitedefaultmidpunct}
{\mcitedefaultendpunct}{\mcitedefaultseppunct}\relax
\EndOfBibitem
\bibitem[Sear(2019)]{Sear2019}
R.~Sear, 2019, \textbf{ArXiv:1901.00802}, \relax
\mciteBstWouldAddEndPuncttrue
\mciteSetBstMidEndSepPunct{\mcitedefaultmidpunct}
{\mcitedefaultendpunct}{\mcitedefaultseppunct}\relax
\EndOfBibitem
\bibitem[Aubret \emph{et~al.}(2017)Aubret, Ramananarivo, and
  Palacci]{aubret2017eppur}
A.~Aubret, S.~Ramananarivo and J.~Palacci, \emph{Current Opinion in Colloid \&
  Interface Science}, 2017, \textbf{30}, 81--89\relax
\mciteBstWouldAddEndPuncttrue
\mciteSetBstMidEndSepPunct{\mcitedefaultmidpunct}
{\mcitedefaultendpunct}{\mcitedefaultseppunct}\relax
\EndOfBibitem
\bibitem[Illien \emph{et~al.}(2017)Illien, Golestanian, and
  Sen]{illien2017fuelled}
P.~Illien, R.~Golestanian and A.~Sen, \emph{Chemical Society Reviews}, 2017,
  \textbf{46}, 5508--5518\relax
\mciteBstWouldAddEndPuncttrue
\mciteSetBstMidEndSepPunct{\mcitedefaultmidpunct}
{\mcitedefaultendpunct}{\mcitedefaultseppunct}\relax
\EndOfBibitem
\bibitem[Bechinger \emph{et~al.}(2016)Bechinger, Di~Leonardo, L{\"o}wen,
  Reichhardt, Volpe, and Volpe]{bechinger2016active}
C.~Bechinger, R.~Di~Leonardo, H.~L{\"o}wen, C.~Reichhardt, G.~Volpe and
  G.~Volpe, \emph{Reviews of Modern Physics}, 2016, \textbf{88}, 045006\relax
\mciteBstWouldAddEndPuncttrue
\mciteSetBstMidEndSepPunct{\mcitedefaultmidpunct}
{\mcitedefaultendpunct}{\mcitedefaultseppunct}\relax
\EndOfBibitem
\bibitem[Colberg \emph{et~al.}(2014)Colberg, Reigh, Robertson, and
  Kapral]{colberg2014chemistry}
P.~H. Colberg, S.~Y. Reigh, B.~Robertson and R.~Kapral, \emph{Accounts of
  chemical research}, 2014, \textbf{47}, 3504--3511\relax
\mciteBstWouldAddEndPuncttrue
\mciteSetBstMidEndSepPunct{\mcitedefaultmidpunct}
{\mcitedefaultendpunct}{\mcitedefaultseppunct}\relax
\EndOfBibitem
\bibitem[Paxton \emph{et~al.}(2004)Paxton, Kistler, Olmeda, Sen, St.~Angelo,
  Cao, Mallouk, Lammert, and Crespi]{paxton2004catalytic}
W.~F. Paxton, K.~C. Kistler, C.~C. Olmeda, A.~Sen, S.~K. St.~Angelo, Y.~Cao,
  T.~E. Mallouk, P.~E. Lammert and V.~H. Crespi, \emph{Journal of the American
  Chemical Society}, 2004, \textbf{126}, 13424--13431\relax
\mciteBstWouldAddEndPuncttrue
\mciteSetBstMidEndSepPunct{\mcitedefaultmidpunct}
{\mcitedefaultendpunct}{\mcitedefaultseppunct}\relax
\EndOfBibitem
\bibitem[Mano and Heller(2005)]{mano2005bioelectrochemical}
N.~Mano and A.~Heller, \emph{Journal of the American Chemical Society}, 2005,
  \textbf{127}, 11574--11575\relax
\mciteBstWouldAddEndPuncttrue
\mciteSetBstMidEndSepPunct{\mcitedefaultmidpunct}
{\mcitedefaultendpunct}{\mcitedefaultseppunct}\relax
\EndOfBibitem
\bibitem[Howse \emph{et~al.}(2007)Howse, Jones, Ryan, Gough, Vafabakhsh, and
  Golestanian]{howse2007self}
J.~R. Howse, R.~A. Jones, A.~J. Ryan, T.~Gough, R.~Vafabakhsh and
  R.~Golestanian, \emph{Physical Review Letters}, 2007, \textbf{99},
  048102\relax
\mciteBstWouldAddEndPuncttrue
\mciteSetBstMidEndSepPunct{\mcitedefaultmidpunct}
{\mcitedefaultendpunct}{\mcitedefaultseppunct}\relax
\EndOfBibitem
\bibitem[Palacci \emph{et~al.}(2010)Palacci, Cottin-Bizonne, Ybert, and
  Bocquet]{palacci2010sedimentation}
J.~Palacci, C.~Cottin-Bizonne, C.~Ybert and L.~Bocquet, \emph{Physical Review
  Letters}, 2010, \textbf{105}, 088304\relax
\mciteBstWouldAddEndPuncttrue
\mciteSetBstMidEndSepPunct{\mcitedefaultmidpunct}
{\mcitedefaultendpunct}{\mcitedefaultseppunct}\relax
\EndOfBibitem
\bibitem[Theurkauff \emph{et~al.}(2012)Theurkauff, Cottin-Bizonne, Palacci,
  Ybert, and Bocquet]{theurkauff2012dynamic}
I.~Theurkauff, C.~Cottin-Bizonne, J.~Palacci, C.~Ybert and L.~Bocquet,
  \emph{Physical Review Letters}, 2012, \textbf{108}, 268303\relax
\mciteBstWouldAddEndPuncttrue
\mciteSetBstMidEndSepPunct{\mcitedefaultmidpunct}
{\mcitedefaultendpunct}{\mcitedefaultseppunct}\relax
\EndOfBibitem
\bibitem[Palacci \emph{et~al.}(2013)Palacci, Sacanna, Steinberg, Pine, and
  Chaikin]{palacci2013living}
J.~Palacci, S.~Sacanna, A.~P. Steinberg, D.~J. Pine and P.~M. Chaikin,
  \emph{Science}, 2013,  1230020\relax
\mciteBstWouldAddEndPuncttrue
\mciteSetBstMidEndSepPunct{\mcitedefaultmidpunct}
{\mcitedefaultendpunct}{\mcitedefaultseppunct}\relax
\EndOfBibitem
\bibitem[Buttinoni \emph{et~al.}(2013)Buttinoni, Bialk{\'e}, K{\"u}mmel,
  L{\"o}wen, Bechinger, and Speck]{buttinoni2013dynamical}
I.~Buttinoni, J.~Bialk{\'e}, F.~K{\"u}mmel, H.~L{\"o}wen, C.~Bechinger and
  T.~Speck, \emph{Physical Review Letters}, 2013, \textbf{110}, 238301\relax
\mciteBstWouldAddEndPuncttrue
\mciteSetBstMidEndSepPunct{\mcitedefaultmidpunct}
{\mcitedefaultendpunct}{\mcitedefaultseppunct}\relax
\EndOfBibitem
\bibitem[Mino \emph{et~al.}(2011)Mino, Mallouk, Darnige, Hoyos, Dauchet,
  Dunstan, Soto, Wang, Rousselet, and Clement]{mino2011enhanced}
G.~Mino, T.~E. Mallouk, T.~Darnige, M.~Hoyos, J.~Dauchet, J.~Dunstan, R.~Soto,
  Y.~Wang, A.~Rousselet and E.~Clement, \emph{Physical Review Letters}, 2011,
  \textbf{106}, 048102\relax
\mciteBstWouldAddEndPuncttrue
\mciteSetBstMidEndSepPunct{\mcitedefaultmidpunct}
{\mcitedefaultendpunct}{\mcitedefaultseppunct}\relax
\EndOfBibitem
\bibitem[Ginot \emph{et~al.}(2015)Ginot, Theurkauff, Levis, Ybert, Bocquet,
  Berthier, and Cottin-Bizonne]{ginot2015nonequilibrium}
F.~Ginot, I.~Theurkauff, D.~Levis, C.~Ybert, L.~Bocquet, L.~Berthier and
  C.~Cottin-Bizonne, \emph{Physical Review X}, 2015, \textbf{5}, 011004\relax
\mciteBstWouldAddEndPuncttrue
\mciteSetBstMidEndSepPunct{\mcitedefaultmidpunct}
{\mcitedefaultendpunct}{\mcitedefaultseppunct}\relax
\EndOfBibitem
\bibitem[Aubret \emph{et~al.}(2018)Aubret, Youssef, Sacanna, and
  Palacci]{aubret2018targeted}
A.~Aubret, M.~Youssef, S.~Sacanna and J.~Palacci, \emph{Nature Physics}, 2018,
  \textbf{14}, 1114--1118\relax
\mciteBstWouldAddEndPuncttrue
\mciteSetBstMidEndSepPunct{\mcitedefaultmidpunct}
{\mcitedefaultendpunct}{\mcitedefaultseppunct}\relax
\EndOfBibitem
\bibitem[Brown and Poon(2014)]{brown2014ionic}
A.~Brown and W.~Poon, \emph{Soft matter}, 2014, \textbf{10}, 4016--4027\relax
\mciteBstWouldAddEndPuncttrue
\mciteSetBstMidEndSepPunct{\mcitedefaultmidpunct}
{\mcitedefaultendpunct}{\mcitedefaultseppunct}\relax
\EndOfBibitem
\bibitem[Brown \emph{et~al.}(2017)Brown, Poon, Holm, and
  de~Graaf]{brown2017ionic}
A.~T. Brown, W.~C. Poon, C.~Holm and J.~de~Graaf, \emph{Soft Matter}, 2017,
  \textbf{13}, 1200--1222\relax
\mciteBstWouldAddEndPuncttrue
\mciteSetBstMidEndSepPunct{\mcitedefaultmidpunct}
{\mcitedefaultendpunct}{\mcitedefaultseppunct}\relax
\EndOfBibitem
\bibitem[Lammert \emph{et~al.}(1996)Lammert, Prost, and
  Bruinsma]{lammert1996ion}
P.~Lammert, J.~Prost and R.~Bruinsma, \emph{Journal of theoretical biology},
  1996, \textbf{178}, 387--391\relax
\mciteBstWouldAddEndPuncttrue
\mciteSetBstMidEndSepPunct{\mcitedefaultmidpunct}
{\mcitedefaultendpunct}{\mcitedefaultseppunct}\relax
\EndOfBibitem
\bibitem[Golestanian \emph{et~al.}(2005)Golestanian, Liverpool, and
  Ajdari]{golestanian2005propulsion}
R.~Golestanian, T.~B. Liverpool and A.~Ajdari, \emph{Physical Review Letters},
  2005, \textbf{94}, 220801\relax
\mciteBstWouldAddEndPuncttrue
\mciteSetBstMidEndSepPunct{\mcitedefaultmidpunct}
{\mcitedefaultendpunct}{\mcitedefaultseppunct}\relax
\EndOfBibitem
\bibitem[Di~Leonardo \emph{et~al.}(2010)Di~Leonardo, Angelani, Dell'Arciprete,
  Ruocco, Iebba, Schippa, Conte, Mecarini, De~Angelis, and
  Di~Fabrizio]{di2010bacterial}
R.~Di~Leonardo, L.~Angelani, D.~Dell'Arciprete, G.~Ruocco, V.~Iebba,
  S.~Schippa, M.~Conte, F.~Mecarini, F.~De~Angelis and E.~Di~Fabrizio,
  \emph{Proceedings of the National Academy of Sciences}, 2010, \textbf{107},
  9541--9545\relax
\mciteBstWouldAddEndPuncttrue
\mciteSetBstMidEndSepPunct{\mcitedefaultmidpunct}
{\mcitedefaultendpunct}{\mcitedefaultseppunct}\relax
\EndOfBibitem
\bibitem[Lion and Allen(2014)]{lion2014osmosis}
T.~W. Lion and R.~J. Allen, \emph{EPL (Europhysics Letters)}, 2014,
  \textbf{106}, 34003\relax
\mciteBstWouldAddEndPuncttrue
\mciteSetBstMidEndSepPunct{\mcitedefaultmidpunct}
{\mcitedefaultendpunct}{\mcitedefaultseppunct}\relax
\EndOfBibitem
\bibitem[Solon \emph{et~al.}(2015)Solon, Fily, Baskaran, Cates, Kafri, Kardar,
  and Tailleur]{solon2015pressure}
A.~P. Solon, Y.~Fily, A.~Baskaran, M.~E. Cates, Y.~Kafri, M.~Kardar and
  J.~Tailleur, \emph{Nature Physics}, 2015, \textbf{11}, 673\relax
\mciteBstWouldAddEndPuncttrue
\mciteSetBstMidEndSepPunct{\mcitedefaultmidpunct}
{\mcitedefaultendpunct}{\mcitedefaultseppunct}\relax
\EndOfBibitem
\bibitem[Rodenburg \emph{et~al.}(2017)Rodenburg, Dijkstra, and van
  Roij]{rodenburg2017van}
J.~Rodenburg, M.~Dijkstra and R.~van Roij, \emph{Soft matter}, 2017,
  \textbf{13}, 8957--8963\relax
\mciteBstWouldAddEndPuncttrue
\mciteSetBstMidEndSepPunct{\mcitedefaultmidpunct}
{\mcitedefaultendpunct}{\mcitedefaultseppunct}\relax
\EndOfBibitem
\bibitem[Takatori \emph{et~al.}(2016)Takatori, De~Dier, Vermant, and
  Brady]{takatori2016acoustic}
S.~C. Takatori, R.~De~Dier, J.~Vermant and J.~F. Brady, \emph{Nature
  communications}, 2016, \textbf{7}, 10694\relax
\mciteBstWouldAddEndPuncttrue
\mciteSetBstMidEndSepPunct{\mcitedefaultmidpunct}
{\mcitedefaultendpunct}{\mcitedefaultseppunct}\relax
\EndOfBibitem
\bibitem[Tailleur and Cates(2008)]{tailleur2008statistical}
J.~Tailleur and M.~Cates, \emph{Physical Review Letters}, 2008, \textbf{100},
  218103\relax
\mciteBstWouldAddEndPuncttrue
\mciteSetBstMidEndSepPunct{\mcitedefaultmidpunct}
{\mcitedefaultendpunct}{\mcitedefaultseppunct}\relax
\EndOfBibitem
\bibitem[Moerman \emph{et~al.}(2017)Moerman, Moyses, Van Der~Wee, Grier,
  Van~Blaaderen, Kegel, Groenewold, and Brujic]{moerman2017solute}
P.~G. Moerman, H.~W. Moyses, E.~B. Van Der~Wee, D.~G. Grier, A.~Van~Blaaderen,
  W.~K. Kegel, J.~Groenewold and J.~Brujic, \emph{Physical Review E}, 2017,
  \textbf{96}, 032607\relax
\mciteBstWouldAddEndPuncttrue
\mciteSetBstMidEndSepPunct{\mcitedefaultmidpunct}
{\mcitedefaultendpunct}{\mcitedefaultseppunct}\relax
\EndOfBibitem
\bibitem[Reigh \emph{et~al.}(2018)Reigh, Chuphal, Thakur, and
  Kapral]{reigh2018diffusiophoretically}
S.~Y. Reigh, P.~Chuphal, S.~Thakur and R.~Kapral, \emph{Soft matter}, 2018,
  \textbf{14}, 6043--6057\relax
\mciteBstWouldAddEndPuncttrue
\mciteSetBstMidEndSepPunct{\mcitedefaultmidpunct}
{\mcitedefaultendpunct}{\mcitedefaultseppunct}\relax
\EndOfBibitem
\bibitem[Ginot \emph{et~al.}(2018)Ginot, Theurkauff, Detcheverry, Ybert, and
  Cottin-Bizonne]{ginot2018aggregation}
F.~Ginot, I.~Theurkauff, F.~Detcheverry, C.~Ybert and C.~Cottin-Bizonne,
  \emph{Nature communications}, 2018, \textbf{9}, 696\relax
\mciteBstWouldAddEndPuncttrue
\mciteSetBstMidEndSepPunct{\mcitedefaultmidpunct}
{\mcitedefaultendpunct}{\mcitedefaultseppunct}\relax
\EndOfBibitem
\bibitem[Keller and Segel(1970)]{keller1970initiation}
E.~F. Keller and L.~A. Segel, \emph{Journal of theoretical biology}, 1970,
  \textbf{26}, 399--415\relax
\mciteBstWouldAddEndPuncttrue
\mciteSetBstMidEndSepPunct{\mcitedefaultmidpunct}
{\mcitedefaultendpunct}{\mcitedefaultseppunct}\relax
\EndOfBibitem
\bibitem[Brenner \emph{et~al.}(1998)Brenner, Levitov, and
  Budrene]{brenner1998physical}
M.~P. Brenner, L.~S. Levitov and E.~O. Budrene, \emph{Biophysical journal},
  1998, \textbf{74}, 1677--1693\relax
\mciteBstWouldAddEndPuncttrue
\mciteSetBstMidEndSepPunct{\mcitedefaultmidpunct}
{\mcitedefaultendpunct}{\mcitedefaultseppunct}\relax
\EndOfBibitem
\bibitem[Golestanian(2012)]{golestanian2012collective}
R.~Golestanian, \emph{Physical Review Letters}, 2012, \textbf{108},
  038303\relax
\mciteBstWouldAddEndPuncttrue
\mciteSetBstMidEndSepPunct{\mcitedefaultmidpunct}
{\mcitedefaultendpunct}{\mcitedefaultseppunct}\relax
\EndOfBibitem
\bibitem[Zhang \emph{et~al.}(2017)Zhang, McMullen, Pontani, He, Sha, Seeman,
  Brujic, and Chaikin]{zhang2017sequential}
Y.~Zhang, A.~McMullen, L.-L. Pontani, X.~He, R.~Sha, N.~C. Seeman, J.~Brujic
  and P.~M. Chaikin, \emph{Nature Communications}, 2017, \textbf{8}, 21\relax
\mciteBstWouldAddEndPuncttrue
\mciteSetBstMidEndSepPunct{\mcitedefaultmidpunct}
{\mcitedefaultendpunct}{\mcitedefaultseppunct}\relax
\EndOfBibitem
\bibitem[Pohl and Stark(2014)]{pohl2014dynamic}
O.~Pohl and H.~Stark, \emph{Physical Review Letters}, 2014, \textbf{112},
  238303\relax
\mciteBstWouldAddEndPuncttrue
\mciteSetBstMidEndSepPunct{\mcitedefaultmidpunct}
{\mcitedefaultendpunct}{\mcitedefaultseppunct}\relax
\EndOfBibitem
\bibitem[Stark(2018)]{stark2018artificial}
H.~Stark, \emph{Accounts of chemical research}, 2018, \textbf{51},
  2681--2688\relax
\mciteBstWouldAddEndPuncttrue
\mciteSetBstMidEndSepPunct{\mcitedefaultmidpunct}
{\mcitedefaultendpunct}{\mcitedefaultseppunct}\relax
\EndOfBibitem
\bibitem[Feng \emph{et~al.}(2013)Feng, Dreyfus, Sha, Seeman, and
  Chaikin]{feng2013dna}
L.~Feng, R.~Dreyfus, R.~Sha, N.~C. Seeman and P.~M. Chaikin, \emph{Advanced
  Materials}, 2013, \textbf{25}, 2779--2783\relax
\mciteBstWouldAddEndPuncttrue
\mciteSetBstMidEndSepPunct{\mcitedefaultmidpunct}
{\mcitedefaultendpunct}{\mcitedefaultseppunct}\relax
\EndOfBibitem
\bibitem[McMullen \emph{et~al.}(2018)McMullen, Holmes-Cerfon, Sciortino,
  Grosberg, and Brujic]{mcmullen2018freely}
A.~McMullen, M.~Holmes-Cerfon, F.~Sciortino, A.~Y. Grosberg and J.~Brujic,
  \emph{Physical Review Letters}, 2018, \textbf{121}, 138002\relax
\mciteBstWouldAddEndPuncttrue
\mciteSetBstMidEndSepPunct{\mcitedefaultmidpunct}
{\mcitedefaultendpunct}{\mcitedefaultseppunct}\relax
\EndOfBibitem
\bibitem[Sheeler and Bianchi(1987)]{sheeler1987cell}
P.~Sheeler and D.~E. Bianchi, \emph{Cell and molecular biology}, Wiley New York
  etc., 1987\relax
\mciteBstWouldAddEndPuncttrue
\mciteSetBstMidEndSepPunct{\mcitedefaultmidpunct}
{\mcitedefaultendpunct}{\mcitedefaultseppunct}\relax
\EndOfBibitem
\bibitem[Jensen \emph{et~al.}(2009)Jensen, Rio, Hansen, Clanet, and
  Bohr]{jensen2009osmotically}
K.~H. Jensen, E.~Rio, R.~Hansen, C.~Clanet and T.~Bohr, \emph{Journal of Fluid
  Mechanics}, 2009, \textbf{636}, 371--396\relax
\mciteBstWouldAddEndPuncttrue
\mciteSetBstMidEndSepPunct{\mcitedefaultmidpunct}
{\mcitedefaultendpunct}{\mcitedefaultseppunct}\relax
\EndOfBibitem
\bibitem[Comtet \emph{et~al.}(2017)Comtet, Jensen, Turgeon, Stroock, and
  Hosoi]{comtet2017passive}
J.~Comtet, K.~H. Jensen, R.~Turgeon, A.~D. Stroock and A.~Hosoi, \emph{Nature
  plants}, 2017, \textbf{3}, 17032\relax
\mciteBstWouldAddEndPuncttrue
\mciteSetBstMidEndSepPunct{\mcitedefaultmidpunct}
{\mcitedefaultendpunct}{\mcitedefaultseppunct}\relax
\EndOfBibitem
\bibitem[Logan and Elimelech(2012)]{logan2012membrane}
B.~E. Logan and M.~Elimelech, \emph{Nature}, 2012, \textbf{488}, 313\relax
\mciteBstWouldAddEndPuncttrue
\mciteSetBstMidEndSepPunct{\mcitedefaultmidpunct}
{\mcitedefaultendpunct}{\mcitedefaultseppunct}\relax
\EndOfBibitem
\bibitem[Werber \emph{et~al.}(2016)Werber, Osuji, and
  Elimelech]{werber2016materials}
J.~R. Werber, C.~O. Osuji and M.~Elimelech, \emph{Nature Reviews Materials},
  2016, \textbf{1}, 16018\relax
\mciteBstWouldAddEndPuncttrue
\mciteSetBstMidEndSepPunct{\mcitedefaultmidpunct}
{\mcitedefaultendpunct}{\mcitedefaultseppunct}\relax
\EndOfBibitem
\bibitem[Jones \emph{et~al.}(2019)Jones, Qadir, van Vliet, Smakhtin, and
  Kang]{jones2018state}
E.~Jones, M.~Qadir, M.~T. van Vliet, V.~Smakhtin and S.-M. Kang, \emph{Science
  of the Total Environment}, 2019, \textbf{657}, 1343--1356\relax
\mciteBstWouldAddEndPuncttrue
\mciteSetBstMidEndSepPunct{\mcitedefaultmidpunct}
{\mcitedefaultendpunct}{\mcitedefaultseppunct}\relax
\EndOfBibitem
\bibitem[Organization \emph{et~al.}(2015)Organization, Supply, and
  Programme]{world2015progress}
W.~H. Organization, W.~J.~W. Supply and S.~M. Programme, \emph{Progress on
  sanitation and drinking water: 2015 update and MDG assessment}, World Health
  Organization, 2015\relax
\mciteBstWouldAddEndPuncttrue
\mciteSetBstMidEndSepPunct{\mcitedefaultmidpunct}
{\mcitedefaultendpunct}{\mcitedefaultseppunct}\relax
\EndOfBibitem
\bibitem[Vidic \emph{et~al.}(2013)Vidic, Brantley, Vandenbossche, Yoxtheimer,
  and Abad]{vidic2013impact}
R.~D. Vidic, S.~L. Brantley, J.~M. Vandenbossche, D.~Yoxtheimer and J.~D. Abad,
  \emph{Science}, 2013, \textbf{340}, 1235009\relax
\mciteBstWouldAddEndPuncttrue
\mciteSetBstMidEndSepPunct{\mcitedefaultmidpunct}
{\mcitedefaultendpunct}{\mcitedefaultseppunct}\relax
\EndOfBibitem
\bibitem[Gregory \emph{et~al.}(2011)Gregory, Vidic, and
  Dzombak]{gregory2011water}
K.~B. Gregory, R.~D. Vidic and D.~A. Dzombak, \emph{Elements}, 2011,
  \textbf{7}, 181--186\relax
\mciteBstWouldAddEndPuncttrue
\mciteSetBstMidEndSepPunct{\mcitedefaultmidpunct}
{\mcitedefaultendpunct}{\mcitedefaultseppunct}\relax
\EndOfBibitem
\bibitem[Onishi \emph{et~al.}(2017)Onishi, Carrero-Parreno, Reyes-Labarta,
  Fraga, and Caballero]{onishi2017desalination}
V.~C. Onishi, A.~Carrero-Parreno, J.~A. Reyes-Labarta, E.~S. Fraga and J.~A.
  Caballero, \emph{Journal of Cleaner Production}, 2017, \textbf{140},
  1399--1414\relax
\mciteBstWouldAddEndPuncttrue
\mciteSetBstMidEndSepPunct{\mcitedefaultmidpunct}
{\mcitedefaultendpunct}{\mcitedefaultseppunct}\relax
\EndOfBibitem
\bibitem[Hoekstra and Mekonnen(2012)]{hoekstra2012water}
A.~Y. Hoekstra and M.~M. Mekonnen, \emph{Proceedings of the national academy of
  sciences}, 2012, \textbf{109}, 3232--3237\relax
\mciteBstWouldAddEndPuncttrue
\mciteSetBstMidEndSepPunct{\mcitedefaultmidpunct}
{\mcitedefaultendpunct}{\mcitedefaultseppunct}\relax
\EndOfBibitem
\bibitem[Alkaisi \emph{et~al.}(2017)Alkaisi, Mossad, and
  Sharifian-Barforoush]{alkaisi2017review}
A.~Alkaisi, R.~Mossad and A.~Sharifian-Barforoush, \emph{Energy Procedia},
  2017, \textbf{110}, 268--274\relax
\mciteBstWouldAddEndPuncttrue
\mciteSetBstMidEndSepPunct{\mcitedefaultmidpunct}
{\mcitedefaultendpunct}{\mcitedefaultseppunct}\relax
\EndOfBibitem
\bibitem[Amy \emph{et~al.}(2017)Amy, Ghaffour, Li, Francis, Linares, Missimer,
  and Lattemann]{amy2017membrane}
G.~Amy, N.~Ghaffour, Z.~Li, L.~Francis, R.~V. Linares, T.~Missimer and
  S.~Lattemann, \emph{Desalination}, 2017, \textbf{401}, 16--21\relax
\mciteBstWouldAddEndPuncttrue
\mciteSetBstMidEndSepPunct{\mcitedefaultmidpunct}
{\mcitedefaultendpunct}{\mcitedefaultseppunct}\relax
\EndOfBibitem
\bibitem[McCutcheon \emph{et~al.}(2005)McCutcheon, McGinnis, and
  Elimelech]{mccutcheon2005novel}
J.~R. McCutcheon, R.~L. McGinnis and M.~Elimelech, \emph{Desalination}, 2005,
  \textbf{174}, 1--11\relax
\mciteBstWouldAddEndPuncttrue
\mciteSetBstMidEndSepPunct{\mcitedefaultmidpunct}
{\mcitedefaultendpunct}{\mcitedefaultseppunct}\relax
\EndOfBibitem
\bibitem[Linares \emph{et~al.}(2016)Linares, Li, Yangali-Quintanilla, Ghaffour,
  Amy, Leiknes, and Vrouwenvelder]{linares2016life}
R.~V. Linares, Z.~Li, V.~Yangali-Quintanilla, N.~Ghaffour, G.~Amy, T.~Leiknes
  and J.~S. Vrouwenvelder, \emph{Water research}, 2016, \textbf{88},
  225--234\relax
\mciteBstWouldAddEndPuncttrue
\mciteSetBstMidEndSepPunct{\mcitedefaultmidpunct}
{\mcitedefaultendpunct}{\mcitedefaultseppunct}\relax
\EndOfBibitem
\bibitem[Chen \emph{et~al.}(2019)Chen, Ge, Xu, and Pan]{chen2019functionalized}
Q.~Chen, Q.~Ge, W.~Xu and W.~Pan, \emph{Journal of Membrane Science}, 2019,
  \textbf{574}, 10--16\relax
\mciteBstWouldAddEndPuncttrue
\mciteSetBstMidEndSepPunct{\mcitedefaultmidpunct}
{\mcitedefaultendpunct}{\mcitedefaultseppunct}\relax
\EndOfBibitem
\bibitem[Wheeler and Stroock(2008)]{wheeler2008transpiration}
T.~D. Wheeler and A.~D. Stroock, \emph{Nature}, 2008, \textbf{455}, 208\relax
\mciteBstWouldAddEndPuncttrue
\mciteSetBstMidEndSepPunct{\mcitedefaultmidpunct}
{\mcitedefaultendpunct}{\mcitedefaultseppunct}\relax
\EndOfBibitem
\bibitem[Semiat(2008)]{semiat2008energy}
R.~Semiat, \emph{Environmental science \& technology}, 2008, \textbf{42},
  8193--8201\relax
\mciteBstWouldAddEndPuncttrue
\mciteSetBstMidEndSepPunct{\mcitedefaultmidpunct}
{\mcitedefaultendpunct}{\mcitedefaultseppunct}\relax
\EndOfBibitem
\bibitem[Chao and Liang(2008)]{chao2008feasibility}
Y.-M. Chao and T.~Liang, \emph{Desalination}, 2008, \textbf{221},
  433--439\relax
\mciteBstWouldAddEndPuncttrue
\mciteSetBstMidEndSepPunct{\mcitedefaultmidpunct}
{\mcitedefaultendpunct}{\mcitedefaultseppunct}\relax
\EndOfBibitem
\bibitem[Oren(2008)]{oren2008capacitive}
Y.~Oren, \emph{Desalination}, 2008, \textbf{228}, 10--29\relax
\mciteBstWouldAddEndPuncttrue
\mciteSetBstMidEndSepPunct{\mcitedefaultmidpunct}
{\mcitedefaultendpunct}{\mcitedefaultseppunct}\relax
\EndOfBibitem
\bibitem[Suss \emph{et~al.}(2012)Suss, Baumann, Bourcier, Spadaccini, Rose,
  Santiago, and Stadermann]{suss2012capacitive}
M.~E. Suss, T.~F. Baumann, W.~L. Bourcier, C.~M. Spadaccini, K.~A. Rose, J.~G.
  Santiago and M.~Stadermann, \emph{Energy \& Environmental Science}, 2012,
  \textbf{5}, 9511--9519\relax
\mciteBstWouldAddEndPuncttrue
\mciteSetBstMidEndSepPunct{\mcitedefaultmidpunct}
{\mcitedefaultendpunct}{\mcitedefaultseppunct}\relax
\EndOfBibitem
\bibitem[Kim \emph{et~al.}(2017)Kim, Gorski, and Logan]{kim2017low}
T.~Kim, C.~A. Gorski and B.~E. Logan, \emph{Environmental Science \& Technology
  Letters}, 2017, \textbf{4}, 444--449\relax
\mciteBstWouldAddEndPuncttrue
\mciteSetBstMidEndSepPunct{\mcitedefaultmidpunct}
{\mcitedefaultendpunct}{\mcitedefaultseppunct}\relax
\EndOfBibitem
\bibitem[Deng \emph{et~al.}(2013)Deng, Dydek, Han, Schlumpberger, Mani,
  Zaltzman, and Bazant]{deng2013overlimiting}
D.~Deng, E.~V. Dydek, J.-H. Han, S.~Schlumpberger, A.~Mani, B.~Zaltzman and
  M.~Z. Bazant, \emph{Langmuir}, 2013, \textbf{29}, 16167--16177\relax
\mciteBstWouldAddEndPuncttrue
\mciteSetBstMidEndSepPunct{\mcitedefaultmidpunct}
{\mcitedefaultendpunct}{\mcitedefaultseppunct}\relax
\EndOfBibitem
\bibitem[Kim \emph{et~al.}(2010)Kim, Ko, Kang, and Han]{kim2010direct}
S.~J. Kim, S.~H. Ko, K.~H. Kang and J.~Han, \emph{Nature Nanotechnology}, 2010,
  \textbf{5}, 297\relax
\mciteBstWouldAddEndPuncttrue
\mciteSetBstMidEndSepPunct{\mcitedefaultmidpunct}
{\mcitedefaultendpunct}{\mcitedefaultseppunct}\relax
\EndOfBibitem
\bibitem[Shahzad \emph{et~al.}(2018)Shahzad, Burhan, Ang, and
  Ng]{shahzad2018adsorption}
M.~W. Shahzad, M.~Burhan, L.~Ang and K.~C. Ng, \emph{Emerging Technologies for
  Sustainable Desalination Handbook}, Elsevier, 2018, pp. 3--34\relax
\mciteBstWouldAddEndPuncttrue
\mciteSetBstMidEndSepPunct{\mcitedefaultmidpunct}
{\mcitedefaultendpunct}{\mcitedefaultseppunct}\relax
\EndOfBibitem
\bibitem[Ar{\'a}mburo-Miranda and
  Ruelas-Ram{\'\i}rez(2017)]{aramburo2017desalination}
I.~V. Ar{\'a}mburo-Miranda and E.~H. Ruelas-Ram{\'\i}rez, \emph{Environmental
  Science and Pollution Research}, 2017, \textbf{24}, 25676--25681\relax
\mciteBstWouldAddEndPuncttrue
\mciteSetBstMidEndSepPunct{\mcitedefaultmidpunct}
{\mcitedefaultendpunct}{\mcitedefaultseppunct}\relax
\EndOfBibitem
\bibitem[Minas \emph{et~al.}(2015)Minas, Karunakaran, Bond, Gandy, Honsbein,
  Madsen, Amezaga, Amtmann, Templeton,
  Biggs,\emph{et~al.}]{minas2015biodesalination}
K.~Minas, E.~Karunakaran, T.~Bond, C.~Gandy, A.~Honsbein, M.~Madsen,
  J.~Amezaga, A.~Amtmann, M.~Templeton, C.~Biggs \emph{et~al.},
  \emph{Desalination and Water Treatment}, 2015, \textbf{55}, 2647--2668\relax
\mciteBstWouldAddEndPuncttrue
\mciteSetBstMidEndSepPunct{\mcitedefaultmidpunct}
{\mcitedefaultendpunct}{\mcitedefaultseppunct}\relax
\EndOfBibitem
\bibitem[Liu \emph{et~al.}(2004)Liu, Ramnarayanan, and
  Logan]{liu2004production}
H.~Liu, R.~Ramnarayanan and B.~E. Logan, \emph{Environmental science \&
  technology}, 2004, \textbf{38}, 2281--2285\relax
\mciteBstWouldAddEndPuncttrue
\mciteSetBstMidEndSepPunct{\mcitedefaultmidpunct}
{\mcitedefaultendpunct}{\mcitedefaultseppunct}\relax
\EndOfBibitem
\bibitem[Elimelech and Phillip(2011)]{elimelech2011future}
M.~Elimelech and W.~A. Phillip, \emph{science}, 2011, \textbf{333},
  712--717\relax
\mciteBstWouldAddEndPuncttrue
\mciteSetBstMidEndSepPunct{\mcitedefaultmidpunct}
{\mcitedefaultendpunct}{\mcitedefaultseppunct}\relax
\EndOfBibitem
\bibitem[Mehta and Zydney(2005)]{mehta2005permeability}
A.~Mehta and A.~L. Zydney, \emph{Journal of Membrane Science}, 2005,
  \textbf{249}, 245--249\relax
\mciteBstWouldAddEndPuncttrue
\mciteSetBstMidEndSepPunct{\mcitedefaultmidpunct}
{\mcitedefaultendpunct}{\mcitedefaultseppunct}\relax
\EndOfBibitem
\bibitem[Park \emph{et~al.}(2017)Park, Kamcev, Robeson, Elimelech, and
  Freeman]{park2017maximizing}
H.~B. Park, J.~Kamcev, L.~M. Robeson, M.~Elimelech and B.~D. Freeman,
  \emph{Science}, 2017, \textbf{356}, eaab0530\relax
\mciteBstWouldAddEndPuncttrue
\mciteSetBstMidEndSepPunct{\mcitedefaultmidpunct}
{\mcitedefaultendpunct}{\mcitedefaultseppunct}\relax
\EndOfBibitem
\bibitem[Werber \emph{et~al.}(2016)Werber, Deshmukh, and
  Elimelech]{werber2016critical}
J.~R. Werber, A.~Deshmukh and M.~Elimelech, \emph{Environmental Science \&
  Technology Letters}, 2016, \textbf{3}, 112--120\relax
\mciteBstWouldAddEndPuncttrue
\mciteSetBstMidEndSepPunct{\mcitedefaultmidpunct}
{\mcitedefaultendpunct}{\mcitedefaultseppunct}\relax
\EndOfBibitem
\bibitem[Stephens \emph{et~al.}(2016)Stephens, Lindsay, Milne, Klein, Fuchs,
  and Rodenburg]{stephens2016pcbs}
L.~W. Stephens, R.~Lindsay, M.~Milne, A.~Klein, V.~Fuchs and L.~Rodenburg,
  \emph{Proceedings of the Water Environment Federation}, 2016, \textbf{2016},
  5423--5429\relax
\mciteBstWouldAddEndPuncttrue
\mciteSetBstMidEndSepPunct{\mcitedefaultmidpunct}
{\mcitedefaultendpunct}{\mcitedefaultseppunct}\relax
\EndOfBibitem
\bibitem[Zeman and Wales(1981)]{zeman1981polymer}
L.~Zeman and M.~Wales, \emph{Synthetic membranes}, 1981, \textbf{2},
  411--434\relax
\mciteBstWouldAddEndPuncttrue
\mciteSetBstMidEndSepPunct{\mcitedefaultmidpunct}
{\mcitedefaultendpunct}{\mcitedefaultseppunct}\relax
\EndOfBibitem
\bibitem[Geise \emph{et~al.}(2011)Geise, Park, Sagle, Freeman, and
  McGrath]{geise2011water}
G.~M. Geise, H.~B. Park, A.~C. Sagle, B.~D. Freeman and J.~E. McGrath,
  \emph{Journal of Membrane Science}, 2011, \textbf{369}, 130--138\relax
\mciteBstWouldAddEndPuncttrue
\mciteSetBstMidEndSepPunct{\mcitedefaultmidpunct}
{\mcitedefaultendpunct}{\mcitedefaultseppunct}\relax
\EndOfBibitem
\bibitem[Barboiu(2016)]{barboiu2016artificial}
M.~Barboiu, \emph{Chemical Communications}, 2016, \textbf{52}, 5657--5665\relax
\mciteBstWouldAddEndPuncttrue
\mciteSetBstMidEndSepPunct{\mcitedefaultmidpunct}
{\mcitedefaultendpunct}{\mcitedefaultseppunct}\relax
\EndOfBibitem
\bibitem[Shen \emph{et~al.}(2018)Shen, Song, Barden, Ren, Lang, Feroz,
  Henderson, Saboe, Tsai, Yan,\emph{et~al.}]{shen2018achieving}
Y.-x. Shen, W.~C. Song, D.~R. Barden, T.~Ren, C.~Lang, H.~Feroz, C.~B.
  Henderson, P.~O. Saboe, D.~Tsai, H.~Yan \emph{et~al.}, \emph{Nature
  communications}, 2018, \textbf{9}, 2294\relax
\mciteBstWouldAddEndPuncttrue
\mciteSetBstMidEndSepPunct{\mcitedefaultmidpunct}
{\mcitedefaultendpunct}{\mcitedefaultseppunct}\relax
\EndOfBibitem
\bibitem[Sun \emph{et~al.}(2018)Sun, Kocsis, Li, Legrand, and
  Barboiu]{sun2018imidazole}
Z.~Sun, I.~Kocsis, Y.~Li, Y.-M. Legrand and M.~Barboiu, \emph{Faraday
  discussions}, 2018, \textbf{209}, 113--124\relax
\mciteBstWouldAddEndPuncttrue
\mciteSetBstMidEndSepPunct{\mcitedefaultmidpunct}
{\mcitedefaultendpunct}{\mcitedefaultseppunct}\relax
\EndOfBibitem
\bibitem[Mauter \emph{et~al.}(2018)Mauter, Zucker, Perreault, Werber, Kim, and
  Elimelech]{mauter2018role}
M.~S. Mauter, I.~Zucker, F.~Perreault, J.~R. Werber, J.-H. Kim and
  M.~Elimelech, \emph{Nature Sustainability}, 2018, \textbf{1}, 166\relax
\mciteBstWouldAddEndPuncttrue
\mciteSetBstMidEndSepPunct{\mcitedefaultmidpunct}
{\mcitedefaultendpunct}{\mcitedefaultseppunct}\relax
\EndOfBibitem
\bibitem[Nednoor \emph{et~al.}(2007)Nednoor, Gavalas, Chopra, Hinds, and
  Bachas]{nednoor2007carbon}
P.~Nednoor, V.~G. Gavalas, N.~Chopra, B.~J. Hinds and L.~G. Bachas,
  \emph{Journal of Materials Chemistry}, 2007, \textbf{17}, 1755--1757\relax
\mciteBstWouldAddEndPuncttrue
\mciteSetBstMidEndSepPunct{\mcitedefaultmidpunct}
{\mcitedefaultendpunct}{\mcitedefaultseppunct}\relax
\EndOfBibitem
\bibitem[Chan \emph{et~al.}(2016)Chan, Marand, and Martin]{chan2016novel}
W.-F. Chan, E.~Marand and S.~M. Martin, \emph{Journal of Membrane Science},
  2016, \textbf{509}, 125--137\relax
\mciteBstWouldAddEndPuncttrue
\mciteSetBstMidEndSepPunct{\mcitedefaultmidpunct}
{\mcitedefaultendpunct}{\mcitedefaultseppunct}\relax
\EndOfBibitem
\bibitem[Yoshida and Bocquet(2016)]{yoshida2016labyrinthine}
H.~Yoshida and L.~Bocquet, \emph{The Journal of chemical physics}, 2016,
  \textbf{144}, 234701\relax
\mciteBstWouldAddEndPuncttrue
\mciteSetBstMidEndSepPunct{\mcitedefaultmidpunct}
{\mcitedefaultendpunct}{\mcitedefaultseppunct}\relax
\EndOfBibitem
\bibitem[Abraham \emph{et~al.}(2017)Abraham, Vasu, Williams, Gopinadhan, Su,
  Cherian, Dix, Prestat, Haigh, Grigorieva,\emph{et~al.}]{abraham2017tunable}
J.~Abraham, K.~S. Vasu, C.~D. Williams, K.~Gopinadhan, Y.~Su, C.~T. Cherian,
  J.~Dix, E.~Prestat, S.~J. Haigh, I.~V. Grigorieva \emph{et~al.}, \emph{Nature
  nanotechnology}, 2017, \textbf{12}, 546\relax
\mciteBstWouldAddEndPuncttrue
\mciteSetBstMidEndSepPunct{\mcitedefaultmidpunct}
{\mcitedefaultendpunct}{\mcitedefaultseppunct}\relax
\EndOfBibitem
\bibitem[Thebo \emph{et~al.}(2018)Thebo, Qian, Zhang, Chen, Cheng, and
  Ren]{thebo2018highly}
K.~H. Thebo, X.~Qian, Q.~Zhang, L.~Chen, H.-M. Cheng and W.~Ren, \emph{Nature
  communications}, 2018, \textbf{9}, 1486\relax
\mciteBstWouldAddEndPuncttrue
\mciteSetBstMidEndSepPunct{\mcitedefaultmidpunct}
{\mcitedefaultendpunct}{\mcitedefaultseppunct}\relax
\EndOfBibitem
\bibitem[Mi(2014)]{mi2014graphene}
B.~Mi, \emph{Science}, 2014, \textbf{343}, 740--742\relax
\mciteBstWouldAddEndPuncttrue
\mciteSetBstMidEndSepPunct{\mcitedefaultmidpunct}
{\mcitedefaultendpunct}{\mcitedefaultseppunct}\relax
\EndOfBibitem
\bibitem[Gravelle \emph{et~al.}(2016)Gravelle, Yoshida, Joly, Ybert, and
  Bocquet]{gravelle2016carbon}
S.~Gravelle, H.~Yoshida, L.~Joly, C.~Ybert and L.~Bocquet, \emph{The Journal of
  chemical physics}, 2016, \textbf{145}, 124708\relax
\mciteBstWouldAddEndPuncttrue
\mciteSetBstMidEndSepPunct{\mcitedefaultmidpunct}
{\mcitedefaultendpunct}{\mcitedefaultseppunct}\relax
\EndOfBibitem
\bibitem[Dumais and Forterre(2012)]{dumais2012vegetable}
J.~Dumais and Y.~Forterre, \emph{Annual Review of Fluid Mechanics}, 2012,
  \textbf{44}, 453--478\relax
\mciteBstWouldAddEndPuncttrue
\mciteSetBstMidEndSepPunct{\mcitedefaultmidpunct}
{\mcitedefaultendpunct}{\mcitedefaultseppunct}\relax
\EndOfBibitem
\bibitem[Broyer(1947)]{broyer1947movement}
T.~Broyer, \emph{The Botanical Review}, 1947, \textbf{13}, 1--58\relax
\mciteBstWouldAddEndPuncttrue
\mciteSetBstMidEndSepPunct{\mcitedefaultmidpunct}
{\mcitedefaultendpunct}{\mcitedefaultseppunct}\relax
\EndOfBibitem
\bibitem[Thompson and Holbrook(2003)]{thompson2003scaling}
M.~V. Thompson and N.~M. Holbrook, \emph{Plant, Cell \& Environment}, 2003,
  \textbf{26}, 1561--1577\relax
\mciteBstWouldAddEndPuncttrue
\mciteSetBstMidEndSepPunct{\mcitedefaultmidpunct}
{\mcitedefaultendpunct}{\mcitedefaultseppunct}\relax
\EndOfBibitem
\bibitem[Jensen \emph{et~al.}(2011)Jensen, Lee, Bohr, Bruus, Holbrook, and
  Zwieniecki]{jensen2011optimality}
K.~H. Jensen, J.~Lee, T.~Bohr, H.~Bruus, N.~M. Holbrook and M.~A. Zwieniecki,
  \emph{Journal of the Royal Society Interface}, 2011, \textbf{8},
  1155--1165\relax
\mciteBstWouldAddEndPuncttrue
\mciteSetBstMidEndSepPunct{\mcitedefaultmidpunct}
{\mcitedefaultendpunct}{\mcitedefaultseppunct}\relax
\EndOfBibitem
\bibitem[Rademaker \emph{et~al.}(2017)Rademaker, Zwieniecki, Bohr, and
  Jensen]{rademaker2017sugar}
H.~Rademaker, M.~A. Zwieniecki, T.~Bohr and K.~H. Jensen, \emph{Physical Review
  E}, 2017, \textbf{95}, 042402\relax
\mciteBstWouldAddEndPuncttrue
\mciteSetBstMidEndSepPunct{\mcitedefaultmidpunct}
{\mcitedefaultendpunct}{\mcitedefaultseppunct}\relax
\EndOfBibitem
\bibitem[Mansfield \emph{et~al.}(1990)Mansfield, Hetherington, and
  Atkinson]{mansfield1990some}
T.~Mansfield, A.~Hetherington and C.~Atkinson, \emph{Annual review of plant
  biology}, 1990, \textbf{41}, 55--75\relax
\mciteBstWouldAddEndPuncttrue
\mciteSetBstMidEndSepPunct{\mcitedefaultmidpunct}
{\mcitedefaultendpunct}{\mcitedefaultseppunct}\relax
\EndOfBibitem
\bibitem[Satter and Galston(1981)]{satter1981mechanisms}
R.~L. Satter and A.~W. Galston, \emph{Annual Review of Plant Physiology}, 1981,
  \textbf{32}, 83--110\relax
\mciteBstWouldAddEndPuncttrue
\mciteSetBstMidEndSepPunct{\mcitedefaultmidpunct}
{\mcitedefaultendpunct}{\mcitedefaultseppunct}\relax
\EndOfBibitem
\bibitem[van Doorn and van Meeteren(2003)]{van2003flower}
W.~G. van Doorn and U.~van Meeteren, \emph{Journal of experimental botany},
  2003, \textbf{54}, 1801--1812\relax
\mciteBstWouldAddEndPuncttrue
\mciteSetBstMidEndSepPunct{\mcitedefaultmidpunct}
{\mcitedefaultendpunct}{\mcitedefaultseppunct}\relax
\EndOfBibitem
\bibitem[Hedrich and Schroeder(1989)]{hedrich1989physiology}
R.~Hedrich and J.~I. Schroeder, \emph{Annual review of plant biology}, 1989,
  \textbf{40}, 539--569\relax
\mciteBstWouldAddEndPuncttrue
\mciteSetBstMidEndSepPunct{\mcitedefaultmidpunct}
{\mcitedefaultendpunct}{\mcitedefaultseppunct}\relax
\EndOfBibitem
\bibitem[Irving \emph{et~al.}(1997)Irving, Ritter, Tomos, and
  Koller]{irving1997phototropic}
M.~Irving, S.~Ritter, A.~Tomos and D.~Koller, \emph{Botanica Acta}, 1997,
  \textbf{110}, 118--126\relax
\mciteBstWouldAddEndPuncttrue
\mciteSetBstMidEndSepPunct{\mcitedefaultmidpunct}
{\mcitedefaultendpunct}{\mcitedefaultseppunct}\relax
\EndOfBibitem
\bibitem[Seminara \emph{et~al.}(2012)Seminara, Angelini, Wilking, Vlamakis,
  Ebrahim, Kolter, Weitz, and Brenner]{seminara2012osmotic}
A.~Seminara, T.~E. Angelini, J.~N. Wilking, H.~Vlamakis, S.~Ebrahim, R.~Kolter,
  D.~A. Weitz and M.~P. Brenner, \emph{Proceedings of the National Academy of
  Sciences}, 2012, \textbf{109}, 1116--1121\relax
\mciteBstWouldAddEndPuncttrue
\mciteSetBstMidEndSepPunct{\mcitedefaultmidpunct}
{\mcitedefaultendpunct}{\mcitedefaultseppunct}\relax
\EndOfBibitem
\bibitem[Marbach and Bocquet(2016)]{marbach2016active}
S.~Marbach and L.~Bocquet, \emph{Physical Review X}, 2016, \textbf{6},
  031008\relax
\mciteBstWouldAddEndPuncttrue
\mciteSetBstMidEndSepPunct{\mcitedefaultmidpunct}
{\mcitedefaultendpunct}{\mcitedefaultseppunct}\relax
\EndOfBibitem
\bibitem[Meyer \emph{et~al.}(2017)Meyer, Ostrenko, Bourantas,
  Morales-Navarrete, Porat-Shliom, Segovia-Miranda, Nonaka, Ghaemi, Verbavatz,
  Brusch,\emph{et~al.}]{meyer2017predictive}
K.~Meyer, O.~Ostrenko, G.~Bourantas, H.~Morales-Navarrete, N.~Porat-Shliom,
  F.~Segovia-Miranda, H.~Nonaka, A.~Ghaemi, J.-M. Verbavatz, L.~Brusch
  \emph{et~al.}, \emph{Cell systems}, 2017, \textbf{4}, 277--290\relax
\mciteBstWouldAddEndPuncttrue
\mciteSetBstMidEndSepPunct{\mcitedefaultmidpunct}
{\mcitedefaultendpunct}{\mcitedefaultseppunct}\relax
\EndOfBibitem
\bibitem[Thiagarajah and Verkman(2018)]{thiagarajah2018water}
J.~R. Thiagarajah and A.~S. Verkman, \emph{Physiology of the Gastrointestinal
  Tract (Sixth Edition)}, Elsevier, 2018, pp. 1249--1272\relax
\mciteBstWouldAddEndPuncttrue
\mciteSetBstMidEndSepPunct{\mcitedefaultmidpunct}
{\mcitedefaultendpunct}{\mcitedefaultseppunct}\relax
\EndOfBibitem
\bibitem[Turner and Sugiya(2002)]{turner2002understanding}
R.~J. Turner and H.~Sugiya, \emph{Oral diseases}, 2002, \textbf{8}, 3--11\relax
\mciteBstWouldAddEndPuncttrue
\mciteSetBstMidEndSepPunct{\mcitedefaultmidpunct}
{\mcitedefaultendpunct}{\mcitedefaultseppunct}\relax
\EndOfBibitem
\bibitem[Gin \emph{et~al.}(2010)Gin, Tanaka, and Brusch]{gin2010model}
E.~Gin, E.~M. Tanaka and L.~Brusch, \emph{Journal of theoretical biology},
  2010, \textbf{264}, 1077--1088\relax
\mciteBstWouldAddEndPuncttrue
\mciteSetBstMidEndSepPunct{\mcitedefaultmidpunct}
{\mcitedefaultendpunct}{\mcitedefaultseppunct}\relax
\EndOfBibitem
\bibitem[Rauch and Farge(2000)]{rauch2000endocytosis}
C.~Rauch and E.~Farge, \emph{Biophysical journal}, 2000, \textbf{78},
  3036--3047\relax
\mciteBstWouldAddEndPuncttrue
\mciteSetBstMidEndSepPunct{\mcitedefaultmidpunct}
{\mcitedefaultendpunct}{\mcitedefaultseppunct}\relax
\EndOfBibitem
\bibitem[Evilevitch \emph{et~al.}(2003)Evilevitch, Lavelle, Knobler, Raspaud,
  and Gelbart]{evilevitch2003osmotic}
A.~Evilevitch, L.~Lavelle, C.~M. Knobler, E.~Raspaud and W.~M. Gelbart,
  \emph{Proceedings of the National Academy of Sciences}, 2003, \textbf{100},
  9292--9295\relax
\mciteBstWouldAddEndPuncttrue
\mciteSetBstMidEndSepPunct{\mcitedefaultmidpunct}
{\mcitedefaultendpunct}{\mcitedefaultseppunct}\relax
\EndOfBibitem
\bibitem[Hubbard \emph{et~al.}(1968)Hubbard, Jones, and
  Landau]{hubbard1968examination}
J.~Hubbard, S.~Jones and E.~Landau, \emph{The Journal of physiology}, 1968,
  \textbf{197}, 639--657\relax
\mciteBstWouldAddEndPuncttrue
\mciteSetBstMidEndSepPunct{\mcitedefaultmidpunct}
{\mcitedefaultendpunct}{\mcitedefaultseppunct}\relax
\EndOfBibitem
\bibitem[Furshpan(1956)]{furshpan1956effects}
E.~Furshpan, \emph{The Journal of physiology}, 1956, \textbf{134},
  689--697\relax
\mciteBstWouldAddEndPuncttrue
\mciteSetBstMidEndSepPunct{\mcitedefaultmidpunct}
{\mcitedefaultendpunct}{\mcitedefaultseppunct}\relax
\EndOfBibitem
\bibitem[K{\"u}cken \emph{et~al.}(2008)K{\"u}cken, Soriano, Pullarkat, Ott, and
  Nicola]{kucken2008osmoregulatory}
M.~K{\"u}cken, J.~Soriano, P.~A. Pullarkat, A.~Ott and E.~M. Nicola,
  \emph{Biophysical journal}, 2008, \textbf{95}, 978--985\relax
\mciteBstWouldAddEndPuncttrue
\mciteSetBstMidEndSepPunct{\mcitedefaultmidpunct}
{\mcitedefaultendpunct}{\mcitedefaultseppunct}\relax
\EndOfBibitem
\bibitem[K{\"u}ppers \emph{et~al.}(1986)K{\"u}ppers, Plagemann, and
  Thurm]{kuppers1986uphill}
J.~K{\"u}ppers, A.~Plagemann and U.~Thurm, \emph{The Journal of Membrane
  Biology}, 1986, \textbf{91}, 107--119\relax
\mciteBstWouldAddEndPuncttrue
\mciteSetBstMidEndSepPunct{\mcitedefaultmidpunct}
{\mcitedefaultendpunct}{\mcitedefaultseppunct}\relax
\EndOfBibitem
\bibitem[Fischbarg \emph{et~al.}(2017)Fischbarg, Hernandez, Rubashkin,
  Iserovich, Cacace, and Kusnier]{fischbarg2017epithelial}
J.~Fischbarg, J.~A. Hernandez, A.~A. Rubashkin, P.~Iserovich, V.~I. Cacace and
  C.~F. Kusnier, \emph{The Journal of membrane biology}, 2017, \textbf{250},
  327--333\relax
\mciteBstWouldAddEndPuncttrue
\mciteSetBstMidEndSepPunct{\mcitedefaultmidpunct}
{\mcitedefaultendpunct}{\mcitedefaultseppunct}\relax
\EndOfBibitem
\bibitem[Dvoriashyna \emph{et~al.}(2018)Dvoriashyna, Foss, Gaffney, Jensen, and
  Repetto]{dvoriashyna2018osmotic}
M.~Dvoriashyna, A.~J. Foss, E.~A. Gaffney, O.~E. Jensen and R.~Repetto,
  \emph{Journal of theoretical biology}, 2018, \textbf{456}, 233--248\relax
\mciteBstWouldAddEndPuncttrue
\mciteSetBstMidEndSepPunct{\mcitedefaultmidpunct}
{\mcitedefaultendpunct}{\mcitedefaultseppunct}\relax
\EndOfBibitem
\bibitem[Preston \emph{et~al.}(1992)Preston, Carroll, Guggino, and
  Agre]{preston1992appearance}
G.~M. Preston, T.~P. Carroll, W.~B. Guggino and P.~Agre, \emph{Science}, 1992,
  \textbf{256}, 385--387\relax
\mciteBstWouldAddEndPuncttrue
\mciteSetBstMidEndSepPunct{\mcitedefaultmidpunct}
{\mcitedefaultendpunct}{\mcitedefaultseppunct}\relax
\EndOfBibitem
\bibitem[Murata \emph{et~al.}(2000)Murata, Mitsuoka, Hirai, Walz, Agre,
  Heymann, Engel, and Fujiyoshi]{murata2000structural}
K.~Murata, K.~Mitsuoka, T.~Hirai, T.~Walz, P.~Agre, J.~B. Heymann, A.~Engel and
  Y.~Fujiyoshi, \emph{Nature}, 2000, \textbf{407}, 599\relax
\mciteBstWouldAddEndPuncttrue
\mciteSetBstMidEndSepPunct{\mcitedefaultmidpunct}
{\mcitedefaultendpunct}{\mcitedefaultseppunct}\relax
\EndOfBibitem
\bibitem[Maurel \emph{et~al.}(2008)Maurel, Verdoucq, Luu, and
  Santoni]{maurel2008plant}
C.~Maurel, L.~Verdoucq, D.-T. Luu and V.~Santoni, \emph{Annu. Rev. Plant
  Biol.}, 2008, \textbf{59}, 595--624\relax
\mciteBstWouldAddEndPuncttrue
\mciteSetBstMidEndSepPunct{\mcitedefaultmidpunct}
{\mcitedefaultendpunct}{\mcitedefaultseppunct}\relax
\EndOfBibitem
\bibitem[Agre(2006)]{agre2006aquaporin}
P.~Agre, \emph{Proceedings of the American Thoracic Society}, 2006, \textbf{3},
  5--13\relax
\mciteBstWouldAddEndPuncttrue
\mciteSetBstMidEndSepPunct{\mcitedefaultmidpunct}
{\mcitedefaultendpunct}{\mcitedefaultseppunct}\relax
\EndOfBibitem
\bibitem[Javot and Maurel(2002)]{javot2002role}
H.~Javot and C.~Maurel, \emph{Annals of Botany}, 2002, \textbf{90},
  301--313\relax
\mciteBstWouldAddEndPuncttrue
\mciteSetBstMidEndSepPunct{\mcitedefaultmidpunct}
{\mcitedefaultendpunct}{\mcitedefaultseppunct}\relax
\EndOfBibitem
\bibitem[Zeidel \emph{et~al.}(1992)Zeidel, Ambudkar, Smith, and
  Agre]{zeidel1992reconstitution}
M.~L. Zeidel, S.~V. Ambudkar, B.~L. Smith and P.~Agre, \emph{Biochemistry},
  1992, \textbf{31}, 7436--7440\relax
\mciteBstWouldAddEndPuncttrue
\mciteSetBstMidEndSepPunct{\mcitedefaultmidpunct}
{\mcitedefaultendpunct}{\mcitedefaultseppunct}\relax
\EndOfBibitem
\bibitem[Roux \emph{et~al.}(2001)Roux, Lapointe, and Bichet]{roux2001structure}
B.~Roux, J.-Y. Lapointe and D.~G. Bichet, \emph{M\'edecine/Science}, 2001,
  \textbf{17}, 115--116\relax
\mciteBstWouldAddEndPuncttrue
\mciteSetBstMidEndSepPunct{\mcitedefaultmidpunct}
{\mcitedefaultendpunct}{\mcitedefaultseppunct}\relax
\EndOfBibitem
\bibitem[Horner \emph{et~al.}(2015)Horner, Zocher, Preiner, Ollinger, Siligan,
  Akimov, and Pohl]{horner2015mobility}
A.~Horner, F.~Zocher, J.~Preiner, N.~Ollinger, C.~Siligan, S.~A. Akimov and
  P.~Pohl, \emph{Science advances}, 2015, \textbf{1}, e1400083\relax
\mciteBstWouldAddEndPuncttrue
\mciteSetBstMidEndSepPunct{\mcitedefaultmidpunct}
{\mcitedefaultendpunct}{\mcitedefaultseppunct}\relax
\EndOfBibitem
\bibitem[Walz \emph{et~al.}(1994)Walz, Smith, Zeidel, Engel, and
  Agre]{walz1994biologically}
T.~Walz, B.~L. Smith, M.~L. Zeidel, A.~Engel and P.~Agre, \emph{Journal of
  Biological Chemistry}, 1994, \textbf{269}, 1583--1586\relax
\mciteBstWouldAddEndPuncttrue
\mciteSetBstMidEndSepPunct{\mcitedefaultmidpunct}
{\mcitedefaultendpunct}{\mcitedefaultseppunct}\relax
\EndOfBibitem
\bibitem[UIUC()]{tajVideo}
T.~UIUC, \emph{Water channels in cell membranes}, \textbf{YouTube},
  \hyperref[Link]{''https://www.youtube.com/watch?v=GSi5--y6NHjY''}\relax
\mciteBstWouldAddEndPuncttrue
\mciteSetBstMidEndSepPunct{\mcitedefaultmidpunct}
{\mcitedefaultendpunct}{\mcitedefaultseppunct}\relax
\EndOfBibitem
\bibitem[Tajkhorshid \emph{et~al.}(2002)Tajkhorshid, Nollert, Jensen, Miercke,
  O'connell, Stroud, and Schulten]{tajkhorshid2002control}
E.~Tajkhorshid, P.~Nollert, M.~{\O}. Jensen, L.~J. Miercke, J.~O'connell, R.~M.
  Stroud and K.~Schulten, \emph{Science}, 2002, \textbf{296}, 525--530\relax
\mciteBstWouldAddEndPuncttrue
\mciteSetBstMidEndSepPunct{\mcitedefaultmidpunct}
{\mcitedefaultendpunct}{\mcitedefaultseppunct}\relax
\EndOfBibitem
\bibitem[Sui \emph{et~al.}(2001)Sui, Han, Lee, Walian, and
  Jap]{sui2001structural}
H.~Sui, B.-G. Han, J.~K. Lee, P.~Walian and B.~K. Jap, \emph{Nature}, 2001,
  \textbf{414}, 872\relax
\mciteBstWouldAddEndPuncttrue
\mciteSetBstMidEndSepPunct{\mcitedefaultmidpunct}
{\mcitedefaultendpunct}{\mcitedefaultseppunct}\relax
\EndOfBibitem
\bibitem[Gravelle \emph{et~al.}(2013)Gravelle, Joly, Detcheverry, Ybert,
  Cottin-Bizonne, and Bocquet]{gravelle2013optimizing}
S.~Gravelle, L.~Joly, F.~Detcheverry, C.~Ybert, C.~Cottin-Bizonne and
  L.~Bocquet, \emph{Proceedings of the National Academy of Sciences}, 2013,
  201306447\relax
\mciteBstWouldAddEndPuncttrue
\mciteSetBstMidEndSepPunct{\mcitedefaultmidpunct}
{\mcitedefaultendpunct}{\mcitedefaultseppunct}\relax
\EndOfBibitem
\bibitem[Pérez-Mitta \emph{et~al.}(2015)Pérez-Mitta, Tuninetti, Knoll,
  Trautmann, Toimil-Molares, and Azzaroni]{perez2015polydopamine}
G.~Pérez-Mitta, J.~S. Tuninetti, W.~Knoll, C.~Trautmann, M.~E. Toimil-Molares
  and O.~Azzaroni, \emph{Journal of the American Chemical Society}, 2015,
  \textbf{137}, 6011--6017\relax
\mciteBstWouldAddEndPuncttrue
\mciteSetBstMidEndSepPunct{\mcitedefaultmidpunct}
{\mcitedefaultendpunct}{\mcitedefaultseppunct}\relax
\EndOfBibitem
\bibitem[Greger and Windhorst(1996)]{greger1996comprehensive}
R.~Greger and U.~Windhorst, \emph{From Cellular Mechanisms to Integration},
  1996, \textbf{2}, year\relax
\mciteBstWouldAddEndPuncttrue
\mciteSetBstMidEndSepPunct{\mcitedefaultmidpunct}
{\mcitedefaultendpunct}{\mcitedefaultseppunct}\relax
\EndOfBibitem
\bibitem[Baylis(1989)]{baylis1989water}
P.~H. Baylis, 1989, \textbf{3}, 229--578\relax
\mciteBstWouldAddEndPuncttrue
\mciteSetBstMidEndSepPunct{\mcitedefaultmidpunct}
{\mcitedefaultendpunct}{\mcitedefaultseppunct}\relax
\EndOfBibitem
\bibitem[Stephenson(1972)]{stephenson1972concentration}
J.~L. Stephenson, \emph{Kidney international}, 1972, \textbf{2}, 85--94\relax
\mciteBstWouldAddEndPuncttrue
\mciteSetBstMidEndSepPunct{\mcitedefaultmidpunct}
{\mcitedefaultendpunct}{\mcitedefaultseppunct}\relax
\EndOfBibitem
\bibitem[Layton(2010)]{layton2010mathematical}
A.~T. Layton, \emph{American Journal of Physiology-Renal Physiology}, 2010,
  \textbf{300}, F356--F371\relax
\mciteBstWouldAddEndPuncttrue
\mciteSetBstMidEndSepPunct{\mcitedefaultmidpunct}
{\mcitedefaultendpunct}{\mcitedefaultseppunct}\relax
\EndOfBibitem
\bibitem[Edwards(2009)]{edwards2009modeling}
A.~Edwards, \emph{American Journal of Physiology-Renal Physiology}, 2009,
  \textbf{298}, F475--F484\relax
\mciteBstWouldAddEndPuncttrue
\mciteSetBstMidEndSepPunct{\mcitedefaultmidpunct}
{\mcitedefaultendpunct}{\mcitedefaultseppunct}\relax
\EndOfBibitem
\bibitem[Gambro(2009)]{gambro}
Gambro, \emph{Phoenix X 36 brochure}, 2009\relax
\mciteBstWouldAddEndPuncttrue
\mciteSetBstMidEndSepPunct{\mcitedefaultmidpunct}
{\mcitedefaultendpunct}{\mcitedefaultseppunct}\relax
\EndOfBibitem
\bibitem[Borenstein \emph{et~al.}(2007)Borenstein, Weinberg, Orrick, Sundback,
  Kaazempur-Mofrad, and Vacanti]{borenstein2007microfabrication}
J.~T. Borenstein, E.~J. Weinberg, B.~K. Orrick, C.~Sundback, M.~R.
  Kaazempur-Mofrad and J.~P. Vacanti, \emph{Tissue engineering}, 2007,
  \textbf{13}, 1837--1844\relax
\mciteBstWouldAddEndPuncttrue
\mciteSetBstMidEndSepPunct{\mcitedefaultmidpunct}
{\mcitedefaultendpunct}{\mcitedefaultseppunct}\relax
\EndOfBibitem
\bibitem[Kim \emph{et~al.}(2011)Kim, Garzotto, Nalesso, Cruz, Kim, Kang, Kim,
  and Ronco]{kim2011wearable}
J.~C. Kim, F.~Garzotto, F.~Nalesso, D.~Cruz, J.~H. Kim, E.~Kang, H.~C. Kim and
  C.~Ronco, \emph{Expert review of medical devices}, 2011, \textbf{8},
  567--579\relax
\mciteBstWouldAddEndPuncttrue
\mciteSetBstMidEndSepPunct{\mcitedefaultmidpunct}
{\mcitedefaultendpunct}{\mcitedefaultseppunct}\relax
\EndOfBibitem
\bibitem[Armignacco \emph{et~al.}(2015)Armignacco, Garzotto, Neri, Lorenzin,
  and Ronco]{armignacco2015wak}
P.~Armignacco, F.~Garzotto, M.~Neri, A.~Lorenzin and C.~Ronco, \emph{Blood
  purification}, 2015, \textbf{39}, 110--114\relax
\mciteBstWouldAddEndPuncttrue
\mciteSetBstMidEndSepPunct{\mcitedefaultmidpunct}
{\mcitedefaultendpunct}{\mcitedefaultseppunct}\relax
\EndOfBibitem
\bibitem[Kaila \emph{et~al.}(2010)Kaila, Verkhovsky, and
  Wikström]{kaila2010proton}
V.~R. Kaila, M.~I. Verkhovsky and M.~Wikström, \emph{Chemical reviews}, 2010,
  \textbf{110}, 7062--7081\relax
\mciteBstWouldAddEndPuncttrue
\mciteSetBstMidEndSepPunct{\mcitedefaultmidpunct}
{\mcitedefaultendpunct}{\mcitedefaultseppunct}\relax
\EndOfBibitem
\bibitem[Pedersen and Carafoli(1987)]{pedersen1987ion}
P.~L. Pedersen and E.~Carafoli, \emph{Trends in Biochemical Sciences}, 1987,
  \textbf{12}, 146--150\relax
\mciteBstWouldAddEndPuncttrue
\mciteSetBstMidEndSepPunct{\mcitedefaultmidpunct}
{\mcitedefaultendpunct}{\mcitedefaultseppunct}\relax
\EndOfBibitem
\bibitem[Mulkidjanian \emph{et~al.}(2007)Mulkidjanian, Makarova, Galperin, and
  Koonin]{mulkidjanian2007inventing}
A.~Y. Mulkidjanian, K.~S. Makarova, M.~Y. Galperin and E.~V. Koonin,
  \emph{Nature Reviews Microbiology}, 2007, \textbf{5}, 892\relax
\mciteBstWouldAddEndPuncttrue
\mciteSetBstMidEndSepPunct{\mcitedefaultmidpunct}
{\mcitedefaultendpunct}{\mcitedefaultseppunct}\relax
\EndOfBibitem
\bibitem[Palmgren(2001)]{palmgren2001plant}
M.~G. Palmgren, \emph{Annual review of plant biology}, 2001, \textbf{52},
  817--845\relax
\mciteBstWouldAddEndPuncttrue
\mciteSetBstMidEndSepPunct{\mcitedefaultmidpunct}
{\mcitedefaultendpunct}{\mcitedefaultseppunct}\relax
\EndOfBibitem
\bibitem[Nishi and Forgac(2002)]{nishi2002vacuolar}
T.~Nishi and M.~Forgac, \emph{Nature reviews Molecular cell biology}, 2002,
  \textbf{3}, 94\relax
\mciteBstWouldAddEndPuncttrue
\mciteSetBstMidEndSepPunct{\mcitedefaultmidpunct}
{\mcitedefaultendpunct}{\mcitedefaultseppunct}\relax
\EndOfBibitem
\bibitem[Murata \emph{et~al.}(2005)Murata, Yamato, Kakinuma, Leslie, and
  Walker]{murata2005structure}
T.~Murata, I.~Yamato, Y.~Kakinuma, A.~G. Leslie and J.~E. Walker,
  \emph{Science}, 2005, \textbf{308}, 654--659\relax
\mciteBstWouldAddEndPuncttrue
\mciteSetBstMidEndSepPunct{\mcitedefaultmidpunct}
{\mcitedefaultendpunct}{\mcitedefaultseppunct}\relax
\EndOfBibitem
\bibitem[Meier \emph{et~al.}(2005)Meier, Polzer, Diederichs, Welte, and
  Dimroth]{meier2005structure}
T.~Meier, P.~Polzer, K.~Diederichs, W.~Welte and P.~Dimroth, \emph{Science},
  2005, \textbf{308}, 659--662\relax
\mciteBstWouldAddEndPuncttrue
\mciteSetBstMidEndSepPunct{\mcitedefaultmidpunct}
{\mcitedefaultendpunct}{\mcitedefaultseppunct}\relax
\EndOfBibitem
\bibitem[Capaldi and Aggeler(2002)]{capaldi2002mechanism}
R.~A. Capaldi and R.~Aggeler, \emph{Trends in biochemical sciences}, 2002,
  \textbf{27}, 154--160\relax
\mciteBstWouldAddEndPuncttrue
\mciteSetBstMidEndSepPunct{\mcitedefaultmidpunct}
{\mcitedefaultendpunct}{\mcitedefaultseppunct}\relax
\EndOfBibitem
\bibitem[Mitchell(1961)]{mitchell1961coupling}
P.~Mitchell, \emph{Nature}, 1961, \textbf{191}, 144--148\relax
\mciteBstWouldAddEndPuncttrue
\mciteSetBstMidEndSepPunct{\mcitedefaultmidpunct}
{\mcitedefaultendpunct}{\mcitedefaultseppunct}\relax
\EndOfBibitem
\bibitem[Junge and Nelson(2015)]{junge2015atp}
W.~Junge and N.~Nelson, \emph{Annual Review of Biochemistry}, 2015,
  \textbf{84}, 631--657\relax
\mciteBstWouldAddEndPuncttrue
\mciteSetBstMidEndSepPunct{\mcitedefaultmidpunct}
{\mcitedefaultendpunct}{\mcitedefaultseppunct}\relax
\EndOfBibitem
\bibitem[Soong \emph{et~al.}(2000)Soong, Bachand, Neves, Olkhovets, Craighead,
  and Montemagno]{soong2000powering}
R.~K. Soong, G.~D. Bachand, H.~P. Neves, A.~G. Olkhovets, H.~G. Craighead and
  C.~D. Montemagno, \emph{Science}, 2000, \textbf{290}, 1555--1558\relax
\mciteBstWouldAddEndPuncttrue
\mciteSetBstMidEndSepPunct{\mcitedefaultmidpunct}
{\mcitedefaultendpunct}{\mcitedefaultseppunct}\relax
\EndOfBibitem
\bibitem[Sowa and Berry(2008)]{sowa2008bacterial}
Y.~Sowa and R.~M. Berry, \emph{Quarterly reviews of biophysics}, 2008,
  \textbf{41}, 103--132\relax
\mciteBstWouldAddEndPuncttrue
\mciteSetBstMidEndSepPunct{\mcitedefaultmidpunct}
{\mcitedefaultendpunct}{\mcitedefaultseppunct}\relax
\EndOfBibitem
\bibitem[Silverman and Simon(1974)]{silverman1974flagellar}
M.~Silverman and M.~Simon, \emph{Nature}, 1974, \textbf{249}, 73\relax
\mciteBstWouldAddEndPuncttrue
\mciteSetBstMidEndSepPunct{\mcitedefaultmidpunct}
{\mcitedefaultendpunct}{\mcitedefaultseppunct}\relax
\EndOfBibitem
\bibitem[Kojima and Blair(2004)]{kojima2004solubilization}
S.~Kojima and D.~F. Blair, \emph{Biochemistry}, 2004, \textbf{43}, 26--34\relax
\mciteBstWouldAddEndPuncttrue
\mciteSetBstMidEndSepPunct{\mcitedefaultmidpunct}
{\mcitedefaultendpunct}{\mcitedefaultseppunct}\relax
\EndOfBibitem
\bibitem[Nishihara and Kitao(2015)]{nishihara2015gate}
Y.~Nishihara and A.~Kitao, \emph{Proceedings of the National Academy of
  Sciences}, 2015, \textbf{112}, 7737--7742\relax
\mciteBstWouldAddEndPuncttrue
\mciteSetBstMidEndSepPunct{\mcitedefaultmidpunct}
{\mcitedefaultendpunct}{\mcitedefaultseppunct}\relax
\EndOfBibitem
\bibitem[Cereijo-Santal\'{o}(1972)]{cereijo1972osmotic}
R.~Cereijo-Santal\'{o}, \emph{Archives of biochemistry and biophysics}, 1972,
  \textbf{152}, 78--82\relax
\mciteBstWouldAddEndPuncttrue
\mciteSetBstMidEndSepPunct{\mcitedefaultmidpunct}
{\mcitedefaultendpunct}{\mcitedefaultseppunct}\relax
\EndOfBibitem
\bibitem[Miller~Jr \emph{et~al.}(2013)Miller~Jr, Rajapakshe, Infante, and
  Claycomb]{miller2013electric}
J.~H. Miller~Jr, K.~I. Rajapakshe, H.~L. Infante and J.~R. Claycomb, \emph{PLoS
  One}, 2013, \textbf{8}, e74978\relax
\mciteBstWouldAddEndPuncttrue
\mciteSetBstMidEndSepPunct{\mcitedefaultmidpunct}
{\mcitedefaultendpunct}{\mcitedefaultseppunct}\relax
\EndOfBibitem
\bibitem[Yip and Elimelech(2012)]{yip2012thermodynamic}
N.~Y. Yip and M.~Elimelech, \emph{Environmental science \& technology}, 2012,
  \textbf{46}, 5230--5239\relax
\mciteBstWouldAddEndPuncttrue
\mciteSetBstMidEndSepPunct{\mcitedefaultmidpunct}
{\mcitedefaultendpunct}{\mcitedefaultseppunct}\relax
\EndOfBibitem
\bibitem[Statistics(2017)]{statistics2017key}
I.~Statistics, \emph{International Energy Agency}, 2017\relax
\mciteBstWouldAddEndPuncttrue
\mciteSetBstMidEndSepPunct{\mcitedefaultmidpunct}
{\mcitedefaultendpunct}{\mcitedefaultseppunct}\relax
\EndOfBibitem
\bibitem[Tufa \emph{et~al.}(2018)Tufa, Pawlowski, Veerman, Bouzek, Fontananova,
  di~Profio, Velizarov, Crespo, Nijmeijer, and Curcio]{tufa2018progress}
R.~A. Tufa, S.~Pawlowski, J.~Veerman, K.~Bouzek, E.~Fontananova, G.~di~Profio,
  S.~Velizarov, J.~G. Crespo, K.~Nijmeijer and E.~Curcio, \emph{Applied
  energy}, 2018, \textbf{225}, 290--331\relax
\mciteBstWouldAddEndPuncttrue
\mciteSetBstMidEndSepPunct{\mcitedefaultmidpunct}
{\mcitedefaultendpunct}{\mcitedefaultseppunct}\relax
\EndOfBibitem
\bibitem[Skilhagen(2010)]{skilhagen2010osmotic}
S.~E. Skilhagen, \emph{Desalination and water treatment}, 2010, \textbf{15},
  271--278\relax
\mciteBstWouldAddEndPuncttrue
\mciteSetBstMidEndSepPunct{\mcitedefaultmidpunct}
{\mcitedefaultendpunct}{\mcitedefaultseppunct}\relax
\EndOfBibitem
\bibitem[MacKay(2008)]{mackay2008sustainable}
D.~MacKay, \emph{Sustainable Energy-without the hot air}, UIT Cambridge,
  2008\relax
\mciteBstWouldAddEndPuncttrue
\mciteSetBstMidEndSepPunct{\mcitedefaultmidpunct}
{\mcitedefaultendpunct}{\mcitedefaultseppunct}\relax
\EndOfBibitem
\bibitem[Rankin and Huang(2016)]{rankin2016effect}
D.~J. Rankin and D.~M. Huang, \emph{Langmuir}, 2016, \textbf{32},
  3420--3432\relax
\mciteBstWouldAddEndPuncttrue
\mciteSetBstMidEndSepPunct{\mcitedefaultmidpunct}
{\mcitedefaultendpunct}{\mcitedefaultseppunct}\relax
\EndOfBibitem
\bibitem[Zhu \emph{et~al.}(2018)Zhu, Hao, Bao, Zhou, Zhang, Pang, Jiang, and
  Jiang]{zhu2018unique}
X.~Zhu, J.~Hao, B.~Bao, Y.~Zhou, H.~Zhang, J.~Pang, Z.~Jiang and L.~Jiang,
  \emph{Science Advances}, 2018, \textbf{4}, eaau1665\relax
\mciteBstWouldAddEndPuncttrue
\mciteSetBstMidEndSepPunct{\mcitedefaultmidpunct}
{\mcitedefaultendpunct}{\mcitedefaultseppunct}\relax
\EndOfBibitem
\bibitem[Lin \emph{et~al.}(2019)Lin, Combs, Su, Yeh, and
  Siwy]{lin2019rectification}
C.-Y. Lin, C.~Combs, Y.-S. Su, L.-H. Yeh and Z.~S. Siwy, \emph{Journal of the
  American Chemical Society}, 2019, \textbf{141}, 3691--3698\relax
\mciteBstWouldAddEndPuncttrue
\mciteSetBstMidEndSepPunct{\mcitedefaultmidpunct}
{\mcitedefaultendpunct}{\mcitedefaultseppunct}\relax
\EndOfBibitem
\bibitem[Brogioli(2009)]{brogioli2009extracting}
D.~Brogioli, \emph{Physical Review Letters}, 2009, \textbf{103}, 058501\relax
\mciteBstWouldAddEndPuncttrue
\mciteSetBstMidEndSepPunct{\mcitedefaultmidpunct}
{\mcitedefaultendpunct}{\mcitedefaultseppunct}\relax
\EndOfBibitem
\bibitem[Brogioli \emph{et~al.}(2011)Brogioli, Zhao, and
  Biesheuvel]{brogioli2011prototype}
D.~Brogioli, R.~Zhao and P.~Biesheuvel, \emph{Energy \& Environmental Science},
  2011, \textbf{4}, 772--777\relax
\mciteBstWouldAddEndPuncttrue
\mciteSetBstMidEndSepPunct{\mcitedefaultmidpunct}
{\mcitedefaultendpunct}{\mcitedefaultseppunct}\relax
\EndOfBibitem
\bibitem[Simoncelli \emph{et~al.}(2018)Simoncelli, Ganfoud, Sene, Haefele,
  Daffos, Taberna, Salanne, Simon, and Rotenberg]{simoncelli2018blue}
M.~Simoncelli, N.~Ganfoud, A.~Sene, M.~Haefele, B.~Daffos, P.-L. Taberna,
  M.~Salanne, P.~Simon and B.~Rotenberg, \emph{Physical Review X}, 2018,
  \textbf{8}, 021024\relax
\mciteBstWouldAddEndPuncttrue
\mciteSetBstMidEndSepPunct{\mcitedefaultmidpunct}
{\mcitedefaultendpunct}{\mcitedefaultseppunct}\relax
\EndOfBibitem
\bibitem[Marino \emph{et~al.}(2016)Marino, Kozynchenko, Tennison, and
  Brogioli]{marino2016capacitive}
M.~Marino, O.~Kozynchenko, S.~Tennison and D.~Brogioli, \emph{Journal of
  Physics: Condensed Matter}, 2016, \textbf{28}, 114004\relax
\mciteBstWouldAddEndPuncttrue
\mciteSetBstMidEndSepPunct{\mcitedefaultmidpunct}
{\mcitedefaultendpunct}{\mcitedefaultseppunct}\relax
\EndOfBibitem
\bibitem[Shin \emph{et~al.}(2018)Shin, Warren, and Stone]{shin2018cleaning}
S.~Shin, P.~B. Warren and H.~A. Stone, \emph{Physical Review Applied}, 2018,
  \textbf{9}, 034012\relax
\mciteBstWouldAddEndPuncttrue
\mciteSetBstMidEndSepPunct{\mcitedefaultmidpunct}
{\mcitedefaultendpunct}{\mcitedefaultseppunct}\relax
\EndOfBibitem
\bibitem[Putnis(2014)]{putnis2014mineral}
A.~Putnis, \emph{Science}, 2014, \textbf{343}, 1441--1442\relax
\mciteBstWouldAddEndPuncttrue
\mciteSetBstMidEndSepPunct{\mcitedefaultmidpunct}
{\mcitedefaultendpunct}{\mcitedefaultseppunct}\relax
\EndOfBibitem
\bibitem[Lager \emph{et~al.}(2008)Lager, Webb, Collins,
  Richmond,\emph{et~al.}]{lager2008losal}
A.~Lager, K.~J. Webb, I.~R. Collins, D.~M. Richmond \emph{et~al.}, SPE
  Symposium on Improved Oil Recovery, 2008\relax
\mciteBstWouldAddEndPuncttrue
\mciteSetBstMidEndSepPunct{\mcitedefaultmidpunct}
{\mcitedefaultendpunct}{\mcitedefaultseppunct}\relax
\EndOfBibitem
\end{mcitethebibliography}
\bibliographystyle{rsc} %the RSC's .bst file

\end{document}